\documentclass[a4paper,english,notitlepage,nofootinbib]{revtex4-1}

\usepackage[T1]{fontenc}
\usepackage[utf8]{inputenc}
\usepackage{color}
\usepackage{babel}
\usepackage{verbatim}
\usepackage{float}
\usepackage{textcomp}
\usepackage{amsmath}
\usepackage{amssymb}
\usepackage{graphicx}
\usepackage{esint}
\usepackage[unicode=true,
 bookmarks=false,
 breaklinks=false,pdfborder={0 0 0},backref=section,colorlinks=true]
 {hyperref}
\hypersetup{pdftitle={MG Forecasts},
 pdfauthor={Santiago Casas},
 pdfsubject={Forecasts for Modified Gravity},
 pdfkeywords={cosmology, forecasts, clustering, lensing},
 pdfborderstyle={},pdfborderstyle={},pdfborderstyle={},pdfborderstyle={},pdfborderstyle={},pdfborderstyle={}}
\usepackage{breakurl}
\usepackage{bbold}

\makeatletter

\providecommand{\tabularnewline}{\\}

\usepackage{babel}
\usepackage{todonotes}
\definecolor{green}{rgb}{0, 0.69, 0.04}
\definecolor{red}{rgb}{1,0,0} 
\definecolor{magenta}{cmyk}{0,1,0,0} 
\definecolor{violet}{cmyk}{0,1,0,0} 
\definecolor{darkgreen}{rgb}{0,0.65,0.05}
\definecolor{antiquefuchsia}{rgb}{0.57, 0.36, 0.61}

\newcommand{\HMG}{\mathrm{HMG}}
\newcommand{\HGR}{\mathrm{HGR}}
\newcommand{\LMG}{\mathrm{LMG}}

\newcommand{\LIN}{\mathrm{L}}
\newcommand{\cnl}{c_{\mathrm{nl}}}
\newcommand{\Pobs}{P_{\mathrm{obs}}}
\newcommand{\lmax}{\ell_{\mathrm{max}}}
\newcommand{\lAs}{\ell \mathcal{A}_{s}}
\newcommand{\planck}{{\it Planck}}

\newcommand{\lcdm}{\Lambda\mathrm{CDM}}
\newcommand{\curH}{\mathcal{H}}

\def\l@section{\@dottedtocline{1}{1em}{2em}}
\def\l@subsection{\@dottedtocline{2}{2em}{4em}}
\def\l@subsubsection{\@dottedtocline{3}{3em}{5em}}

\usepackage[titletoc]{appendix}

\usepackage{colortbl}

\newcommand{\Tstrut}{\rule{0pt}{2.6ex}}       % "top" strut
\newcommand{\Bstrut}{\rule[-0.9ex]{0pt}{0pt}} % "bottom" strut
\newcommand{\TBstrut}{\Tstrut\Bstrut} % top&bottom struts
\usepackage{tikz}
\usepackage{collcell}

\newcommand\ColCell[1]{%
	\pgfmathparse{#1>0?1:0}%
	\ifnum\pgfmathresult=1\relax\color{red}\fi#1}

\newcolumntype{E}{>{\collectcell\ColCell}c<{\endcollectcell}}

\makeatother

\begin{document}

\title{Linear and non-linear Modified Gravity forecasts with future surveys}

\author{Santiago Casas$^{1}$, Martin Kunz$^{2}$, Matteo Martinelli$^{1,3}$, Valeria Pettorino$^{1,4}$}
\affiliation{$^{1}$ Institut fuer Theoretische Physik, Universitaet Heidelberg,
Philosophenweg 16, D-69120 Heidelberg, Germany.\\
 $^{2}$ D\'epartement de Physique Th\'eorique and Center for Astroparticle Physics, Universit\'e de Gen\`eve, 24 quai Ansermet, CH--1211 Gen\`eve 4, Switzerland. \\
 $^{3}$ Institute Lorentz, Leiden University, PO Box 9506, Leiden 2300 RA, The Netherlands, \\
 $^{4}$ Laboratoire AIM, UMR CEA-CNRS-Paris 7, Irfu, Service d'Astrophysique CEA Saclay, F-91191 Gif-sur-Yvette, France}

\begin{abstract}
Modified Gravity theories generally affect the Poisson equation and
the gravitational slip (effective anisotropic stress) in an observable way, that can be parameterized
by two generic functions ($\eta$ and $\mu$) of time and space. We
bin the time dependence of these functions in redshift and present forecasts on each
bin for future surveys like Euclid. We consider
both Galaxy Clustering and Weak Lensing surveys, showing the impact
of the non-linear regime, treated with two different semi-analytical
approximations. In addition to these future observables, 
we use a prior covariance matrix derived from the \planck\ observations of the Cosmic Microwave Background.
Our results show that $\eta$ and $\mu$ in different redshift bins are
significantly correlated, but including non-linear scales reduces or even eliminates the correlation, breaking the degeneracy between Modified Gravity parameters and the overall amplitude of the matter power spectrum. 
We further decorrelate parameters with a Zero-phase Component Analysis and 
identify which combinations of the Modified Gravity parameter amplitudes, in different redshift bins, are best constrained by future surveys. 
We also extend the analysis to two particular parameterizations of the time evolution of $\mu$
and $\eta$ and consider, in addition to Euclid, also SKA1, SKA2, DESI: we find in this case that future surveys will be able to 
constrain the current values of $\eta$ and $\mu$ at the 2-5\% level when using only linear scales (wavevector k < 0.15 h/Mpc), depending on the specific time parameterization; sensitivity improves to about $1\%$ when non-linearities are included. 
\end{abstract}
\maketitle
\pdfbookmark{\contentsname}{toc}
\tableofcontents{}

\section{Introduction}

Future large scale structure surveys will be able to measure with
percent precision the parameters governing the evolution of matter
perturbations. While we have the tools to investigate the standard model, the next challenge is to be able to compare those data with cosmologies that go beyond General Relativity, in order to test whether a fluid component like Dark Energy or similarly a Modified Gravity scenario can better fit the data. On the theoretical side, while many Modified Gravity models are still allowed by type Ia supernova (SNIa) and
Cosmic Microwave Background (CMB) data \cite{planck_collaboration_planck_2016}; structure formation can help us to distinguish among them and the standard scenario, thanks to their signatures on the matter power spectrum,
in the linear and mildly non-linear regimes (for some examples of forecasts, see \cite{amendola_cosmology_2013, casas_fitting_2015,  bielefeld_cosmological_2014}).

The evolution of matter perturbations can be fully described by two generic functions of time and space \cite{kunz_phenomenological_2012, amendola_observables_2013}, which can be measured via Galaxy Clustering and Weak Lensing surveys. In this work we want to forecast how well we can measure those functions, in different redshift bins.

While any two independent functions of the gravitational potentials would do, we follow the notation of \cite{planck_collaboration_planck_2016} and consider $\mu$ and $\eta$: the first modifies the Poisson equation for $\Psi$ while the second is equal to the ratio of the gravitational potentials (and is therefore also a direct observable \cite{amendola_observables_2013}). We will consider forecasts for the planned surveys Euclid, SKA1 and SKA2 and a subset of DESI, DESI-ELG, using as priors the
constraints from recent \planck\ data (see also \cite{Alonso2016, hojjati_cosmological_2012, baker_observational_2015, bull_extending_2015, Gleyzes2016} for previous works that address forecasts in Modified Gravity.

In section \ref{sec:Parameterizing-Modified-Gravity} we define $\mu$ and $\eta$ and parameterize them in three different ways. First, in a general manner, we let these functions vary freely in different  redshift bins. Complementarily, we also consider two specific parameterizations of the time evolution proposed in \cite{planck_collaboration_planck_2016}. Here, we also specify the fiducial values of our cosmology for each of the
parameterizations considered.
Section \ref{sec:The-non-linear-power} discusses our treatment for the linear and mildly non-linear regime. Linear spectra are obtained from a modified Boltzmann
code \cite{hojjati_testing_2011}; the mild non-linear regime (up to k $\sim$ 0.5 h/Mpc) compares two methods to emulate the non-linear power spectrum: the commonly used Halofit \cite{smith_stable_2003, takahashi_revising_2012}, and a semi-analytic prescription to model
the screening mechanisms present in Modified Gravity models \cite{hu_parameterized_2007}.
In section \ref{sec:Fisher-Matrix-method} we explain the method used
to produce the Fisher forecasts both for Weak Lensing and Galaxy
Clustering. We explain how we compute and add the CMB \planck\ priors to our Fisher matrices.
Section \ref{sec:Results:-Redshift-Binned} discusses the results
obtained for the redshift binned parameterization both for Galaxy Clustering
and for Weak Lensing in the linear and non-linear cases. We describe our method to decorrelate the errors in section
\ref{sub:Decorrelation-of-covariance}. The results for the other two time parameterizations are instead discussed in sections \ref{sub:MG-DE} and \ref{sub:MG-TR}, both for Weak Lensing and Galaxy Clustering in the linear and mildly non-linear
regimes. To test the effect of our non-linear prescription, we show in section \ref{sub:Testing-the-effect-of-Zhao} the impact of different choices of the non-linear prescription parameters on the cosmological parameter estimation.

\section{\label{sec:Parameterizing-Modified-Gravity}Parameterizing Modified
Gravity}

In linear perturbation theory, scalar, vector and tensor perturbations
do not mix, which allows us to consider only the scalar perturbations
in this paper. We work in the conformal Newtonian gauge, with the
line element given by 
\begin{equation}
ds^{2}=-(1+2\Psi)dt^{2}+a^{2}(1-2\Phi)dx^{2}\,\,\,.
\end{equation}
Here $\Phi$ and $\Psi$ are two functions of time and scale that coincide
with the gauge-invariant Bardeen potentials in the Newtonian
gauge.

In theories with extra degrees of freedom (Dark Energy, DE) or modifications
of General Relativity (MG) the normal linear perturbation equations
are no longer valid, so that for a given matter source the values
of $\Phi$ and $\Psi$ will differ from their usual values. We can
parameterize this change generally with the help of two new functions
that encode the modifications. Many different choices are possible
and have been adopted in the literature, see e.g.\
\cite{planck_collaboration_planck_2016} for a limited overview. In this paper
we introduce the two functions through a gravitational slip (leading
to $\Phi\neq\Psi$ also at linear order and for pure cold dark matter)
and as a modification of the Poisson equation for $\Psi$, 
\begin{eqnarray}
-k^{2}\Psi(a,k) & \equiv & 4\pi
Ga^{2}\mu(a,k)\rho(a)\delta(a,k)\,\,\,;\label{eq: mu_def}\\
\eta(a,k) & \equiv & \Phi(a,k)/\Psi(a,k)\,\,\,.\label{eq: eta_def}
\end{eqnarray}
These expressions define $\mu$ and $\eta$. Here $\rho(a)$ is the average dark matter density and $\delta(a,k)$
the comoving matter density contrast -- we will neglect relativistic
particles and radiation as we are only interested in modeling the
perturbation behaviour at late times. In that situation, $\eta$,
which is effectively an observable \cite{amendola_observables_2013}, is closely related to
modifications of GR \cite{saltas_anisotropic_2014,sawicki_non-standard_2016}, while $\mu$ encodes for example deviations in
gravitational clustering, especially in redshift-space distortions
as non-relativistic particles are accelerated by the gradient of $\Psi$.

When considering Weak Lensing observations then it is also natural
to parameterize deviations in the lensing or Weyl potential $\Phi+\Psi$,
since it is this combination that affects null-geodesics (relativistic particles).
To this end we introduce a function $\Sigma(t,k)$ so that
\begin{equation}
-k^{2}(\Phi(a,k)+\Psi(a,k))\equiv8\pi
Ga^{2}\Sigma(a,k)\rho(a)\delta(a,k)\,\,\,.\label{eq:Sigma-def}
\end{equation}
Since metric perturbations are fully specified by two functions of
time and scale, $\Sigma$ is not independent from $\mu$ and $\eta$,
and can be obtained from the latter as follows: 
\begin{equation}
\Sigma(a,k)=(\mu(a,k)/2)(1+\eta(a,k))\,\,\,.\label{eq:SigmaofMuEta}
\end{equation}

Throughout this work, we will denote the standard Lambda-Cold-Dark-Matter ($\Lambda$CDM) model,
defined through the Einstein-Hilbert action with a cosmological constant, simply as GR. For this case
we have that $\mu=\eta=\Sigma=1$. All other cases in which these functions are not unity will be labeled
as Modified Gravity (MG) models.

Using effective quantities like $\mu$ and $\eta$ has the advantage
that they are able to model {\em any} deviations of the perturbation
behaviour from $\Lambda$CDM expectations, they are relatively close 
to observations, and they can also be related to other commonly used 
parameterization \cite{pogosian_how_2010}
On the other hand, they are not
easy to map to an action (as opposed to approaches like effective
field theories that are based on an explicit action) and in addition they
contain so much freedom that we normally restrict their parameterisation
to a subset of possible functions.

This has however
the disadvantage of loosing generality and making our constraints
on $\mu$ and $\eta$ parameterization-dependent. In this
paper, we prefer to complement specific choices of parameterizations
adopted in the literature (we will use the choice made in \cite{planck_collaboration_planck_2016}) 
with a more general approach: we will bin
the functions $\mu(a)$ and $\eta(a)$ in redshift bins with index
$i$ and we will treat each $\mu_{i}$ and $\eta_{i}$ as independent
parameters in our forecast; we will then apply a variation of Principal Component Analysis (PCA), called Zero-phase Component Analysis (ZCA).
This approach has been taken previously
in the literature by \cite{hojjati_cosmological_2012}, where they
bin $\mu$ and $\eta$ in several redshift and $k$-$scale$ bins
together with a binning of $w(z)$ and cross correlate large scale
structure observations with CMB temperature, E-modes and polarization
data together with Integrated Sachs-Wolfe (ISW) observations to forecast
the sensitivity of future surveys to modifications in $\mu$ and $\eta$.
In the present work, we will neglect a possible $k$-dependence, 
we will focus on Galaxy Clustering (GC) and weak
lensing (WL) surveys and we will show that there are important differences
between the linear and non-linear cases; including the non-linear regime generally reduces correlations
among the cosmological parameters. In the remainder of this section we will introduce the parameterizations that we will use.

\subsection{Parameterizing gravitational potentials in discrete redshift bins \label{sub:param-z-bins-th}}

As a first approach we neglect scale dependence and bin
the time evolution of the functions $\mu$ and $\eta$ without specifying
any parameterized evolution. To this purpose we divide the redshift
range $0\leq z\leq3$ in 6 redshift bins and we consider the values
$\mu(z_{i})$ and $\eta(z_{i})$ at the right limiting redshift $z_{i}$
of each bin as free parameters, thus with the $i$ index spanning the values
$\{0.5,1.0,1.5,2.0,2.5,3.0\}$. The first bin is assumed to have a constant 
value, coinciding with the one at $z_1=0.5$, i.e. $\mu(z<0.5)=\mu(z_{1})$ and $\eta(z<0.5)=\eta(z_{1})$.
The $\mu(z)$ function (and analogously $\eta(z)$) is then reconstructed
as 
\begin{equation}\label{eq:MGbin-mu-parametrization}
\mu(z)=\mu(z_{1})+\sum_{i=1}^{N-1}{\frac{\mu(z_{i+1})-\mu(z_{i})}{2}\left[1+\tanh{\left(s\frac{z-z_{i+1}}{z_{i+1}-z_{i}}\right)}\right]},
\end{equation}
where $s=10$ is a smoothing parameter and $N$ is the number of binned
values. We assume that both $\mu$ and $\eta$ reach the GR limit
at high redshifts: to realize this, the last $\mu(z_{6})$ and $\eta(z_{6})$
values assume the standard $\lcdm$ value $\mu=\eta=1$ and both functions
are kept constant at higher redshifts $z>3$.

Similarly, the derivatives of these functions are obtained by computing
\begin{equation}
\mu'({\bar{z}_{j}})=\frac{\mu(z_{i+1})-\mu(z_{i})}{z_{i+1}-z_{i}},
\end{equation}
with $\bar{z}_{j}=(z_{i+1}+z_{i})/2$, using the same $\tanh(x)$
smoothing function: 
\begin{equation}\label{eq:MGbin-muderiv-parametrization}
\frac{d\mu(z)}{dz}=\mu'(\bar{z}_{1})+\sum_{j=1}^{N-2}{\frac{\mu'(\bar{z}_{j+1})-\mu'(\bar{z}_{j})}{2}\left[1+\tanh{\left(s\frac{z-\bar{z}_{j+1}}{\bar{z}_{j+1}-\bar{z}_{j}}\right)}\right]}\,\,\,.
\end{equation}
In particular we assume $\mu'=\eta'=0$ for $z<0.5$ and for $z>3$.
We set the first five amplitudes of $\mu_{i}$ and $\eta_{i}$ as
free parameters, thus the set we consider is:
$\theta=\{\Omega_{m},\Omega_{b},h,\ln10^{10} A_{s},n_{s},\{\mu_{i}\},\{\eta_{i}\}\}$,
with $i$ an index going from 1 to 5. We take as fiducial cosmology
the values shown in Tab. \ref{tab:fiducial-MG-AllCases} columns 5
and 6. We only modify the evolution of perturbations and assume that
the background expansion is well described by the standard $\Lambda$CDM
expansion law for a flat universe with given values of $\Omega_{m}$,
$\Omega_{b}$ and $h$.

\subsection{Parameterizing gravitational potentials with simple smooth functions of the scale factor \label{sub:param-smooth-funct}}
As an alternative approach, we assume simple specific, time parameterizations for
the $\mu$ and $\eta$ MG functions, adopting the ones used in the
\planck\ analysis \cite{planck_collaboration_planck_2016}. 
We neglect here as well any scale dependence:
\begin{itemize}
\item a parameterization in which the time evolution is related to the dark
energy density fraction, to which we refer as `late-time' parameterization:
\begin{eqnarray}
\mu(a,k)\equiv1+E_{{\rm 11}}\Omega_{{\rm
DE}}(a)\,\,\,,\label{eq:DE-mu-parametrization}\\
\eta(a,k)\equiv1+E_{{\rm 22}}\Omega_{{\rm
DE}}(a)\,\,\,;\label{eq:DE-eta-parametrization}
\end{eqnarray}

\item a parameterization in which the time evolution is the simplest first
order Taylor expansion of a general function of the scale factor $a$
(and closely resembles the $w_{0}-w_{a}$ parametrization for the
equation of state of DE), referred to as `early-time' parameterization, because it
allows departures from GR also at high redshifts
\footnote{Notice that our early-time parametrization is called
`time-related' in \cite{planck_collaboration_planck_2016}.}:
\begin{eqnarray}
\mu(a,k)\equiv1+E_{{\rm 11}}+E_{{\rm
12}}(1-a)\,\,\,,\label{eq:TR-mu-parametrization}\\
\eta(a,k)\equiv1+E_{{\rm 21}}+E_{{\rm
22}}(1-a)\,\,\,.\label{eq:TR-eta-parametrization}
\end{eqnarray}

\end{itemize}

The late-time parameterization is forced to behave as GR ($\mu=\eta=1$)
at high redshift when $\Omega_{{\rm DE}}(a)$ becomes negligible;
the early time one allows more freedom as the amplitude of the deviations
from GR do not necessarily reduce to zero at high redshifts. Both parameterizations have been used in \cite{planck_collaboration_planck_2016}. In \cite{bull_extending_2015, Gleyzes2016, Alonso2016} the authors used a similar time parameterization in which the Modified Gravity
parameters depend on the time evolution of the dark energy fraction. In \cite{bull_extending_2015} an extra parameter accounts for a scale-dependent $\mu$: their treatment keeps $\eta$ (called $\gamma$ in their paper) fixed and equal to 1; it uses linear power spectra up to $k_{\mathrm{max}}(z)$ with $k_{\rm max}(z=0)=0.14$/Mpc $\approx$ 0.2 h/Mpc. 
In \cite{Gleyzes2016} the authors also use a combination of Galaxy Clustering, Weak Lensing and ISW cross-correlation to constrain
Modified Gravity in the Effective Field Theory formalism \cite{Gubitosi2013}. In \cite{Bellini2016} and \cite{Alonso2016} a similar parameterization was used to constrain the Horndeski functions \cite{Bellini2014} with present data and future forecasts respectively, in the linear regime.

For the late-time parameterization, the set of free parameters we
consider is: $\theta=\{\Omega_{m},\Omega_{b},h,\ln10^{10}A_{s},n_{s},E_{11},E_{22}\}$,
where $E_{11}$ and $E_{22}$ determine the amplitude of the variation
with respect to $\lcdm$. As fiducial cosmology we use the values
shown in Table \ref{tab:fiducial-MG-AllCases}, columns 1 and 2, i.e.
the marginalized parameter values obtained fitting these models with
recent \planck\ data; notice that these results differ slightly from
the \planck\ analysis in \cite{planck_collaboration_planck_2016} for
the same parameterization, because we don't consider here the effect
of massive neutrinos.

For the early time parameterization we have $E_{11}$ and $E_{21}$
which determine the amplitude of the deviation from GR at present
time ($a=0$) and 2 additional parameters ($E_{12},\ E_{22}$), which
determine the time dependence of the $\mu(a)$ and $\eta(a)$ functions
for earlier times. The
fiducial values for this model, obtained from the {\it Planck}+BSH best
fit is given in columns 3 and 4 of Table \ref{tab:fiducial-MG-AllCases}.

\begin{table}[htbp]
\centering{}
\begin{tabular}{|cc|cc|cc|}
\hline 
\multicolumn{2}{|c|}{\textbf{Late time}} &
\multicolumn{2}{c|}{\textbf{Early time}} &
\multicolumn{2}{c|}{\Tstrut \textbf{Redshift Binned}}\tabularnewline
\multicolumn{1}{|c}{Parameter } & \multicolumn{1}{c|}{Fiducial } &
\multicolumn{1}{c}{Parameter } & \multicolumn{1}{c|}{Fiducial } &
\multicolumn{1}{c}{Parameter } & \multicolumn{1}{c|}{Fiducial }\tabularnewline
\hline 
$\Omega_{c}$  & \multicolumn{1}{c|}{$0.254$ } & $\Omega_{c}$  &
\multicolumn{1}{c|}{$0.256$} & $\Omega_{c}$  & $0.254$ \tabularnewline
$\Omega_{b}$  & \multicolumn{1}{c|}{$0.048$ } & $\Omega_{b}$  &
\multicolumn{1}{c|}{$0.048$} & $\Omega_{b}$  & $0.048$ \tabularnewline
$n_{s}$  & \multicolumn{1}{c|}{$0.969$ } & $n_{s}$  &
\multicolumn{1}{c|}{$0.969$} & $n_{s}$  & $0.969$ \tabularnewline
$\ln10^{10}A_{s}$  & \multicolumn{1}{c|}{$3.063$ } & $\ln10^{10}A_{s}$  &
\multicolumn{1}{c|}{$3.091$} & $\ln10^{10}A_{s}$  & $3.057$ \tabularnewline
$h$  & \multicolumn{1}{c|}{$0.682$ } & $h$  & \multicolumn{1}{c|}{$0.682$} &
$h$  & $0.682$ \tabularnewline
$E_{11}$  & \multicolumn{1}{c|}{$0.100$ } & $E_{11}$  &
\multicolumn{1}{c|}{$-0.098$} & $\mu_{1}$  & $1.108$\tabularnewline
$E_{22}$  & \multicolumn{1}{c|}{$0.829$ } & $E_{12}$  &
\multicolumn{1}{c|}{$0.096$} & $\mu_{2}$  & $1.027$\tabularnewline
\cline{1-2} 
\multicolumn{1}{c}{} & \multicolumn{1}{c|}{} & $E_{21}$  &
\multicolumn{1}{c|}{$0.940$} & $\mu_{3}$  & $0.973$\tabularnewline
\multicolumn{1}{c}{} & \multicolumn{1}{c|}{} & $E_{22}$  &
\multicolumn{1}{c|}{$-0.894$} & $\mu_{4}$  & $0.952$\tabularnewline
\cline{3-4} 
\multicolumn{1}{c}{} & \multicolumn{1}{c}{} & \multicolumn{1}{c}{} &
\multicolumn{1}{c|}{} & $\mu_{5}$  & $0.962$\tabularnewline
\multicolumn{1}{c}{} & \multicolumn{1}{c}{} & \multicolumn{1}{c}{} &
\multicolumn{1}{c|}{} & $\eta_{1}$  & $1.135$\tabularnewline
\multicolumn{1}{c}{} & \multicolumn{1}{c}{} & \multicolumn{1}{c}{} &
\multicolumn{1}{c|}{} & $\eta_{2}$  & $1.160$\tabularnewline
\multicolumn{1}{c}{} & \multicolumn{1}{c}{} & \multicolumn{1}{c}{} &
\multicolumn{1}{c|}{} & $\eta_{3}$  & $1.219$\tabularnewline
\multicolumn{1}{c}{} & \multicolumn{1}{c}{} & \multicolumn{1}{c}{} &
\multicolumn{1}{c|}{} & $\eta_{4}$  & $1.226$\tabularnewline
\multicolumn{1}{c}{} & \multicolumn{1}{c}{} & \multicolumn{1}{c}{} &
\multicolumn{1}{c|}{} & $\eta_{5}$  & $1.164$\tabularnewline
\cline{5-6} 
\end{tabular}\protect\caption{\label{tab:fiducial-MG-AllCases} Fiducial values
for the Modified
Gravity parameterizations and the redshift-binned model of $\mu$
and $\eta$ used in this work. The DE related parameterization contains
two extra parameters $E_{11}$ and $E_{22}$ with respect to GR; the
early-time parametrization depends on 4 extra parameters
$E_{11},\,E_{12},\,E_{22}$
and $E_{21}$ with respect to GR; the redshift-binned model contains
10 extra parameters, corresponding to the amplitudes $\mu_{i}$ and
$\eta_{i}$ in five redshift bins. In this work we will
use alternatively and for simplicity the notation $\ell \mathcal{A}_s \equiv \ln(10^{10} A_{s})$.
The fiducial values are obtained
performing a Monte Carlo analysis of {\it Planck}+BAO+SNe+H$_{0}$ (BSH) data
\cite{planck_collaboration_planck_2016}.}
\label{tab:DEfid} 
\end{table}

\begin{figure}[htbp]
	\centering{}\begin{center}
		\includegraphics[width=0.45\linewidth]{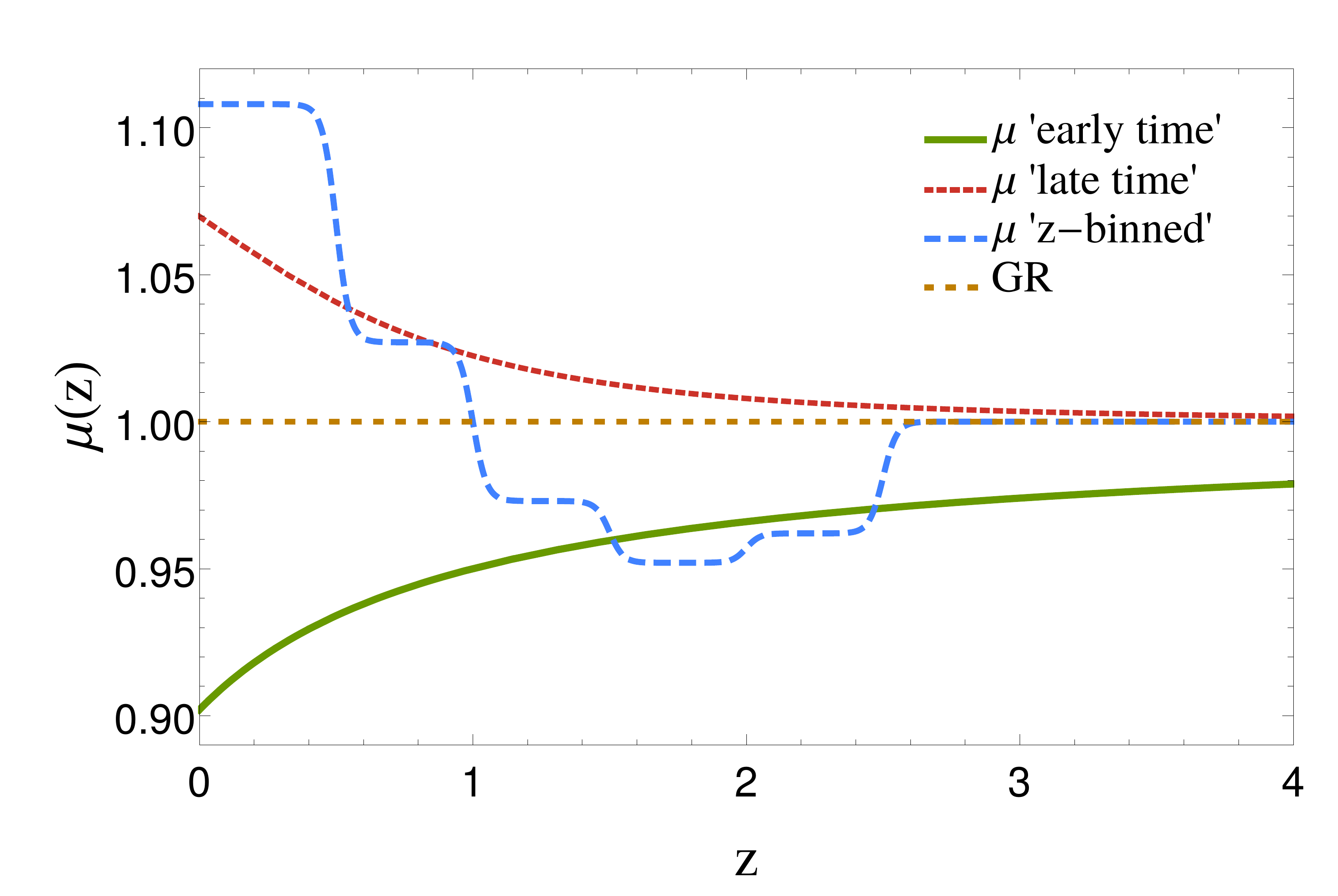}
		\includegraphics[width=0.45\linewidth]{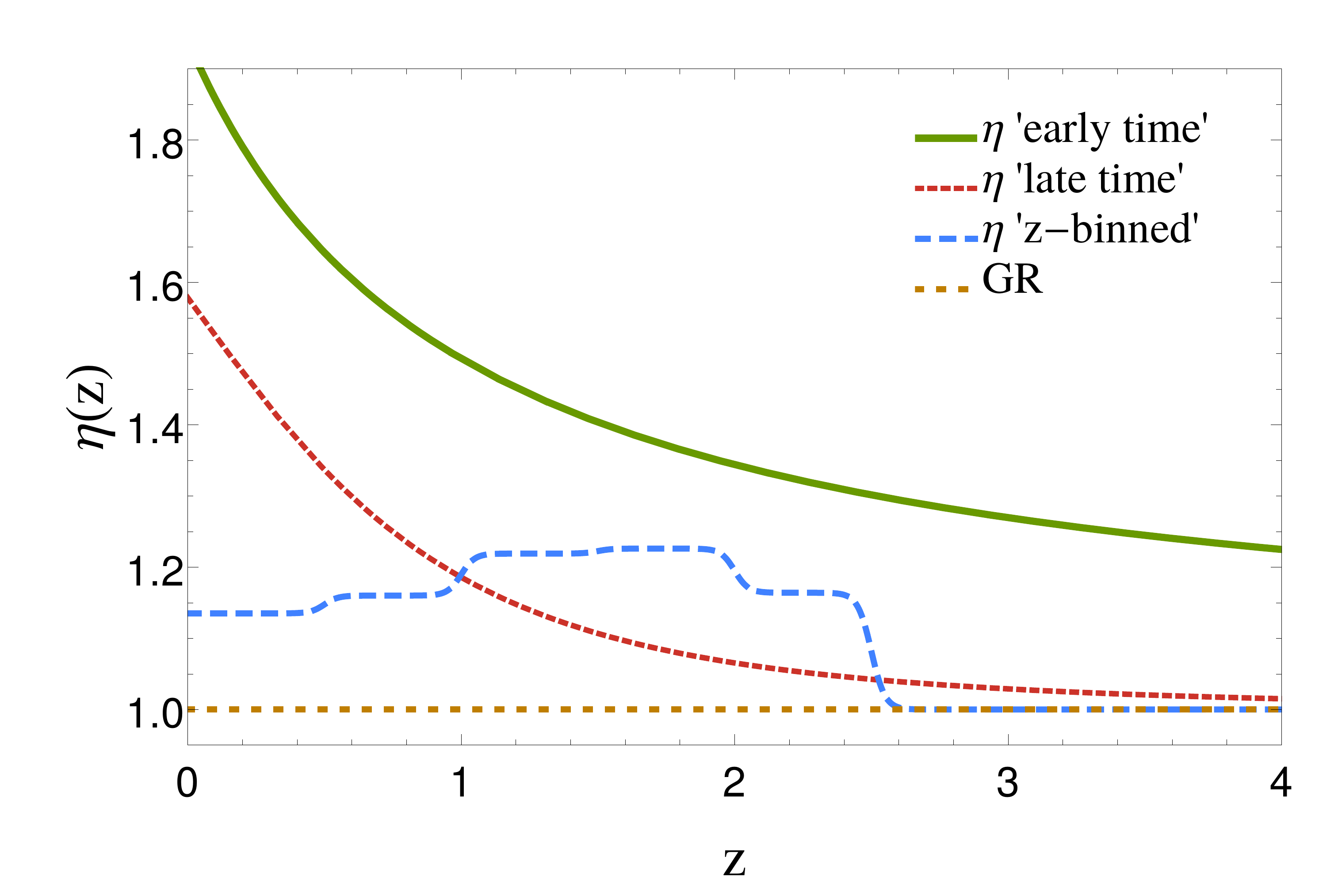}
	\end{center}
	\caption{\label{fig:fidplot}
The Modified Gravity functions $\mu$ and $\eta$ as a function of redshift $z$ for each of the models considered in this work, evaluated at the fiducials specified inTable \ref{tab:fiducial-MG-AllCases}. 
In light long-dashed blue lines, the `redshift binned model' (Eqns.\ \ref{eq:MGbin-mu-parametrization}-\ref{eq:MGbin-muderiv-parametrization}). In short-dashed red lines the late-time parameterization (Eqns.\ \ref{eq:DE-mu-parametrization}-\ref{eq:DE-eta-parametrization}) and in green solid lines the early-time parameterization (Eqns.\ \ref{eq:TR-mu-parametrization}-\ref{eq:TR-eta-parametrization}). Finally the medium-dashed orange line represents the standard $\lcdm$ model (GR) for reference.
		}
\end{figure}

\section{\label{sec:The-non-linear-power}The power spectrum in Modified
Gravity}

\subsection{The linear power spectrum}

In this work we will use linear power spectra calculated with MGCAMB
\cite{zhao_searching_2009,hojjati_testing_2011}, a modified version
of the Boltzmann code CAMB \cite{lewis_efficient_2000}. We do so,
as MGCAMB offers the possibility to input directly any parameterization
of $\mu$ and $\eta$ without requiring further assumptions:
MGCAMB uses our Eqns.\ (\ref{eq: mu_def}) and (\ref{eq: eta_def}) in the Einstein-Boltzmann
system of equations, providing the modified evolution of matter perturbations,
corresponding to our choice of the gravitational potential functions.
Non-relativistic particles like cold dark matter are accelerated by the
gradient of $\Psi$, so that especially the redshift space distortions are
sensitive to the modification given by $\mu(a,k)$. For relativistic particles
like photons and neutrinos on the other hand, the combination of $\Phi+\Psi$ (and therefore $\Sigma$)
enters the equations of motion. The impact on the matter power spectrum
is more complicated, as the dark matter density contrast is linked via
the relativistic Poisson equation to $\Phi$. In addition, an early-time
modification of $\Phi$ and $\Psi$ can also affect the baryon distribution through
their coupling to radiation during that period.
As already mentioned above, we will not consider
the $k-$dependence of $\mu$ and $\eta$ in this work and our modifications
with respect to standard GR will be only functions of the scale factor
$a$.

\subsection{Non-linear power spectra}

As the Universe evolves, matter density fluctuations
($\delta_{m}$) on small scales (k > 0.1 h/Mpc) become larger than unity and shell-crossing will eventually occur. The usual continuity and
Euler equations, together with the Poisson equation become singular
\cite{bernardeau_large-scale_2001,bernardeau_evolution_2013},
therefore making a computation of the matter power spectrum in the
highly non-linear regime practically impossible under the standard
perturbation theory approach. However, at intermediate scales of around
$k\approx0.1-0.2$ h/Mpc, there is the possibility of calculating
semi-analytically the effects of non-linearities on the oscillation
patterns of the baryon acoustic oscillations (BAO). There are many
approaches in this direction, from renormalized perturbation theories
\cite{crocce_renormalized_2006,blas_time-sliced_2016,taruya_closure_2008}
to time flow equations \cite{pietroni_flowing_2008,anselmi_nonlinear_2012}, effective or coarse grained theories
\cite{carrasco_effective_2012,baumann_cosmological_2012,pietroni_coarse-grained_2011,manzotti_coarse_2014} and many others.
 This direction of research is important, since from the
amplitude and widths of the first few BAO peaks, one can extract more information from data and break degeneracies among parameters, especially in Modified Gravity.

Computing the non-linear power spectrum in standard GR is still an
open question, and even more so when the Poisson equations are modified,
as it is in the case in Modified Gravity theories. A solution to this problem
is to calculate the evolution of matter perturbations in an N-body
simulation
\cite{springel_cosmological_2005,fosalba_mice_2013,takahashi_revising_2012,lawrence_coyote_2010,heitmann_coyote_2014},
however, this procedure is time-consuming and computationally expensive.

Because of these issues, several previous analyses
have been done with a conservative removal of the information at small
scales (see for example the \planck\ Dark Energy paper
\cite{planck_collaboration_planck_2016},
several CFHTLenS analysis \cite{heymans_cfhtlens_2013,kitching_3d_2014}
or the previous PCA analysis by \cite{hojjati_cosmological_2012}).
However, future surveys will probe an extended range of scales, therefore
removing non-linear scales from the analysis would strongly reduce
the constraining power of these surveys. Moreover, at small scales
we also expect to find means of discriminating between different Modified
Gravity models, such as the onset of screening mechanisms needed to
recover GR at small scales where experiments strongly constrain
deviations from it. For these reasons it is crucial to find methods
which will allow us to investigate, at least approximately, the non-linear
power spectrum.

Attempts to model the non-linear power spectrum semi-analytically in Modified
Gravity have been investigated for f(R) theories in
\cite{zhao_modeling_2014,taruya_regularized_2014},
for coupled dark energy in
\cite{casas_fitting_2015,saracco_non-linear_2010,vollmer_efficient_2014}
and for growing neutrino models in \cite{brouzakis_nonlinear_2011}.
Typically they rely on non-linear expansions of the perturbations
using resummation techniques based on
\cite{pietroni_flowing_2008,taruya_closure_2008}
or on fitting formulae based on N-Body simulations
\cite{casas_fitting_2015,takahashi_revising_2012,bird_massive_2011}.
A similar analysis is not available for the model-independent approach considered
in this paper. 
In order to give at least a qualitative estimate
of what the importance of non-linearities would be for constraining these
Modified Gravity models, we will adopt in the rest of the paper a method 
which interpolates between the standard approach to non linear scales in GR
and the same applied to MG theories.

\subsubsection{Halofit}
\begin{figure}[htbp]
\begin{centering}
\includegraphics[width=0.45\textwidth]{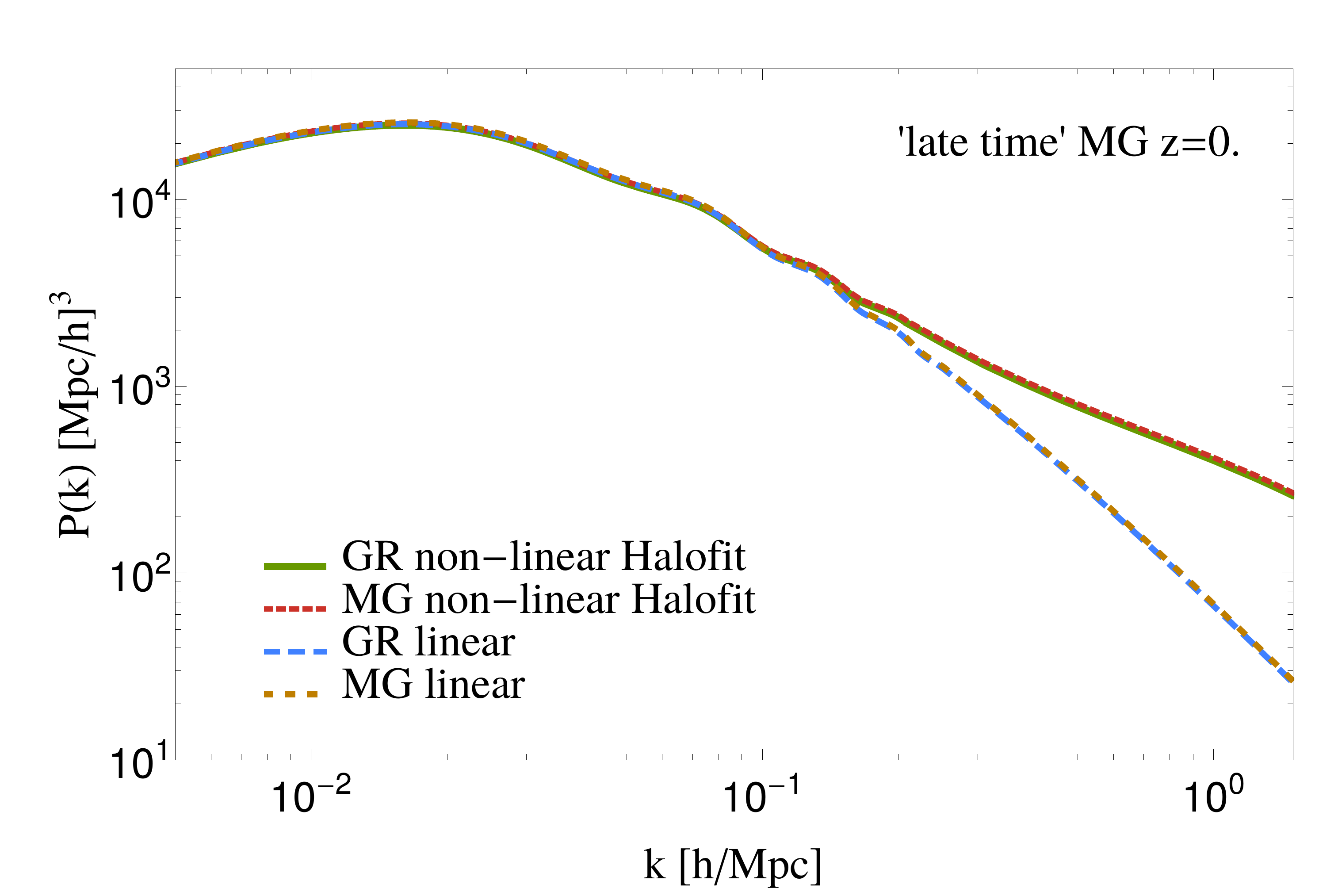}
\includegraphics[width=0.45\textwidth]{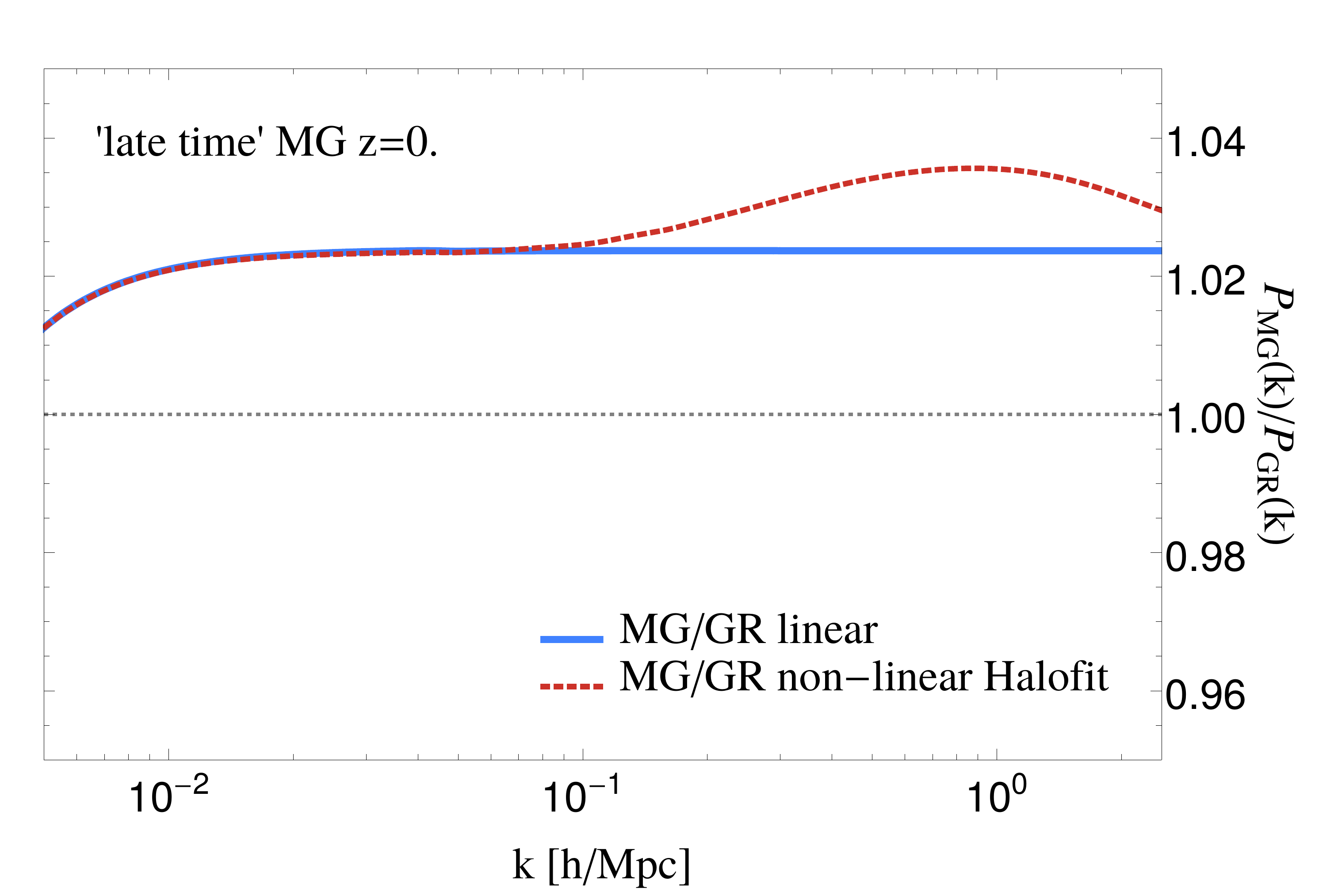} 
\par\end{centering}

\protect\caption{\label{fig:lin-non-pk-mg} \textbf{Left: }matter power
spectra
computed with MGCAMB (linear) and MGCAMB+Halofit (non-linear), illustrating the impact of non-linearities at different scales. As an illustrative example, MG in this plot corresponds to the fiducial model in the late-time parametrization defined in Eq.\ (\ref{eq:DE-mu-parametrization}).
All curves are computed at $z=0$. The green solid line is the GR
fiducial in the non-linear case, the blue long-dashed line is also
GR but in the linear case. The short-dashed red line is the MG fiducial
in the non-linear case and the medium-dashed brown line the MG fiducial
in the linear case.
\textbf{Right:
}in order to have a closer look at small scales, we plot here the
ratio of the MG power spectrum to the GR power spectrum for the linear
(blue solid) and non-linear (red short-dashed) cases separately. The blue solid line compared to the horizontal grey dashed line,
shows the effect of Modified Gravity when taking only linear spectra into account. While the red dashed line, which
represents the non-linear case, shows that the ratio to GR presents clearly a bump that peaks around
$k\approx1.0$ h/Mpc, meaning that the power spectrum in MG differs
at most 4\% from the non-linear power spectrum in GR. We will see later that we are able to 
discriminate between these two models using future surveys, especially when non-linear scales ($k \gtrsim 0.1$ h/Mpc) are included.
} 
\end{figure}

We describe here the effect of applying the standard 
approach to non linearities to MG theories. This is done using
the revised Halofit \cite{takahashi_revising_2012}, based on
\cite{smith_stable_2003},
which is a fitting function of tens of numerical parameters that reproduces
the output of a certain set of $\lcdm$ N-body simulations in a specific
range in parameter space as a function of the linear power spectrum.
This fitting function is reliable with an accuracy of better than
10\% at scales larger than $k\lesssim1$h/Mpc and redshifts in between
$0\leq z\leq10$ (see \cite{takahashi_revising_2012} for more details).
This fitting function can be used within Boltzmann codes to estimate
the non-linear contribution which corrects the linear power spectrum
as a function of scale and time. We will use the Halofit fitting function
 as a way of approximating the non-linear power spectrum in
our models even though it is really only valid for $\lcdm$. 
In Fig.\ \ref{fig:lin-non-pk-mg}, the left panel
shows a comparison between the linear and non-linear power spectra
calculated by MGCAMB in two different models, our fiducial late-time
model (as from Table \ref{tab:fiducial-MG-AllCases}) and GR, both sharing
the same $\lcdm$ parameters. At small length scales (large $k$),
the non-linear deviation is clearly visible at scales $k\gtrsim0.3$ h/Mpc
and both MG and GR seem to overlap due to the logarithmic scale used.
In the right panel, we can see the ratio between MG and GR for both linear
and non-linear power spectra, using the same 5 $\lcdm$ parameters
$\{\Omega_{m},\Omega_{b},h,\,A_{s},n_{s}\}$. We can see clearly that
MG in the non-linear regime, using the standard Halofit, shows a distinctive
feature at scales in between $0.2\lesssim k\lesssim2$. This feature
however, does not come from higher order perturbations induced by
the modified Poisson equations (\ref{eq: mu_def}, \ref{eq:Sigma-def}),
because Halofit, as explained above, is calibrated with simulations
within the $\lcdm$ model and does not contain any information from
Modified Gravity. The feature seen here is caused by the different
growth rate of perturbations in Modified Gravity, that yields then
a different evolution of non-linear structures. 

\subsubsection{Prescription for mildly non-linear scales including screening}
\label{sub:Prescription-HS}

\begin{figure}[htbp]
\begin{centering}
\includegraphics[width=0.45\textwidth]{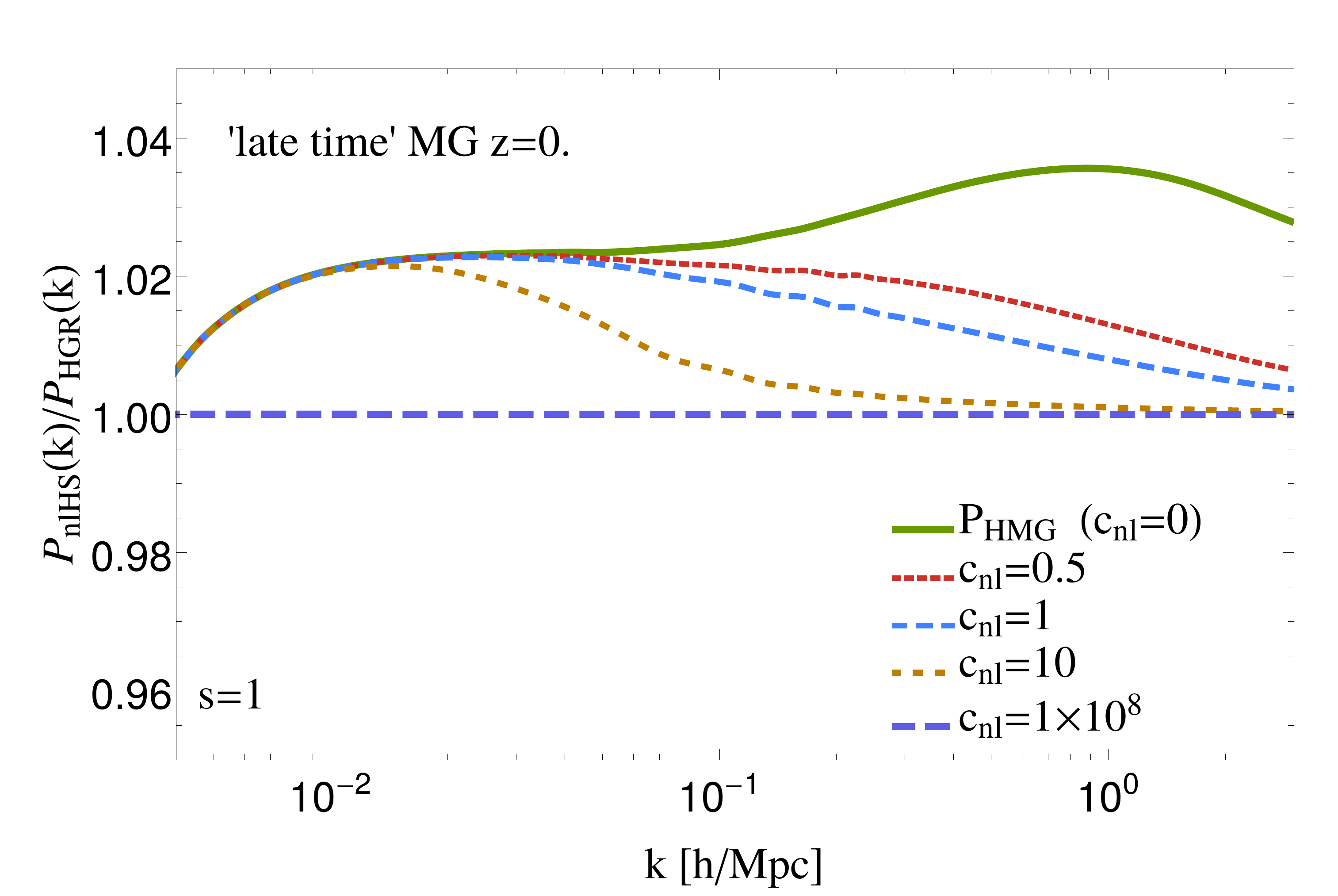}
\includegraphics[width=0.45\textwidth]{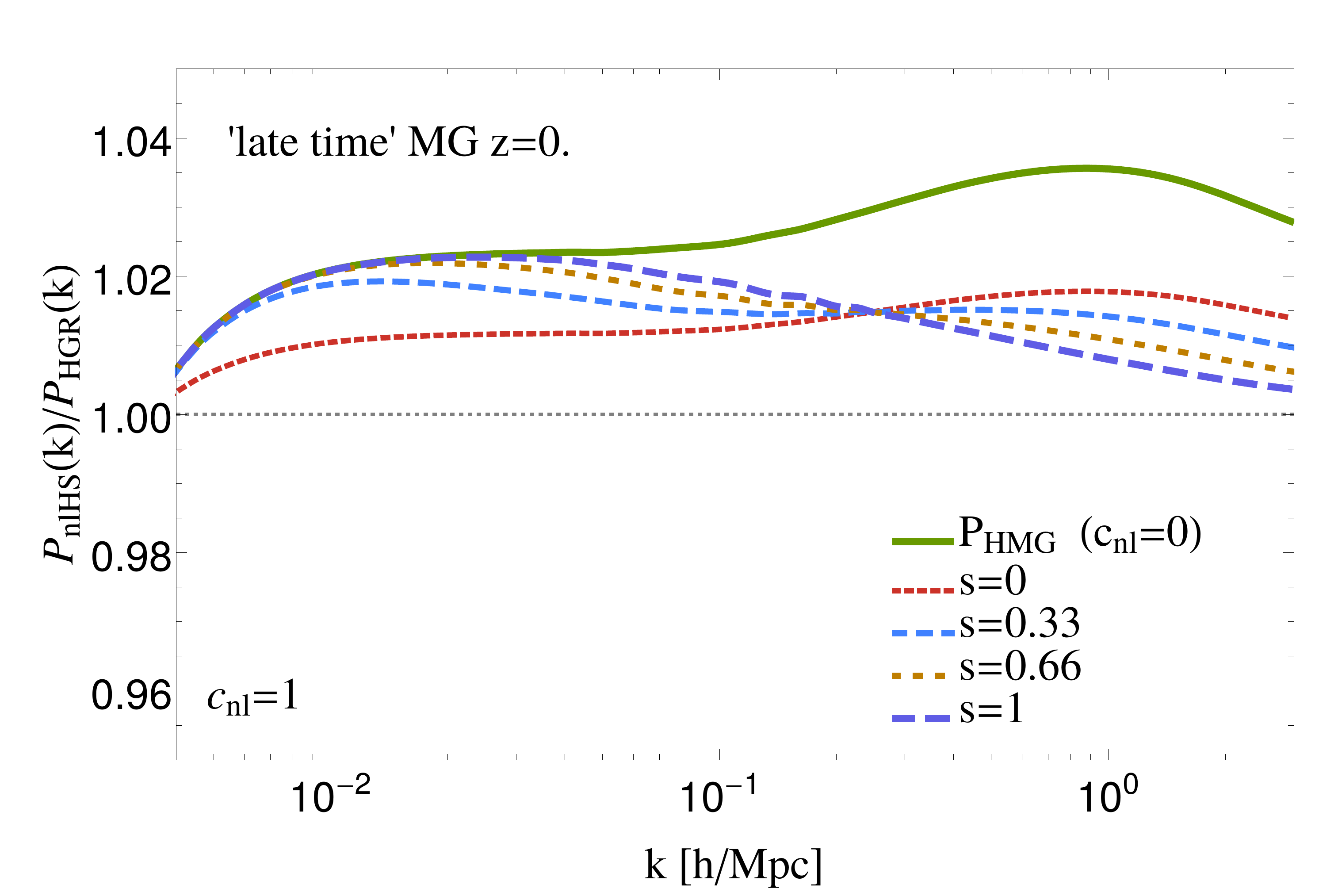} 
\par\end{centering}
\protect\caption{\label{fig:lin-nonlin-Zhao-MG} The ratio of the Modified Gravity non-linear
power spectrum using the HS prescription by \cite{hu_parameterized_2007}
($P_{\mathrm{nlHS}}$) with respect to the GR+Halofit fiducial non-linear power spectrum $P_{\mathrm{HGR}}$, 
for different values of $\cnl$ (left panel) and $s$ (right panel),
illustrated in Eqns.(\ref{eq:PHSDefinition}, \ref{eq:prescription_sigma_def}).
The value $\cnl=0$ (green solid line) corresponds
to MG+Halofit $P_{\HMG}$.
All curves are calculated at $z=0$.
\textbf{Left: } We show the ratio for $\cnl=\{0.5,1.0,10,10^8\}$,
plotted as short-dashed red, medium-dashed blue, short-dashed brown and medium-dashed purple
lines respectively. When $\cnl\rightarrow\infty$, Eqn.\ (\ref{eq:PHSDefinition})
corresponds to the limit of $P_{\HGR}$ and therefore the ratio is
just 1. The effect of the HS prescription
is to grasp some of the features of the non-linear power spectrum
at mildly non-linear scales induced by Modified Gravity, taking into
account that at very small scales, a screening mechanism might yield
again just a purely GR non-linear power spectrum. The parameter $\cnl$
interpolates between these two cases. \textbf{Right: }in this panel
we show the effect of the parameter $s$, for $s=\{0,0.33,0.66,1\}$
(short-dashed red, medium-dashed blue, short-dashed brown and long-dashed
purple, respectively). Both parameters need to be fitted with simulations
in order to yield a reliable match with the shape of the non-linear
power spectrum in Modified Gravity, as it was done in \cite{zhao_n-body_2011}
and references therein. The grey dashed line marks the constant value of 1.}
\end{figure}

As discussed above, modifications to the $\Phi$ and $\Psi$ 
potentials make the use of Halofit to compute the evolution at non linear scales 
unreliable. In order to
take into account the non-linear contribution to the power spectrum
in Modified Gravity, we investigate here a different method, which
starts from the consideration that whenever we
modify $\mu$ and $\eta$ with respect to GR, we modify the strength
of gravitational attraction in a way universal to all species: this
means that, similarly to the case of scalar-tensor or $f(R)$ theories,
we need to assume the existence of a non-perturbative screening mechanism, acting at
small scales, that guarantees agreement with solar system experiments.
In other words, it is reasonable to think that the non-linear power
spectrum will have to match GR at sufficiently small scales, while
at large scales it is modified. Of course, without having a specific model in mind, it remains arbitrary
how the interpolation between the small scale regime and the large
scale regime is done. In this
paper, we adopt the Hu \& Sawicki (HS) Parametrized Post-Friedmann prescription proposed
in \cite{hu_parameterized_2007}, which was used for the case of $f(R)$
theories previously by \cite{zhao_modeling_2014}. Given a MG model,
this prescription interpolates between the non-linear power spectrum
in Modified Gravity (which is in our case just the linear MG power
spectrum corrected with standard Halofit, $P_{\HMG}$) and the non-linear
power spectrum in GR calculated with Halofit ($P_{\mathrm{HGR}}$). The resulting
power spectrum will be denoted as $P_{\mathrm{nlHS}}$
\begin{equation}
P_{\mathrm{nlHS}}(k,z)=\frac{P_{\HMG}(k,z)+\cnl S_{\LIN}^{2}(k,z)P_{\mathrm{HGR}}(k,z)}{1+\cnl S_{\LIN}^{2}(k,z)} \, ,\label{eq:PHSDefinition}
\end{equation}
with 
\begin{equation}
S_{\LIN}^{2}(k,z)=\left[\frac{k^{3}}{2\pi^{2}}P_{\LMG}(k,z)\right]^{s} \, . \label{eq:prescription_sigma_def}
\end{equation}
The weighting function $S_{\LIN}$ used in the interpolation quantifies the onset of non-linear clustering and it is constructed
using the linear power spectrum in Modified Gravity ($P_{\LMG}$).
The constant $\cnl$ and the constant exponent $s$ are free parameters.
In Figure \ref{fig:lin-nonlin-Zhao-MG} we show the ratio $P_{\mathrm{nlHS}}/P_{\HGR}$,
which illustrates the relative difference between the non-linear HS prescription
in MG and the Halofit non-linear power spectrum in GR, for different
values of $\cnl$ (left panel) and different values of $s$ (right
panel). The parameter $\cnl$ controls at which scale there is a
transition into a non-linear regime in which standard GR is valid
(this can be the case when a screening mechanism is activated);
$s$ controls the smoothness of the transition and is in principle
a model and redshift dependent quantity. When $\cnl=0$ we recover
the Modified Gravity power spectrum with Halofit $P_{\HMG}$; when
$\cnl\rightarrow\infty$ we recover the non-linear power spectrum
in GR calculated with Halofit $P_{\mathrm{HGR}}$. In
\cite{zhao_modeling_2014,zhao_n-body_2011,koyama_non-linear_2009},
the $\cnl$ and $s$ constants were obtained fitting expression (\ref{eq:PHSDefinition}) to N-Body
simulations or to a semi-analytic perturbative approach. In the case
of $f(R)$, $s=1/3$ seems to match very well the result from simulations
up to a scale of $k=0.5$h/Mpc \cite{koyama_non-linear_2009}. A relatively
good agreement up to such small scales is enough for our purposes.
In the absence of N-Body simulations or semi-analytic methods available
for the models investigated in this work, we will assume unity for
both parameters, which is a natural choice, and we will test in Section \ref{sub:Testing-the-effect-of-Zhao}
how our results vary for different values of these parameters, namely
$\cnl=\{0.1,0.5,\,1,3\}$ and $s=\{0,\,1/3,\,2/3,\,1\}$. This will
give a qualitative estimate of the impact of non-linearities on the
determination of cosmological parameters.

\section{\label{sec:Fisher-Matrix-method}Fisher Matrix forecasts}

The Fisher matrix formalism
(\cite{tegmark_measuring_1998,seo_improved_2007,seo_baryonic_2005})
is one of the most popular tools to forecast the outcome of an experiment,
because of its speed and its versatility when the likelihood is approximately
Gaussian. Here we apply the Fisher matrix formalism to two different
probes, Galaxy Clustering (GC) and Weak Lensing (WL), which are the
main cosmological probes for the future Euclid satellite \cite{mukherjee_planck_2008}.
The background and perturbations quantities we
use in the following equations are computed with a version of
\texttt{MGCAMB} \cite{zhao_searching_2009,hojjati_testing_2011}
modified in order to account for the binning and the parameterizations
described in Section \ref{sec:Parameterizing-Modified-Gravity}.

\subsection{Future large scale galaxy redshift surveys \label{sub:FutureSurveys}}

In this work we choose to present results on some of the future galaxy redshift surveys, which are planned to be started and analyzed within the next decade.
Our baseline survey will be the Euclid satellite \cite{amendola_cosmology_2013, laureijs_euclid_2011}. Euclid\footnote{http://www.euclid-ec.org/} is a European Space Agency medium-class mission scheduled for launch in 2020. Its main goal is to explore the expansion history of the Universe and the evolution of large scale cosmic structures by measuring shapes and redshifts of galaxies, covering 15000$\text{deg}^2$ of the sky, up to redshifts of about $z\sim2$. It will be able to measure up to 100 million spectroscopic redshifts which can be used for Galaxy Clustering measurements and 2 billion photometric galaxy images, which can be used for Weak Lensing observations (for more details, see \cite{amendola_cosmology_2013, laureijs_euclid_2011}). We will use in this work the  Euclid Redbook specifications for Galaxy Clustering and Weak Lensing forecasts \cite{laureijs_euclid_2011}, some of which are listed in Tables \ref{tab:GC-specifications} and \ref{tab:WL-specifications} and the rest can be found in the above cited references.

Another important future survey will be the Square Kilometer Array (SKA)\footnote{https://www.skatelescope.org/}, which is planned to become the world's largest radiotelescope. It will be built in two phases, phase 1 split into SKA1-SUR in Australia and SKA1-MID in South Africa and SKA2 which will be at least 10 times as sensitive. The first stage is due to finish observations around the year 2023 and the second phase is scheduled for 2030 (for more details, see \cite{yahya_cosmological_2015,santos_hi_2015,raccanelli_measuring_2015,bull_measuring_2015}). The first phase SKA1, will be able to measure in an area of 5000$\text{deg}^2$ of the sky and a redshift of up to $z\sim0.8$  an estimated number of about $5\times10^6$ galaxies; SKA2 is expected to cover a much larger fraction of the sky ($\sim$30000$\text{deg}^2$), will yield much deeper redshifts (up to $z\sim2.5$) and is expected to detect about $10^9$ galaxies with spectroscopic redshifts \cite{santos_hi_2015}.
SKA1 and SKA2 will also be capable of performing 
radio Weak Lensing experiments, which are very promising, since they are expected to be less sensitive to systematic effects in the instruments, related to residual point spread function (PSF) anisotropies \cite{harrison_ska_2016}.
In this work we will use for our forecasts of SKA1 and SKA2, the specifications computed by \cite{santos_hi_2015} for GC and by \cite{harrison_ska_2016} for WL. The numerical survey parameters are listed in Tables \ref{tab:GC-specifications} and \ref{tab:WL-specifications}, while the galaxy bias $b(z)$ and the number density of galaxies $n(z)$, can be found in the references mentioned above.

We will also forecast the results from DESI\footnote{http://desi.lbl.gov/}, a stage IV, ground-based dark energy experiment, that will study large scale structure formation in the Universe through baryon acoustic oscillations (BAO) and redshift space distortions (RSD), using redshifts and positions from galaxies and quasars \cite{desi_collaboration_desi_2016-1,desi_collaboration_desi_2016,levi_desi_2013}.
It is scheduled to start in 2018 
and will cover an estimated area in the sky of about 
14000$\text{deg}^2$. It will measure spectroscopic redshifts for four different classes of objects, luminous red galaxies (LRGs) up to a redshift of $z=1.0$, bright [O II] emission line galaxies (ELGs) up to $z=1.7$, quasars (QSOs) up to $z\sim3.5$ and at low redshifts ($z\sim0.2$) magnitude-limited bright galaxies (BLGs). In total, 
DESI will be able to measure more than 30 million spectroscopic redshifts.
In this paper we will use for our forecasts only the specifications for the ELGs, as found in \cite{desi_collaboration_desi_2016-1}, since this observation provides the largest number density of galaxies in the redshift range of our interest. We cite the geometry and redshift binning specifications in Table \ref{tab:GC-specifications}, while the galaxy number density and bias can be found in \cite{desi_collaboration_desi_2016-1}.

\subsection{Galaxy Clustering\label{sub:Fisher-Galaxy-Clustering}}

The distribution of galaxies in space is not perfectly uniform. Instead it follows, up to a bias, the underlying
matter power spectrum so that the observed power spectrum $\Pobs$ (the Fourier transform of the real-space two point correlation
function of the galaxy number counts) is closely linked to the dark matter power spectrum $P(k)$. 
The observed power spectrum however also contains
additional effects like redshift-space distortions due to velocities and a suppression of power due to redshift-uncertainties. Here
we follow \cite{seo_improved_2007}, neglecting further relativistic and observational effects, and write the observed power spectrum as
\begin{equation}
\Pobs (k,\mu,z)=\frac{D_{A,f}^{2}(z)H(z)}{D_{A}^{2}(z)H_{f}(z)}b^{2}(z)(1+\beta_{d}(z)\mu^{2})^{2}e^{-k^{2}\mu^{2}(\sigma_{r}^{2}+\sigma_{v}^{2})}P(k,z)\,\,\,.\label{eq:observed-Pk}
\end{equation}
$\Pobs (k,\mu,z)$ is the observed power spectrum as a
function of the redshift $z$, the wavenumber $k$ and of $\mu\equiv\cos\alpha$,
where $\alpha$ is the angle between the line of sight and the 3D-wavevector
$\vec{k}$.
This observed power spectrum contains all the cosmological information
about the background and the matter perturbations as well as corrections
due to redshift-space distortions, geometry and observational uncertainties.
In the formula, the subscript $f$ denotes the fiducial value
of each quantity, $P(k,z)$ is the matter power spectrum, $D_{A}(z)$
is the angular diameter distance, $H(z)$ the Hubble function and
$\beta_{d}(z)$ is the redshift space distortion factor, which in
linear theory is given by $\beta_{d}(z)=f(z)/b(z)$, with $f(z)\equiv d\ln
G/d\ln a$
representing the linear growth rate of matter perturbations and $b(z)$ the galaxy bias as a function of redshift,
which we assume to be local and scale-independent. The exponential
factor represents the damping of the observed power spectrum, due
to two different effects: $\sigma_{z}$ an error induced by spectroscopic
redshift measurement errors, which translates into an uncertainty in the position of galaxies at a scale of
$\sigma_{r}=\sigma_{z}/H(z)$ and $\sigma_{v}$ which is the dispersion
of pairwise peculiar velocities which are present at non-linear scales
and also introduces a damping scale in the mapping between real and
redshift space. We marginalize over this parameter, similarly to what \cite{bull_extending_2015} and others have done, and we take as fiducial value $\sigma_{v} = 300$km/s, compatible with the estimates by \cite{de_la_torre_modelling_2012}. We also include the Alcock-Paczynski effect \cite{alcock_evolution_1979}, by which the $k$ modes and the angle cosine $\mu$ perpendicular and parallel to the line-of-sight get distorted by geometrical factors related to the Hubble function and the angular diameter distance \cite{ballinger_measuring_1996, feldman_power_1994}. 

We can then write the
Fisher matrix for the galaxy power spectrum in the following
form \citep{seo_improved_2007, amendola_testing_2012}: 
\begin{equation}
F_{ij}=\frac{V_{\rm survey}}{8\pi^{2}}\int_{-1}^{+1}\mbox{d}\mu\int_{k_{\rm min}}^{k_{\rm max}}\mbox{d}k\,k^{2}\frac{\partial\ln
\Pobs (k,\mu,z)}{\partial\theta_{i}}\frac{\partial\ln
\Pobs (k,\mu,z)}{\partial\theta_{j}}\left[\frac{n(z)\Pobs (k,\mu,z)}{n(z)\Pobs (k,\mu,z)+1}\right]^{2}\,\,.\label{eq:fisher-matrix-gc}
\end{equation}
Here $V_{\rm survey}$
is the volume covered by the survey and contained in a redshift slice
$\Delta z$ and $n(z)$ is the galaxy number density as a function of
redshift.
We will consider the smallest wavenumber $k_{\rm min}$ to be $k_{\rm min}=0.0079\textrm{h/Mpc}$ , while the maximum wavenumber will be 
$k_{\rm max}=0.15$h/Mpc for the linear forecasts and $k_{\rm max}=0.5$h/Mpc for the non-linear forecasts.
In the above formulation of the Galaxy Clustering Fisher matrix, we neglect the correlation among different redshift bins and possible
redshift bin uncertainties as was explored recently in \cite{bailoni_improving_2016}, we will use for our forecasts the 
more standard recipe specified in the Euclid Redbook \cite{laureijs_euclid_2011}.

\begin{table}[h]
\centering{}
\begin{tabular}{|c|cccc|c|}
\hline 
\Tstrut \textbf{Parameter}  & \textbf{Euclid}  & \textbf{DESI-ELG}  & \textbf{SKA1-SUR}
& \textbf{SKA2}  & \textbf{Description}\tabularnewline
\hline 
\Tstrut $A_{\rm survey}$  & 15000 $\mbox{deg}^{2}$  & 14000 $\mbox{deg}^{2}$  & 5000
$\mbox{deg}^{2}$  & 30000 $\mbox{deg}^{2}$  & Survey area in the
sky\tabularnewline
$\sigma_{z}$  & 0.001  & 0.001  & 0.0001  & 0.0001  & Spectroscopic redshift
error\tabularnewline
$\{z_{\rm min},\ z_{\rm max}\}$  & \{0.65, 2.05\}  & \{0.65, 1.65\}  & \{0.05, 0.85\}
& \{0.15, 2.05\}  & Min. and max. limits for redshift bins \tabularnewline
$\Delta z$  & 0.1  & 0.1  & 0.1  & 0.1  & Redshift bin width\tabularnewline
\hline 
\end{tabular}\protect\caption{\label{tab:GC-specifications} Specifications for
the spectroscopic
galaxy redshift surveys used in this work. The number density of tracers $n(z)$ and the galaxy bias $b(z)$, can be found for SKA in \cite{santos_hi_2015} and for DESI in reference \cite{desi_collaboration_desi_2016-1}.}
\end{table}

\subsection{Weak Lensing \label{sub:Fisher-Weak-Lensing}}

Light propagating through the universe is deflected by variations in the Weyl potential $\Phi+\Psi$,
leading to distortions in the images of galaxies. In the regime of small deflections (Weak Lensing) we
can write the power spectrum of the shear field as
\begin{equation}
\label{def_shear}
C_{ij}(\ell)=\frac{9}{4}\int_{0}^{\infty}\mbox{d}z\:\frac{W_{i}(z)W_{j}(z)H^{3}(z)\Omega_{m}^{2}(z)}{(1+z)^{4}}\left[\Sigma(\ell/r(z),z)\right]P_{m}(\ell/r(z)) \, .
\end{equation}
In this expression we used Eqn.\ (\ref{eq:Sigma-def}) to relate the Weyl potential to $\Sigma$ and
to the matter power spectrum $P_m$ and we use the Limber approximation to write down the conversion $k=\ell/r(z)$, where $r(z)$ is the comoving
distance given by
\begin{equation}
r(z) = c\int_0^z \frac{d\tilde{z}}{H(\tilde{z})} \, .
\end{equation}
The indices $i,\:j$ stand for each of the $\mathcal{N}_{bin}$
redshift bins, such that $C_{ij}$ is a matrix of dimensions $\mathcal{N}_{bin}\times\mathcal{N}_{bin}$. The window functions $W_i$ are given by
\begin{equation}
W(z)=\int_{z}^{\infty}\mbox{d}\tilde{z}\left(1-\frac{r(z)}{r(\tilde{z})}\right)n(\tilde{z})
\end{equation}
where the normalized galaxy distribution function is

\begin{equation}
n(z)\propto z^{2}\exp\left(-(z/z_{0})^{3/2}\right) \, . \label{eq:ngal dist}
\end{equation}

Here the median redshift $z_{\rm med}$ and $z_{0}$ are related by $z_{\rm med}=\sqrt{2}z_{0}$.
The Weak Lensing Fisher matrix is then 
given by a sum over all possible correlations at different redshift bins
\citep{tegmark_measuring_1998},
\begin{equation}
F_{\alpha\beta}=f_{\rm sky}\sum_{\ell}^{\lmax}\sum_{i,j,k,l}\frac{(2\ell+1)\Delta\ell}{2}\frac{\partial
C_{ij}(\ell)}{\partial\theta_{\alpha}}\textrm{Cov}_{jk}^{-1}\frac{\partial
C_{kl}(\ell)}{\partial\theta_{\beta}}\textrm{Cov}_{li}^{-1} \, . \label{eq:FisherSum-WL}
\end{equation}
The prefactor $f_{\rm sky}$ is the fraction of the sky covered by the survey. The upper limit of the sum, $\lmax$, is a high-multipole cutoff due to our ignorance of clustering and baryon physics on small
scales, similar to the role of $k_{\rm max}$ in Galaxy Clustering. In this work we choose $\lmax = 1000$ for the linear forecasts and $\lmax = 5000$ for the
non-linear forecasts (this cutoff is not necessarily reached at all multipoles $\ell$, as what matters is the minimum scale between $\ell_{\rm max}$ and $k_{\rm max}$, as we discuss below; see also \cite{casas_fitting_2015}).
In Eqn.\ (\ref{eq:FisherSum-WL}), $\textrm{Cov}_{ij}$ is the corresponding covariance matrix of the shear power spectrum and it has the following form:
\begin{equation}
\textrm{Cov}_{ij}(\ell)=C_{ij}(\ell)+\delta_{ij}\gamma_{\rm int}^{2}n_{i}^{-1}+K_{ij}(\ell)
\end{equation}
where $\gamma_{\rm int}$ is the intrinsic galaxy ellipticity, whose value can be seen in Table \ref{tab:WL-specifications} for each survey. The
shot noise term $n_{i}^{-1}$ is expressed as
\begin{equation}
n_{i}=3600\left(\frac{180}{\pi}\right)^{2}n_{\theta}/\mathcal{N}_{bin}
\end{equation}
with $n_{\theta}$ the total number of galaxies per $\text{arcmin}^2$ and the index $i$ standing for each redshift bin.
Since the redshift bins have been chosen such that each of them contains
the same amount of galaxies (equi-populated redshift bins), the shot noise term is equal for each bin.
The matrix $K_{ij}(\ell)$
is a diagonal ``cutoff'' matrix, discussed for the first time in \cite{casas_fitting_2015} whose entries increase to very high
values at the scale where the power spectrum $P(k)$ has to be cut
to avoid the inclusion of uncertain or unresolved non-linear scales. We choose to add this matrix to have 
further control on the inclusion of non-linearities. Without this matrix, due to the redshift-dependent relation between
$k$ and $\ell$, a very high $\lmax$ would correspond at low redshifts, to a very high $k_{\rm max}$ where we do not longer
trust the accuracy of the non-linear power spectrum. Therefore, the sum in Eqn.\ (\ref{eq:FisherSum-WL}) is limited by the minimum scale imposed
either by $\lmax$ or by $k_{\rm max}$, which is the maximum wavenumber considered in the matter power spectrum $P(k,z)$. As we did for Galaxy Clustering, we use for linear forecasts $k_{\rm max}=0.15$ and for non-linear forecasts $k_{\rm max}=0.5$.

\begin{table}[h]
\centering{}
\begin{tabular}{|c|ccc|c|}
\hline 
\Tstrut \textbf{Parameter}  & \textbf{Euclid}  & \textbf{SKA1}  & \textbf{SKA2}  &
\textbf{Description}\tabularnewline
\hline 
\Tstrut $f_{\rm sky}$  & 0.364  & 0.121  & 0.75 & Fraction of the sky covered\tabularnewline
$\sigma_{z}$  & 0.05  & 0.05  & 0.05  & Photometric redshift
error\tabularnewline
$n_{\theta}$  & 30  & 10  & 2.7  & Number of galaxies per arcmin\tabularnewline 
$\gamma_{\rm int}$  & 0.22  & 0.3  & 0.3  & Intrinsic galaxy ellipticity \tabularnewline
$z_{0}$  & 0.9  & 1.0  & 1.6  & Median redshift over $\sqrt{2}$ \tabularnewline
$\mathcal{N}_{bin}$ & 12 & 12 & 12 & Total number of tomographic redshift bins
\tabularnewline
\hline 
\end{tabular}\protect\caption{\label{tab:WL-specifications} Specifications for
the Weak Lensing
surveys Euclid, SKA1 and SKA2 used in this work. Other needed quantities can be found in the references cited in section \ref{sub:FutureSurveys}.
For all WL surveys we use a redshift range between $z=0.5$ and $z=3.0$, using 6 equi-populated redshift bins.}
\end{table}

\subsection{Covariance and correlation matrix and the Figure of Merit\label{sec:covcorr}}

The covariance matrix is defined for a \emph{d-dimensional }vector
$p$ of random variables as
\begin{equation}
\mathbf{C}=\langle \Delta p \Delta p^{T} \rangle\label{eq:covariance_def}
\end{equation}
with $\Delta p = p - \langle p \rangle$ and the angular brackets $\langle \, \rangle$ representing an expectation value. 
The matrix $\mathbf{C}$, with all its off-diagonal elements set to zero, is called the variance matrix $V$ and contains the
square of the errors $\sigma_{i}$ for each parameter
$p_{i}$
\begin{equation}
\mathbf{V} \equiv diag(\sigma_{1}^{2},...,\sigma_{d}^{2})\,\,\,.
\end{equation}
The Fisher matrix $\mathbf{F}$ is the inverse of the covariance matrix
\begin{equation}
\mathbf F= \mathbf C^{-1}\,\,\,.
\end{equation}
The correlation matrix $\mathbf P$ is obtained from the covariance matrix
$\mathbf C$, in the following way
\begin{equation}
P_{ij}=\frac{C_{ij}}{\sqrt{C_{ii}C_{jj}}} \, . \label{eq:correlation_def}
\end{equation}
If the covariance matrix is non-diagonal, then there are correlations
among some elements of $p$. We can observe this also by plotting
the marginalized error ellipsoidal contours. The orientation of the ellipses can
tell us if two variables $p_{i}$ and $p_{j}$ are correlated ($P_{ij}>0$),
corresponding to ellipses with 45 degree orientation to the right of the vertical
line or if they are anti-correlated ($P_{ij}<0$), corresponding to ellipses oriented 45
degrees to the left of the vertical line.

To summarize the information contained in the Fisher/covariance matrices we can
define a Figure of Merit (FoM). Here we choose the logarithm of the determinant, while another possibility would
be the Kullback-Leibler divergence, which is a measure of the information entropy gain, see Appendix \ref{sec:KL}.

The square-root $\sqrt{\det(\mathbf{C})}$ of the determinant of the covariance matrix is proportional to the volume of the error
ellipsoid. We can see this if we rotate our coordinate system so that the 
covariance matrix is diagonal, $\mathbf C = {\rm diag}(\sigma_1^2, \sigma_2^2, \ldots \sigma_{d}^{2})$, then $\det(\mathbf C) = \prod_i \sigma_i^2$ and $(1/2)\ln(\det(\mathbf C)) = \ln \prod_i\sigma_i$ would indeed represent the logarithm of an error volume. Thus, the smaller the determinant (and therefore also $\ln(\det(\mathbf{C}))$), the smaller is the ellipse and the stronger are the constraints on the parameters. We define
\begin{equation}\label{eq:FoM}
\mathrm{FoM} = -\frac{1}{2} \ln(\det(\mathbf{C})) \, ,
\end{equation}
with a negative sign in front such that stronger constraints lead to a higher Figure of Merit. In the following, the value of the FoM reported in all tables will be obtained including only the dark energy parameters (i.e.\ the $(\mu_i,\eta_i)$ sub-block for the binned case and the $(\mu,\eta)$ sub-block in the smooth functions case), after marginalizing over
all other parameters. 
The FoM allows us to compare not only the constraining power of different probes but also of the different experiments. 
As the absolute value depends on the details of the setup, we define the relative figure of merit between probe $a$ and probe $b$: $\mathrm{FoM}_{a,b} = -1/2 \ln(\det(\mathbf{C_a})/\det(\mathbf{C_b})) = \mathrm{FoM}_{a}-\mathrm{FoM}_{b}$ and we fix our reference case (probe $b$), for each parametrization, to the Galaxy Clustering observation using linear power spectra with the Euclid survey (labeled as `Euclid Redbook GC-lin' in all figures and tables). The FoM has units of `nits', since we are using the natural logarithm. These are similar to `bits', but `nits' are counted in base $e$ instead of base 2.

An analogous construction allows us to study quantitatively the strength of the correlations encoded by the correlation matrix $\mathbf P$.
We define the `Figure of Correlation' (FoC) as:
\begin{equation}\label{eq:FoC}
\mathrm{FoC} = -\frac{1}{2} \ln(\det(\mathbf{P})) \, .
\end{equation}
If the parameters are independent, i.e.\ fully decorrelated, then $\mathbf{P}$ is just the unit matrix and $\ln(\det(\mathbf{P}))=0$. 
Off-diagonal correlations will decrease the logarithm of the determinant, therefore making the FoC larger. 
From a geometrical point of view, the determinant expresses a volume spanned by the vector of (normalized) variables. 
If these variables are independent, the volume would be maximal and equal to one, while if they are strongly linearly dependent, the volume would be squeezed 
and in the limit where all variables are effectively the same, the volume would be reduced to zero. 
Hence, a more positive FoC indicates a stronger correlation of the parameters.

\subsection{CMB \planck\ priors}
\label{sub:Fisher-Planck}
Alongside the information brought by LSS probes,
we also include CMB priors on the parameterizations considered. In
order to obtain these, we analyze the binned and parameterized approaches
described in Section \ref{sec:Parameterizing-Modified-Gravity} with
the {\it Planck}+BSH combination of CMB and background (BAO+SN-Ia+$H_{0}$)
datasets discussed in the \planck\ Dark Energy and Modified Gravity paper
\cite{planck_collaboration_planck_2016}.
We use a Markov Chain Monte Carlo (MCMC) approach, using the publicly
available code \texttt{COSMOMC} \cite{lewis_cosmological_2002,lewis_efficient_2013},
interfaced with our modified version of \texttt{MGCAMB}. The MCMC
chains sample the parameter vector $\Theta$ which contains the standard
cosmological parameters
$\{\omega_{b}\equiv\Omega_{b}h^{2},\,\omega_{c}\equiv\Omega_{c}h^{2},\,\theta_{\rm MC},\,\tau,n_{s},\ln{10^{10}A_{s}}\}$
to which we add the $E_{ij}$ parameters when we parameterize the time
evolution of $\mu$ and $\eta$ with continuous functions of the scale factor, 
and the $\mu_{i},\ \eta_{i}$ parameters in the binned
case. On top of these, also the 17 nuisance parameters of the \planck\
analysis are included. From the MCMC analysis of the \planck\ likelihood
we obtain a covariance matrix in terms of the parameters $\Theta$.
We marginalize over the nuisance parameters and over the optical
depth $\tau$ since this parameter does not enter into
the physics of large scale structure formation.

$\theta$ is usually the ratio of sound horizon
to the angular diameter distance at the time of decoupling. Since
calculating the decoupling time $z_{\rm CMB}$ is relatively time consuming,
as it involves the minimization of the optical transfer function,
\texttt{COSMOMC} uses instead an approximate quantity $\theta_{\rm MC}$
based on the following fitting formula from \cite{hu_small_1996}
\begin{align}
z_{\rm CMB} & =1048\times(1+0.00124\omega_{b}^{-0.738})\nonumber \\
 &
\times\left(1+0.0783\omega_{b}^{-0.238}/(1+39.5\omega_{b}^{0.763})\right.\nonumber
\\
 & \times\left.(\omega_{d}+\omega_{b})^{0.560/(1+21.1\omega_{b}^{1.81})}\right)
\end{align}
where $\omega_{d}\equiv(\Omega_{c}+\Omega_{\nu})h^{2}$.
The sound horizon is defined as 
\begin{equation}
r_{s}(z_{\rm CMB})=cH_{0}^{-1}\int_{z_{\rm CMB}}^{\infty}\mbox{d}z\frac{c_{s}}{E(z)}
\end{equation}
where the sound speed is $c_{s}=1/\sqrt{3(1+\overline{R}_{b}a)}$with
the baryon-radiation ratio being $\overline{R}_{b}a=3\rho_{b}/4\rho_{\gamma}$.
$\overline{R}_{b}=31500\Omega_{b}h^{2}(T_{\rm CMB}/2.7\mbox{K})^{-4}$.
However, \texttt{CAMB }approximates it as
$\overline{R}_{b}a=30000a\Omega_{b}h^{2}$.

Therefore we first marginalize the covariance matrix over the nuisance
parameters and the parameter $\tau$, which cannot be constrained
by LSS observations. Then, we invert the resulting matrix, to obtain a \planck\
prior Fisher matrix and then use a Jacobian to convert between the
MCMC parameter basis $\Theta_{i}$ and the GC-WL parameter basis $\theta_{i}$.
We use the formulas above for the sound horizon $r_{s}$ and the angular
diameter distance $d_{A}$ to calculate the derivatives of $\theta_{\rm MC}$
with respect to the parameters of interest. Our Jacobian is then simply (see Appendix \ref{sec:appjac} for details)
\begin{equation}
J_{ij}=\frac{\partial\Theta_{i}}{\partial\theta_{j}} \, .
\end{equation}

\section{\label{sec:Results:-Redshift-Binned}Results: Euclid forecasts for redshift binned parameters}
In this section we analyze the Modified Gravity functions $\mu(a)$ and $\eta(a)$,
described in Section \ref{sec:Parameterizing-Modified-Gravity}, when they are allowed
to vary freely in five redshift bins.

For this purpose, we calculate a Fisher matrix of fifteen parameters:
five for the standard $\lcdm$ parameters $\{\Omega_{m},\Omega_{b},h,\ln10^{10}A_{s},n_{s}\}$,
five for $\mu$ (one
for each bin amplitude $\mu_{i}$) and five for $\eta$ (one for each
bin amplitude $\eta_{i}$), corresponding to the 5 redshift bins z=\{0-0.5, 0.5-1.0, 1.0-1.5, 1.5-2.0, 2.0-2.5\}. 
The fiducial values for all fifteen parameters
were calculated running a Markov-Chain-Monte-Carlo with \planck\ likelihood
data and can be found in Table \ref{tab:fiducial-MG-AllCases}.

We first show the constraints on our 15 parameters for Galaxy Clustering (GC) forecasts in subsection \ref{sub:GC-Correlations},
while in subsection \ref{sub:Weak-Lensing} we report results for Weak Lensing (WL). In subsection \ref{sub:Combined-GC-WL-Planck-Binned}, we
comment on the combination of forecasts for GC+WL together with \planck\ data.
All forecasts are performed using Euclid Redbook specifications. 
Other surveys 
will be considered for the other two time parameterizations in section \ref{subsub: other-surveys-late-time} and 
\ref{subsub: other-surveys-early-time}. 
For each case, we show the correlation matrix obtained from the covariance matrix and argue that 
the redshift-binned
parameters show a strong correlation, therefore we illustrate the decorrelation
procedure for the covariance matrix in subsection \ref{sub:Decorrelation-of-covariance}
where we also include combined GC+WL and GC+WL+{\it Planck} cases.

\subsection{\label{sub:GC-Correlations}Euclid Galaxy Clustering Survey}

For the Galaxy Clustering survey, we give results for two cases, one
using only linear power spectra up to a maximum wavevector of $k_{\rm max}=0.15$
h/Mpc and another one using non-linear power spectra up to $k_{\rm max}=0.5$
h/Mpc, as obtained by using the HS parameterization of
Eqn.\ (\ref{eq:PHSDefinition}).
For the redshift-binned case, we will report forecasts only for a Euclid survey, 
using Euclid Redbook specifications which
are detailed in section \ref{sub:Fisher-Galaxy-Clustering}.

\begin{figure}[H]
\centering
\includegraphics[width=0.4\textwidth]{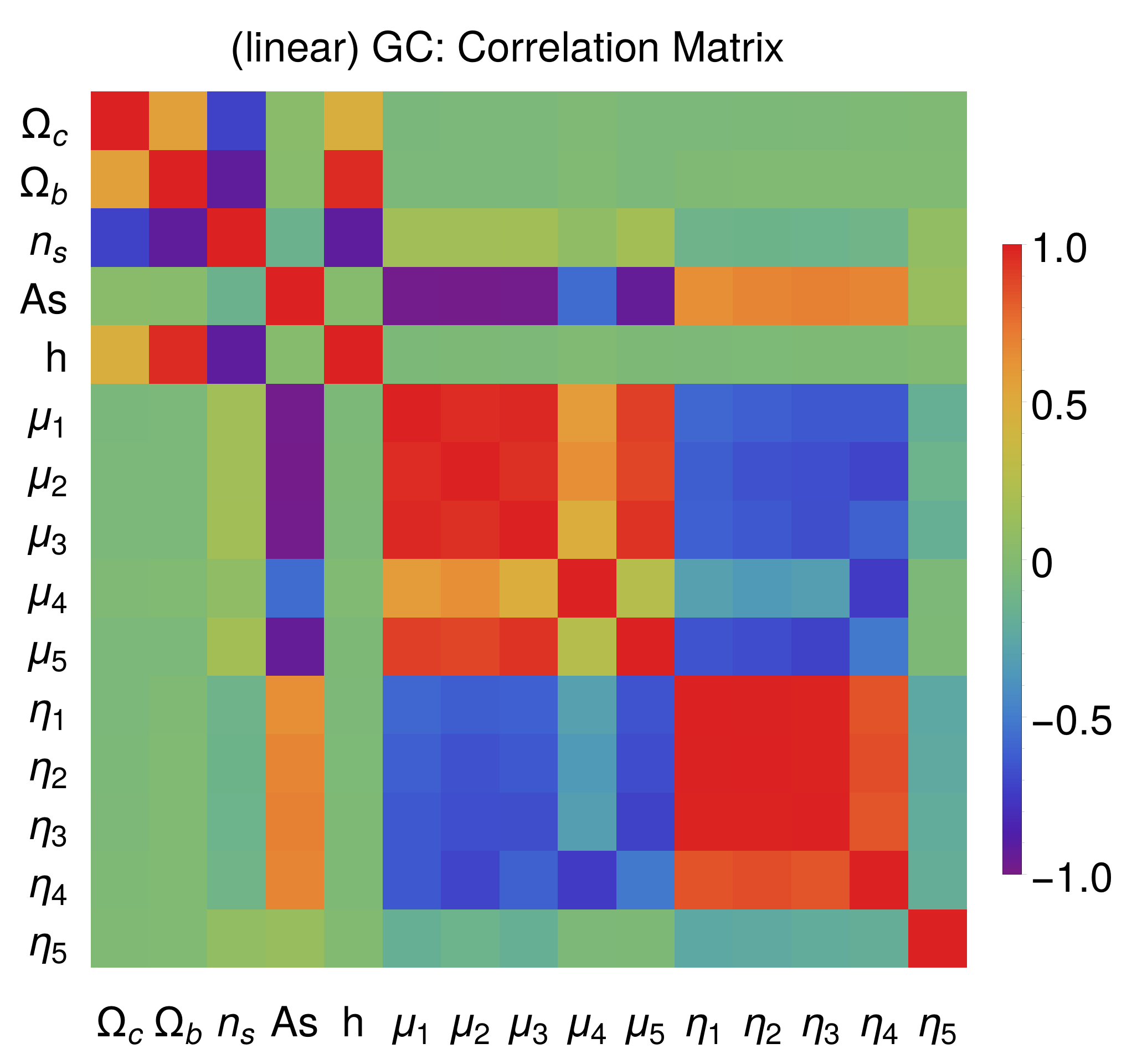}
\includegraphics[width=0.4\textwidth]{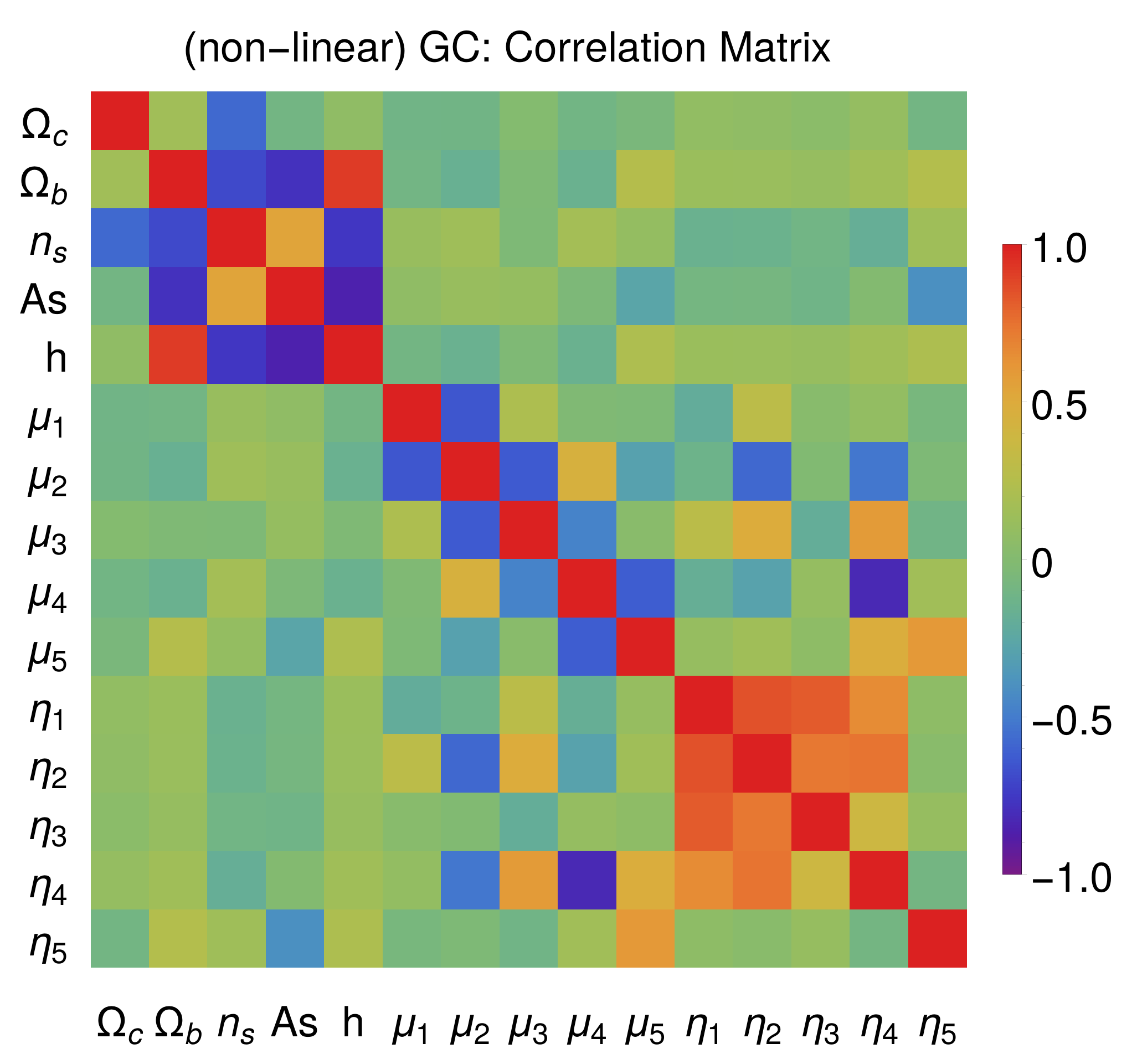}
\caption{\label{fig:GCcorr}
Correlation matrix $\mathbf P$ defined in (\ref{eq:correlation_def}) obtained from the covariance matrix in the MG-binning case, for a Galaxy Clustering Fisher forecast using Euclid
Redbook specifications. \textbf{Left panel:}
Linear forecasts. Here there are strong positive correlations among the $\mu_i$ and $\eta_i$ parameters and anti-correlations between
 $\ln10^{10}A_{s}$  and the $\mu_i$ parameters, as well as between $\mu_i$ and $\eta_i$. The FoC in this case is $\approx 65$. (see Eqn.\ (\ref{eq:FoC}) for its definition).
\textbf{Right panel: } Non-linear 
forecasts using the HS prescription. Interestingly, the anti-correlations between  $\ln10^{10}A_{s}$  and $\mu_i$ 
have disappeared, as well as the correlations among the  $\mu_i$ parameters. The FoC is in this case   $\approx 32$, meaning that the variables are much less correlated than in the linear case.
This is due to the fact that taking into account non-linear structure formation breaks degeneracies between the primordial amplitude parameter and the modifications
to the Poisson equation.}
\end{figure}

We calculate the Fisher matrix for the 15 parameters
$\theta=\{\Omega_{m},\Omega_{b},h,\ln10^{10}A_{s},n_{s},\mu_{i},\eta_{i}\}$
where $\eta_{i}$ and $\mu_{i}$ represent ten independent parameters, one for each function
at each of the 5 redshift bins corresponding to the redshifts z=\{0-0.5, 0.5-1.0, 1.0-1.5, 1.5-2.0, 2.0-2.5\}. As a standard procedure, we marginalize over the unknown bias parameters.
From the covariance matrix, defined
previously in Eqn.\ (\ref{eq:covariance_def}), we obtain the correlation
matrix $P_{ij}$ defined in Eqn.\ (\ref{eq:correlation_def}) for the
set of parameters $\theta_{i}$. In Figure \ref{fig:GCcorr} we show
the matrix $P_{ij}$ in the linear (left panel) and non-linear-HS
(right panel) cases. Redder (bluer) colors signal stronger correlations
(anti-correlations). 

A covariance matrix that contains strong correlations among parameter A and B, means that the 
experimental or observational setting has difficulties distinguishing between A and B for the assumed theoretical model, i.e.\ this represents a parameter degeneracy.
Therefore if for example parameter A is poorly constrained, then parameter B will be badly constrained as well.
The appearance of correlations among parameters is linked to the non-diagonal elements of the covariance matrix. Subsequently, this means that
the fully marginalized errors on a single parameter, will be larger if there are strong correlations and will be smaller (closer to the value of the
fully maximized errors) if the correlations are negligible.

In the linear case, $\mu_{i}$ and $\eta_{i}$ parameters show correlations
among each other, while the primordial
 amplitude parameter $\ln10^{10}A_{s}$ exhibits a strong anti-correlation with all the $\mu_{i}$.
This can be explained considering that a larger growth of
structures in linear theory can also be mimicked with a larger initial
amplitude of density fluctuations.

Interestingly, including non-linear scales in the analysis (right panel of Fig.\ \ref{fig:GCcorr})
leads to a strong suppression of the correlations among the $\mu_i$.
Also the correlation between these and $\ln10^{10}A_{s}$ is suppressed
as a change in the initial amplitude of the power spectrum is not able
to compensate for a modified Poisson equation when non-linear evolution 
is considered.

As discussed in Section \ref{sec:covcorr}, we can also express the difference between the correlation matrix of the linear forecast and the non-linear forecast in a more quantitative way, by computing the
determinant of the correlation matrix, or equivalently the FoC (\ref{eq:FoC}). 
If the correlations were negligible, this determinant would be equal to one (and therefore its FoC would be 0), while if the correlations were strong, the determinant
would be closer to zero with a corresponding large positive value of the FoC.
For the linear forecast, the FoC is about $62$, while for the non-linear forecast, it is much smaller at approximately $35$. 
In Table \ref{tab:errors-all-MGBin3} we show the 1$\sigma$ constraints obtained on $\ln{(10^{10}A_s)}$
and on the $\mu_i$ and $\eta_i$ parameters, both in the linear and 
non-linear cases for a Euclid Redbook GC survey (top rows).
While linear GC alone ($k_{\rm max}$ = 0.15 h/Mpc) is not very constraining in any bin, the inclusion of non-linear scales ($k_{\rm max} $= 0.5 h/Mpc) drastically reduces errors on the $\mu_{i}$ parameters: the first three bins in $\mu_i$ (0. < z < 1.5 ) are the best constrained, to less than $10\%$, with the corresponding $\eta_i$ constrained at 20$\%$ by non-linear GC alone. This is also visible in the FoM which increases by 19 nits (`natural units', similar to bits but using base $e$ instead of base 2), nearly 4 nits per redshift bin on average, when including the non-linear scales. The fact that the error on $\ln10^{10}A_{s}$ improves from 90\% to 0.68\% shows that the decorrelation induced by the non-linearities breaks the degeneracy with the amplitude and therefore improves considerably the determination of cosmological parameters. This shows that it is important to include non-linear scales in GC surveys (and not only in Weak Lensing ones, which is usually more expected and will be shown in the next subsection).

\begin{table}[htbp]
\centering{}%
\begin{tabular}{|l|c|c|c|c|c|c|c|c|c|c|c||c|}
\hline 
\Tstrut \textbf{Euclid} (Redbook)  & $\ell \mathcal{A}_{s}$  & $\mu_{1}$  & $\mu_{2}$  & $\mu_{3}$  & $\mu_{4}$  & $\mu_{5}$  & $\eta_{1}$  
& $\eta_{2}$  & $\eta_{3}$  & $\eta_{4}$ & $\eta_{5}$ & MG FoM
\tabularnewline
\hline 
\Tstrut Fiducial  & 3.057  & 1.108  & 1.027  & 0.973  & 0.952  & 0.962  & 1.135  & 1.160  & 1.219  & 1.226 & 1.164 &  relative\tabularnewline
\hline 
\Tstrut \textbf{GC (lin)}  \Tstrut & 160\% & 119\% & 159\% & 183\% & 450\% & 1470\% & 509\% & 570\% & 586\% & 728\% & 3390\%  & 0 \tabularnewline
\Tstrut \textbf{GC (nl-HS)} \Tstrut & 0.8\% & 7.0\% & 6.7\% & 10.9\% & 27.4\% & 41.1\% & 20\% & 24.3\% & 19.9\% & 38.2\% & 930\% & 19  \tabularnewline
\hline
\hline
\Tstrut \textbf{WL (lin)}  & 640\% & 165\% & 2210\% & 4150\% & 13100\% & 22500\% & 2840\% & 3140\% & 8020\% & 29300\% & 39000\%  
& -27\tabularnewline
\Tstrut \textbf{WL (nl-HS)}  & 7.3\% & 188\% & 255\% & 419\% & 222\% & 206\% & 330\% & 488\% & 775\% & 8300\% & 9380\% 
& -10  \tabularnewline
\hline
\hline
\Tstrut \textbf{GC+WL (lin)} & 11.3\% & 5.8\% & 10\% & 19.2\% & 282\% & 469\% & 7.9\% & 9.6\% & 16.1\% & 276\% & 2520\%
& 12  \tabularnewline
\Tstrut \textbf{GC+WL+{\it Planck} (lin)} & 1.1\% & 3.4\% & 4.8\% & 7.8\% & 9.3\% & 13.1\% & 6.2\% & 7.7\% & 9.1\% & 12.7\% & 23.6\%
& 27 \tabularnewline
\hline
\hline
\Tstrut \textbf{GC+WL (nl-HS)} & 0.8\% & 2.2\% & 3.3\% & 8.2\% & 24.8\% & 34.1\% & 3.6\% & 5.1\% & 8.1\% & 25.4\% & 812\%
& 24 \tabularnewline
\Tstrut \textbf{GC+WL+{\it Planck}} &  &   &  &   &  &   &   &  &   &  & & 
\tabularnewline \textbf{(nl-HS)}  
& 0.3\% & 1.8\% & 2.5\% & 5.8\% & 7.8\% & 10.3\% & 3.2\% & 4.1\% & 5.9\% & 9.6\% & 19.5\% 
& 33 \tabularnewline
\Tstrut \textbf{GC+WL+{\it Planck}} &  &   &  &   &  &   &   &  &   &  & & 
\tabularnewline \textbf{(nl-Halofit)}  
& 0.4\% & 2.0\% & 2.4\% & 5.1\% & 7.4\% & 10.2\% & 3.5\% & 4.1\% & 5.8\% & 9.2\% & 18.9\% 
& 33 \tabularnewline
\hline
\end{tabular}\protect\caption{\label{tab:errors-all-MGBin3}
1$\sigma$
fully marginalized errors (as a percentage of the corresponding fiducial) on cosmological parameters for Euclid (Redbook) Galaxy Clustering and Weak Lensing surveys, alone and combining the two probes. We compare forecasts using linear spectra (lin) and forecasts using the nonlinear HS prescription (nl-HS). In Galaxy Clustering, the cutoff is set to $k_{\rm max}=0.15$ h/Mpc in the linear case and $k_{\rm max}=0.5$ h/Mpc in the non-linear case.
For WL, the maximum cutoff in the linear case is at $\ell_{\rm max} = 1000$, 
while in the nl-HS case it is $\ell_{\rm max} = 5000$.
At the bottom, we add on top a \planck\ prior (see section \ref{sub:Fisher-Planck}).
For comparison, we also show in the last row the combined GC+WL+{\it Planck}, using just Halofit power spectra. 
The last column indicates the relative Figure of Merit ($\text{FoM}_{a,b}$) of the MG parameters in nits (`natural units', i.e.\ using the natural logarithm), with respect to our reference GC linear case, see (\ref{eq:FoM}) and surrounding text. A larger FoM indicates a more constraining probe. We notice a considerable improvement, in both GC and WL, when non-linearities are included.
The combination GC+WL in the linear case 
constrains the MG parameters in the first two bins (z < 1.0 ) to less than 10$\%$ and including \planck\ priors allows to access higher redshifts with the same accuracy. A significant improvement in the constraints is obtained when adding the non-linear regime, in agreement with the observed reduction in correlation seen in Figs.\ \ref{fig:GCcorr} and \ref{fig:WLcorr}. This is especially well exemplified by the error on $\ell \mathcal{A}_s \equiv \ln(10^{10} A_{s})$, which reduces from 160\%  to 0.82\%, from the linear to the non-linear forecast in the GC case
and from 640\%  to 7.3\% in the WL case. 
Finally, we note that since we are showing errors on $\mu$ and $\eta$, WL seems to be unfairly poor at constraining parameters. However, when converting this errors into errors on $\Sigma$, which is directly measured by WL, the constrains on $\Sigma_{1,2,3}$ are slightly better, of the order of 40\% for WL(nl-HS) as can be guess from the degeneracy directions shown in Fig.\ \ref{fig:DE+Planck-ellipses-mu-sig-eta}. The FoM itself is nearly unaffected by the choice of $\{\mu,\eta\}$ vs $\{\mu,\Sigma\}$ as it is rotationally invariant.
}
\end{table}

\subsection{\label{sub:Weak-Lensing}Euclid Weak Lensing Survey}

\begin{figure}[htbp]
\centering
\includegraphics[width=0.4\textwidth]{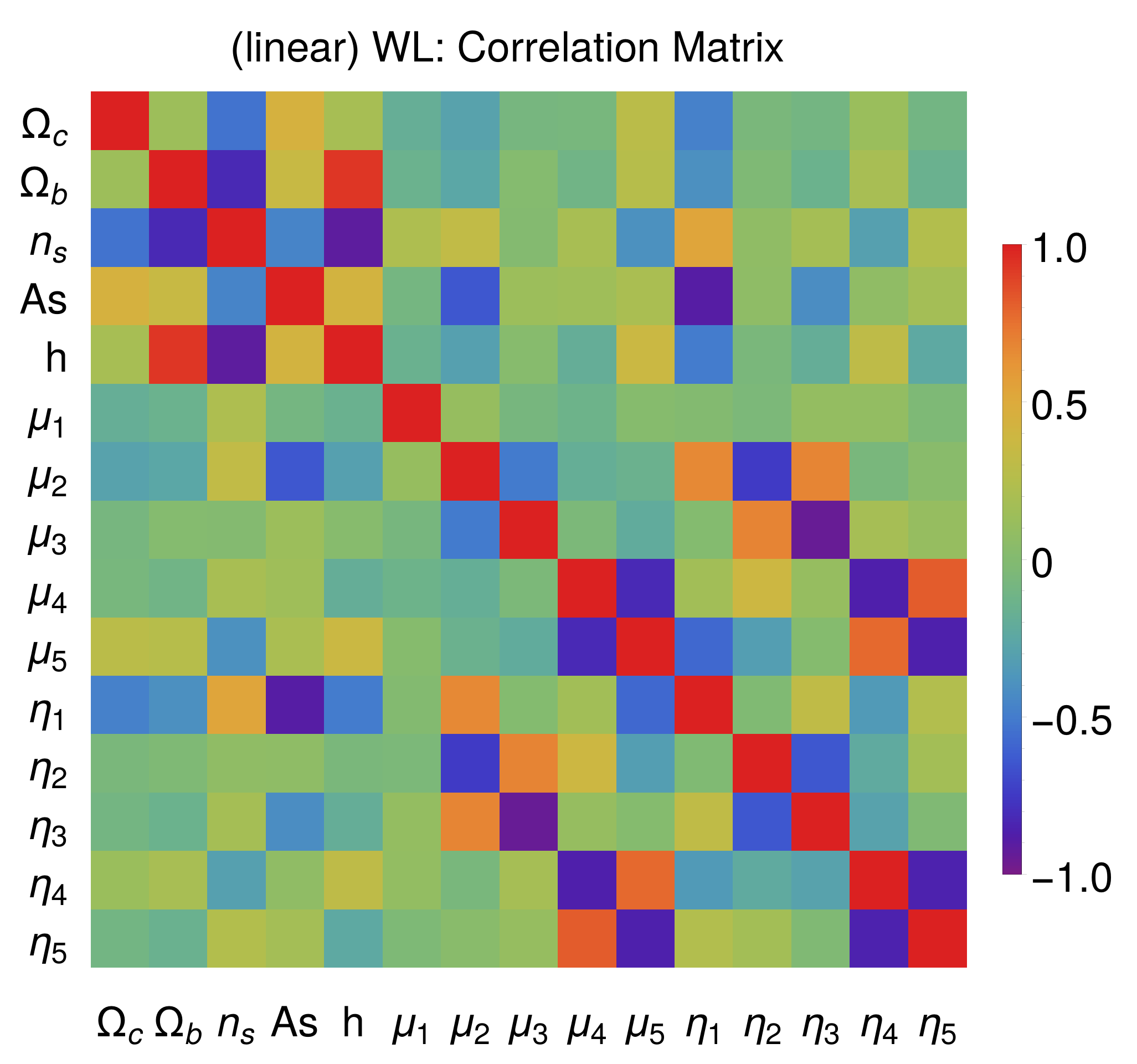}
\includegraphics[width=0.4\textwidth]{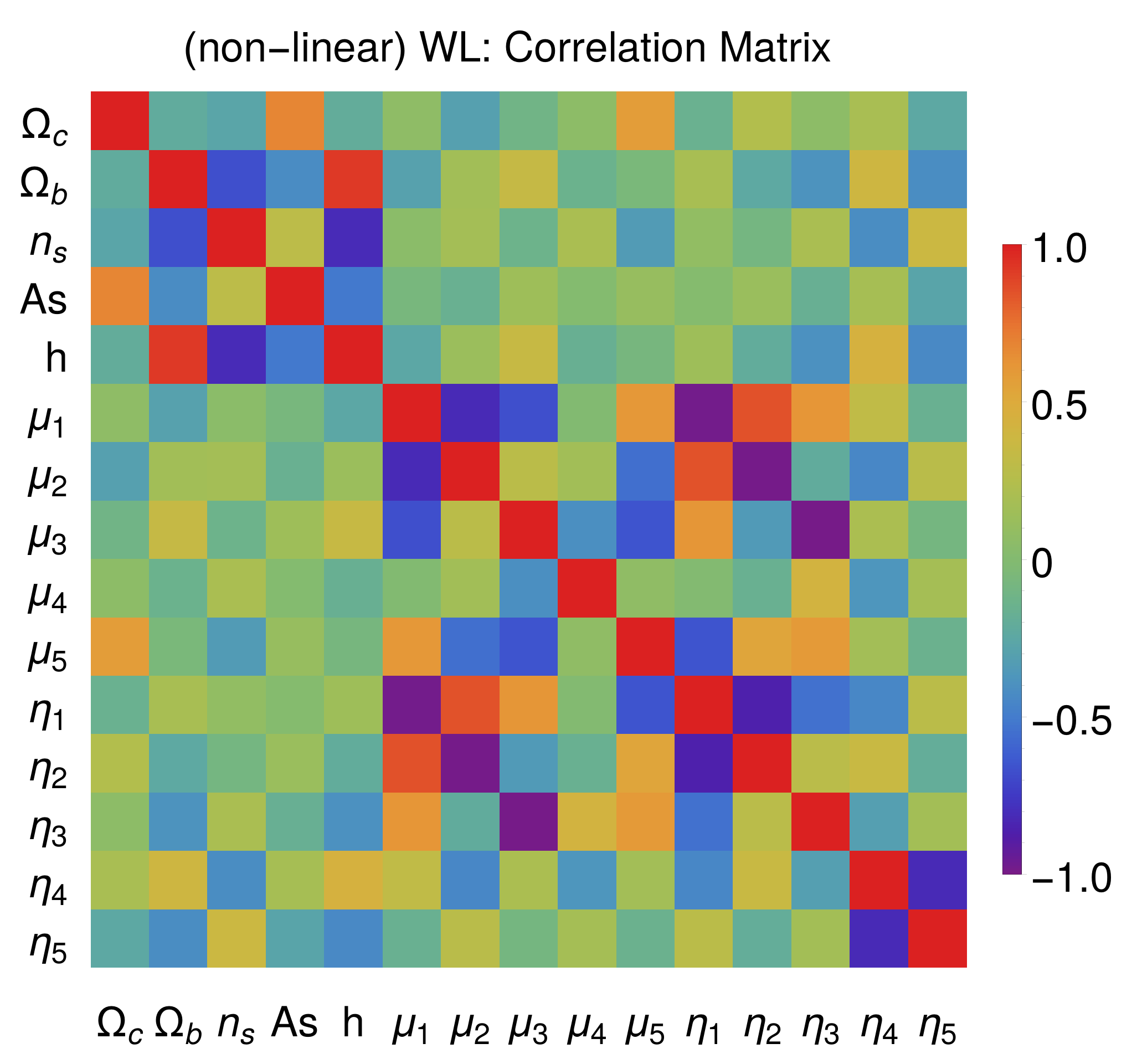}
\caption{\label{fig:WLcorr}
Correlation matrix obtained from the covariance matrix in the MG-binning case, for a Weak Lensing Fisher Matrix forecast using Euclid
Redbook specifications. \textbf{Left panel:} linear forecasts. Strong anti-correlations are present between the $\mu_i$ and the $\eta_i$ parameters for the same value of i in {2,3,4,5}. The amplitude $\ln(10^{10}A_{s})$ parameter is mostly uncorrelated except with $\eta_1$. 
The FoC (\ref{eq:FoC}) in this case is approximately $69$.
\textbf{Right panel:} non-linear 
forecasts using the HS prescription. Here the same trend as in the linear case is present, with just subtle changes. 
The FoC in this case is about $32$, meaning that the variables are indeed less correlated than in the linear case. 
The parameter $\ln(10^{10}A_{s})$ is effectively not correlated to other parameters, 
and the anti-correlation of $\mu_{i}$ and the $\eta_{i}$ for the same value of the index i is present in the first three bins. 
The anti-correlation between these two sets of parameters
is expected, since WL is sensitive to the Weyl potential $\Sigma$, which is a product of $\mu$ and $\eta$.}
\end{figure}

In the case of Weak Lensing, the linear forecast is performed with
linear power spectra up to a maximum multipole $\ell_{\rm max}=1000$,
while the non-linear forecast is performed with non-linear spectra
up to a maximum multipole of $\ell_{\rm max}=5000$, as explained in
section \ref{sec:Fisher-Matrix-method}. Since we limit 
our power spectrum to a maximum in $k$-space, as explained
in section \ref{sub:Fisher-Weak-Lensing}, these multipole values are not reached at every redshift.
Like for GC, also for WL it is very important
to include information from the non-linear power spectrum, since in
that range will lie most of the constraining power of next-generation
surveys like Euclid. In Figure \ref{fig:WLcorr} we show the correlation
matrices for the linear (left panel) and non-linear (right panel)
Fisher Matrix forecasts. In this case, as opposed to the GC correlation matrices, it is not visually 
clear which case is more correlated than the other.
At a closer look, in the linear case we can observe strong anti-correlations between the $\mu_{j}$ and $\eta_{j}$ parameters for 
$j={2,3,4,5}$ and an anti-correlation between $\eta_{1}$ and the primordial amplitude $\ln10^{10}A_{s}$.
In the non-linear case, the primordial amplitude parameter is effectively decorrelated from the Modified Gravity parameters,
and the anti correlation between $\mu_{j}$ and the $\eta_{j}$ affects the first three bins (effectively increasing degeneracies in the first bin).
The anti-correlation between these two sets of parameters
is expected, since WL is sensitive to $\Sigma$, which is a product of $\eta$ and $\mu$, c.f.\ Eqn.\ (\ref{eq:SigmaofMuEta}).
The decrease of correlation when going from the linear to the non-linear case is confirmed, also quantitatively, by the FoC: the one for the linear case is of $\approx 69$, larger than the one for the non-linear case $\approx 32$. Once again, the inclusion of non-linear scales breaks degeneracies, especially among the linear amplitude of the power spectrum and the MG parameters. 

Table \ref{tab:errors-all-MGBin3} shows the corresponding 1$\sigma$ marginalized errors on $\ln{(10^{10}A_s)}$
and on the $\mu_i$ and $\eta_i$, both in the linear and 
non-linear cases for a Euclid Redbook WL survey. 
As in the case of GC, linear WL cannot constrain alone any of the amplitudes of 
the Modified Gravity parameters $\mu_i$ and $\eta_i$ for any redshift bin. 
Being able to include non-linear scales improves constraints on the amplitude $\ln(10^{10}A_{s})$ by a factor 100. 
The 1$\sigma$ errors on $\mu_{i}$ and $\eta_{i}$ 
improve up to one order of magnitude, with the FoM increasing by 17 nits, although remaining quite unconstraining for Modified Gravity parameters in all redshift bins.
Notice, however, that the 1$\sigma$ error on $\mu_1$ from WL in the linear case is slightly smaller than in the non-linear case.
This can be attributed to the fact that in the linear case, $\mu_1$ is uncorrelated to any other parameter, as 
shown in Figure \ref{fig:WLcorr} and on that specific bin (0 < z < 0.5) non-linearities don't seem to improve the constraints on this parameter.

\subsection{Combining Euclid Galaxy Clustering and Weak Lensing, with \planck\ data \label{sub:Combined-GC-WL-Planck-Binned}}

The combination of Galaxy Clustering and Weak Lensing is expected to be very powerful for 
Modified Gravity parameters as they measure two different combinations of $\mu$ and $\eta$, 
thus breaking their degeneracy as illustrated in Fig. \ref{fig:DE+Planck-ellipses-mu-sig-eta}. 
This is shown in Table \ref{tab:errors-all-MGBin3}, where the sensitivity drastically increases with respect to the two separate probes, 
especially in the low redshifts bins $(0.<z<1.5)$, where the lensing signal is dominant. 
Adding non-linearities further doubles the FoM. The \planck\ 
data constrains mostly the standard $\lcdm$ parameters and has only a limited ability to constrain the MG sector. 
However, the additional information breaks parameter degeneracies and in this way significantly decreases
the uncertainties on all parameters, so that the linear GC+WL+{\it Planck} is comparable to the non-linear GC+WL. 
Also, quantitatively, the correlation among parameters is reduced by combining GC+WL with \planck\ data. The FoC in this case is of $\approx 22$.
We also see in Table \ref{tab:errors-all-MGBin3} that the differences between the non-linear prescription
adopted here (nl-HS) and a straightforward application of Halofit to the MG case (nl-Halofit) is not
very large. We will investigate further the impact of the parameters used in the non-linear prescription in section \ref{sub:Testing-the-effect-of-Zhao}.

\subsection{\label{sub:Decorrelation-of-covariance}Decorrelation of covariance
matrices and the Zero-phase Component Analysis}

In the previous subsections we highlighted how the MG parameters and the 
amplitude of the primordial power spectrum exhibit significant correlations and showed how including non-linearities helps to decorrelate them. Even without including non-linearities, however, it is interesting to investigate how we can completely decorrelate the parameters, identifying in this way those parameter combinations which are best constrained by data.

Given again a \emph{d-dimensional} vector
$p$ of random variables (our originally correlated parameters), we can calculate its covariance matrix $\mathbf C$ defined in Eqn.\ (\ref{eq:covariance_def}). The process of decorrelation is the process of making the matrix $\mathbf C$ 
a diagonal matrix.

Let us define some important identities. The covariance
matrix can be decomposed in its eigenvalues (the elements of a diagonal matrix $\Lambda$) and eigenvectors
(the rows of an orthogonal matrix $U$).
\begin{equation}
\mathbf C=U\Lambda U^{T} \quad \Leftrightarrow \quad 
\mathbf F=U\Lambda^{-1}U^{T} \,\,\, ,
\label{eq:eigensystemofC}
\end{equation}
where $\mathbf F$ is the Fisher Matrix.

It is possible to show that applying a transformation matrix $W$ to the $p$ parameter vector, thus obtaining a new vector of variables $q=Wp$, the covariance matrix of the transformed vector $q$ is whitened (i.e. it is the identity matrix, and whitening is defined as the process
of converting the covariance matrix into an identity matrix)
\begin{align}
\mathbf {\tilde{C}} & =W\langle\Delta\hat{p}\Delta\hat{p}^{T}\rangle W^{T} =\langle\Delta\hat{q}\Delta\hat{q}^{T}\rangle \label{eq:Cwhitened}\\
 & = \mathbb{1} \,\,\, . \nonumber
\end{align} 
This means that the transformed $q$ parameters are decorrelated, since their correlation matrix is diagonal. The choice of $W$ 
is not unique as several  possibilities
exist; we focus on a particular choice in the rest of the paper referred to as Zero-phase Component Analysis (ZCA, first 
introduced by \cite{Bell19973327} in the context of image processing), but we show
two other possible choices and their effect on the analysis in Appendix
\ref{sec:appdec}.

Zero-phase Component Analysis (sometimes also called Mahalanobis transformation \cite{kessy_optimal_2015}) is a specific choice of decorrelation method that 
minimizes the squared norm of the difference between the $q_i$
and the $p_i$ vector $\|\vec{p}-\vec{q}\|$, under the constraint that the vector $\vec{q}$ should be uncorrelated \cite{kessy_optimal_2015}. 
In this way the uncorrelated variables $q$ will be as close 
as possible to the original variables $p$ in a least squares sense. 
This is achieved by using the $W$ matrix:
\begin{equation}
W \equiv F^{1/2} \,\,\, .
\end{equation}
Then in this case, the covariance matrix is whitened, by following Eqn.\ (\ref{eq:Cwhitened}):
\begin{equation}
\tilde{C} =F^{1/2}F^{-1}F^{1/2}=\mathbb{1} \,\,\, .
\end{equation}
In our case, since we do not want to whiten, but just decorrelate the covariance matrix,
we renormalize the $W$ matrix by multiplying with a diagonal matrix $N_{jj}\equiv\sum_j(W_{ij}^{-2})$, such that the sum of the square of the 
elements on each row of the new weighting matrix 
$\tilde{W} \equiv N W$, is equal to unity; therefore the final
transformed covariance is still diagonal but is not the identity matrix: 
\begin{equation}
\tilde{C}=\tilde{W}C\tilde{W}=N^{2} \mathbb{1} \label{eq:Ctilde-decor}
\end{equation}
and at the same time we ensure that 
the vector of new variables $q_i$ will have the same norm as the old vector of variables $p_i$.

\subsubsection{ZCA for Galaxy Clustering \label{subsub:ZCA-GC}}

From Fig.\ \ref{fig:GCcorr} we can see that the correlations are present in sub-blocks, one for 
the standard $\lcdm$ parameters and another one for the Modified Gravity parameters. The exception lies in the linear case
where $\lAs \equiv \ln{(10^{10}A_s)}$ is strongly anti-correlated with all the $\mu_i$ and positively correlated with the $\eta_i$. 
To use a more objective criterion, we choose the $10\times 10$ block of MG parameters $\mu_i$ and $\eta_i$ with parameter indices 6 to 15, and only add to this block a $\lcdm$ parameter with index $a$
if the following condition is satisfied:
\[
 \sum_{i=6}^{i=15}(P_{ai})^2 \geq 1
\]
where the index $a$ corresponds to one of the first five standard parameters. For Galaxy Clustering, the only index satisfying this condition is $a=4$ in the linear case, corresponding to $\lAs \equiv \ln{(10^{10}A_s)}$: i.e. the standard parameter corresponding to the amplitude is, as said, degenerate with Modified Gravity parameters $\mu_i$ and $\eta_i$.
In the non-linear case no parameter satisfies this condition (because, as we have seen, non-linearities are able to eliminate correlation with the amplitude), but for consistency we will use the same vector of 11 parameters
$p_i = \left\{ \lAs, \mu_i, \eta_i \right\}$ for our decorrelation procedure.
Therefore we will also have 11 transformed uncorrelated $q_i$ parameters, function of the original $p_i$ parameters, in all the cases presented below.

Figure \ref{fig:Wmat-ZCA-GC} shows the coefficients that relate the 
$q_i$ parameters to the original $p_i$ ones, in the linear (left panel) and the non-linear (right panel) cases, also shown explicitly in Tables
\ref{tab:Wcoeff-lin-GC} and \ref{tab:Wcoeff-nlHS-GC} of Appendix \ref{sec:Wmatrices}.
We plot in Figure \ref{fig:GCbinerrs} a comparison between the 1$\sigma$ errors on the primary parameters $p_i$ 
(represented by circles connected with dark green dashed lines) and the decorrelated parameters $q_i$ 
(represented by squares connected with orange solid lines). 
In the linear case (left panel), we can see that the errors on the $q_i$ parameters are 2 orders of magnitude 
better than the errors on the $p_i$ parameters. In the non-linear case (right panel) the improvement is of at most 1 order of magnitude 
and that for a completely decorrelated parameter like $\lAs$, 
the error on its corresponding $q_i$ is exactly the same. 
This shows that a decorrelation procedure is still worth to do, even when including the non-linear regime, 
even if the degeneracy with the amplitude is already completely broken thanks to the non-linear prescription.
The fact that the curve of 1$\sigma$ errors for the $q_i$ follows the same pattern as the curve for the $p_i$ errors, 
is due to the fact that we have used a ZCA decomposition 
(see Section \ref{sub:Decorrelation-of-covariance}) and therefore the $q_i$ are as similar as possible to the $p_i$.

\begin{figure}[htbp]
	\centering
	\includegraphics[width=0.4\textwidth]{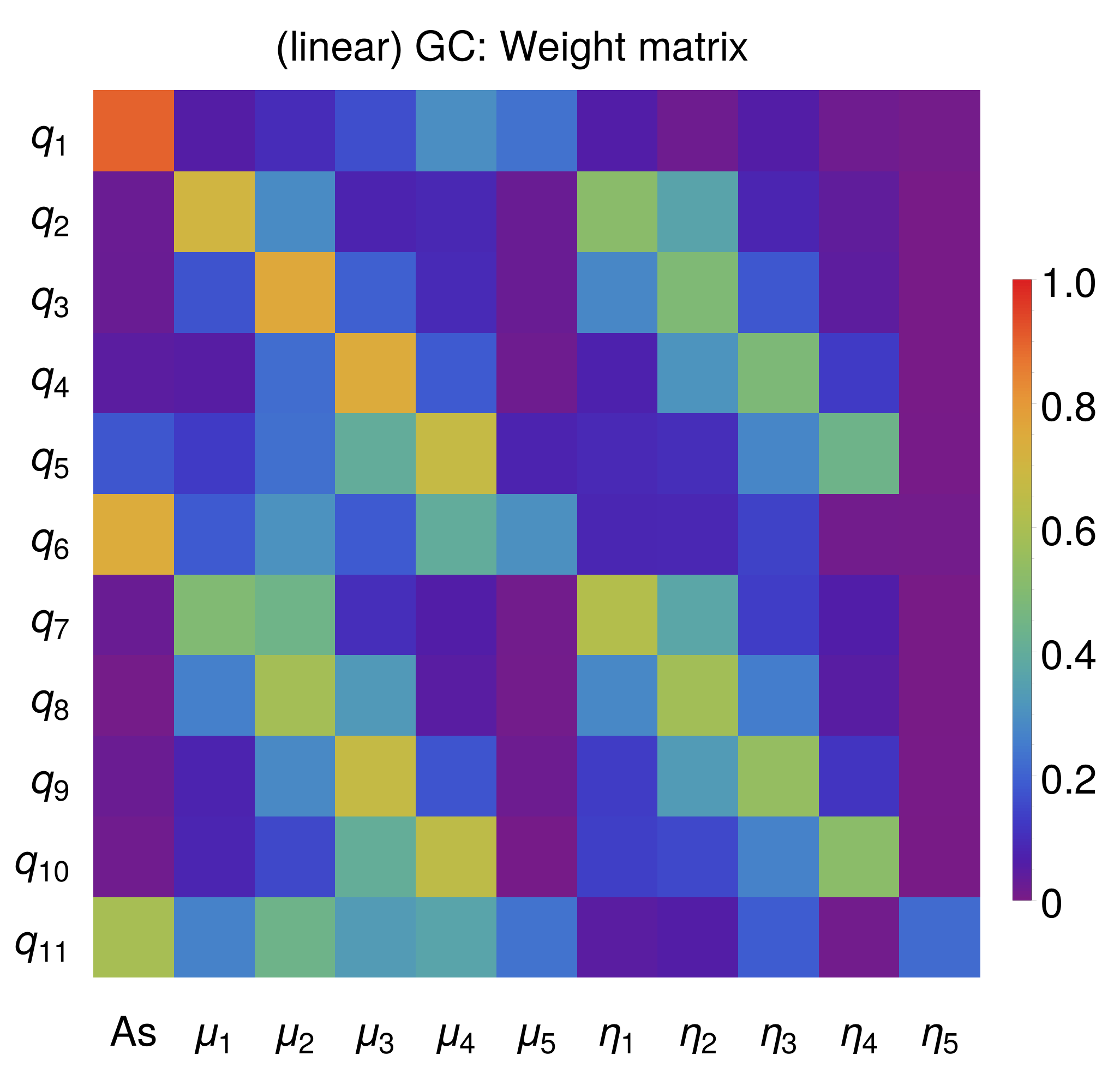}
	\includegraphics[width=0.4\textwidth]{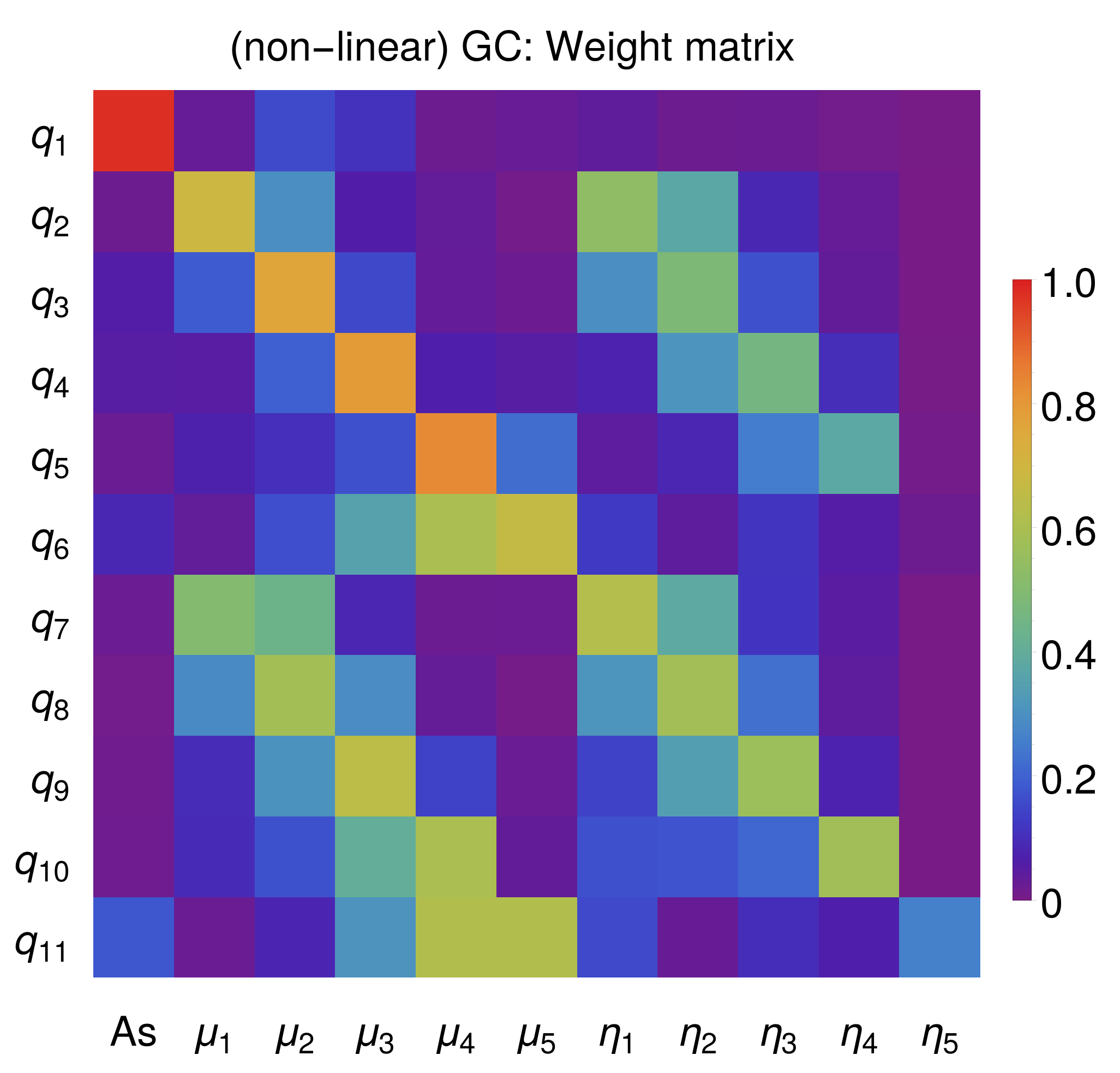}
	\caption{\label{fig:Wmat-ZCA-GC}
	Entries of the matrix $W$ that relates the $q_i$ parameters to the original $p_i$ ones, after applying the ZCA decorrelation of the covariance matrix in the linear and non-linear GC cases. This matrix shows for each new variable $q_i$ on the vertical axis, the coefficients of the linear combination of parameters $\mu_i$, $\eta_i$ and $A_s$ that give rise to that variable $q_i$. The red (blue) colors, indicate a large (small) contribution of the respective variable on the horizontal axis. \textbf{Left panel:} linear forecast for Euclid Redbook specifications.
\textbf{Right panel:} non-linear forecast for Euclid Redbook specifications, using the HS prescription.
In both cases one can observe that most $q_i$ parameters have only small or negligible contributions from $\mu_5$  and $\eta_5$, which are found to be the less constrained bins.
    }
\end{figure}

\begin{figure}[htbp]
\centering{}\includegraphics[width=0.4\linewidth]{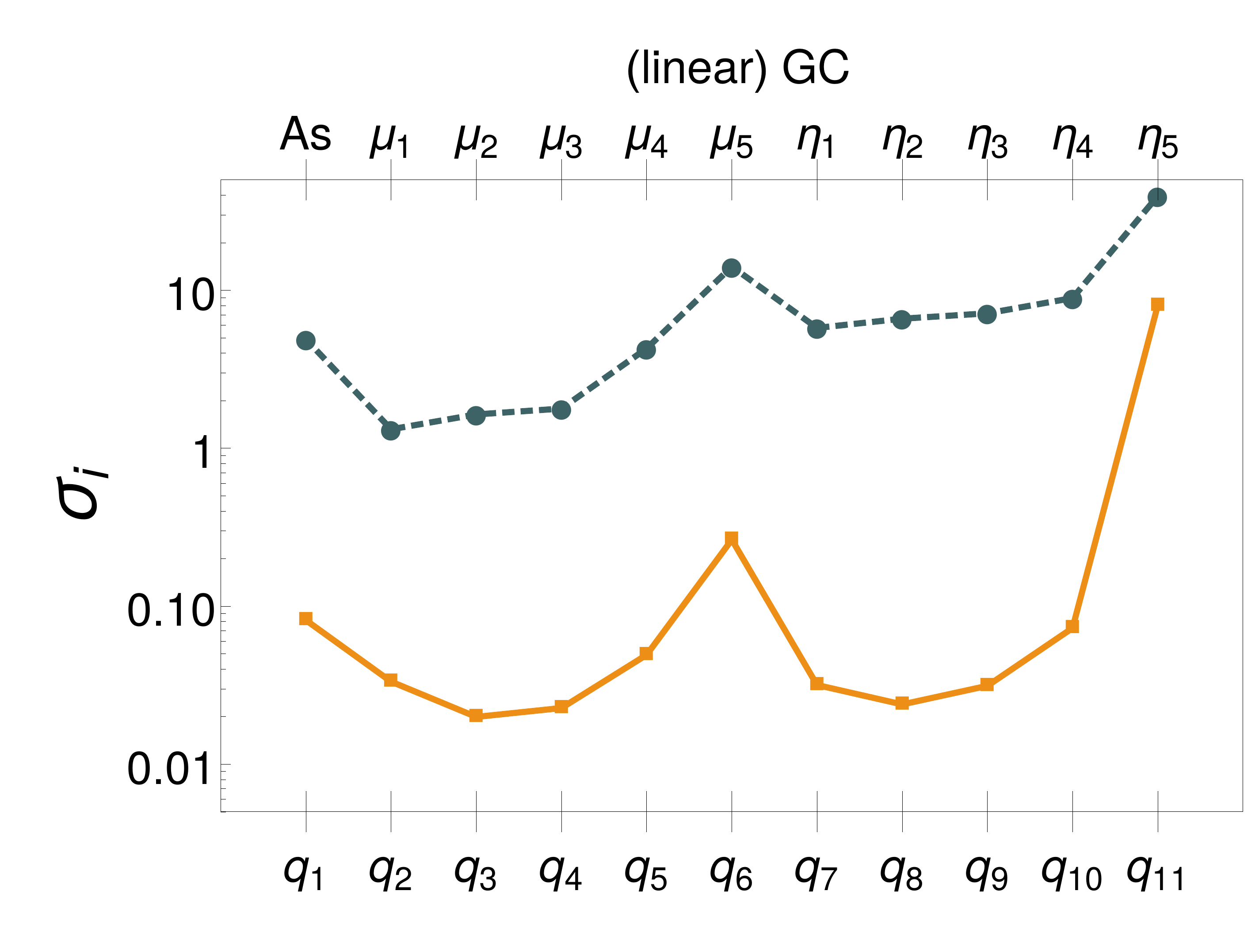}
\includegraphics[width=0.4\linewidth]{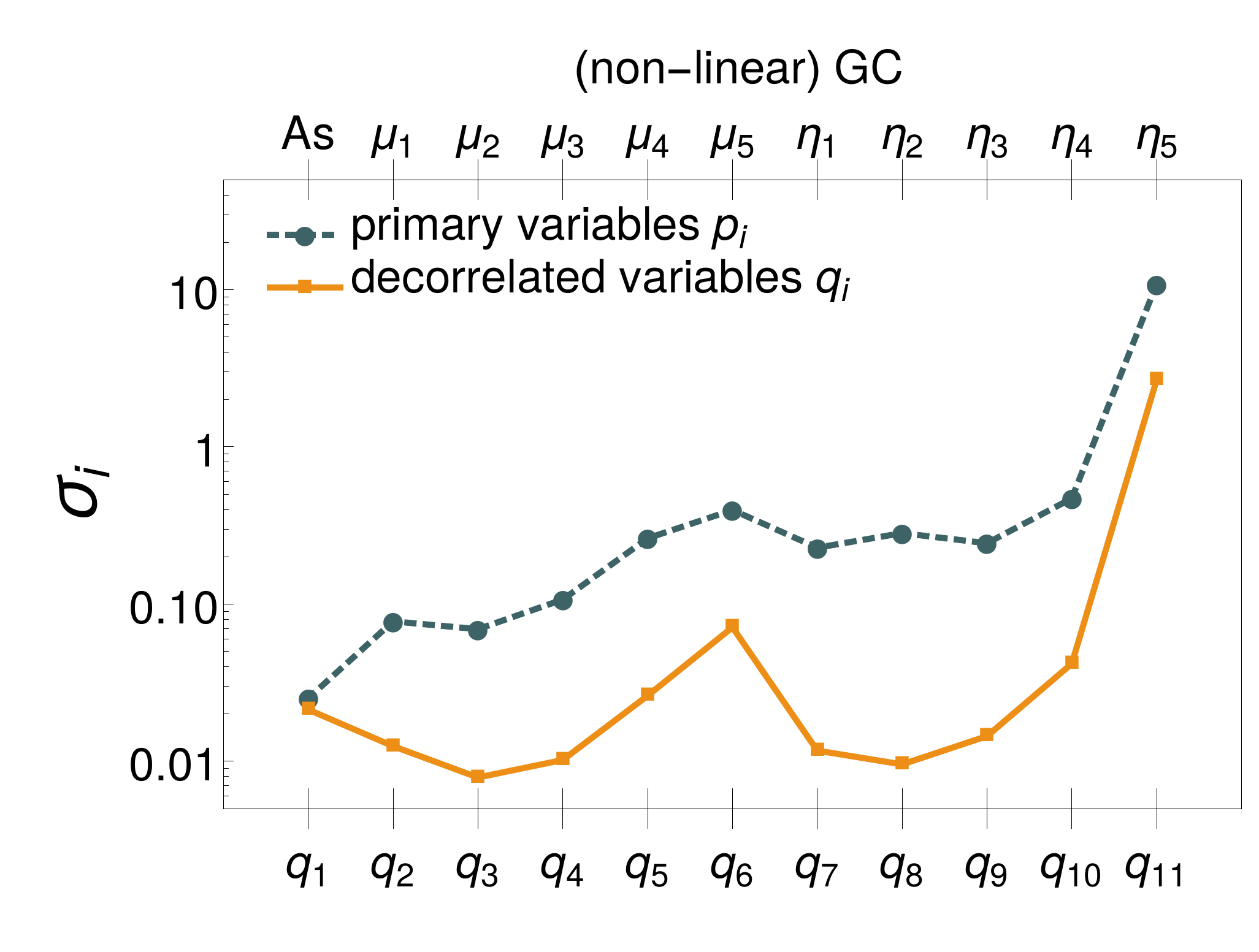}
\caption{\label{fig:GCbinerrs}
Results for a Euclid Redbook GC survey, with redshift-binned parameters, before and after applying the ZCA decorrelation.
Each panel shows the 1$\sigma$ fully marginalized errors on the primary parameters $p_i$ (green dashed
lines), and the 1$\sigma$ errors on the decorrelated 
parameters $q_i$ (orange solid lines). \textbf{Left: } linear forecasts,
performed using linear power spectra up to a maximum wavenumber $k_{\rm max}=0.15$h/Mpc.
\textbf{Right: }non-linear forecasts using non-linear spectra with the HS prescription up to a maximum wavenumber $k_{\rm max}=0.5$h/Mpc.
In the linear case, the errors on the decorrelated $q_i$ parameters are about 2 orders of magnitude smaller than for the primary parameters, 
while in the non-linear HS case, the improvement in the errors is of one order of magnitude. This means that applying a decorrelation procedure is worth even when non-linearities are considered.
In both cases for GC, the least constrained parameters are $\mu_5$ and $\eta_5$, corresponding to $2.0 < z < 2.5$.}
\end{figure}

\begin{figure}[htbp]
	\centering{}\includegraphics[width=0.4\linewidth]{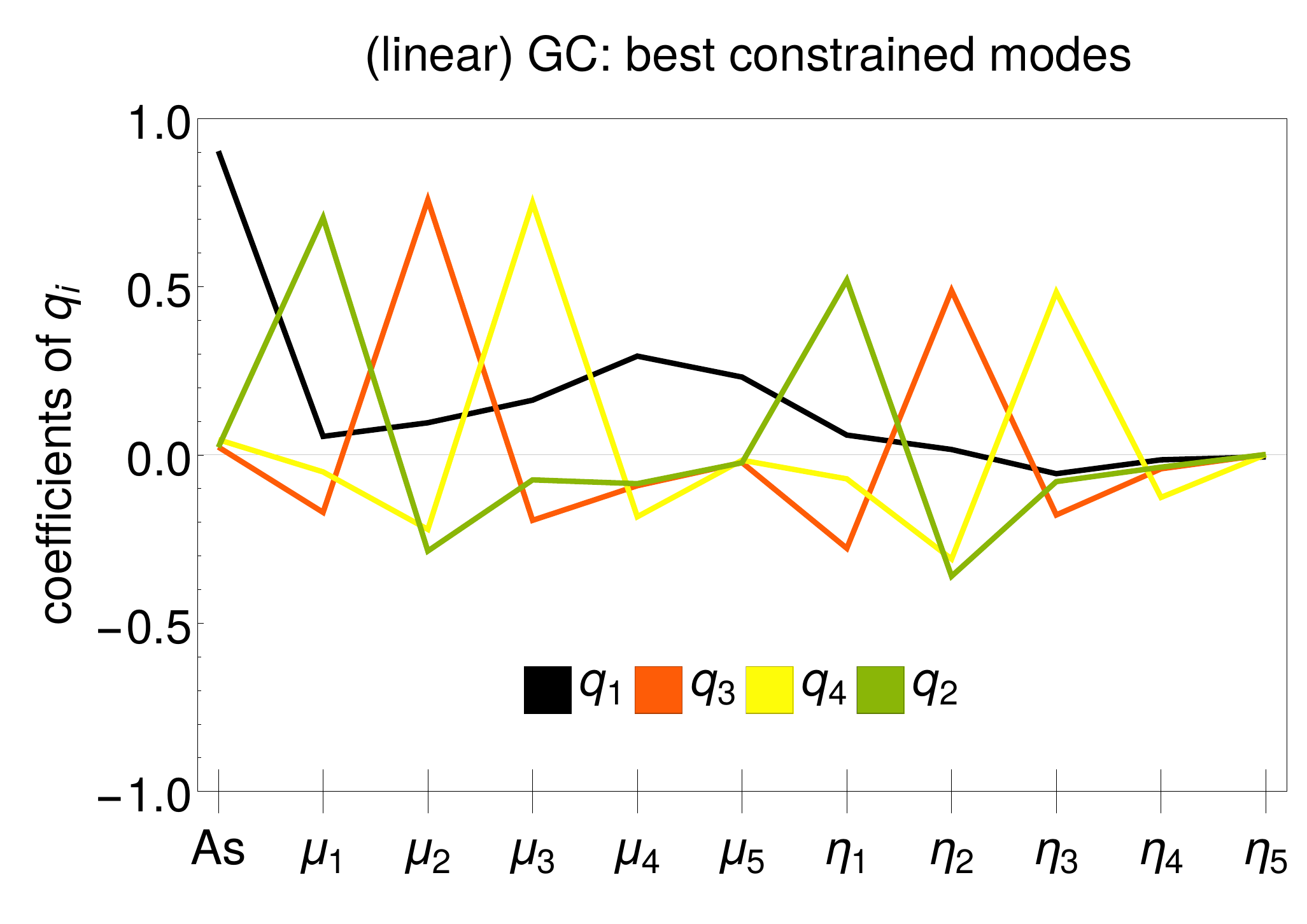}
	\includegraphics[width=0.4\linewidth]{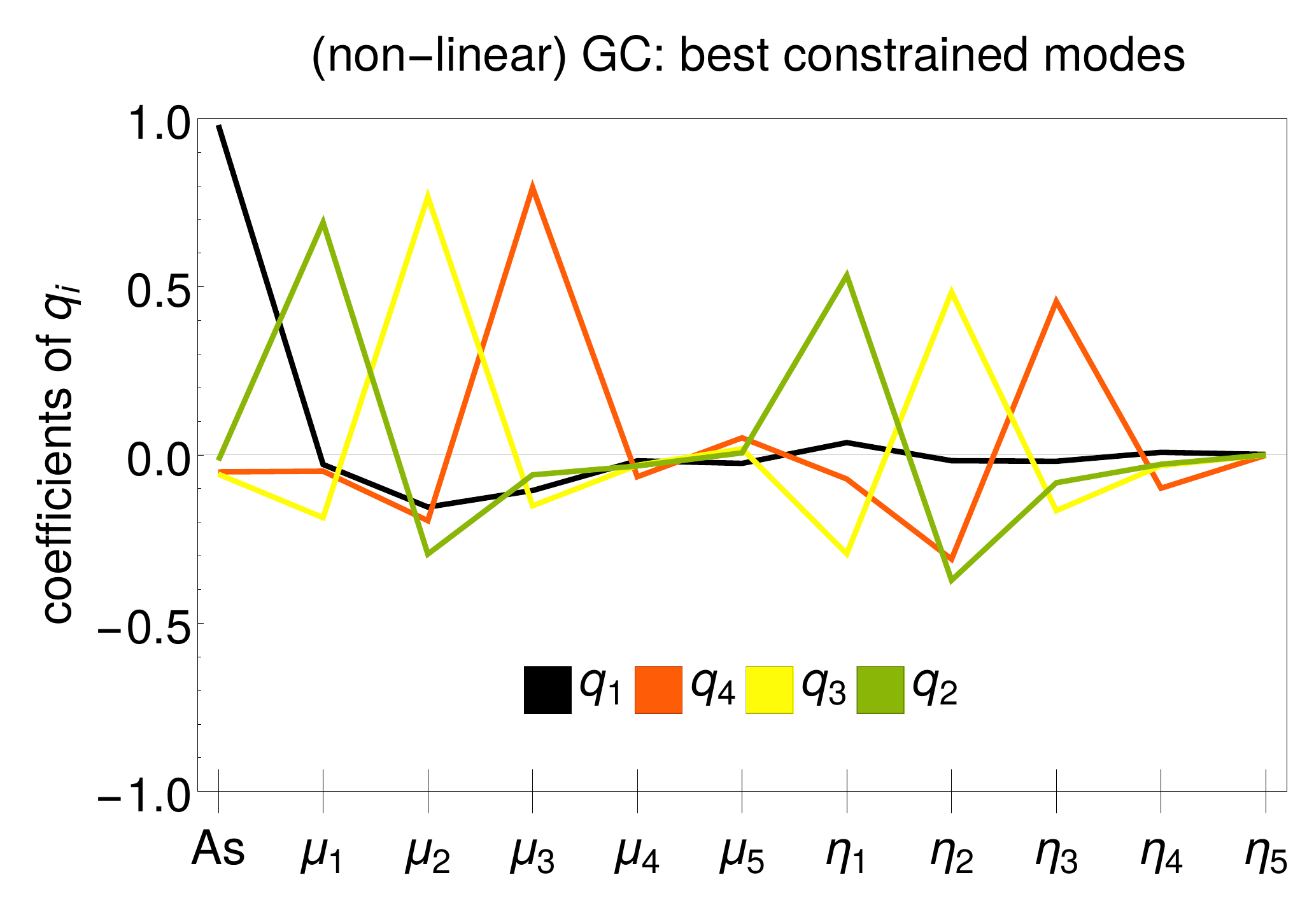}
	\caption{\label{fig:GCbestconst}
Best constrained modes for a Euclid Redbook GC survey, with $\mu$ and $\eta$ binned in redshift, 
after transforming into uncorrelated $q$ parameters via ZCA. 
Each of the four best constrained parameters $q_i$, shown in the panels, 
is a linear combination of the primary parameters $p_i$. The $q_i$ in the legends are ordered from left to right, from the best constrained to the least constrained.}
\end{figure}

We are interested in finding the best combination of primary parameters $p_i$ giving rise
to the best constrained uncorrelated parameters $q_i$. In order to find the errors 
on the parameters $q_i$, we need to look at the diagonal of the decorrelated covariance matrix $\mathbf{\tilde{C}}$ expressed in Eqn.\ (\ref{eq:Ctilde-decor}) and 
identify the $q_i$ parameters with the smallest relative errors ($\sigma_{q_i}/q_i$): we find than in the linear GC case, the best 
constrained combinations of primary parameters (ordered from most to least constrained) are given approximately by:

\begin{equation} \label{eq:bestCombined-GClin}
\begin{aligned}
	q_1  &= +0.9 \lAs  +  0.32 \mu_4 \\
	q_3  &= +0.75\mu_2 - 0.29\eta_1 + 0.50 \eta_2\\
	q_4  &= -0.25\mu_2 + 0.74\mu_3  - 0.32 \eta_2 + 0.49 \eta_3 \\
	q_2  &= +0.70\mu_1 - 0.30\mu_2  + 0.52 \eta_1 - 0.36 \eta_2 \,\,\, .
\end{aligned}
\end{equation}
In contrast, for the non-linear GC case, the parameter $\lAs \equiv \ln{(10^{10}A_s)}$ is not correlated to any other, and therefore it 
is well constrained on its own. The best 4 constrained parameters (ordered from most to least constrained) in the non-linear case, are:
\begin{equation} \label{eq:bestCombined-GCnonlin}
\begin{aligned}
	q_1  &= +0.99\lAs \\
	q_4  &= -0.28\mu_2 + 0.76\mu_3 -0.33 \eta_2 + 0.47 \eta_3\\
	q_3  &= +0.73\mu_2 -0.32 \eta_1 + 0.49 \eta_2\\
	q_2  &= +0.68\mu_1 -0.35\mu_2 + 0.52 \eta_1 -0.37 \eta_2 \,\,\, .
\end{aligned}
\end{equation}
The best constrained decorrelated parameters $q_i$ for a Euclid GC survey, 
expressed in the set of Equations (\ref{eq:bestCombined-GClin}) (linear) and (\ref{eq:bestCombined-GCnonlin}) (non-linear HS), 
can be seen graphically in Fig.\ \ref{fig:GCbestconst} for the linear (left panel) 
and non-linear HS (right panel) cases respectively. 
From these combinations we see that a survey like Euclid, using GC only, will be sensitive to Modified Gravity parameters $\mu$ and $\eta$ mainly in the first three redshift bins, corresponding to a range $0. < z < 1.5$.
The complete matrix $W$ of coefficients relating the $q_i$ to the $p_i$ parameters, 
can be found in Tables \ref{tab:Wcoeff-lin-GC} and \ref{tab:Wcoeff-nlHS-GC} of Appendix \ref{sec:Wmatrices}.

\subsubsection{ZCA for Weak Lensing}

\begin{figure}[htbp]
	\centering{}\includegraphics[width=0.4\linewidth]{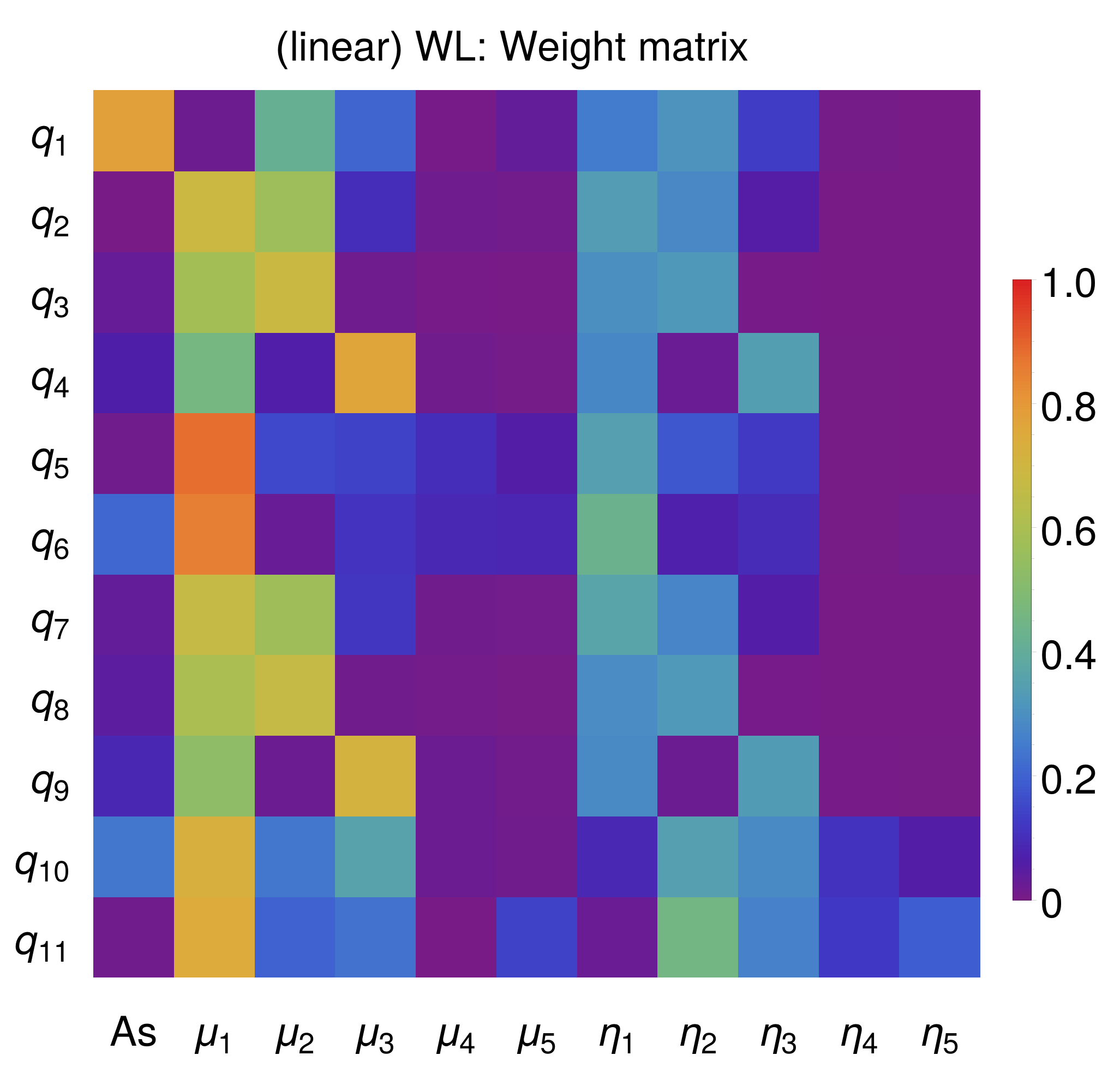}\includegraphics[width=0.4\linewidth]{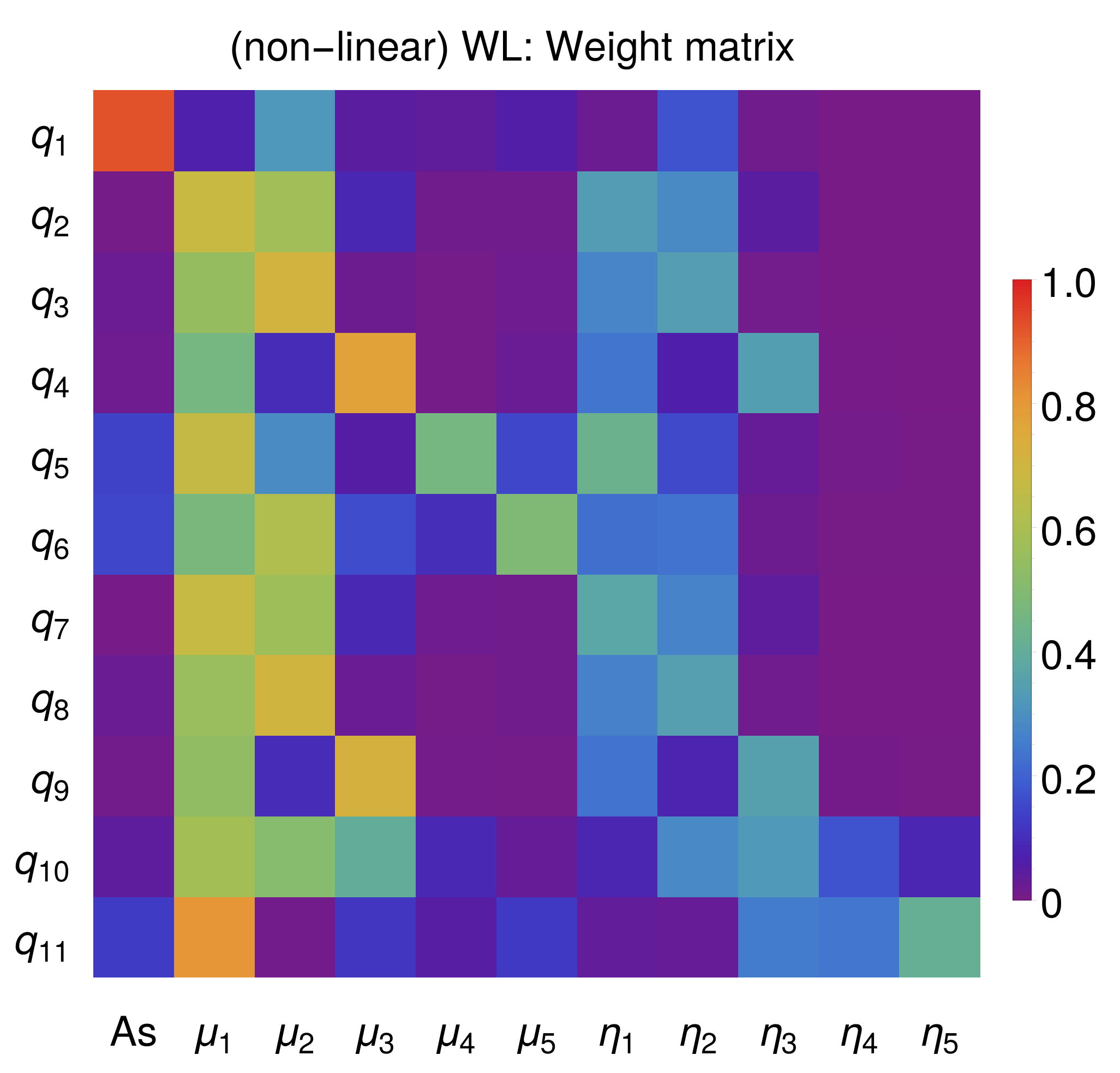}
	\caption{
Entries of the matrix $W$ that relates the $q_i$ parameters to the original $p_i$ ones, after applying the ZCA decorrelation of the covariance matrix in the linear and non-linear WL cases. This matrix shows for each new variable $q_i$ on the vertical axis, the coefficients of the linear combination of parameters $\mu_i$, $\eta_i$ and $A_s$ that give rise to that variable $q_i$. The red (blue) colors, indicate a large (small) contribution of the respective variable on the horizontal axis. \textbf{Left panel:} linear forecast for Weak Lensing Euclid Redbook specifications.
\textbf{Right panel:} non-linear forecast for Weak Lensing Euclid Redbook specifications, using the HS prescription.
As for GC, most $q_i$ parameters have only small or negligible contributions from $\mu_5$  and $\eta_5$, which are found to be the less constrained bins.
	\label{fig:Wmat-ZCA-WL}}
\end{figure}

We apply the same decorrelation procedure to the WL case, obtaining the $q$ vectors
shown in the weight matrix of Figure \ref{fig:Wmat-ZCA-WL}
again reported explicitly in Tables \ref{tab:Wcoeff-lin-WL} and \ref{tab:Wcoeff-nlHS-WL}
in Appendix \ref{sec:Wmatrices}. 

\begin{figure}[htbp]
	\centering{}\begin{center}
		\includegraphics[width=0.4\linewidth]{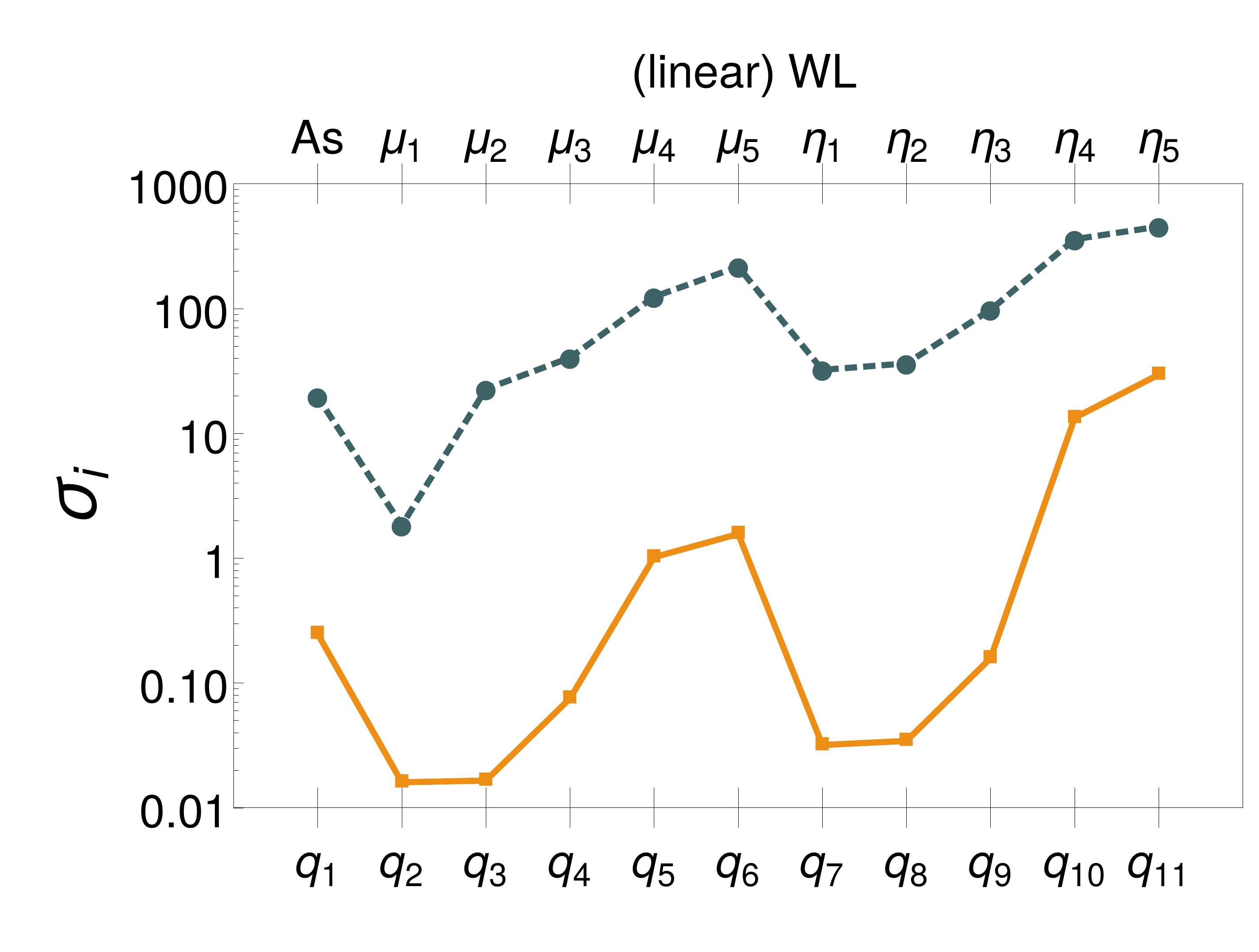}
		\includegraphics[width=0.4\linewidth]{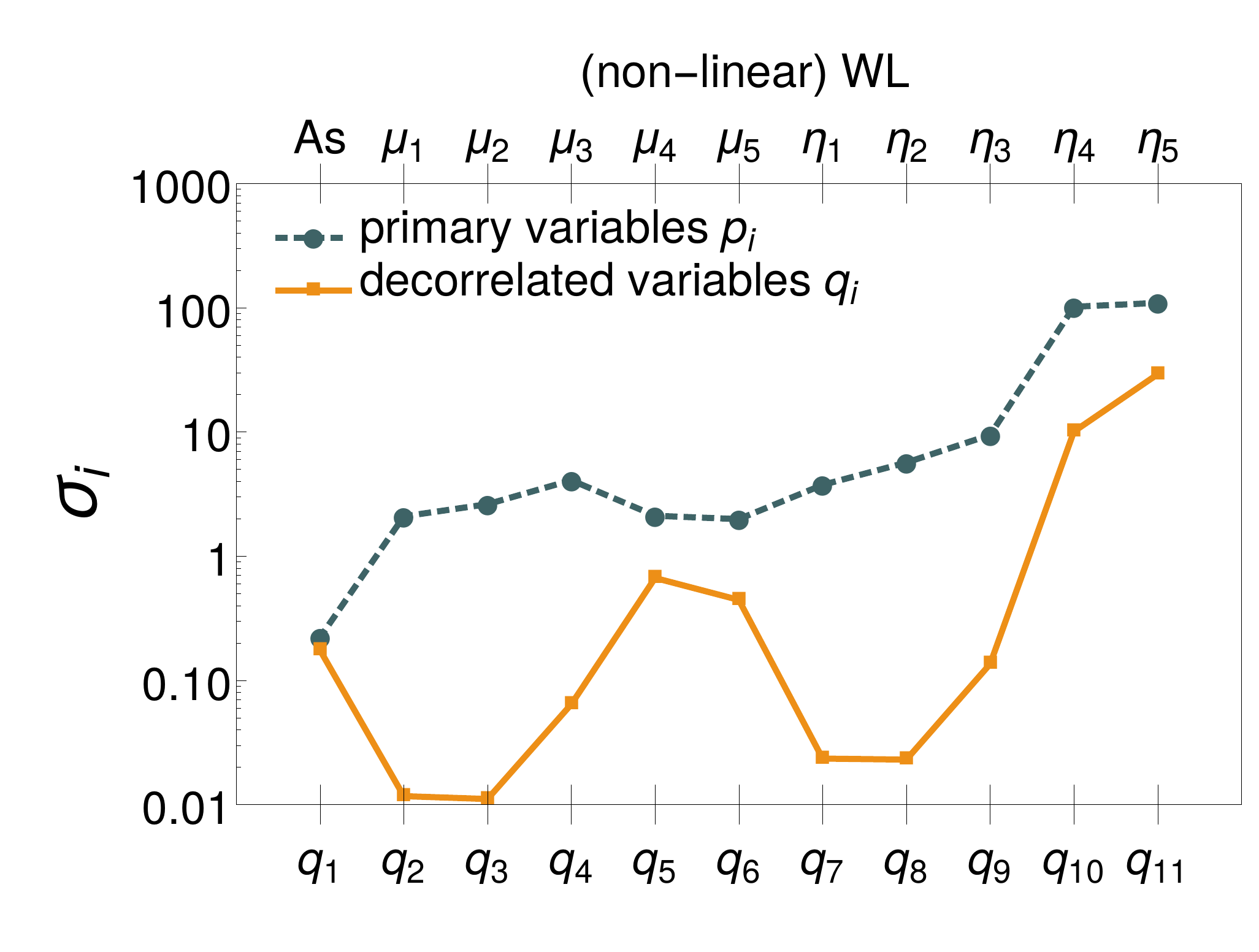}
	\end{center}
	\caption{\label{fig:WLbinerrs} 
	Results for a Euclid Redbook WL survey, with redshift-binned parameters, 
	before and after applying the ZCA decorrelation.
	Each panel shows the 1$\sigma$ fully marginalized errors on the primary parameters $p_i$ (green dashed
		lines), and the 1$\sigma$ errors on the decorrelated
		parameters $q_i$ (orange solid lines). \textbf{Left: }Linear forecasts,
		performed with an $\ell_{\rm max}=1000$ and linear matter power spectra.
		\textbf{Right: }Non-linear forecasts using the non-linear spectra with the HS prescription, up to an $\ell_{\rm max}=5000$.
		The errors in the non-linear HS case, are about 1 order of magnitude smaller than in the linear case. 
		For the best constrained $q_i$ parameters, the 
		decorrelated errors are up to 2 orders of magnitude smaller than the corresponding fully marginalized parameters on the parameters $p_i$.}
\end{figure}

In Figure \ref{fig:WLbinerrs} we show the comparison between the errors on the primary
parameters $p_i$ and the de-correlated ones $q_i$. As for the GC case, the errors in the linear case improve by 2 orders of magnitude after applying the decorrelation procedure (left panel). In the non-linear case (right panel) the improvement is smaller, but still worth to do, especially to constrain $q_2,q_3,q_7,q_8$.

\begin{figure}[htbp]
	\centering{}\includegraphics[width=0.4\linewidth]{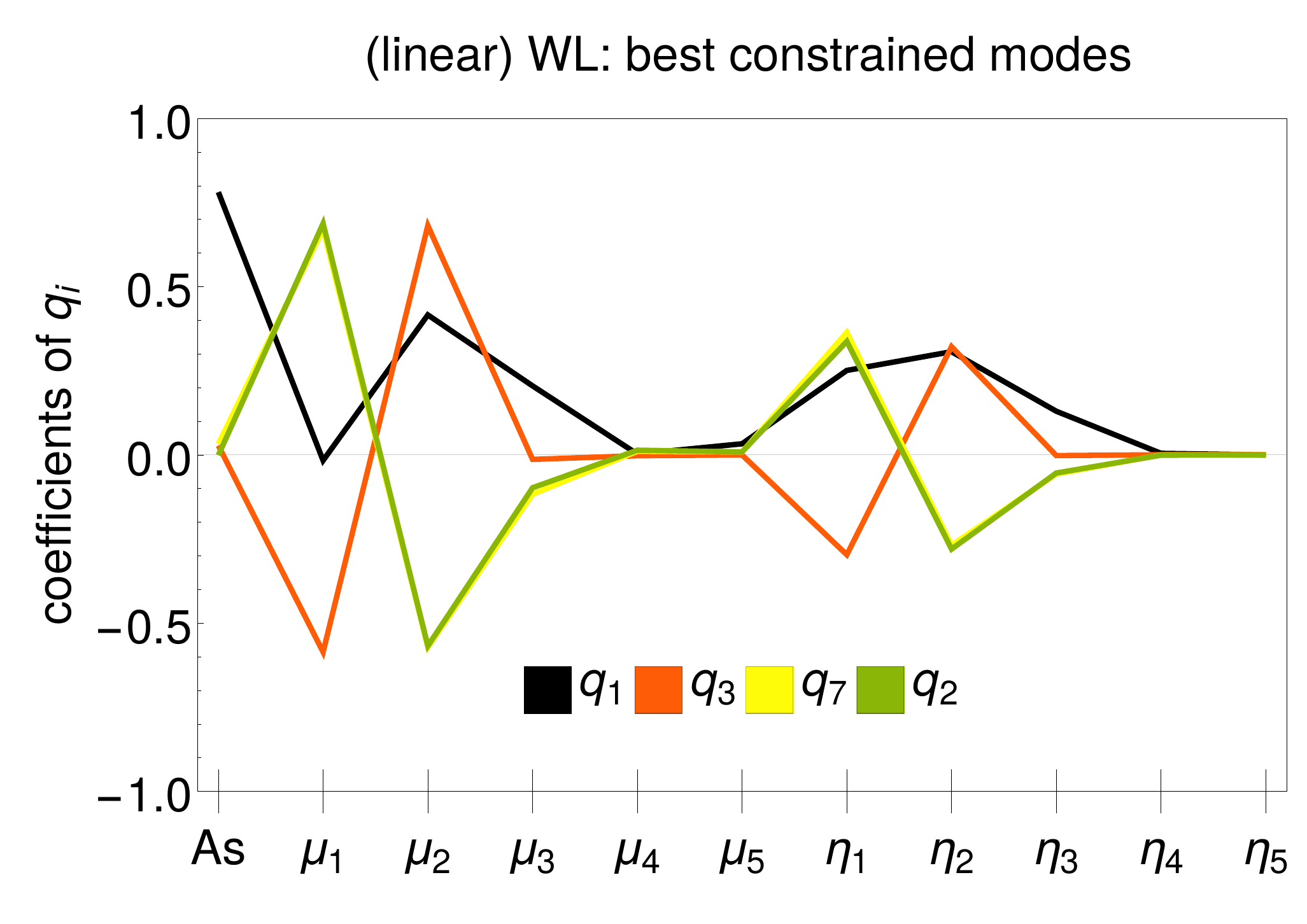}
	\includegraphics[width=0.4\linewidth]{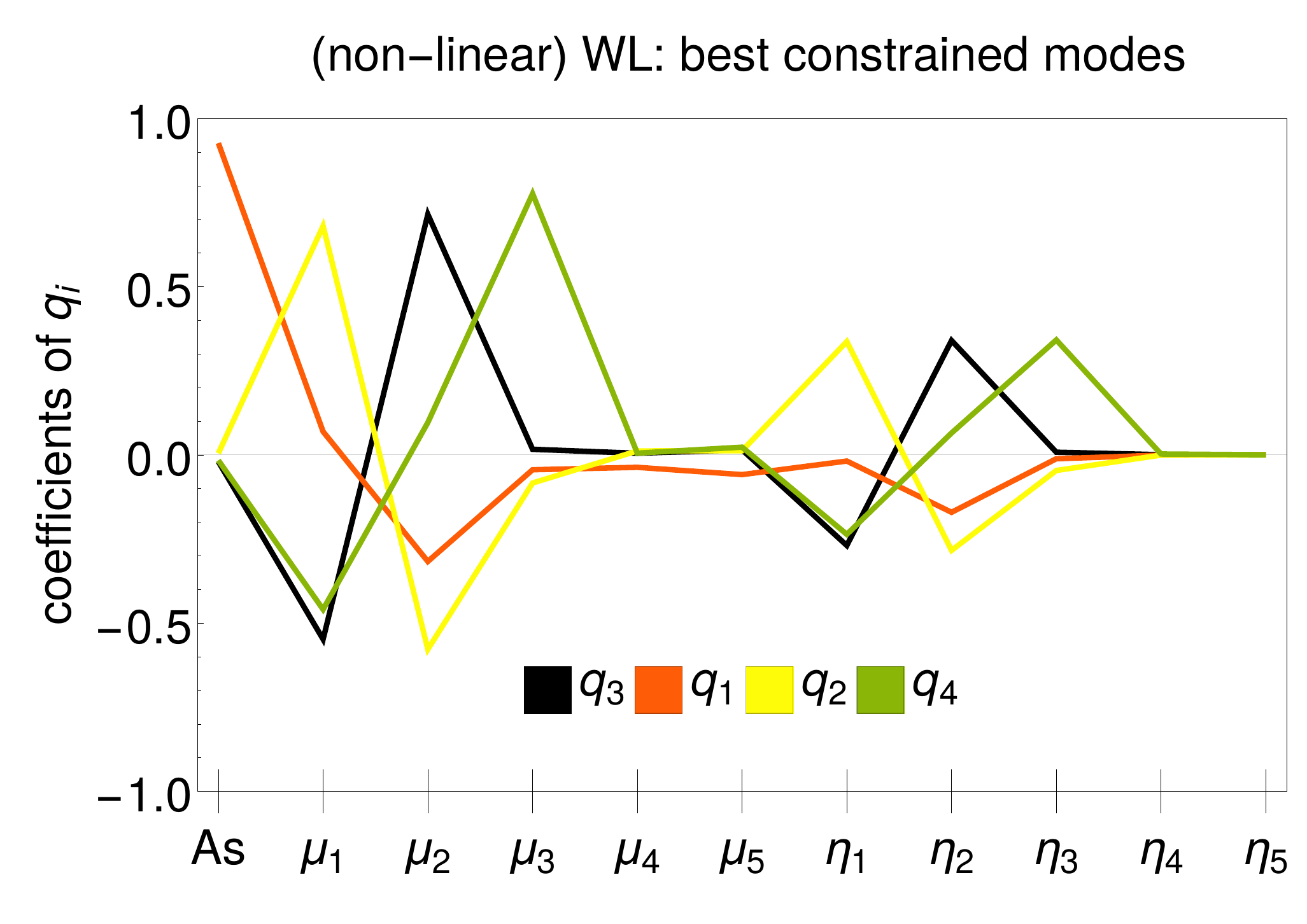}
	\caption{\label{fig:WLbestconst}
Best constrained modes for a Euclid Redbook WL survey, 
with $\mu$ and $\eta$ binned in redshift, after transforming into uncorrelated $q$ parameters via ZCA.
Each of the four best constrained parameters $q_i$, shown in the panels, is a linear combination of the primary parameters $p_i$. $q_i$ in the label are ordered from the best constrained to the least constrained.}
\end{figure}

More generally, as we did for the GC case in the previous section, we look for the $q_i$ 
parameters with the smallest relative errors ($\sigma_{q_i}/q_i$) and find in the linear WL case, that the best 
constrained combinations (ordered from most to least constrained) of primary parameters are given approximately by: 

\begin{equation} \label{eq:bestCombined-WLlin}
\begin{aligned}
	q_1  &= +0.76\lAs + 0.48 \mu_2 + 0.33\eta_2 \\
	q_3  &= -0.59\mu_1 + 0.67 \mu_2 - 0.30\eta_1 + 0.32 \eta_2\\
	q_7  &= +0.65\mu_1 - 0.60 \mu_2 + 0.36\eta_1 - 0.28 \eta_2\\
	q_2  &= +0.67\mu_1 - 0.59 \mu_2 + 0.33\eta_1 - 0.29 \eta_2 \,\,\, .
\end{aligned}
\end{equation}
This means that WL in the linear case will only be able to constrain 
combinations of the first two redshift bins in $\mu$ and $\eta$ (corresponding to  $ 0. < z < 1.0 $).
This can also be observed graphically in the left panel of Figure \ref{fig:WLbestconst}.
For the non-linear WL case, the combinations remain practically the same, except for $q_1$, which will depend much more strongly on the parameter $\lAs$. The best 4 constrained parameters in this case, are (ordered from most to least constrained):
\begin{equation} \label{eq:bestCombined-WLnonlin}
\begin{aligned}
	q_3  &= -0.55\mu_1 + 0.71 \mu_2 + -0.27\eta_1 + 0.34\eta_2\\            
	q_1  &= +0.93\lAs - 0.32\mu_2 \\ 
	q_2  &= +0.67\mu_1 - 0.60 \mu_2 + 0.33\eta_1 - 0.29 \eta_2\\
	q_4  &= -0.46\mu_1 + 0.29 \mu_2 + 0.73\mu_3 + 0.31 \eta_3 \,\,\, .
\end{aligned}
\end{equation}
These combinations can also be visualized in the right panel of Figure \ref{fig:WLbestconst}.
The complete matrix $W$ of coefficients relating the $q_i$ to the $p_i$ parameters, can be found in Tables \ref{tab:Wcoeff-lin-GC} and \ref{tab:Wcoeff-nlHS-GC} of Appendix \ref{sec:Wmatrices}.

\subsubsection{ZCA for Weak Lensing + Galaxy Clustering +  CMB \planck\ priors}

\begin{figure}[htbp]
	\centering
    \includegraphics[width=0.4\linewidth]{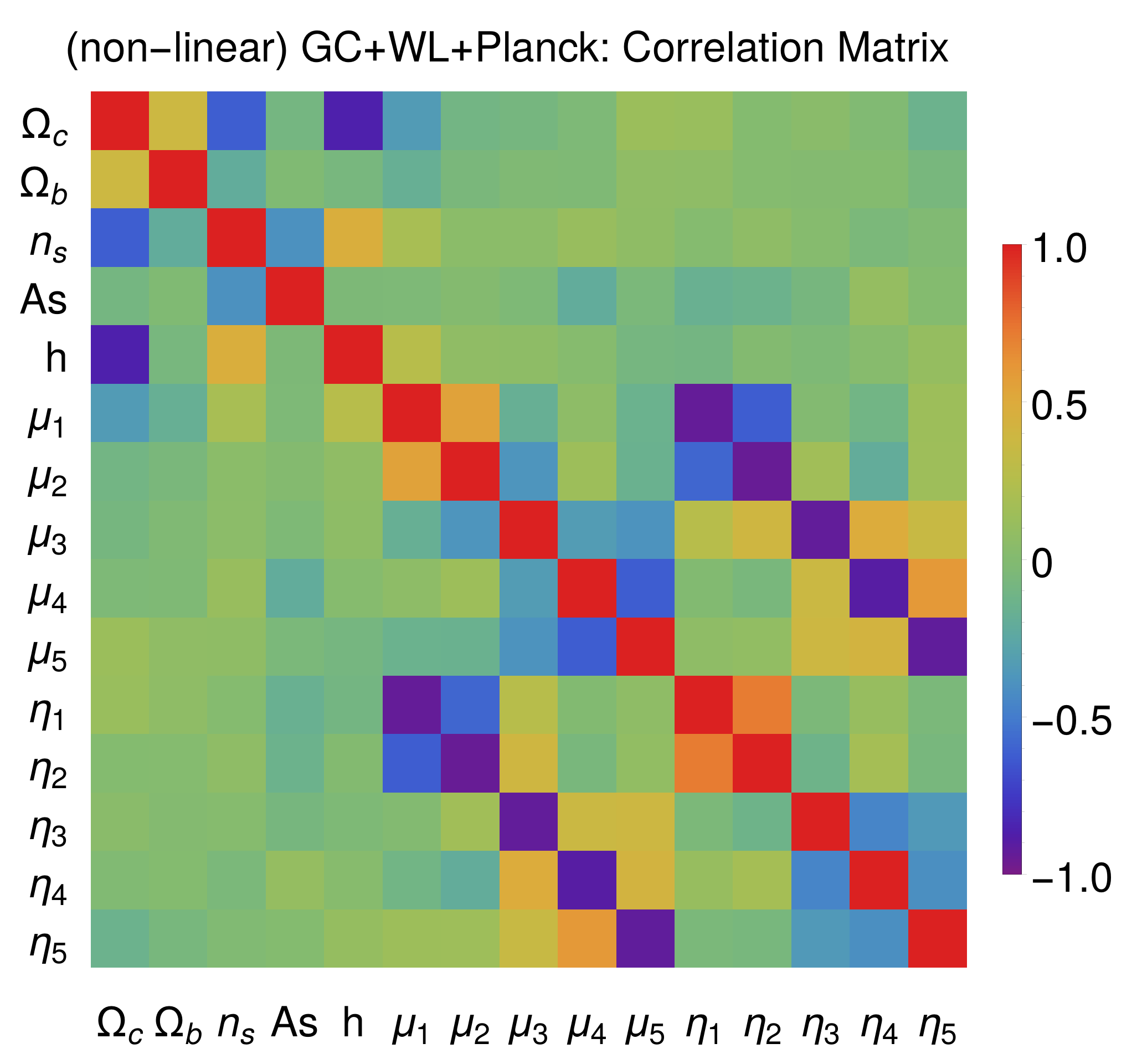}
    \includegraphics[width=0.4\linewidth]{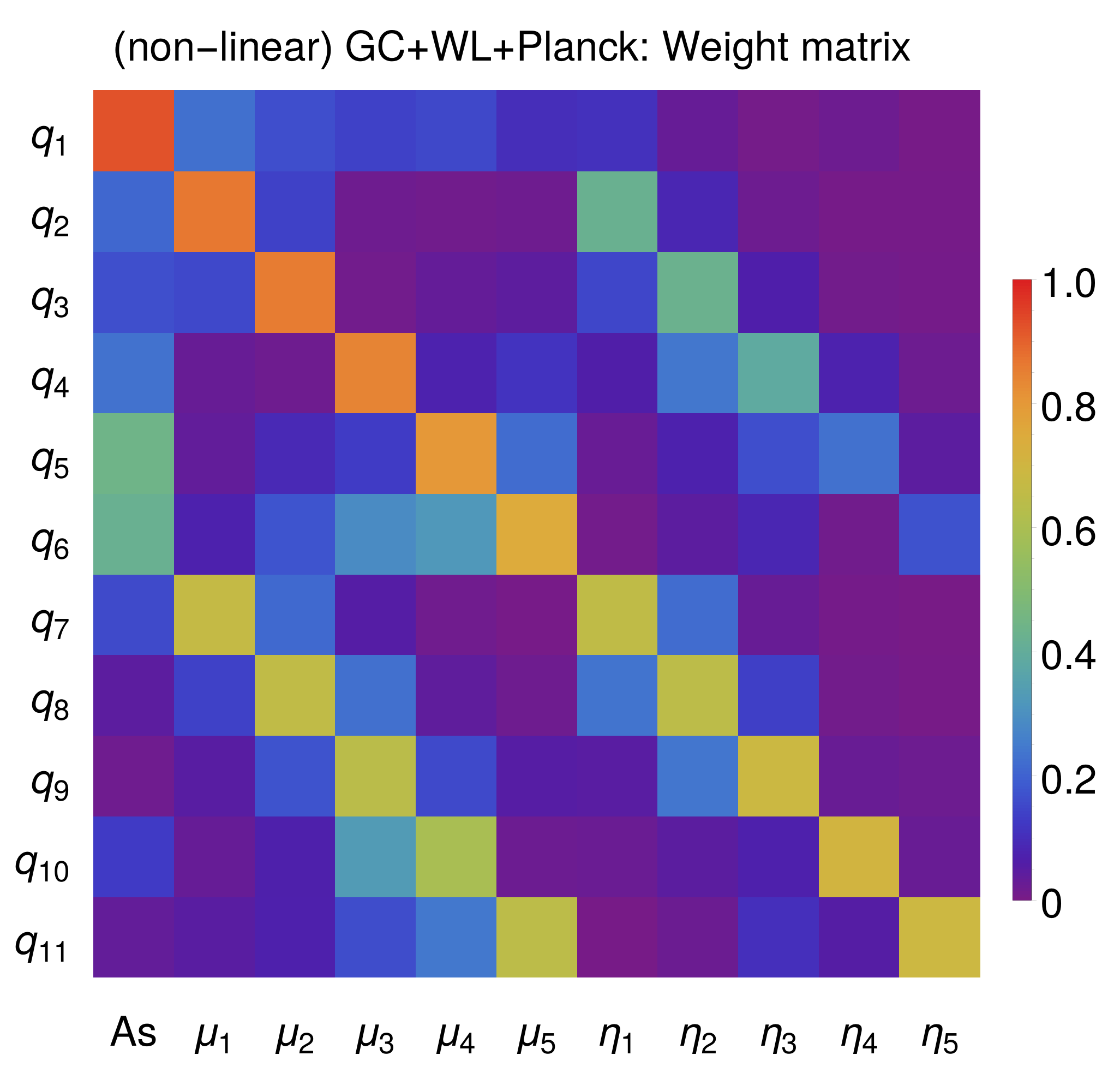}
	\caption{\label{fig:GC+WL+Planck-corr-Wmat}
Results for the combined forecasts of Euclid Redbook GC+WL using the non-linear HS prescription together with the addition 
of \planck\ CMB priors.
\textbf{Left panel:} correlation matrix obtained from the covariance matrix in the MG-binning case. 
Red (purple blue) colors represent strong positive (negative) correlations. 
The structure of this matrix is considerably diagonal, except for the strong anti-correlations of the pair $(\mu_i,\eta_i)$ 
for $i=\{1,2,3,4,5\}$, which resembles the correlations found for the WL case alone (see Figure \ref{fig:WLcorr}). 
However, the sub-block of standard cosmological parameters is now  much more diagonal 
and shows less correlations than in the GC (Fig.\ \ref{fig:GCcorr}) or WL cases.
The natural FoC (defined in \ref{eq:FoC}) in this case is $\approx 22$, showing that the variables are much less correlated 
than in the two previous cases.
\textbf{Right panel:} entries of the matrix $W$ for the ZCA decorrelation of the covariance matrix. 
This matrix shows for each new variable $q_i$ on the vertical axis, the coefficients of the 
linear combination of parameters $\mu_i$ and $\eta_i$ that give rise to that variable $q_i$. 
The red (blue) colors, indicate a large (small) contribution of the respective variable on the horizontal axis.}
\end{figure}

As mentioned earlier, Galaxy Clustering and Weak Lensing are particularly important, combined together, to constrain Modified Gravity parameters, as they probe two independent combinations of the gravitational potentials. We now show results for their combination, using for both the non-linear HS prescription, together with a \planck\ prior
 (which was obtained by performing an MCMC analysis on {\it Planck}+BSH background data, as specified in Section \ref{sub:Fisher-Planck}). Notice that we neglect here any
 information coming from the cross correlation of the two probes; we therefore assume that these two observables are independent of each other
and we simply add the GC and WL Fisher matrices to obtain our combined results; this appears to be a conservative (pessimistic) choice 
\cite{Lacasa2016}.
 In Table \ref{tab:errors-all-MGBin3}, we can see that the inclusion of the \planck\ prior improves considerably certain parameters, especially
 the less constrained ones by GC+WL, namely $\mu_{4,5}$ and $\eta_{4,5}$.
 In terms of correlations, we can observe in the left panel of Fig.\ \ref{fig:GC+WL+Planck-corr-Wmat}, that the structure of the correlation matrix resembles the one of the linear WL case (Fig.\ \ref{fig:WLcorr}), except that the block of standard $\lcdm$ parameters is much less correlated and that the anti-correlation among $\mu_i$ and $\eta_i$ is much stronger now.
On the other hand, applying the decorrelation procedure (Section \ref{sub:Decorrelation-of-covariance}), the weight matrix $W$ (right panel of Fig.\ \ref{fig:GC+WL+Planck-corr-Wmat}), resembles the $W$ matrix observed in the non-linear GC case, illustrated in Figure \ref{fig:Wmat-ZCA-GC}). Notice that now the variables $q_i$ depend quite strongly on only one of the $\mu_i$ for $i=\{1,2,3,4\}$, while the $q_i$ for $i=\{6,7,8,9,10\}$ depend on a balanced sum of $\mu_i$ and $\eta_i$ for $i=\{6,7,8,9,10\}$.

\begin{figure}[htbp]
	\centering
	\includegraphics[width=0.4\linewidth]{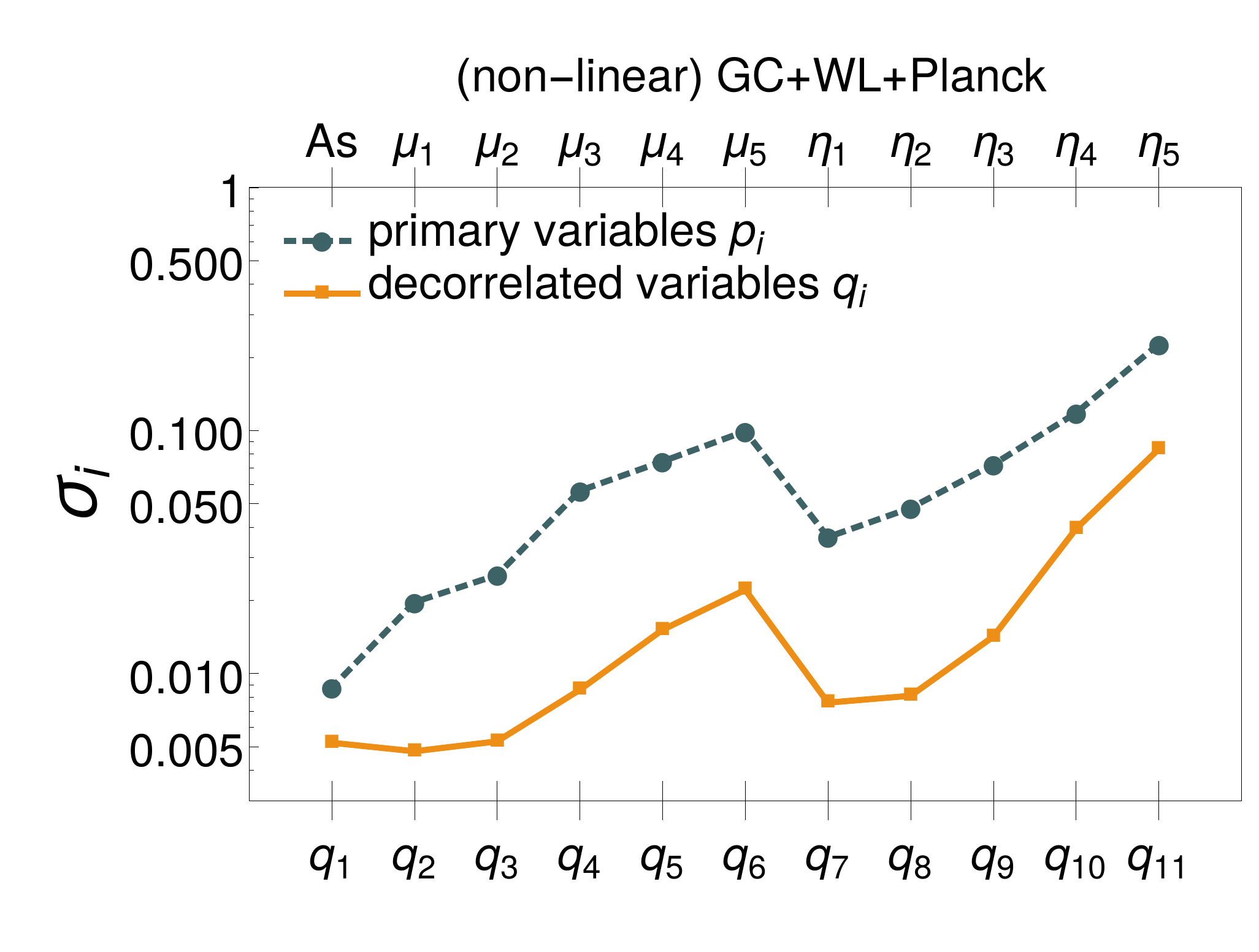}
	\includegraphics[width=0.4\linewidth]{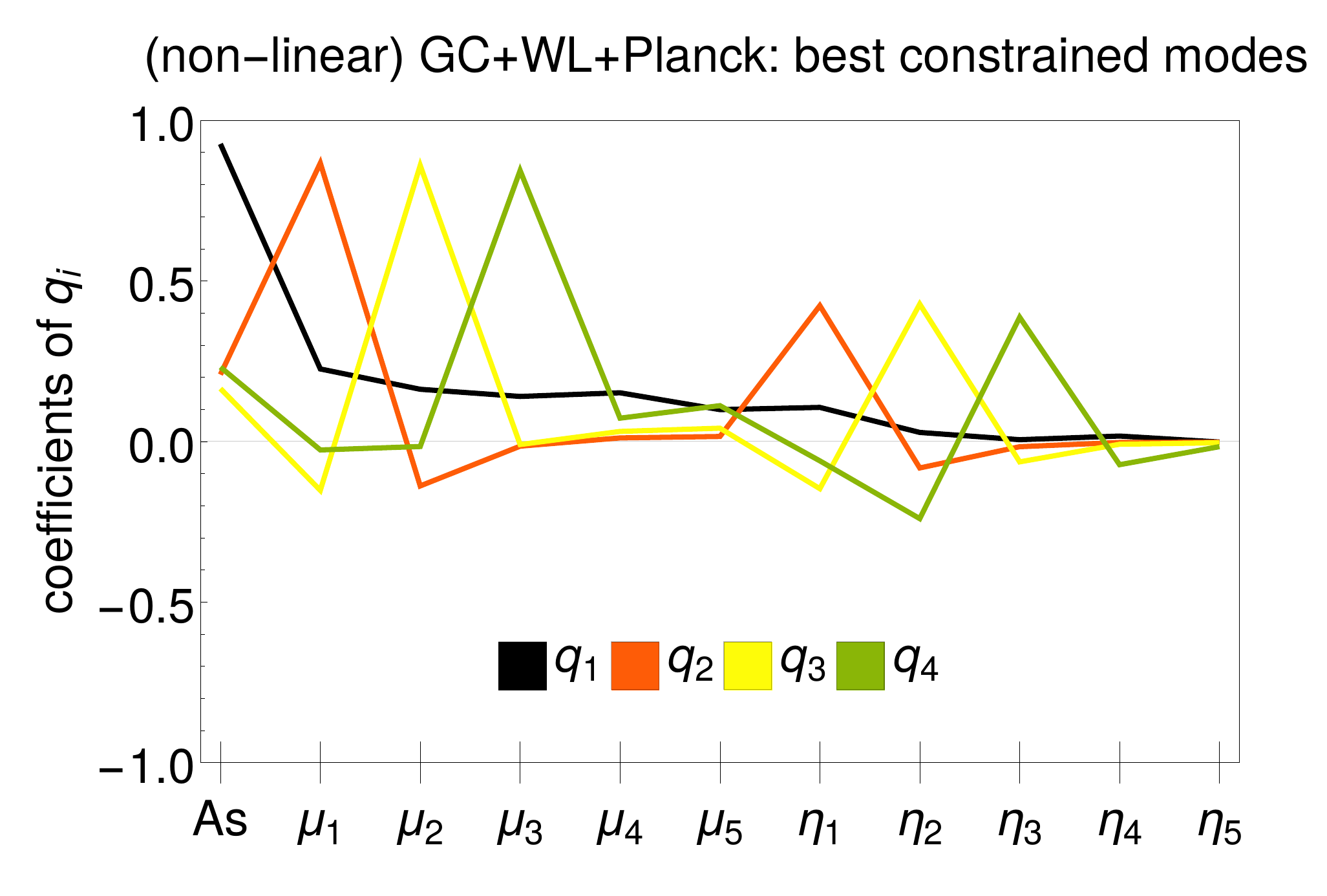}
	\caption{\label{fig:GC+WL+Planck-bestconst-errspq}
		\textbf{Left:} the 1$\sigma$ fully marginalized errors on the primary parameters $p_i$ (green dashed
		lines), and the 1$\sigma$ errors on the decorrelated derived
		parameters $q_i$ (yellow solid lines). As opposed to the GC or WL cases (figs.\ref{fig:GCbinerrs},\ref{fig:WLbinerrs}), here the decorrelated errors are much more similar to the standard errors.
		This is due to the fact that in the GC+WL+{\it Planck} combination, the cosmological parameters are not so strongly correlated.
		\textbf{Right:} best constrained modes for a Euclid Redbook GC+WL case using the non-linear HS prescription and adding a CMB \planck\ prior.
		Each panel shows the four best constrained parameters $q_i$. Each of them is a linear combination of the primary parameters $p_i$. 
		The best constrained modes are sums $a\mu_i + b\eta_i$ for $i=\{1,2,3\}$ and positive values $a$ and $b$.}
\end{figure}

Finally, in this combined case the best constrained $q_i$ variables, are $q_1,\;q_2,\;q_3,\;$, $q_4$ approximately given by:
\begin{equation} \label{eq:bestCombined-GCWLnonlinPlanck}
\begin{aligned}
	q_1  &= +0.93\lAs\\            
	q_2  &= +0.84\mu_1 + 0.48\eta_1 \\ 
	q_3  &= +0.80\mu_2 - 0.26\eta_1 + 0.45 \eta_2\\
	q_4  &= +0.28\lAs + 0.79\mu_3 - 0.29 \eta_2 + 0.39 \eta_3 \,\,\, .
\end{aligned}
\end{equation}
These combinations of primary parameters are illustrated in the right panel of Figure \ref{fig:GC+WL+Planck-bestconst-errspq}.
The combination $q_2$ is similar to the combination $2\mu+\eta$ that was also identified in \cite{planck_collaboration_planck_2016} as being well-constrained. The best constrained modes $q_2$, $q_3$ and $q_4$ all contain terms of the form $a\mu_i + b\eta_i$ for $i=\{1,2,3\}$, with positive coefficients $a$ and $b$, where $a \approx 2b$.

All errors are shown in the left panel of Figure \ref{fig:GC+WL+Planck-bestconst-errspq}. Notice how in this case, the improvement on the errors of the $q_i$ variables is less than an order of magnitude, thus smaller than what found in GC and WL separately; this is due to combination of GC and WL which, together with the inclusion of the CMB prior, lead to smaller correlations among the parameters. 
When combining GC+WL in the non-linear HS case, the FoC (defined in \ref{eq:FoC}) is $\approx 31$, 
showing that there is not much gain in decorrelation, compared to GC or WL alone, where this quantity was approximately $32$.
However, combining GC+WL (non-linear HS) with \planck\ priors yields $\textrm{FoC} \approx 22$, showing that correlations among
parameters are drastically reduced.
The fact that the curve of 1$\sigma$ errors for the $q_i$ (orange line, marked with circles) 
follows the same pattern as the curve for the $p_i$ errors (green dashed line, marled with circles), 
is due to the fact that we have used a ZCA decomposition and therefore the $q_i$ are as close as possible to the $p_i$.
The complete matrix of coefficients $W$ can be found in Table \ref{tab:Wcoeff-nlHS-WL+GC+Planck} in appendix \ref{sec:Wmatrices}.

\section{Modified gravity with simple smooth functions of the scale factor}

As discussed in Sec.\ \ref{sec:Parameterizing-Modified-Gravity}, $\mu$ and $\eta$ (or an equivalent pair of functions of the gravitational potentials) depend in general on time and space. We will now investigate the time dependence further, starting from the two parameterizations proposed in \cite{planck_collaboration_planck_2016} and recalled in Eqns.\ (\ref{eq:DE-mu-parametrization}-\ref{eq:TR-eta-parametrization}) in this paper. We extend the analysis of the \planck\ paper by testing different prescriptions for the non-linear regime in Modified Gravity (as illustrated in Section \ref{sec:The-non-linear-power}) and further investigate forecasts for future experiments like Euclid, SKA, DESI. In the following subsections we first give results for the late-time parameterization of Eqns.\ (\ref{eq:DE-mu-parametrization},\ref{eq:DE-eta-parametrization}) and then for the early time parameterization of Eqns.\ (\ref{eq:TR-mu-parametrization},\ref{eq:TR-eta-parametrization}). In both cases we consider Galaxy Clustering and Weak Lensing, neglecting, as in Section \ref{sec:Results:-Redshift-Binned} any information coming from the cross correlation of the two probes.

\subsection{\label{sub:MG-DE}Modified Gravity in the late-time parameterization}
The late-time parameterization is defined in Eqns.\ (\ref{eq:DE-mu-parametrization},\ref{eq:DE-eta-parametrization}). We now calculate forecasts for Galaxy Clustering and Weak Lensing, with future surveys, in the linear and mildly non-linear regimes.
We also include prior information obtained from
the analysis of the {\it Planck}+BSH datasets (where we recall that BSH stands for BAO + SNe + $H_0$ prior),
as discussed in Section \ref{sub:Fisher-Planck}.

\subsubsection{Galaxy Clustering in the linear and mildly non-linear regime}
In Table \ref{tab:errors-Euclid-GC-WL-late_time} we show forecasts for the Euclid survey \cite{laureijs_euclid_2011} for Galaxy Clustering (top part of the table) and three different cases: using only linear scales with a cutoff at $k_{\rm max}=0.15$h/Mpc, labeled GC(lin); extending forecasts in the mildly non-linear regime, obtained using the prescription described in Sec.\ \ref{sub:Prescription-HS}, with a cutoff at $k_{\rm max}=0.5$ h/Mpc, labeled GC(nl-HS); combining the mildly non-linear case with \planck\ priors, as described in Sec.\ \ref{sub:Fisher-Planck}. We take into account
the BAO features, redshift space distortions and the full shape of
the power spectrum, as well as the survey specifications of the Euclid Redbook, recalled
in Section \ref{sub:Fisher-Galaxy-Clustering} for convenience. The columns correspond to the marginalized errors on five standard cosmological parameters $\{\Omega_c, \Omega_b, n_s, \ln (10^{10} A_s), h\}$ and three combinations of the gravitational potentials $\{\mu, \eta, \Sigma\}$: the latter are reconstructed in time, according to the late-time parameterization, as defined in Eqns.\ (\ref{eq:DE-mu-parametrization},\ref{eq:DE-eta-parametrization}). We recall that only two of these three functions are independent and fully determine cosmological linear perturbations.

In the late-time scenario, for a Galaxy Clustering survey,  
neither $\eta$ nor $\Sigma$ are actually
constrained by a linear forecast, while $\mu$ is mildly constrained.  
Adding the non-linear regime improves constraints on $\mu$, while the other parameters remain unconstrained, 
unless we also include \planck\ priors, which yields an improvement in the FoM of 6.3 nits. In general the observable power spectrum may depend on both $\mu$ (explicitely appearing in the last term of the equation for the density perturbation (cf. Eqn.(\ref{eq-delta})) and on $\eta$ (implicitly contained in the derivatives of the gravitational potential in the same equation). The contribution of the derivative of the potentials is larger in the early-time parameterization, with respect to the late-time one by construction. This is due to the fact that in the late-time case deviations from GR go to zero at large redshifts. In this sense, in the specific case of the late time parameterization, the observed power spectrum mainly depends on $\mu$ only, which explains why this is the only quantity (mildly) constrained by GC alone. In the early-time parameterization, though, modifications can appear also at earlier times, so that both $\eta$ and $\mu$ effectively affect the power spectrum, which explains why they can both be constrained with a smaller uncertainty, as we will discuss in Section \ref{sub:MG-TR}.

In Appendix \ref{sec:appder} we review the equation governing the evolution of cold dark matter density fluctuations, 
as a function of the Modified Gravity functions $\mu(a)$ and $\eta(a)$ as they are implemented in the code MGCAMB \cite{hojjati_testing_2011}.
The inability of GC to constrain $\eta$ in this parametrization, is also visible in Fig.\ \ref{fig:DE+Planck-ellipses-mu-sig-eta} which shows that the GC constraints are
degenerate along the $\eta$ or $\Sigma$ directions.
Therefore, we show how with this parameterization choice Euclid GC will be extremely sensitive to
modifications of the Poisson equation for $\Psi$, while it would require additional
information to constrain departures from the standard Weyl potential.

\begin{table}[htbp]
\centering{}%
\begin{tabular}{|l|c|c|c|c|c||c|c|c|c|}
\hline 
\textbf{Euclid} (Redbook) & $\Omega_{c}$  & $\Omega_{b}$  & $n_{s}$  &
$\ell\mathcal{A}_{s}$  & $h$  & $\mu$  & $\eta$  & $\Sigma$ & MG FoM
\Tstrut\tabularnewline
\hline 
\multicolumn{1}{|l|}{Fiducial}   & {0.254}  & {0.048}  & {0.969}  & { 3.060}  & {0.682 } & {1.042}  & {1.719}  & {1.416} & relative \Tstrut\tabularnewline
\hline
\hline
\Tstrut \textbf{GC(lin)
}  
& 1.9\% & 6.4\% & 3\% & 2.8\% & 4.5\% & 17.1\% & 1030\% & 641\%
& 0\tabularnewline
\Tstrut\textbf{GC(nl-HS) 
}  
& 0.9\% & 2.5\% & 1.3\% & 0.8\% & 1.7\% & 1.7\% & 475\% & 291\%
& 2.9 \tabularnewline
\Tstrut\textbf{GC(nl-HS)+{\it Planck} 
}  
& 0.7\% & 0.6\% & 0.3\% & 0.2\% & 0.3\% & 1.7\% & 16.8\% & 10.3\%
& 6.3 \tabularnewline
\hline 
\hline
\Tstrut \textbf{WL(lin) 
}  
& 7.8\% & 25.7\% & 9.9\% & 10.3\% & 19.1\% & 58.2\% & 106\% & 9.3\%
& 3.2\tabularnewline
\Tstrut \textbf{WL(nl-HS) 
}    
& 6.3\% & 20.7\% & 4.6\% & 5.8\% & 13.8\% & 23.3\% & 40.9\% & 4.6\%
& 4.5 \tabularnewline
\Tstrut \textbf{WL(nl-HS)+{\it Planck} 
}  
& 2.1\% & 1.1\% & 0.4\% & 0.7\% & 0.7\% & 11.8\% & 21.8\% & 2.8\%
& 5.7 \tabularnewline
\hline
\hline
 \Tstrut \textbf{GC+WL(lin)}  
 & 1.8\% & 5.9\% & 2.8\% & 2.3\% & 4.2\% & 7.1\% & 10.6\% & 2\%
& 6.6 \tabularnewline
\Tstrut \textbf{GC+WL(lin)+{\it Planck}} $\;$  
 & 1.0\% & 0.7\% & 0.4\% & 0.4\% & 0.4\% & 6.2\% & 9.8\% & 1.5\%
& 7.0  \tabularnewline
\hline
\hline
 \Tstrut \textbf{GC+WL(nl-HS)}  
 & 0.8\% & 2.2\% & 0.8\% & 0.7\% & 1.5\% & 1.6\% & 2.4\% & 1.0\%
& 8.8 \tabularnewline
\Tstrut \textbf{GC+WL(nl-HS)+{\it Planck}} $\;$  
 & 0.7\% & 0.6\% & 0.2\% & 0.2\% & 0.3\% & 1.6\% & 2.4\% & 0.9\%
& 8.9  \tabularnewline
\Tstrut \textbf{GC+WL(nl-Halofit)+{\it Planck}} $\;$ 
& 0.6\% & 0.5\% & 0.2\% & 0.2\% & 0.2\% & 0.8\% & 1.7\% & 0.8\%
& 9.6  \tabularnewline
\hline  
\end{tabular}\protect
\caption{\label{tab:errors-Euclid-GC-WL-late_time}
1$\sigma$
fully marginalized errors on the cosmological parameters in the late-time parameterization of Modified Gravity for a Euclid
Galaxy Clustering forecast (top), a Weak Lensing forecast (middle) and the combination
of both probes (bottom): 
Modified Gravity is encoded in two of the three functions $\mu$, $\eta$, $\Sigma$, 
which are reconstructed in the late-time parameterization defined in 
Eqns.\ (\ref{eq:DE-mu-parametrization},\ref{eq:DE-eta-parametrization}). For each case, we also list
the forecasted errors using a {\it Planck}+BSH prior.
Linear forecasts are labeled by ``lin'', and correspond to a cutoff $k_{\rm max}=0.15$h/Mpc for GC and $\ell_{\rm max}=5000$ for WL; 
non-linear forecasts use the prescription described in 
sec.\ \ref{sub:Prescription-HS}, are labeled by ``nl-HS'' and correspond to a cutoff of $k_{\rm max}=0.5$h/Mpc for GC and $\ell_{\rm max}=5000$ for WL.
in both cases, power spectra have been computed using the MGCAMB Boltzmann code. For completeness,
we also show the GC+WL+{\it Planck} case using the non-linear power spectra computed using Halofit only (nl-Halofit).
In the last column, we show for each observation the Figure of Merit (FoM) 
relative to our base observable in this parametrization, which is GC(linear).
GC(linear) has an absolute FoM of -0.94 in `nits'.
We can see that in both GC and WL there is a considerable gain when including non-linear scales and \planck\ priors.
The difference in the FoM between the non-linear HS prescription and the standard Halofit approach
is quite small. 
}
\end{table}

\begin{table}[htbp]
\centering{}%
\begin{tabular}{|l|c|c|c|c|c||c|c|c|c|}
\hline
\Tstrut \textbf{Euclid} (Redbook) & $\Omega_{c}$ & $\Omega_{b}$ & $n_{s}$ & $\ell\mathcal{A}_{s}$ & $h$ & $\mu$ & $\eta$ & $\Sigma$
& MG FoM  \tabularnewline
\hline 
\Tstrut Fiducial & {0.256} & {0.048} & {0.969} & {3.091} & {0.682} & {0.902} & {1.939} & {1.326} & relative\tabularnewline
\hline
\hline 
\Tstrut\textbf{GC(lin)} & 3.7\% & 19.2\% & 9.1\% & 16.6\% & 16.5\% & 16.7\% & 758\% & 489\%  
& 0 \tabularnewline
\Tstrut \textbf{GC(nl-HS)} & 1.1\% & 2.3\% & 1.3\% & 0.7\% & 1.6\% & 1.8\% & 7.9\% & 4.8\%
& 6.6 \tabularnewline 
\Tstrut \textbf{GC(nl-HS)+{\it Planck}}  & 0.8\% & 0.7\% & 0.3\% & 0.3\% & 0.3\% & 1.7\% & 7.6\% & 4.6\%
& 6.7 \tabularnewline
\hline 
\hline 
\Tstrut \textbf{WL(lin)}  & 12.1\% & 28.9\% & 11.3\% & 13.3\% & 24\% & 6.8\% & 11.1\% & 11.9\%
& 4.9 \tabularnewline 
\Tstrut \textbf{WL(nl-HS)}  & 6.5\% & 21.9\% & 6.6\% & 5.9\% & 15.8\% & 2.8\% & 8.0\% & 3.4\%
& 6.6  \tabularnewline
\Tstrut \textbf{WL(nl-HS)+{\it Planck}}  & 2.1\% & 1.3\% & 0.5\% & 0.9\% & 0.7\% & 2.2\% & 7.2\% & 2.9\%
& 7.2 \tabularnewline
\hline
\hline 
\Tstrut \textbf{GC+WL(lin)}  & 1.8\% & 6.6\% & 3.4\% & 5.6\% & 5.2\% & 3.0\% & 6.8\% & 3.4\%  
& 6.4  \tabularnewline
\Tstrut \textbf{GC+WL(lin)+{\it Planck}}  & 1.2\% & 0.9\% & 0.6\% & 2.3\% & 0.4\% & 2.4\% & 6.5\% & 2.8\%  
& 6.9  \tabularnewline
\hline
\hline 
\Tstrut \textbf{GC+WL(nl-HS)}  & 1.0\% & 2.2\% & 1.2\% & 0.7\% & 1.6\% & 1.3\% & 4.4\% & 1.9\% 
& 8.0   \tabularnewline
\Tstrut \textbf{GC+WL(nl-HS)+{\it Planck}}  & 0.8\% & 0.7\% & 0.3\% & 0.3\% & 0.3\% & 1.3\% & 4.4\% & 1.9\% 
& 8.2  \tabularnewline
\Tstrut \textbf{GC+WL(nl-Halofit)+{\it Planck} $\;$}  & 0.7\% & 0.7\% & 0.3\% & 1.3\% & 0.3\% & 0.9\% & 2.3\% & 1\% 
& 8.9  \tabularnewline
\hline
\end{tabular}\protect\caption{\label{tab:errors-Euclid-GC-WL-early_time}
Same as Table (\ref{tab:errors-Euclid-GC-WL-late_time}) but for the  early-time parameterization. Note that the last column (MG FoM) cannot be compared to the one for a different parameterization (Table \ref{tab:errors-Euclid-GC-WL-late_time}) since the reference value is different (GC(lin)) and the two parameterizations have a different number of primary parameters). In this case the absolute MG FoM of GC(lin) is $\approx -0.47$ nits.
}
\end{table}

\begin{table}[htbp]
\centering{}%
\begin{tabular}{|l|c|c|c|c|c||c|c|c|c|}
	\hline
	 & $\Omega_{c}$ & $\Omega_{b}$ & $n_{s}$ & $\ell\mathcal{A}_{s}$ & $h$ & $\mu$ & $\eta$ & $\Sigma$ & MG FoM \TBstrut\tabularnewline
\hline 
Fiducial & {0.254 } & {0.048 } & {0.969 } & {3.060 } & {0.682 } & {1.042 } & {1.719 } & {1.416 } & relative\TBstrut\tabularnewline
\hline 
\hline 
\textbf{GC(nl-HS)} &  &  &  &  &  &  &  & & \TBstrut\tabularnewline
Euclid   
& 0.9\% & 2.5\% & 1.3\% & 0.8\% & 1.7\% & 1.7\% & 475\% & 291\%
& 2.9
\tabularnewline
SKA1-SUR  
& 5\% & 15.3\% & 8.7\% & 3.8\% & 10.8\% & 18.1\% & 165\% & 108\%
& 1.7
\tabularnewline
SKA2  
& 0.5\% & 1.3\% & 0.4\% & 0.4\% & 0.8\% & 0.7\% & 86.8\% & 53.2\%
& 5.5
\tabularnewline
DESI-ELG   
& 1.6\% & 4.1\% & 2.3\% & 1.3\% & 2.9\% & 3.3\% & 899\% & 552\%
& 1.8
\tabularnewline
\hline 
\hline 
\textbf{WL(nl-HS)} &  &  &  &  &  &  &  &  & \TBstrut\tabularnewline
Euclid   
& 6.3\% & 20.7\% & 4.6\% & 5.8\% & 13.8\% & 23.3\% & 40.9\% & 4.6\%
& 4.5
\tabularnewline
SKA1  
& 30.8\% & 109\% & 35\% & 36.5\% & 77.6\% & 220\% & 405\% & 36.8\%
& 0.5
\tabularnewline
SKA2  
& 6\% & 22.5\% & 5.9\% & 6.8\% & 15.9\% & 19\% & 33.2\% & 3.7\%
& 4.9
\tabularnewline
\hline 
\hline 
\textbf{GC+WL(lin)} &  &  &  &  &  &  &  & & \Tstrut\tabularnewline
Euclid  
& 1.8\% & 5.9\% & 2.8\% & 2.3\% & 4.2\% & 7.1\% & 10.6\% & 2\%
& 6.6
\tabularnewline
SKA1  
& 10.1\% & 47.6\% & 25.4\% & 21.7\% & 40.4\% & 26.4\% & 28.8\% & 13.6\%
& 3.7
\tabularnewline
SKA2 
& 1.2\% & 4.5\% & 2.2\% & 1.9\% & 3.3\% & 4.1\% & 5.5\% & 1.6\%
& 7.5
\tabularnewline
\hline 
\hline 
\textbf{GC+WL(lin)+{\it Planck}} &  &  &  &  &  &  &  & & \Tstrut\tabularnewline
Euclid  
& 1.0\% & 0.7\% & 0.4\% & 0.4\% & 0.4\% & 6.2\% & 9.8\% & 1.5\%
& 6.9
\tabularnewline
SKA1  
& 2.4\% & 1.2\% & 0.4\% & 1.2\% & 0.7\% & 12\% & 19.8\% & 3.8\%
& 5.3
\tabularnewline
SKA2 
& 0.7\% & 0.6\% & 0.3\% & 0.4\% & 0.3\% & 3.6\% & 5.2\% & 1.2\%
& 7.8
\tabularnewline
\hline 
\hline 
\textbf{GC+WL(nl-HS)} &  &  &  &  &  &  &  & & \Tstrut\tabularnewline
Euclid  
& 0.8\% & 2.2\% & 0.8\% & 0.7\% & 1.5\% & 1.6\% & 2.4\% & 1.0\%
& 8.7
\tabularnewline
SKA1  
& 4.7\% & 14.3\% & 6.2\% & 3.6\% & 9.6\% & 12.8\% & 11\% & 7.3\%
& 5.5
\tabularnewline
SKA2 
& 0.4\% & 1.3\% & 0.3\% & 0.4\% & 0.8\% & 0.7\% & 0.9\% & 0.6\%
& 10.3
\tabularnewline
\hline
\hline
\textbf{GC+WL(nl-HS)+{\it Planck}} &  &  &  &  &  &  &  & & \Tstrut\tabularnewline
Euclid  
& 0.7\% & 0.6\% & 0.2\% & 0.2\% & 0.3\% & 1.6\% & 2.4\% & 0.9\%
& 8.9
\tabularnewline 
SKA1  
& 2.0\% & 1.0\% & 0.4\% & 0.8\% & 0.6\% & 3.5\% & 6\% & 2.7\%
& 6.9
\tabularnewline 
SKA2 
& 0.4\% & 0.5\% & 0.2\% & 0.1\% & 0.2\% & 0.6\% & 0.9\% & 0.5\%
& 10.3
\tabularnewline
\hline
\end{tabular}\protect\caption{\label{tab:errors-GC-SKAcompare-MG-DE-mu-eta-sigma}
1$\sigma$ fully
marginalized errors on the cosmological parameters
$\{\Omega_{m},\Omega_{b},h,\ell \mathcal{A}_{s},n_{s},\mu,\eta, \Sigma\}$
in the late-time parameterization comparing different surveys in the linear and non-linear case. 
In the last column, we show for each observation the Modified Gravity Figure of Merit (MG FoM) relative to our base observable, which is the Euclid Redbook GC(linear), see Table \ref{tab:errors-Euclid-GC-WL-late_time}.
We can see that in general terms, SKA2 is the most powerful survey, followed by Euclid and SKA1.
In the case of GC alone, DESI-ELG is more constraining than SKA1-SUR.
Notice that in this parameterization, a GC survey would only constrain $\mu$ with
a high accuracy, while a WL survey would constrain $\Sigma$ with
a very good accuracy. The combination of both is much more
powerful than the single probes. Adding \planck\ priors (last row) improves considerably the
constraints on the $\lcdm$ parameters but has an almost negligible effect on the MG parameters (MG FoM remain almost constant when adding \planck\ priors to the GC+WL (non-linear HS) case. The marginalized contours for the $\mu$-$\eta$ plane , comparing these surveys, can be seen in the left panel ofFig.\ref{fig:combined_surveys}.
}
\end{table}

\begin{table}[htbp]
\centering{}%
\small
\begin{tabular}{|l|c|c|c|c|c||c|c|c|c|}
\hline
	 & $\Omega_{c}$ & $\Omega_{b}$ & $n_{s}$ & $\ell\mathcal{A}_{s}$ & $h$ & $\mu$ & $\eta$ & $\Sigma$  & MG FoM \TBstrut\tabularnewline
\hline 
Fiducial & {0.256} & {0.0485} & {0.969} & {3.091} & {0.682} & {0.902} & {1.939} & {1.326} & relative \TBstrut\tabularnewline 
\hline 
\hline 
\textbf{GC(nl-HS)} &  &  &  &  &  &  &  & & \TBstrut\tabularnewline
Euclid   
& 1.1\% & 2.3\% & 1.3\% & 0.7\% & 1.6\% & 1.8\% & 7.9\% & 4.8\%
& 6.6
\tabularnewline
SKA1-SUR  
& 7.9\% & 14.2\% & 13.4\% & 4.2\% & 11\% & 12.6\% & 82.7\% & 52.6\%
& 2.2
\tabularnewline
SKA2  
& 0.6\% & 1.3\% & 0.7\% & 0.4\% & 0.9\% & 0.9\% & 3.4\% & 1.8\%
& 8.3
\tabularnewline
DESI-ELG   
& 2.0\% & 4.3\% & 2.7\% & 1.4\% & 3.0\% & 8.2\% & 32\% & 28.6\%
& 4.3
\tabularnewline
\hline 
\hline 
\textbf{WL(nl-HS)} &  &  &  &  &  &  &  &  & \TBstrut\tabularnewline
Euclid   
& 6.5\% & 21.9\% & 6.6\% & 5.9\% & 15.8\% & 2.8\% & 8.0\% & 3.4\%
& 6.6
\tabularnewline
SKA1  
& 32\% & 106\% & 37.2\% & 33\% & 79.3\% & 13.1\% & 37.1\% & 16.4\%
& 3.4
\tabularnewline
SKA2  
& 5.9\% & 22.1\% & 6.7\% & 6.1\% & 16.1\% & 2.4\% & 7.0\% & 2.9\%
& 6.9
\tabularnewline
\hline 
\hline 
\textbf{GC+WL(lin)} &  &  &  &  &  &  &  & & \Tstrut\tabularnewline
Euclid  
& 1.8\% & 6.6\% & 3.4\% & 5.6\% & 5.2\% & 3.0\% & 6.8\% & 3.4\%
& 6.4
\tabularnewline
SKA1  
& 10.3\% & 46.4\% & 24.2\% & 33.6\% & 40.2\% & 14.4\% & 29.6\% & 15.5\%
& 3.3
\tabularnewline
SKA2 
& 1.3\% & 4.9\% & 2.5\% & 4.2\% & 3.9\% & 2.5\% & 5.7\% & 2.7\%
& 6.8
\tabularnewline
\hline 
\hline 
\textbf{GC+WL(lin)+{\it Planck}} &  &  &  &  &  &  &  & & \Tstrut\tabularnewline
Euclid  
& 1.2\% & 0.9\% & 0.6\% & 2.3\% & 0.4\% & 2.4\% & 6.5\% & 2.8\%
& 6.8
\tabularnewline
SKA1  
& 2.5\% & 1.5\% & 0.8\% & 2.9\% & 0.8\% & 8.8\% & 22.2\% & 8.5\%
& 4.5
\tabularnewline
SKA2 
& 0.9\% & 0.7\% & 0.6\% & 2.1\% & 0.3\% & 2.1\% & 5.4\% & 2.3\%
& 7.2
\tabularnewline
\hline 
\hline 
\textbf{GC+WL(nl-HS)} &  &  &  &  &  &  &  & & \Tstrut\tabularnewline
Euclid  
& 1.0\% & 2.2\% & 1.2\% & 0.7\% & 1.6\% & 1.3\% & 4.4\% & 1.9\%
& 8.1
\tabularnewline
SKA1  
& 7.1\% & 13.4\% & 10.7\% & 4\% & 10\% & 8.2\% & 24.4\% & 10.5\%
& 4.4
\tabularnewline
SKA2 
& 0.6\% & 1.3\% & 0.7\% & 0.4\% & 0.9\% & 0.8\% & 2.7\% & 1.3\%
& 8.8
\tabularnewline
\hline
\hline
\textbf{GC+WL(nl-HS)+{\it Planck}} &  &  &  &  &  &  &  & & \Tstrut\tabularnewline
Euclid  
& 0.8\% & 0.7\% & 0.3\% & 0.3\% & 0.3\% & 1.3\% & 4.4\% & 1.9\%
& 8.1
\tabularnewline 
SKA1  
& 2.1\% & 1.3\% & 0.5\% & 0.9\% & 0.7\% & 7\% & 20.8\% & 8.2\%
& 4.9
\tabularnewline 
SKA2 
& 0.5\% & 0.5\% & 0.3\% & 0.2\% & 0.2\% & 0.8\% & 2.7\% & 1.3\%
& 8.8
\tabularnewline
\hline
\end{tabular}
\caption{\label{tab:errors-GC-SKAcompare-MG-TR-mu-eta-sigma-Zhao-1}
Same as Table \ref{tab:errors-GC-SKAcompare-MG-DE-mu-eta-sigma} but for the early-time parameterization.
The last 4 columns correspond to the
projection of the errors on $E_{11}$ and $E_{22}$ onto $\mu,\;\eta$ and $\Sigma$, respectively. 
We have marginalized over $E_{12}$
and $E_{21}$ since at $z=0$ they don't contribute to the Modified
Gravity parameters. Notice that in this
parameterization, a GC survey alone is able to constrain both $\mu$ and $\Sigma$ to a good level for all surveys, better than with the late time parameterization, more often used in literature. The combination of GC+WL is however less constraining in the early time parametrization than in late time parameterization one. The reference case for the MG FoM is the Euclid (Redbook) GC linear forecast (Table \ref{tab:errors-Euclid-GC-WL-early_time}).
The non-linear forecast for GC+WL+{\it Planck} would yield, for Euclid and SKA2, contraints at the 1-2\% accuracy on $\mu$, $\Sigma$, while
for SKA1 the contraints would be at the 8\% level. The marginalized contours for the $\mu$-$\eta$ plane , comparing these surveys, can be seen in the right panel ofFig.\ref{fig:combined_surveys}.
}
\end{table}

\subsubsection{Weak Lensing in the linear and mildly non-linear regime}

Using the Euclid Weak Lensing specifications described in Section
sub:Fisher-Weak-Lensing, we obtain the results displayed in
the middle panel of Table \ref{tab:errors-Euclid-GC-WL-late_time}. 
Also in this case, we use the late-time parameterization, in three different cases: 
the first uses only linear quantities, with a maximum multipole of $\ell_{\rm max}=1000$; 
the second case uses the non-linear HS prescription of section \ref{sub:Prescription-HS} 
up to a maximum multipole of $\ell_{\rm max}=5000$; the third case combines Weak Lensing with \planck\ priors, 
as described in \ref{sub:Fisher-Planck}.

In the linear case, WL forecast yields constraints on the standard $\lcdm$ parameters
at around 10\% of accuracy, with the exception of $\Omega_{b}$ which
is poorly constrained at around 26\%, and the expansion rate h (19\%). This is likely due to the fact
that WL is only directly sensitive to the total matter distribution
in the Universe and cannot differentiate baryons from dark matter. All constraints improve when adding non-linear information, with $\Omega_b$ and h still constrained only at the level of about $20\%$. When the \planck\ priors are included, though, constraints shrink down to about 1$\%$ for all cosmological parameters.
The Modified Gravity parameters show the expected
trend; in the linear case, only $\Sigma$ is constrained, at 11$\%$, as this parameter is directly defined in terms of the lensing potential $\Phi+\Psi$; a Weak Lensing
probe is however not directly sensitive to $\mu$ and $\eta$ separately, as can be seen in Fig.\ \ref{fig:DE+Planck-ellipses-mu-sig-eta}.
The linear FoM is only slightly weaker than the one from GC for this parameterization, probably because both probes have an effectively unconstrained degeneracy direction. When adding non-linear information, errors on $\mu$ and $\eta$ improve, though still remaining in a poorly constrained interval (25$\%$-44$\%$).
Already on its own, however, Weak Lensing could rule out many
models of Modified Gravity that change $\Sigma$ at more than $5\%$ (or even $2.9\%$, if we include \planck\ priors). The combination GC+{\it Planck} 
and WL+{\it Planck} has a comparable overall constraining power, with GC+{\it Planck} being about 1 nit stronger and providing smaller errors on $\mu$.

\subsubsection{Combining Weak Lensing and Galaxy Clustering}\label{subsub:late-time-comb-GC+WL}

\begin{figure}[htbp]
\begin{centering}
\includegraphics[width=0.3\textwidth]{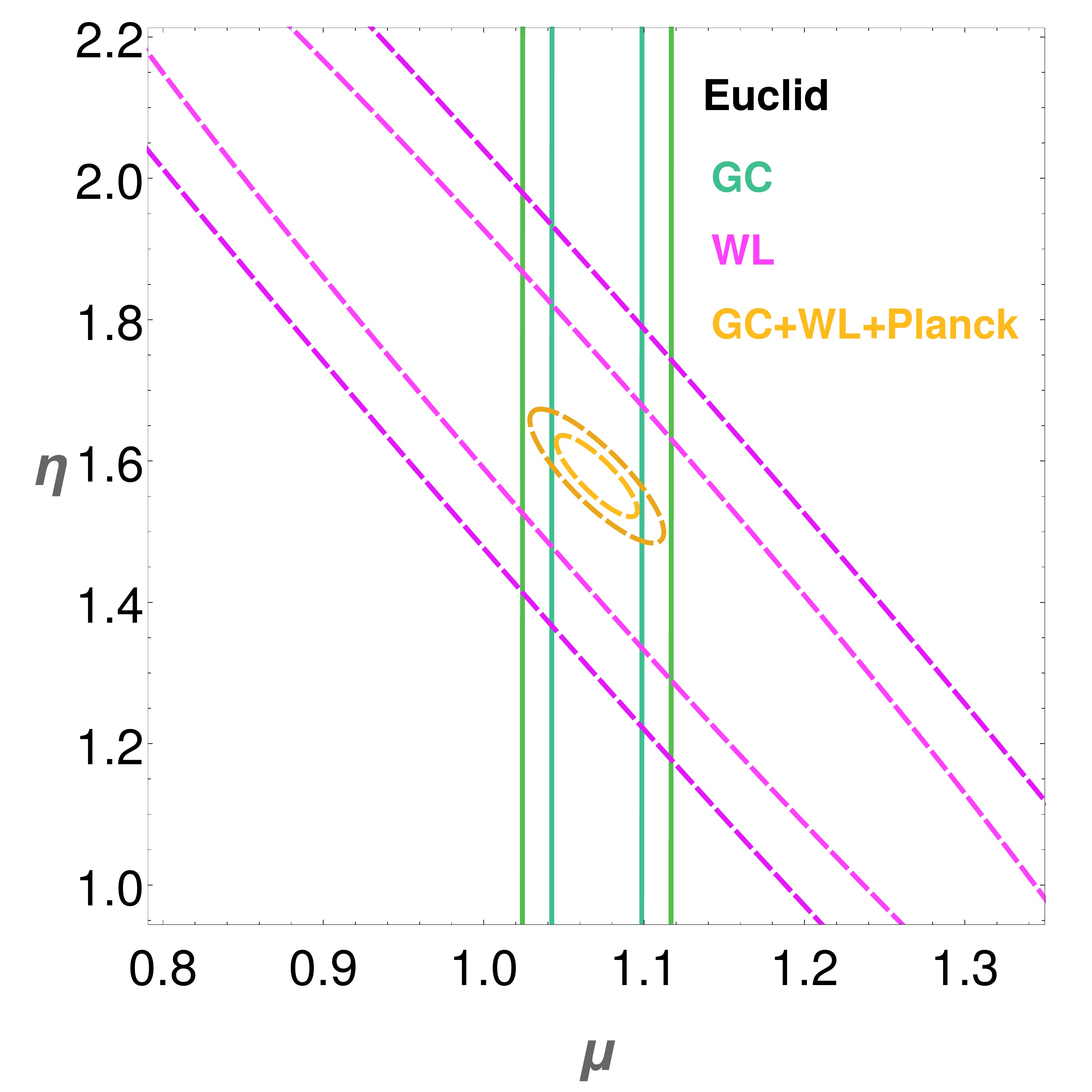}
\includegraphics[width=0.3\textwidth]{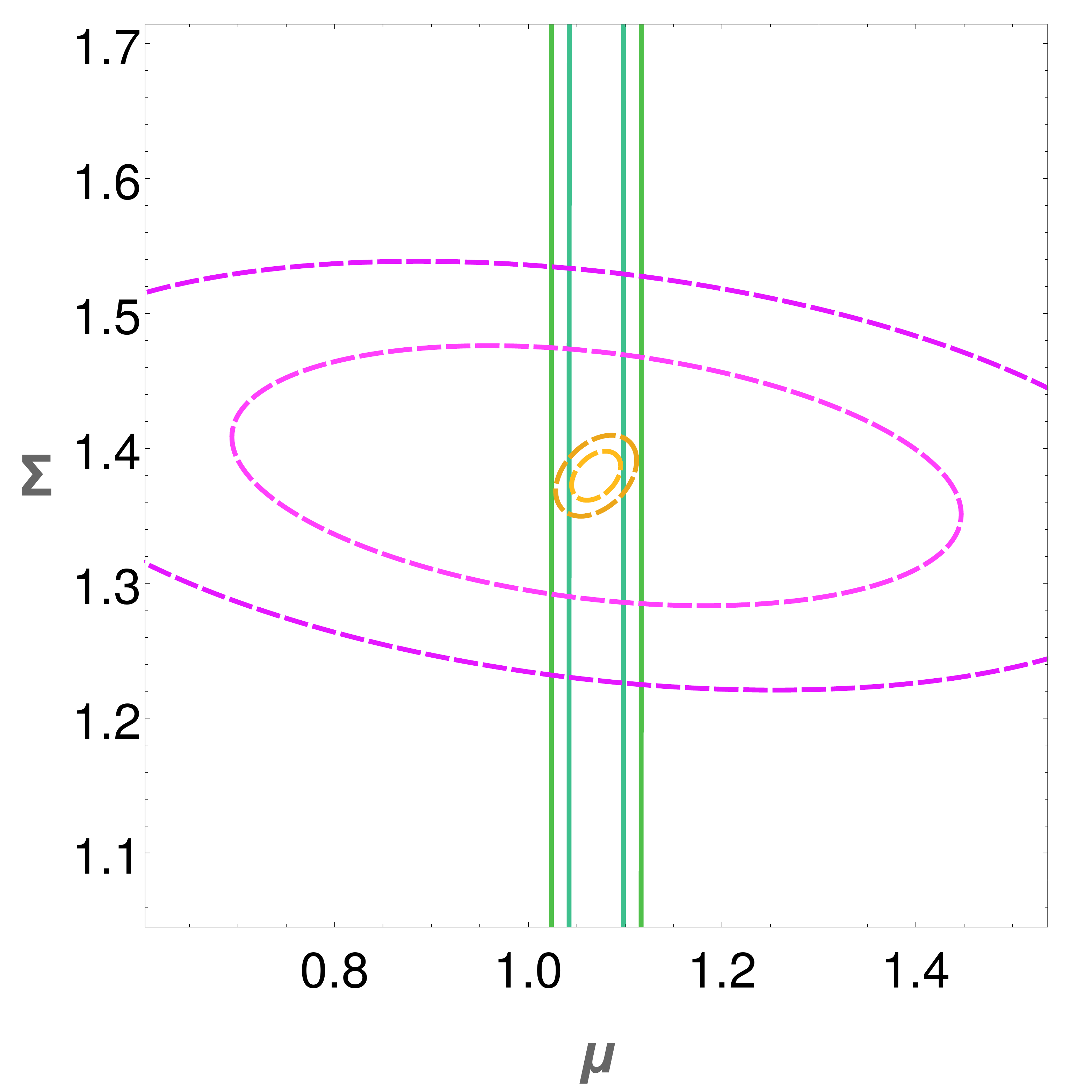} 
\end{centering}

\begin{centering}
\includegraphics[width=0.3\textwidth]{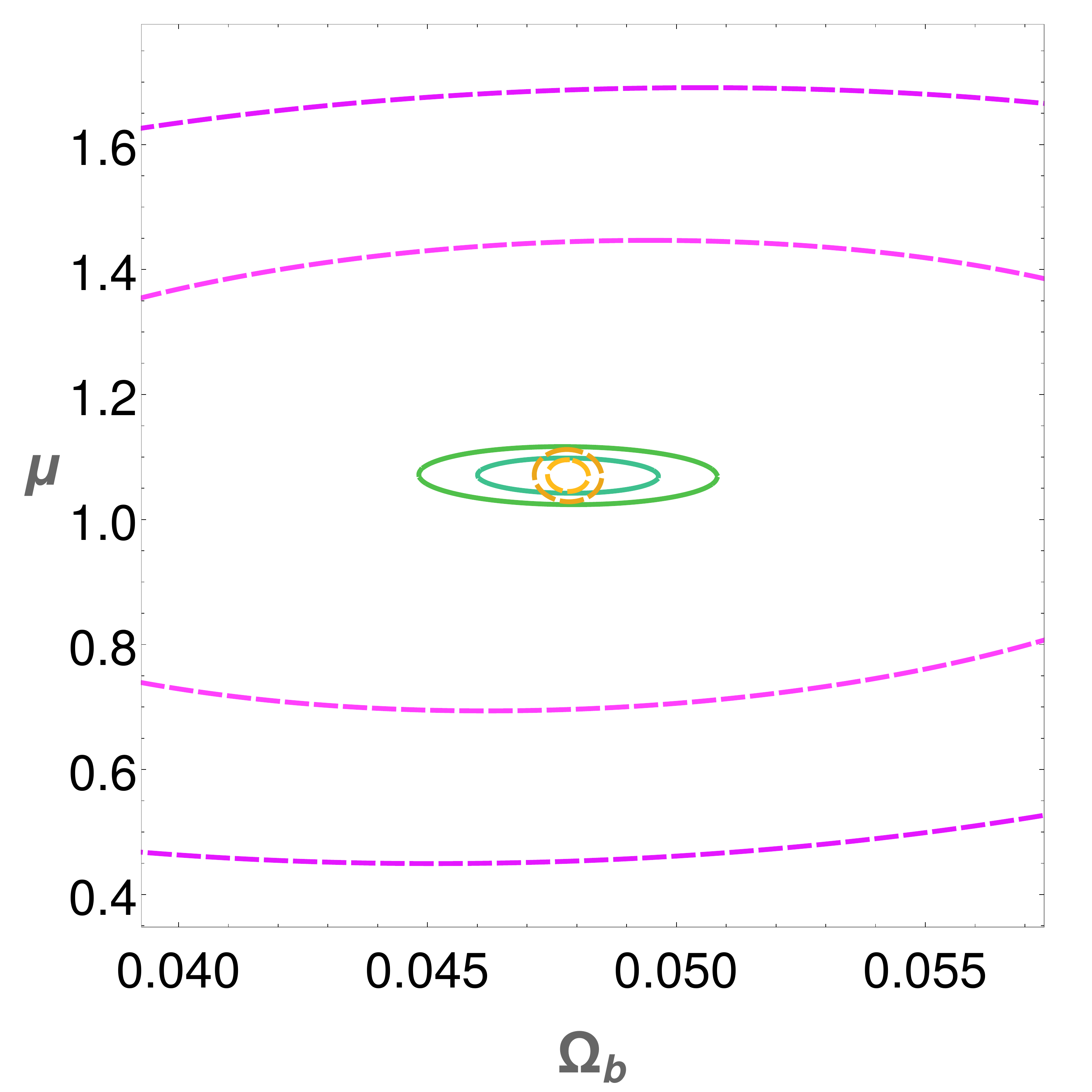}
\includegraphics[width=0.3\textwidth]{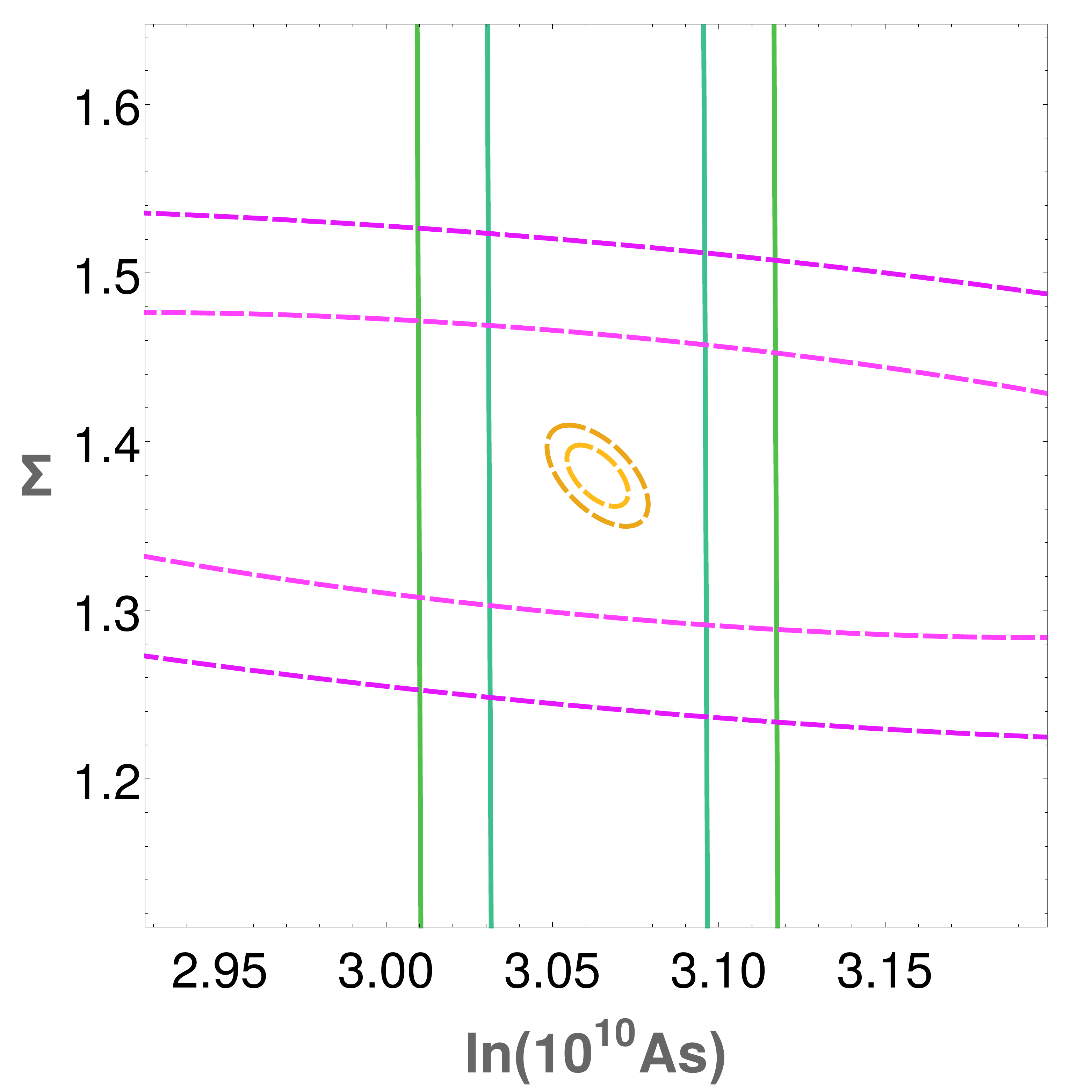} 
\end{centering}

\caption{\label{fig:DE+Planck-ellipses-mu-sig-eta} 
Fisher Matrix marginalized contours
(1, 2 $\sigma$) for the Euclid space mission in the late-time parameterization
using mildly non-linear scales and the HS prescription. Green lines
represent constraints from a Galaxy Clustering survey, pink lines
stand for the Weak Lensing observables, and orange lines represent the GC+WL+{\it Planck} combined confidence
regions. \textbf{Upper Left: }contours for the fully marginalized
errors on $\eta$ and $\mu$. \textbf{Upper Right: }contours for the
fully marginalized errors on $\Sigma$ and $\mu$.\textbf{ Lower Left:
}contours for the fully marginalized errors on $\mu$ and $\Omega_{b}$.
\textbf{Lower Right: }contours for the fully marginalized errors on
$\Sigma$ and $\ln(10^10 A_{s})$. The fact that the combination of GC, WL and \planck\
breaks many degeneracies in the 7-dimensional parameter space, explains why
the combined contours (yellow) have a much smaller area. Notice that in this parametrization, 
GC measures mostly $\mu$ and WL constrains mostly just $\Sigma$.}
\end{figure}

After using the two primary probes from Euclid
separately, we discuss here the constraints obtained combining Weak
Lensing and Galaxy Clustering. 
The combination between GC and
WL can be seen in the bottom panel of Table
\ref{tab:errors-Euclid-GC-WL-late_time}.
In the late-time parameterization, in the linear case, Weak Lensing combined with a galaxy
clustering for a Euclid survey (Redbook specifications) constrains the standard $\Lambda$CDM parameters
in the range $2 \% - 6 \%$, and below 1$\%$ when \planck\ priors are included. Modified Gravity parameters $\mu$ and $\eta$ are now also constrained below 10$\%$, reaching $1\%$ when adding non-linear scales.
The remarkable improvement can be attributed
to the fact that the combination of GC and WL Fisher matrices breaks
many degeneracies in the parameter space. This is shown in Figure
\ref{fig:DE+Planck-ellipses-mu-sig-eta}, where it is possible to notice
how the two probes are almost orthogonal both in the $\mu$-$\eta$-
and $\mu$-$\Sigma$-plane. Weak Lensing measures
the changes in the Weyl potential, parametrized by $\Sigma$, while $\mu$ is related to the Poisson equation, and therefore to the potential $\Psi$, modified by peculiar velocities and sensitive to Galaxy Clustering; $\eta$ can also be written as a combination of $\mu$ and $\Sigma$ (see Eqn.\ \ref{eq:SigmaofMuEta}). 

Further improvement is brought by the sensitivity of Galaxy Clustering to standard $\Lambda$CDM parameters; even
though GC constraints on $\Sigma$ and $\eta$ are not as good as the ones for Weak Lensing, the better measurement of standard parameters provided by Galaxy Clustering breaks degeneracies in the Modified Gravity sector of the parameter space, leading to
narrower bounds for $\eta$ and $\Sigma$ with respect to both probes taken separately. WL is instead not sensitive to modifications of the Poisson equation for matter and this explains why constraints on $\mu$ are not improved
by the combination of the two probes, but are rather dominated by
GC. 
The correlation among parameters can also help us explain the observed results. 
The Figure of Correlation, defined in Eq.\ (\ref{eq:FoC}), for GC (non-linear HS) alone is 4.9, while
for WL (non-linear HS) the correlation is higher, with $\textrm{FoC}=16.9$.
When combining both probes (GC+WL (non-linear HS)) the FoC goes to an intermediate point of 7.6.

Given the constraining power of the GC+WL combination on MG functions, adding the \planck\ priors does not lead to significant improvements on the dark energy related parameters. On the other hand, standard parameters significantly benefit
from the inclusion of CMB and background priors and we can expect
this to be a relevant factor for MG models with degeneracies
with $\Lambda$CDM parameters, e.g. models affecting also the expansion
history of the universe.
An overview of the constraints on Modified Gravity described in this section is shown in Fig.\ref{fig:DE+Planck-ellipses-mu-sig-eta}, with Euclid GC, Euclid WL and Euclid GC+WL combined with \planck\ priors.

\subsubsection{Forecasts in Modified Gravity for SKA1, SKA2 and DESI \label{subsub: other-surveys-late-time}}

\begin{figure}[htbp]
\begin{centering}
\includegraphics[width=0.45\textwidth]{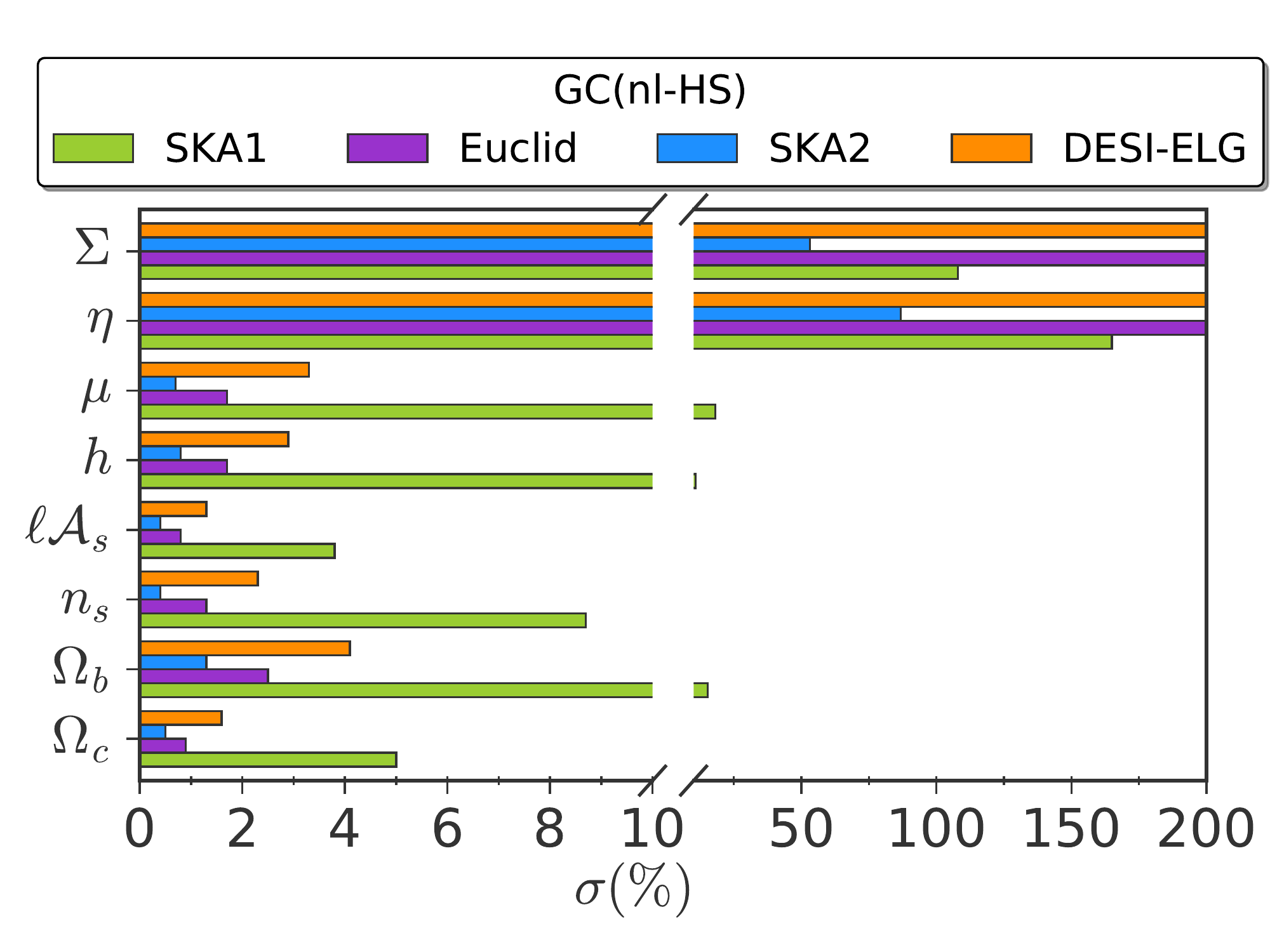}\hspace{-0.5pt}
\includegraphics[width=0.45\textwidth]{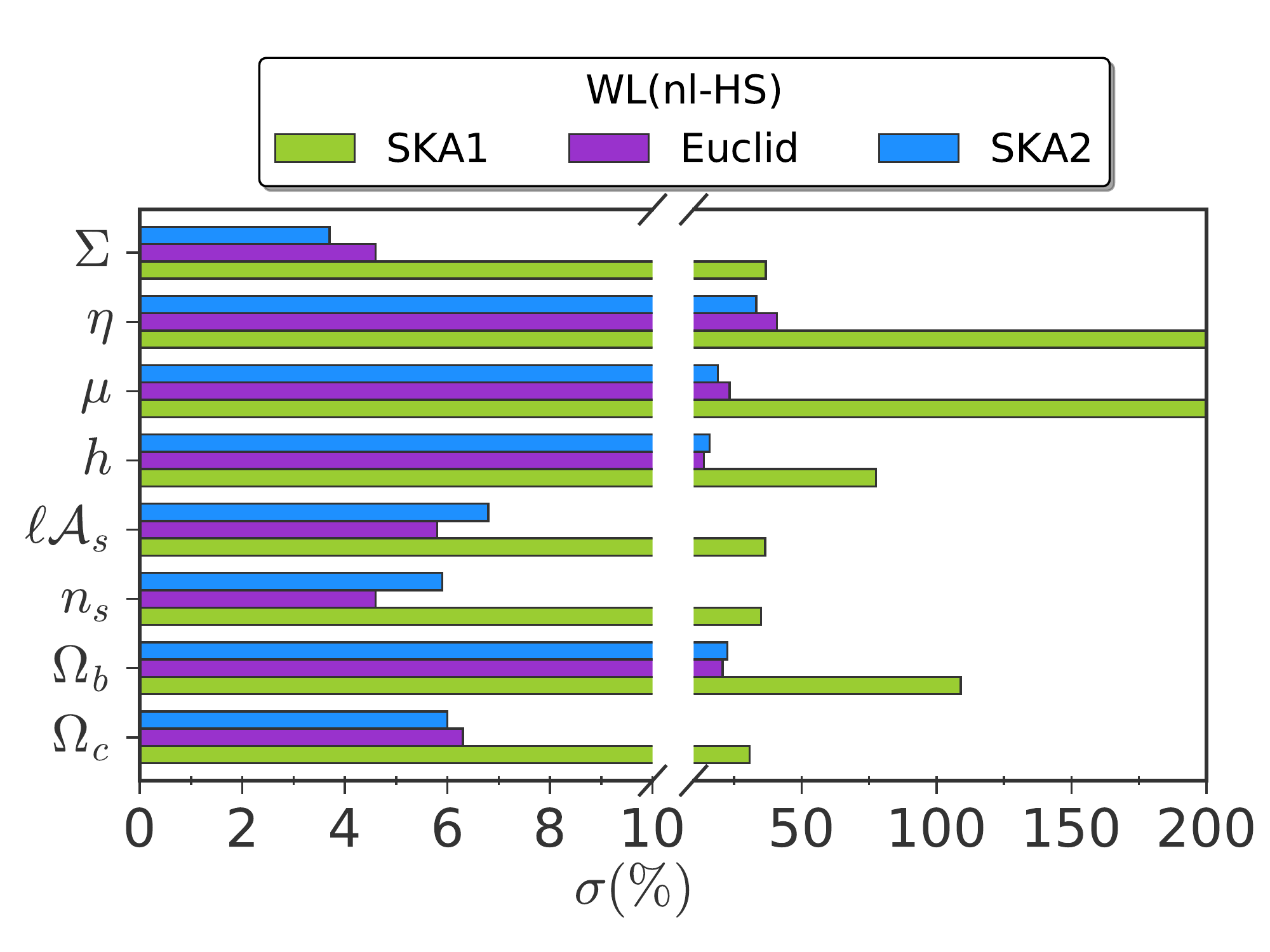}

\includegraphics[width=0.45\textwidth]{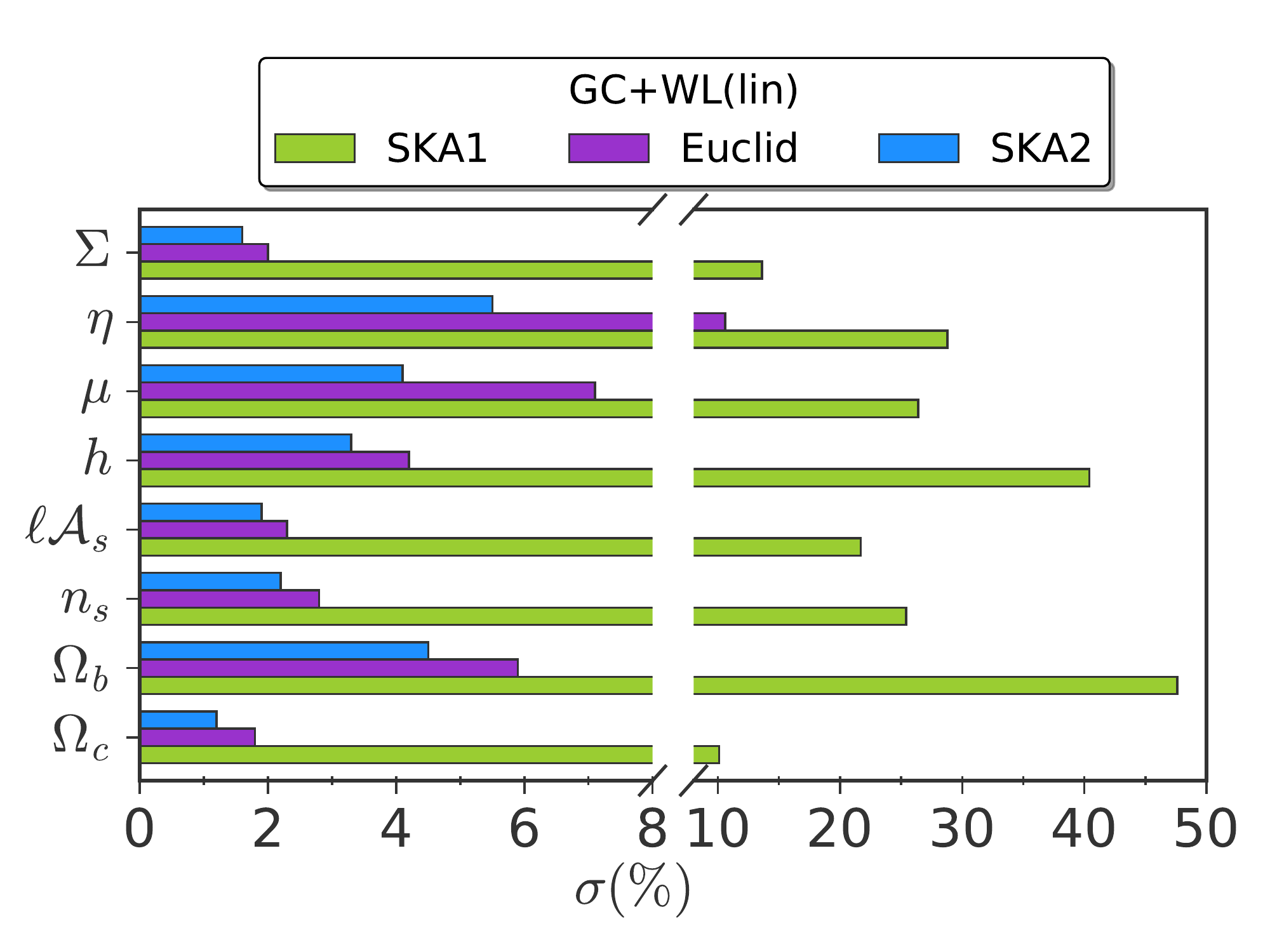}\hspace{-0.5pt}
\includegraphics[width=0.45\textwidth]{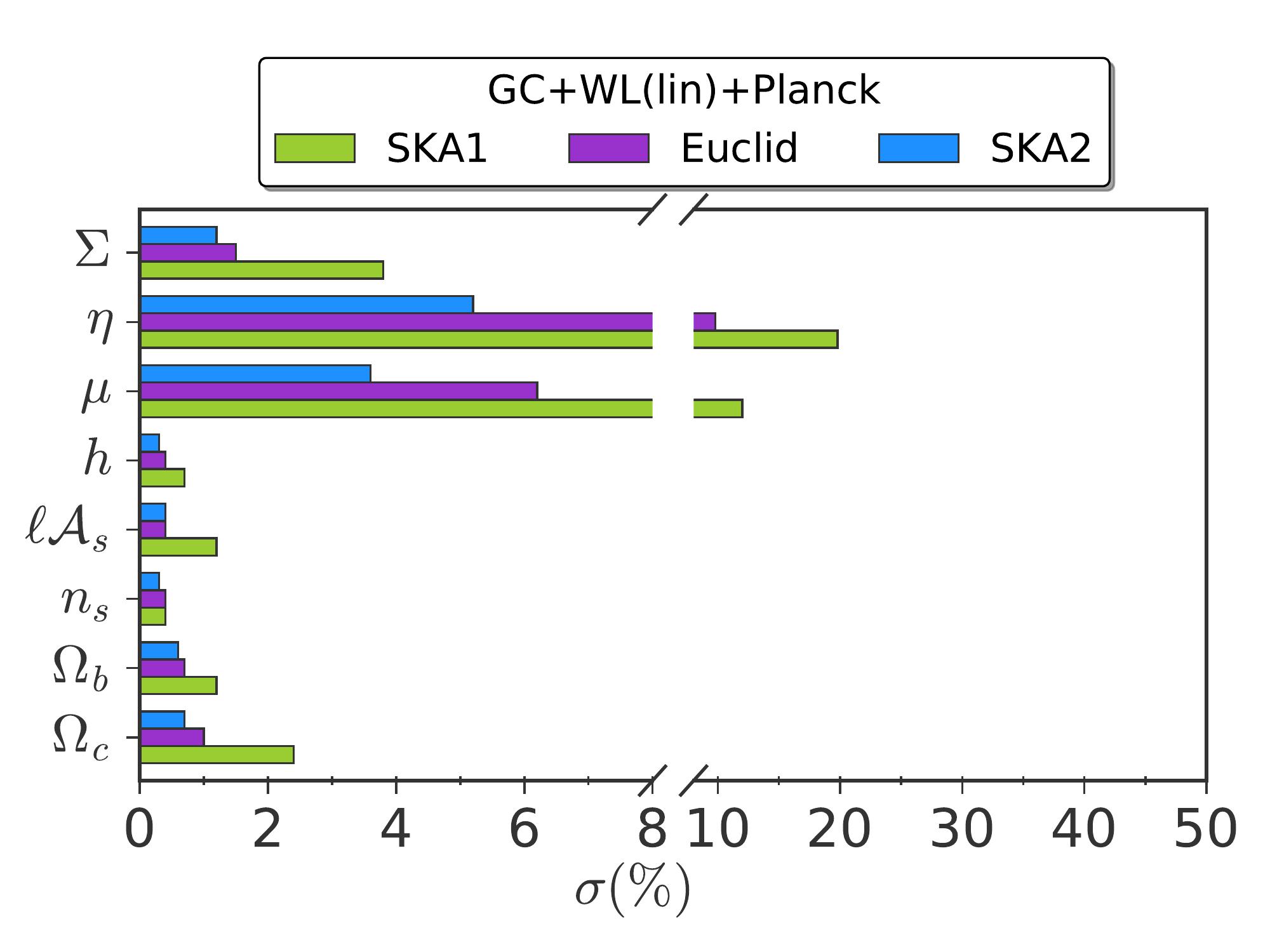}

\includegraphics[width=0.45\textwidth]{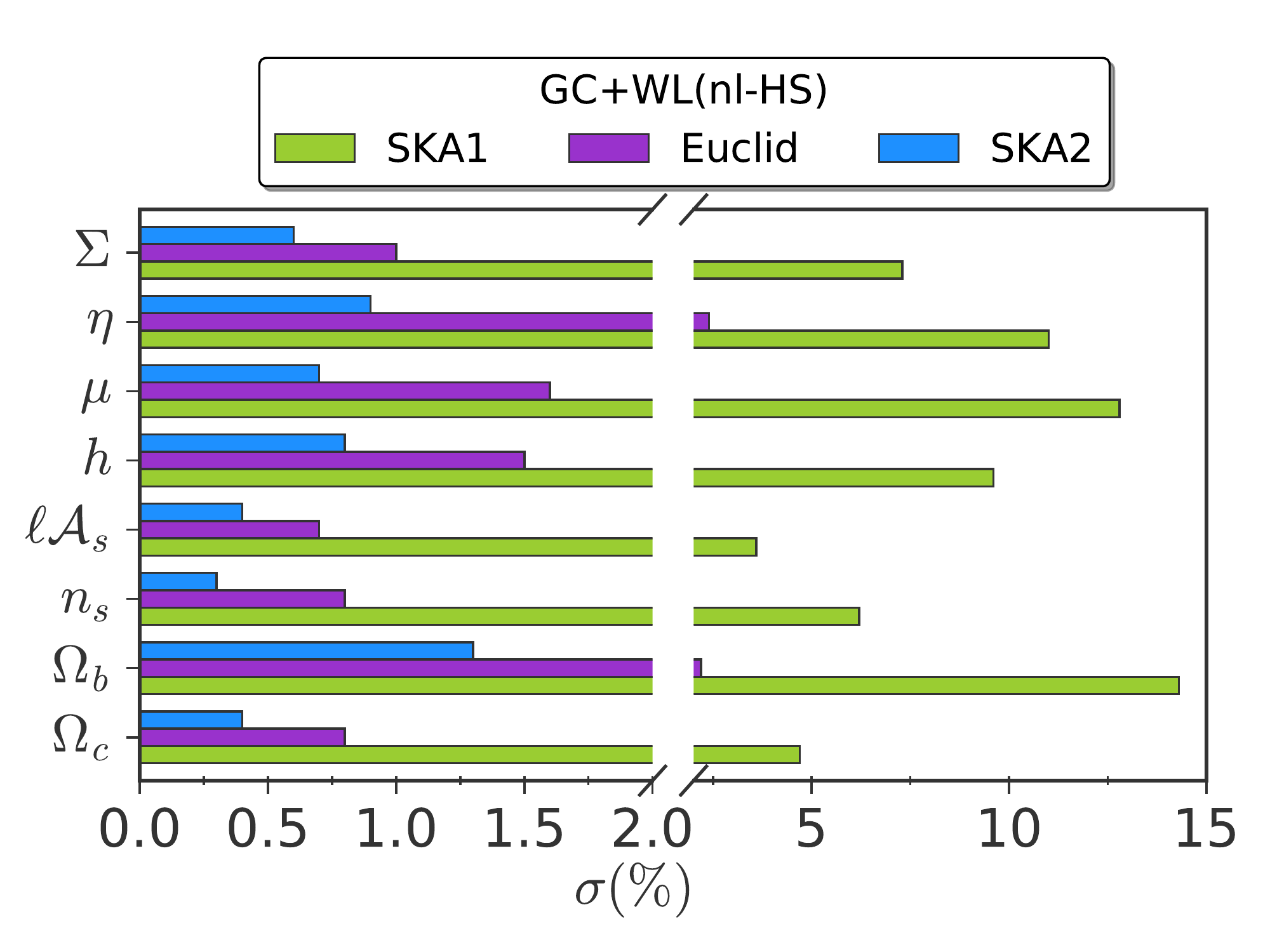}\hspace{-0.5pt}
\includegraphics[width=0.45\textwidth]{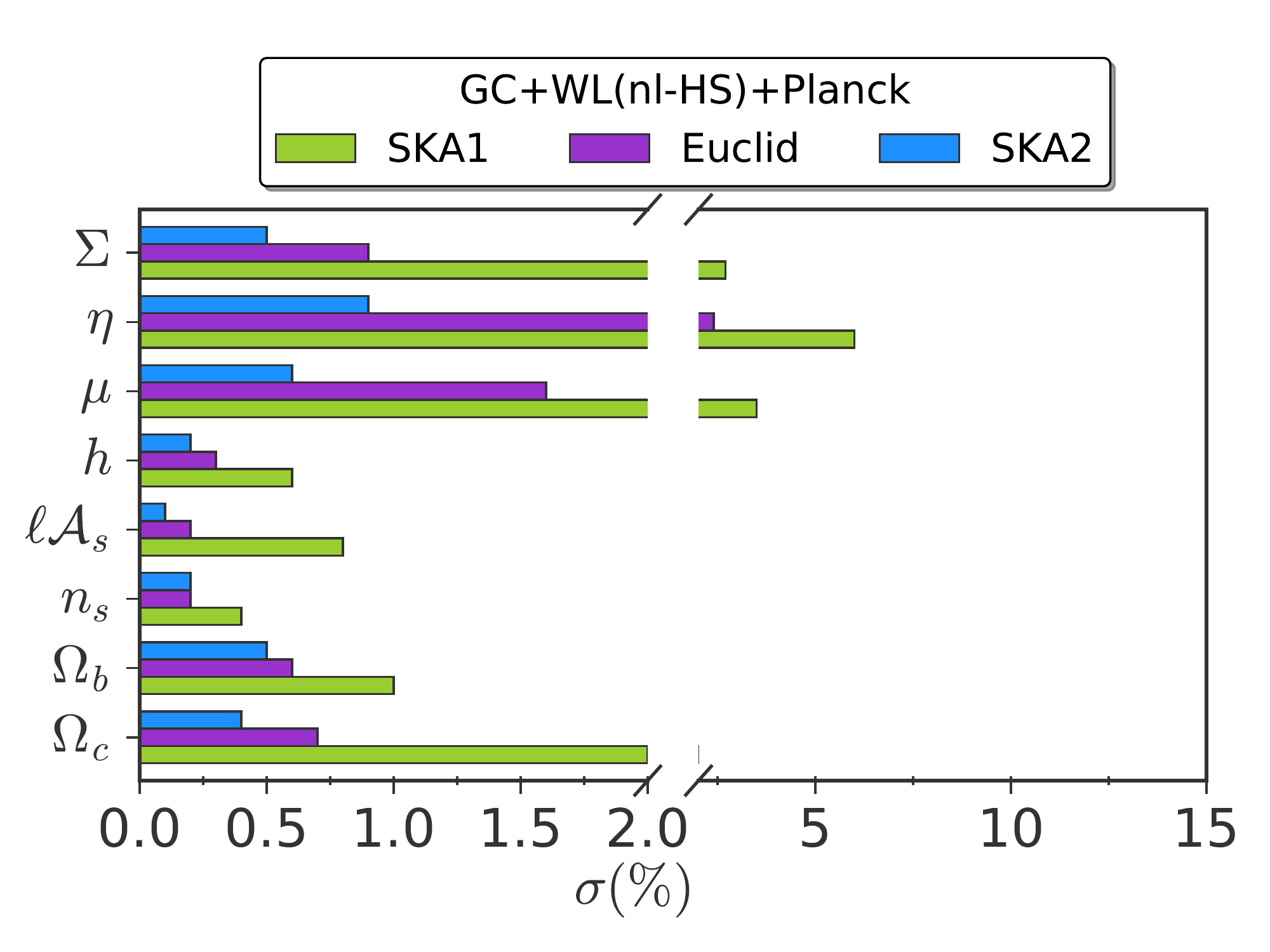}
\par\end{centering}
\caption{\label{fig:BarPlot-DE-GC-1}1$\sigma$ fully marginalized errors
on the parameters $\{\Omega_{m},\Omega_{b},h,\ell\mathcal{A}_{s},n_{s},\mu,\eta,\Sigma\}$ 
for the late-time parameterization of MG obtained by forecasts on
Galaxy Clustering (non-linear HS) (top left panel), Weak Lensing (non-linear HS) (top right panel), the 
combinations GC+WL (linear) (middle left) and GC+WL+{\it Planck} (linear) (middle right) and
the combinations GC+WL (non-linear HS) (bottom left) and GC+WL+{\it Planck} (non-linear) (bottom right).
In the GC case, the surveys
considered are SKA2 (blue), SKA1-SUR (green), Euclid Redbook (purple) and DESI-ELG (orange). For forecasts including WL, only Euclid, SKA1 and SKA2 are included.
Although the 1$\sigma$ constraints on the standard parameters are overall weaker for WL than for GC, Weak Lensing surveys perform better on Modified Gravity parameters. Comparing the different surveys,
Euclid and SKA2 perform similarly well for the WL observable alone, if non-linearities are included.
Notice that SKA1-SUR performs better than Euclid on the $\eta$ and $\Sigma$ parameters, because it can measure better at lower redshifts.
Including the \planck\ prior, the GC+WL combination for Euclid and SKA2 constrains all parameters at much better than percent accuracy. 
Detailed specifications of the different surveys are explained in the text.
}
\end{figure}

For the SKA1 and SKA2 surveys (whose specifications are explained in detail in Section \ref{sub:FutureSurveys}), 
previous work on forecasting cosmological parameters has been done, among others, by \cite{baker_observational_2015} and \cite{bull_extending_2015}. 
In the latter work, the author parameterizes the evolution of $\mu(a)$ using the late-time parameterization, but also adds an extra parameter
allowing for a scale dependence in $\mu(a)$ and including a \planck\ prior. For a fixed scale,
the 1$\sigma$ errors on the amplitude of $\mu$ lie between 0.045 and 0.095, depending on the details of the SKA1 specifications, while  
for SKA2, this error is of about 0.017. This setting would correspond to our GC+WL(linear) + \planck\ case 
(see Table \ref{tab:errors-GC-SKAcompare-MG-DE-mu-eta-sigma}) where we find for SKA1 a 1$\sigma$ error on $\mu$ of 0.12 and for SKA2 the forecasted error is 0.036.
Our errors are somewhat larger, but we also have extended our analysis to let the gravitational slip $\eta$ be different from 1 at present time, our departure from $\mu=1$ at present time is larger by a factor 4 and 
our linear forecast is conservative in the sense that it includes less wavenumbers $k$ at higher redshifts, compared to theirs.

\begin{figure}[htbp]
\centering
\includegraphics[width=0.35\linewidth]{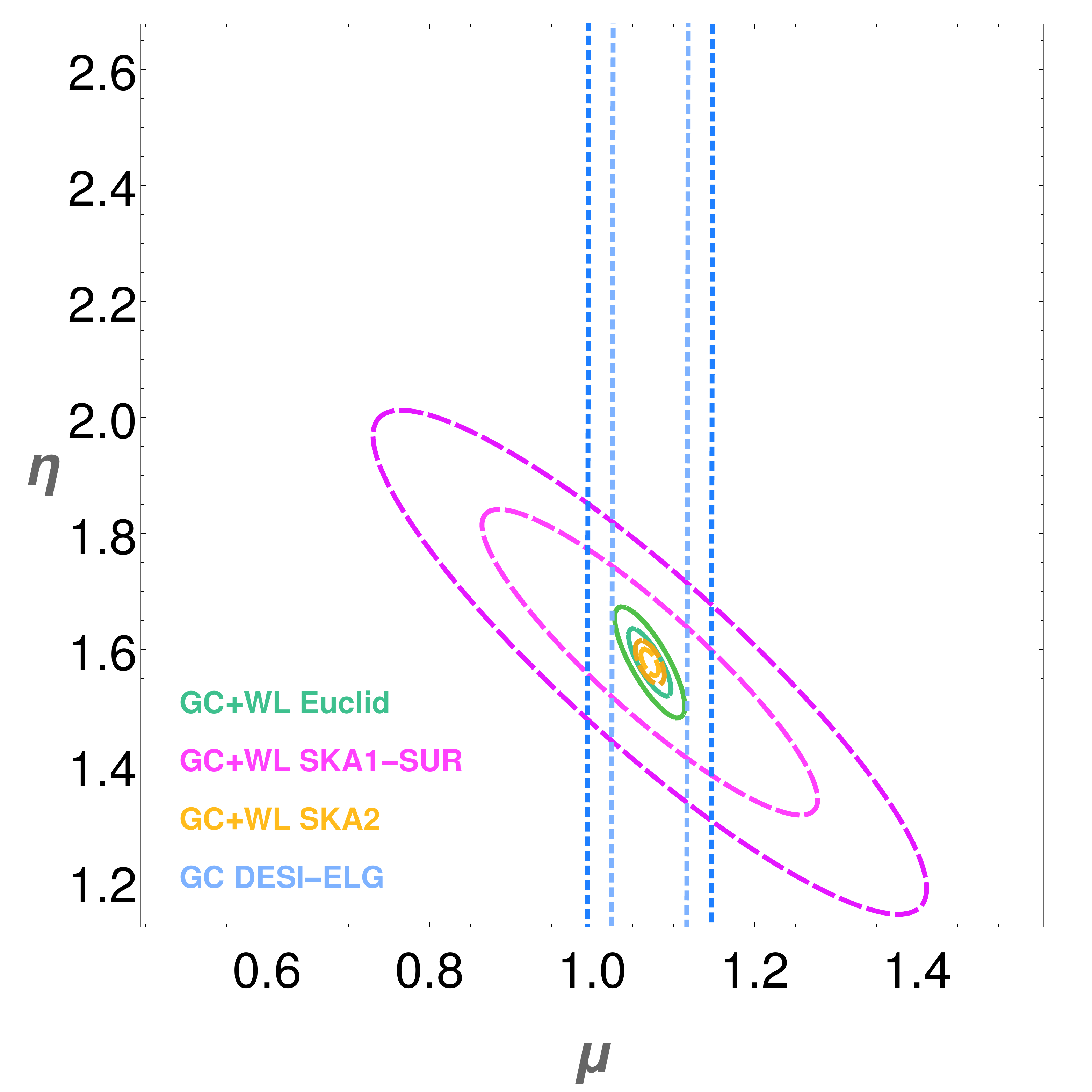}
\includegraphics[width=0.35\linewidth]{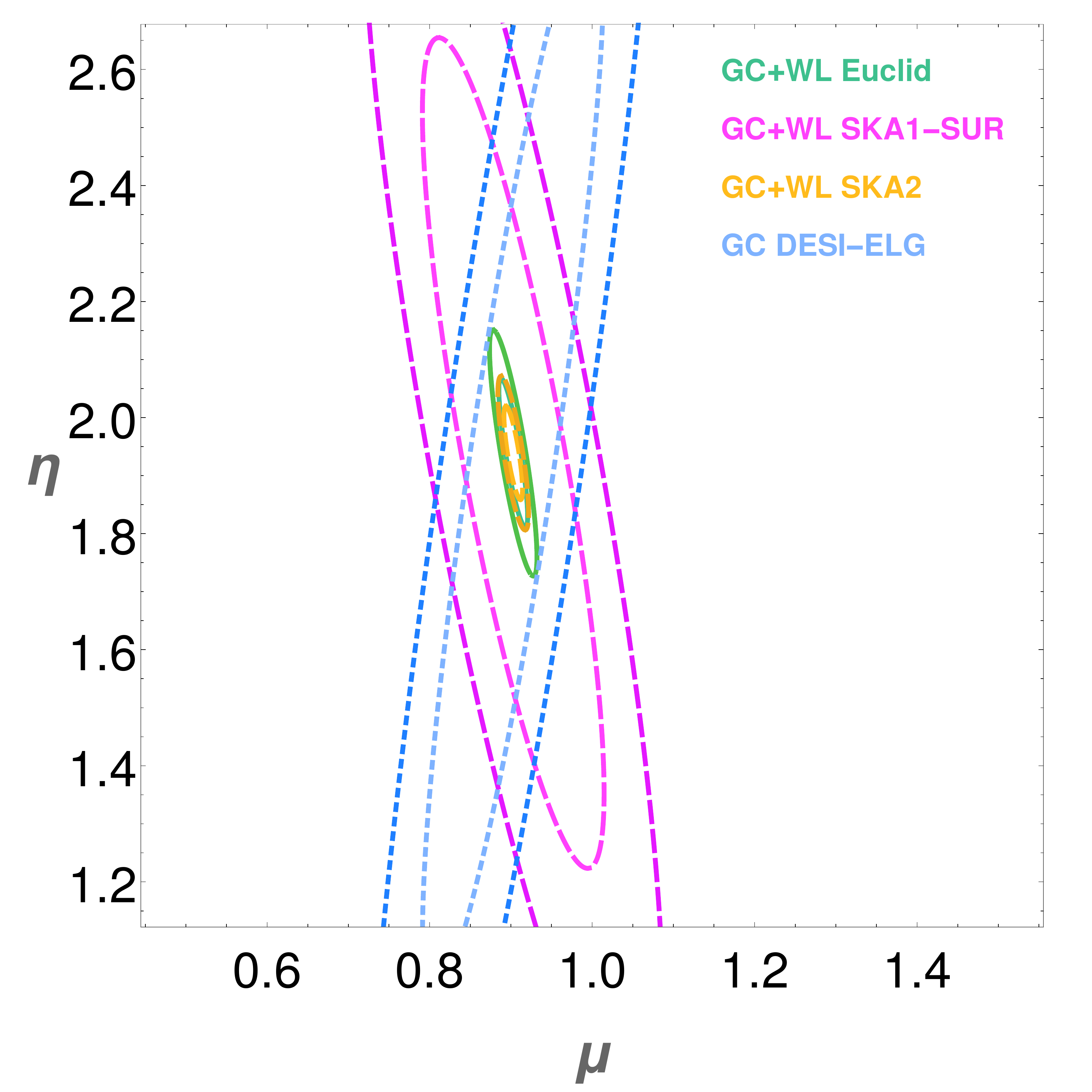}
\caption{\label{fig:combined_surveys}
1$\sigma$ and 2$\sigma$
fully marginalized confidence contours on the parameters
$\mu$ and $\eta$, for 3 different surveys combining Galaxy Clustering (GC) and Weak Lensing (WL): Euclid, SKA1-SUR
and SKA2 and for GC only: DESI-ELG, all in the late-time (left panel) and early-time (right panel) parameterizations
of sections \ref{sub:MG-DE} and \ref{sub:MG-TR}, respectively.
As explained in the main text, the constraints are parameterization-dependent, especially on $\eta$, where in
the late-time scenario GC alone is not able to constrain it, while in the early-time scenario GC can constrain both $\mu$ and $\eta$.
}
\end{figure}

In Figure \ref{fig:BarPlot-DE-GC-1} we show the 1$\sigma$ fully
marginalized forecasted errors on the parameters 
$\{\Omega_{m},\Omega_{b},h,\ell \mathcal{A}_{s},n_{s},\mu,\Sigma\}$
for different Weak Lensing (left panel) and Galaxy Clustering (right panel)
surveys in the late-time parameterization. In the GC case, the surveys
considered are DESI-ELG (yellow), SKA2 (green), SKA1-SUR (orange)
and Euclid (blue). For the WL forecast, we considered Euclid (blue),
SKA1 (orange) and SKA2 (green). These constraints correspond to the ones
listed in Table \ref{tab:errors-GC-SKAcompare-MG-DE-mu-eta-sigma}. 
The marginalized confidence contours for the $\mu$-$\eta$ plane , 
comparing all these surveys, can be seen in the left panel ofFig.\ref{fig:combined_surveys}.
The 1$\sigma$ fully marginalized
constraints on the parameters are weaker for WL than for GC,
which may be a consequence of the higher correlation among variables
for the Weak Lensing observable, described in the previous section \ref{subsub:late-time-comb-GC+WL}.
Comparing the different surveys,
the general trend is that Euclid and SKA2 perform at a similar level for WL at both linear and non-linear level; for GC and when combining both probes, SKA2
gives the strongest constraints, followed by Euclid, SKA1 and DESI (for GC).
Notice that in this parameterization, a SKA1-SUR GC survey constrains
the $\Sigma$ parameter alone better than a Euclid Galaxy Clustering
survey (although Euclid is overall much stronger as can be seen with the FoM). This is due to the fact that SKA1-SUR probes much lower
redshifts (from $z=0.05-0.85$) than Euclid and is therefore suitable to better constrain those parameterizations in which the effect of the Modified Gravity parameters is stronger at lower redshifts; this is the case of the late-time parameterization, which is proportional
to the dark energy density, dominating at low redshifts only. This result is reversed in the early time parameterization, in which Modified Gravity can play a role also at earlier redshifts.

\subsection{\label{sub:MG-TR} Modified Gravity in the early-time parameterization}

\subsubsection{Galaxy Clustering, Weak Lensing and its combination}

We extend our analysis now to an alternative choice, the early time parameterization specified in Eqns.\ (\ref{eq:TR-mu-parametrization}) and (\ref{eq:TR-eta-parametrization}). As before, we use
Euclid Redbook specifications for WL and GC
and the cut in scales discussed previously for the two observables, i.e. 
a maximum wavelength cutoff at $k_{\rm max}=0.15$ for GC and a maximum multipole of $\ell_{\rm max}=1000$ for WL in the 
linear case, and a cutoff $k_{\rm max}=0.5$h/Mpc and a maximum multipole of $\ell_{\rm max}=5000$ for WL in the non-linear regime,
which is analyzed using the prescription described in Sec.\ \ref{sub:Prescription-HS}.
We use the two observables both separately and in combination, without accounting for cross correlation of the two (as discussed in section \ref{sec:Results:-Redshift-Binned} this seems to correspond to a conservative choice), 
with and without \planck\ priors.

\begin{figure}[htbp]
\includegraphics[width=0.3\textwidth]{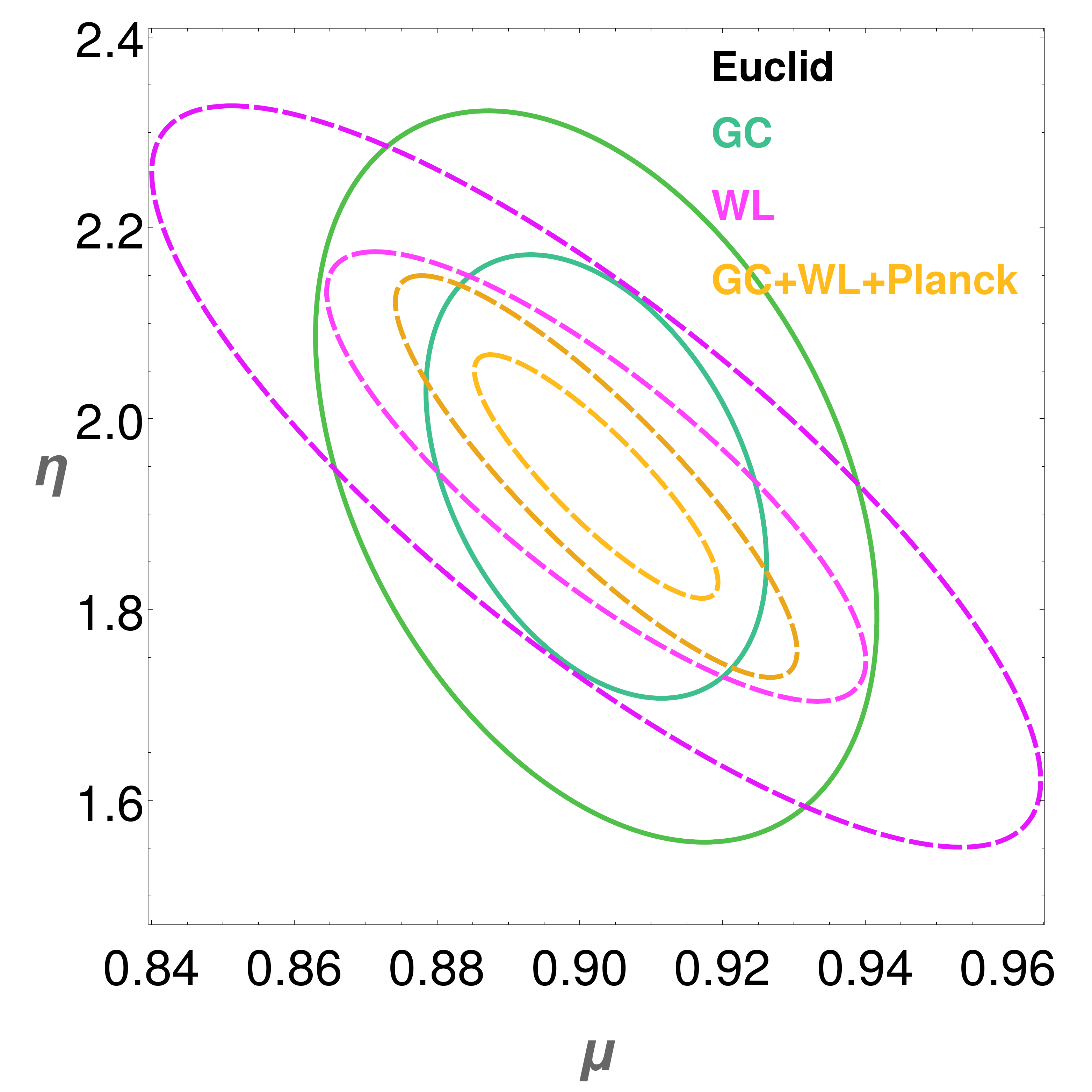}
\includegraphics[width=0.3\textwidth]{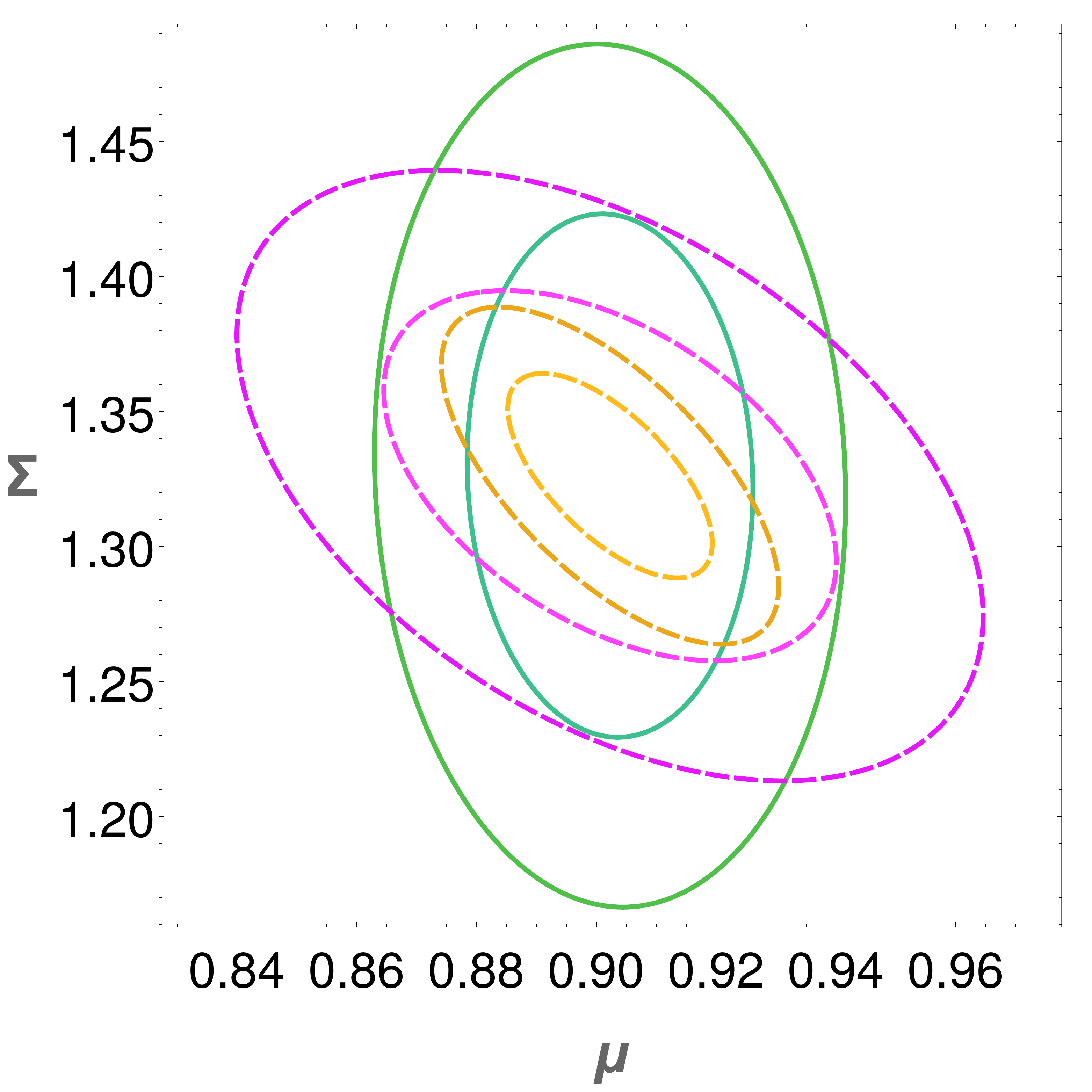}
\\
\includegraphics[width=0.3\textwidth]{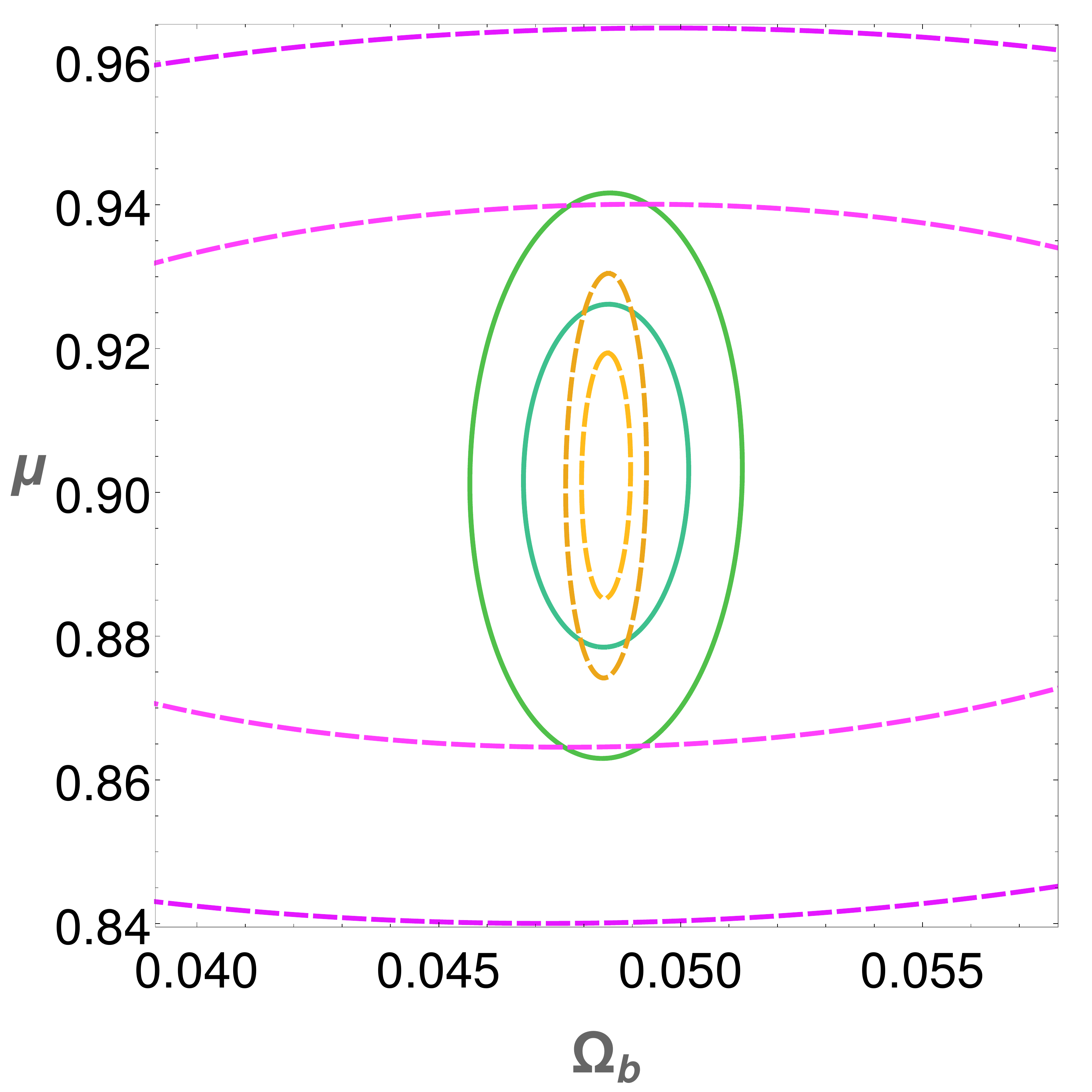}
\includegraphics[width=0.3\textwidth]{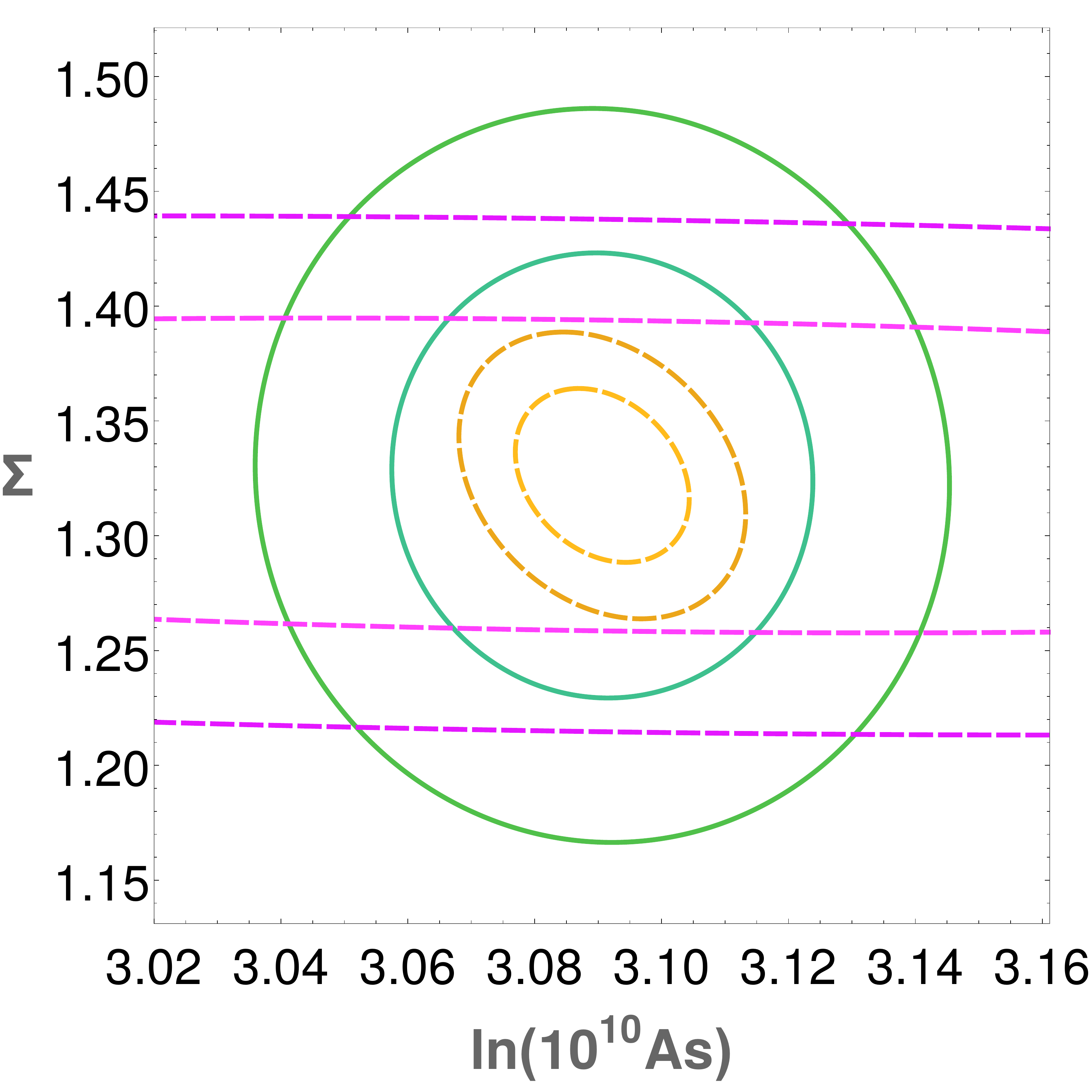}
\caption{\label{fig:T-related-ellipses-mu-omegac} Fisher Matrix marginalized forecasted contours
(1$\sigma$, 2$\sigma$) for the Euclid Redbook satellite in the early-time
parameterization using mildly non-linear scales and the HS prescription.
Green lines represent constraints from the Galaxy Clustering survey,
pink lines stand for the Weak Lensing observables, and orange lines represent the
 GC+WL+{\it Planck} combined confidence
regions. \textbf{Upper left: }contours for the fully marginalized
errors on $\eta$ and $\mu$. \textbf{Upper right: }contours for the
fully marginalized errors on $\Sigma$ and $\mu$.\textbf{ Lower left:
}contours for the fully marginalized errors on $\mu$ and $\Omega_{b}$.
\textbf{Lower right: }contours for the fully marginalized errors on
$\Sigma$ and $\ln 10^{10} A_s$. Notice that in this parameterization, GC and WL are able to constrain both $\mu$ and
$\eta$ or $\Sigma$ on their own.}
\end{figure}

Results are shown in Table \ref{tab:errors-Euclid-GC-WL-early_time} and 
Figure \ref{fig:T-related-ellipses-mu-omegac}.
The general behaviour of the constraints is similar to the one in the late-time parameterization,
with the combination of GC and WL able to break the degeneracies with standard cosmological
parameters, leading to a significant improvement of the constraints on MG parameters, constraining $\mu$ and $\Sigma$ at the 1-2\% level. 
There are some other interesting differences with the late-time scenario. 
First, the addition of \planck\ priors does not really improve much the constraints obtained by GC or WL alone, which was not the case in the 
late-time parametrization. This is related to the fact that in the early-time parameterization, GC and WL (non-linear) on their own are already good at constraining both
$\mu$ (at 2-3 \%) and $\eta$ (at around 8\%), with consequently small errors on $\Sigma$ ($\approx 3$\%).
In Appendix \ref{sec:appder} we show the derivatives of the matter power spectrum with respect to the MG parameters $\mu$ and $\eta$ in both
parameterizations. We can observe that in the early-time scenario, the derivative  $dP(k)/d\eta$ is larger than in the late-time
parameterization, leading therefore to better constraints.
Another difference lies in the correlation among parameters, which for WL and GC+WL is considerably smaller than in the late-time scenario. 
The Figure of Correlation (defined in Eqn.\ \ref{eq:FoC}) for GC (non-linear HS) alone is 4.7, while
for WL (non-linear HS) the correlation is somewhat higher, with $\textrm{FoC}=7.3$.
When combining both probes (GC+WL (non-linear HS)) the FoC goes to an intermediate point of 5.2.

\subsubsection{Other Surveys: DESI-ELG, SKA1 and SKA2 \label{subsub: other-surveys-early-time}}

\begin{figure}[htbp]
\begin{centering}
\includegraphics[width=0.45\textwidth]{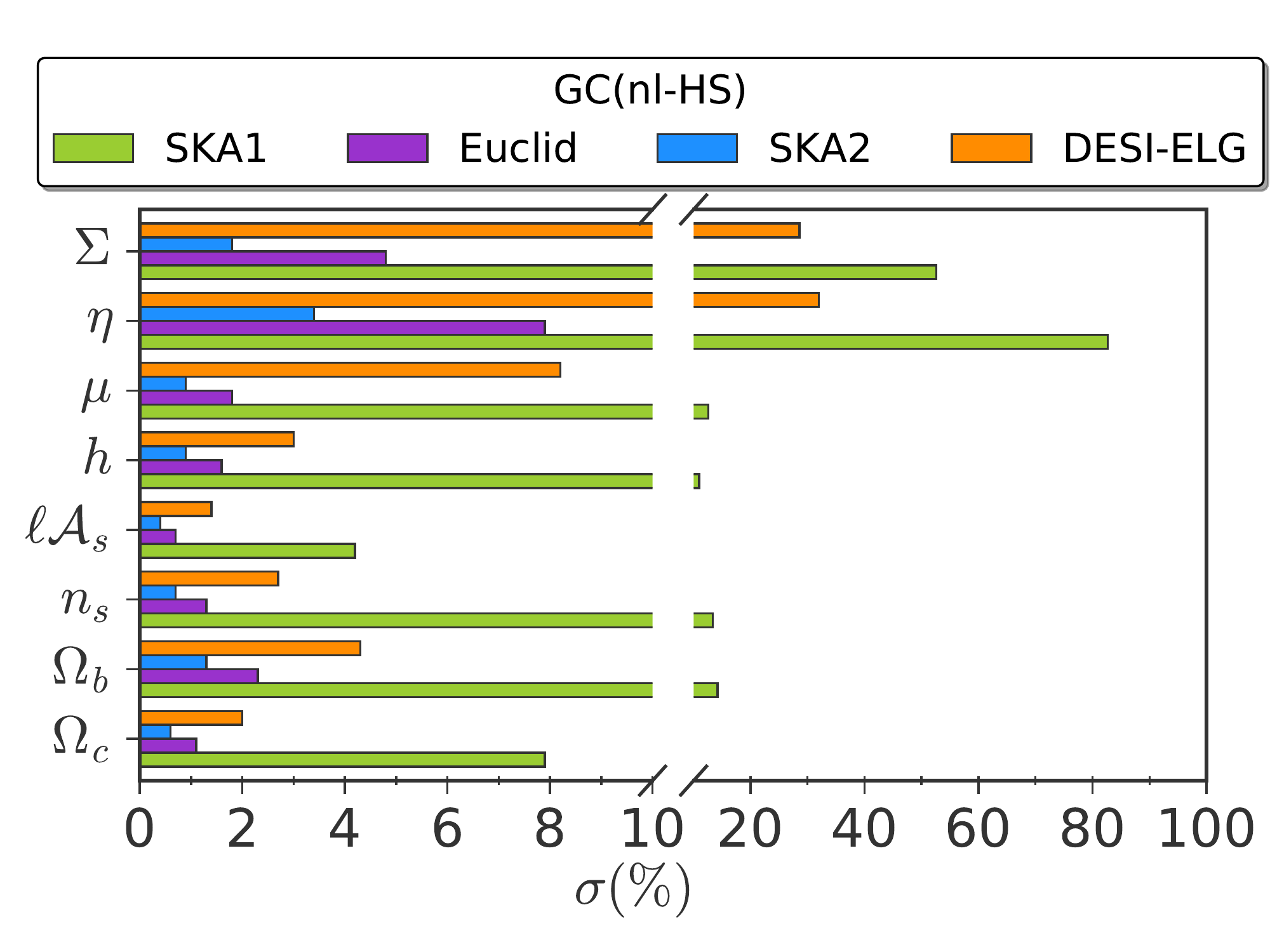}\hspace{-0.5pt}
\includegraphics[width=0.45\textwidth]{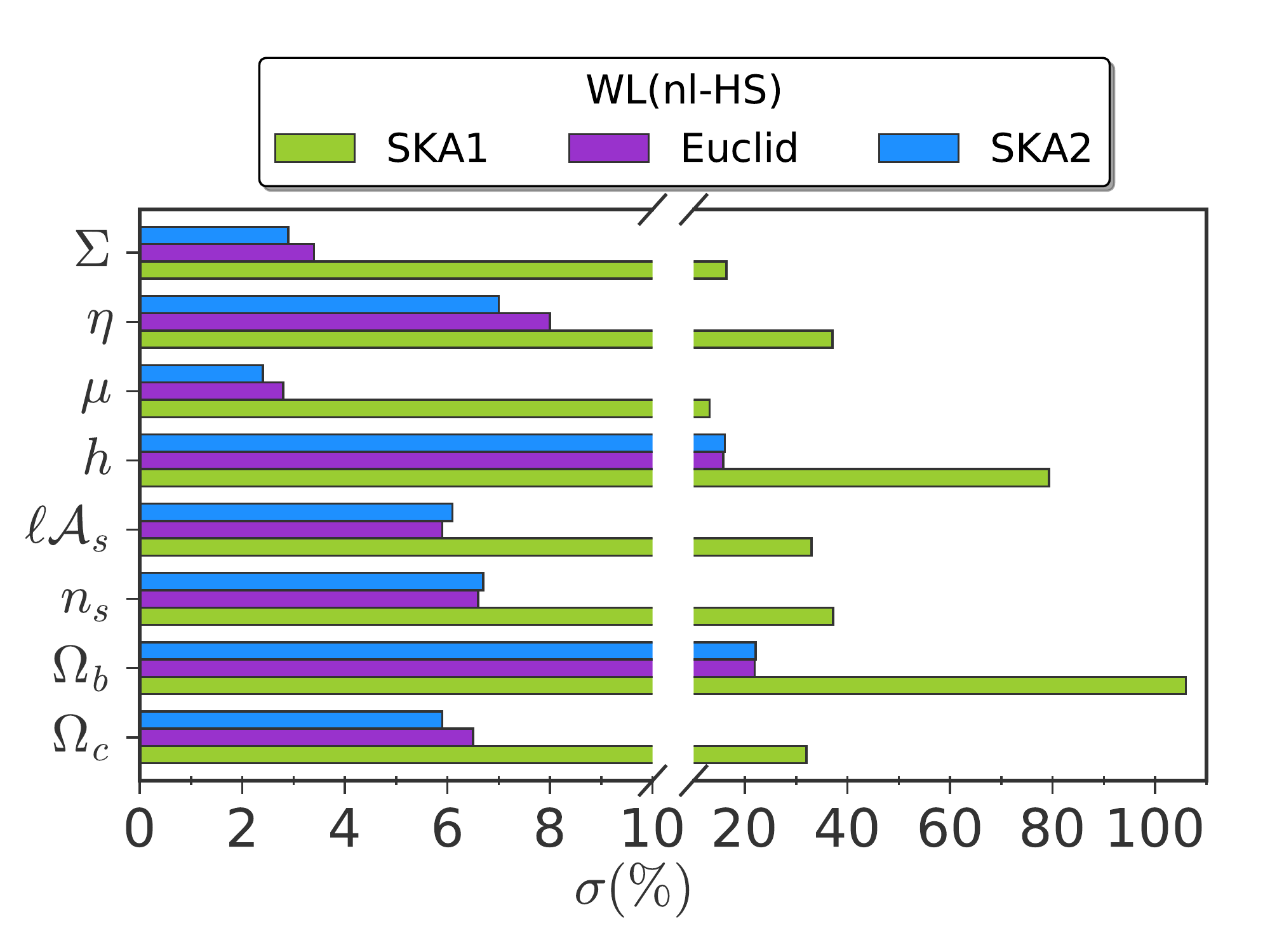}

\includegraphics[width=0.45\textwidth]{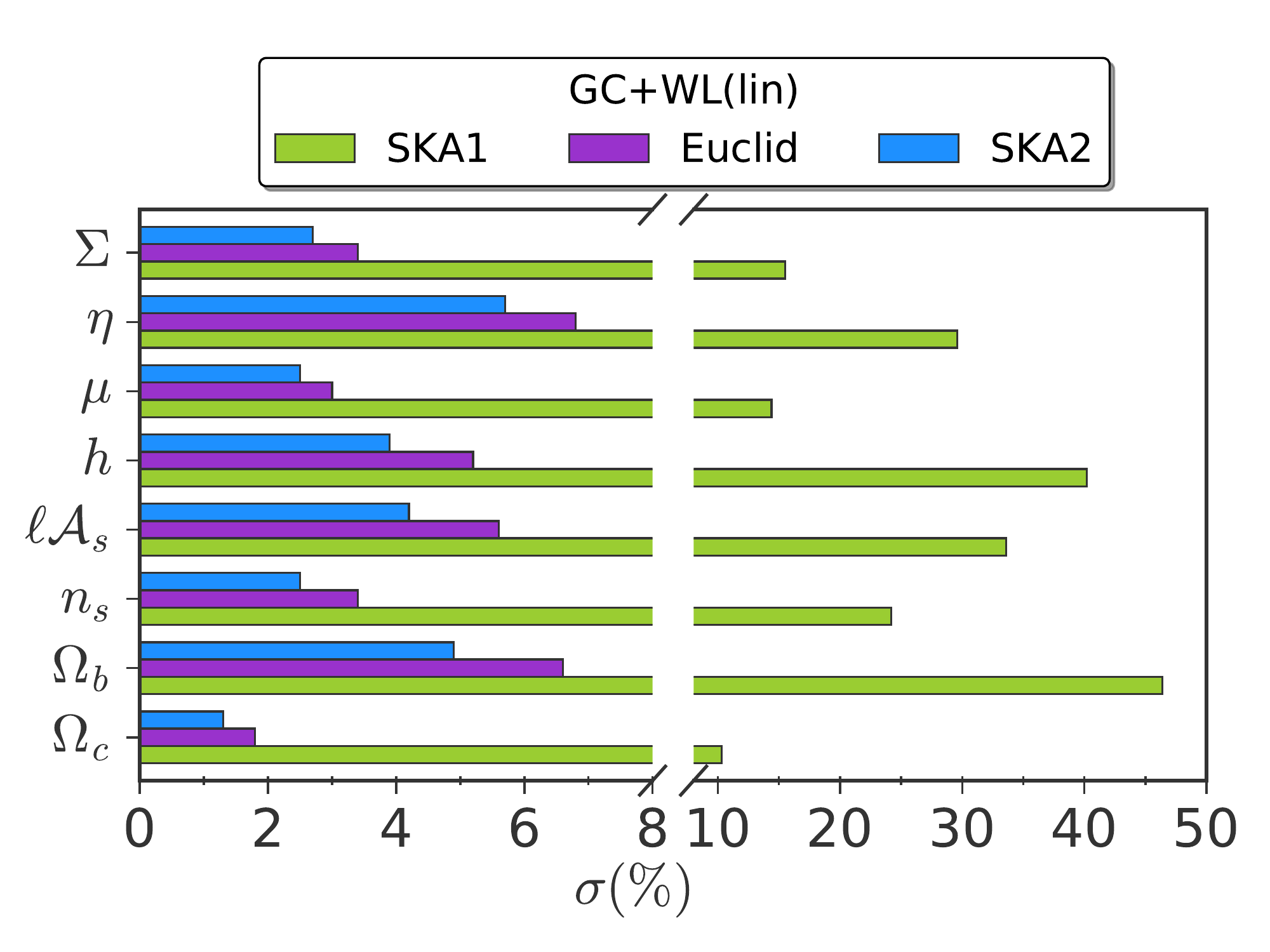}\hspace{-0.5pt}
\includegraphics[width=0.45\textwidth]{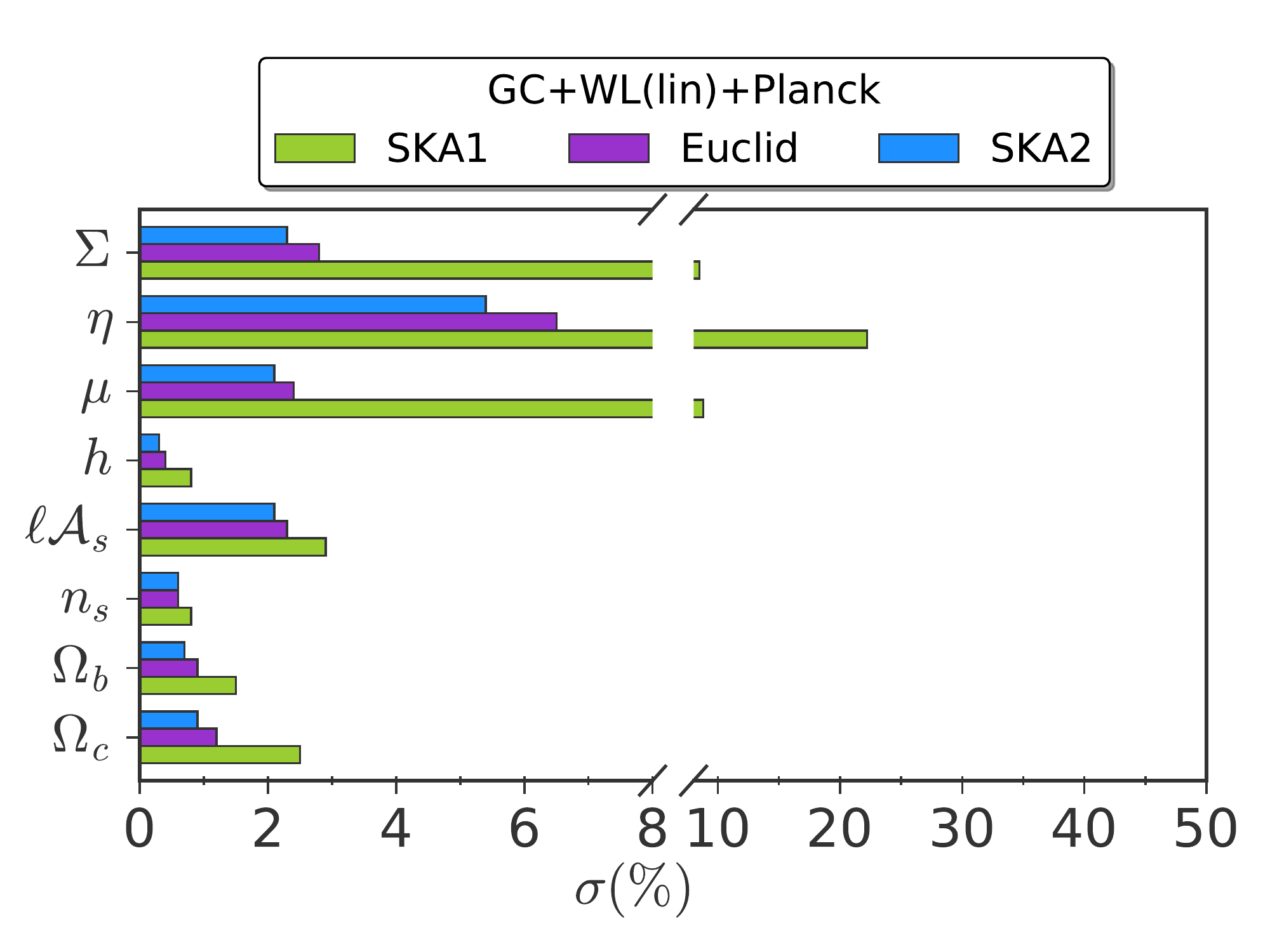}

\includegraphics[width=0.45\textwidth]{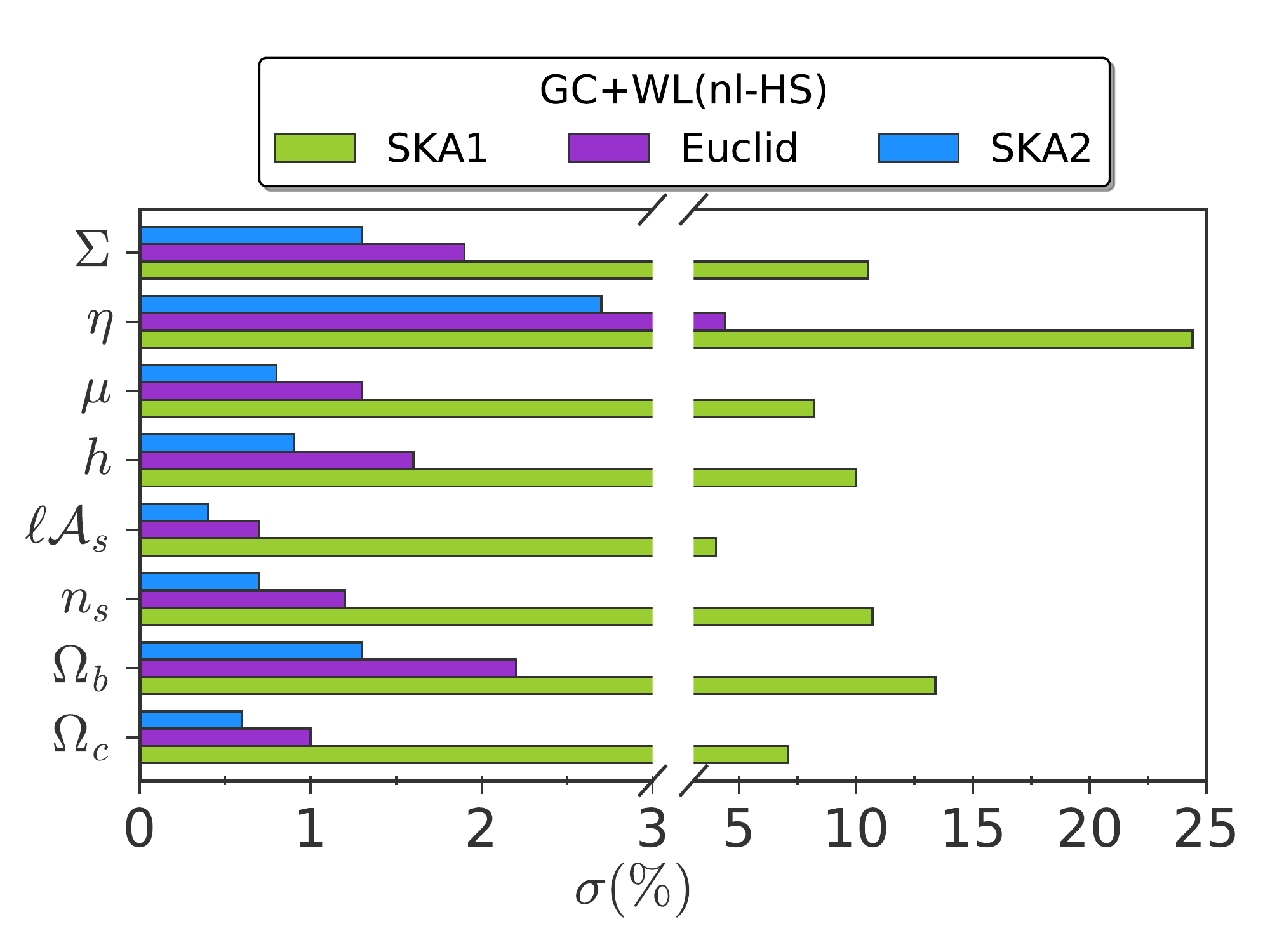}\hspace{-0.5pt}
\includegraphics[width=0.45\textwidth]{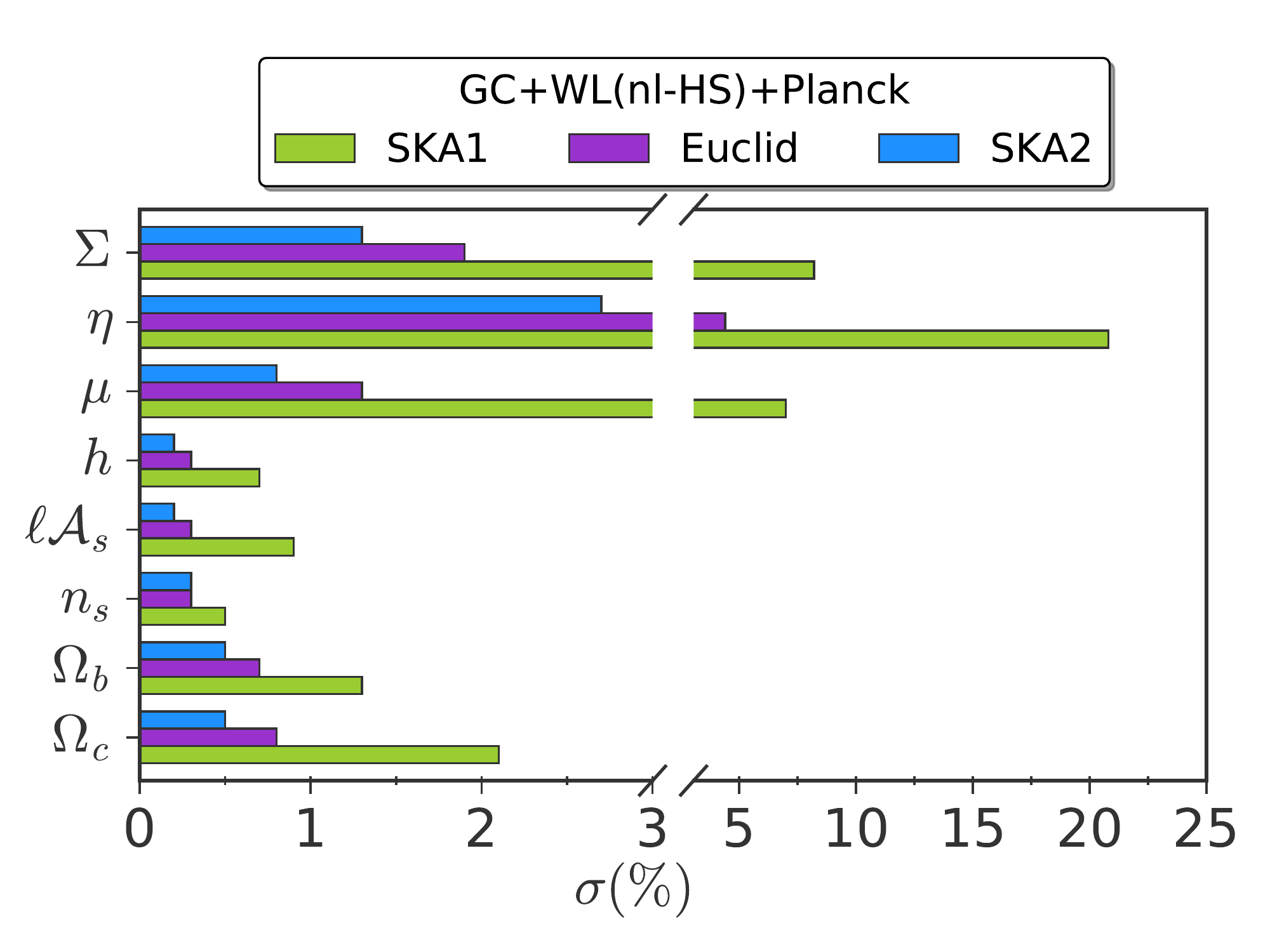}
\par\end{centering}

\caption{\label{fig:BarPlot-MGTR-Surveys} Same as Fig.\ref{fig:BarPlot-DE-GC-1} but for the early-time parameterization (Eqns. \ref{eq:TR-mu-parametrization},\ref{eq:TR-eta-parametrization})
The 1$\sigma$ fully marginalized
constraints on the parameters are weaker for WL than for GC,
which is a consequence of the higher correlation among variables
for the Weak Lensing observable. 
}
\end{figure}

Also in the early time parameterization we obtain
the 1-$\sigma$ fully marginalized errors for Galaxy Clustering (top left
panel of Figure \ref{fig:BarPlot-MGTR-Surveys}) considering DESI-ELG
(yellow), SKA2 (green), SKA1-SUR (orange) and Euclid (blue), and for
Weak Lensing (top right panel of Figure \ref{fig:BarPlot-MGTR-Surveys})
using Euclid (blue), SKA1 (orange) and SKA2 (green). We also report
the results in Table \ref{tab:errors-GC-SKAcompare-MG-TR-mu-eta-sigma-Zhao-1},
where it is possible to notice how the conclusions drawn on the hierarchy
 of the considered experiments do not change
with respect to the late-time parameterization. The main difference
is that in this case the full SKA1-SUR GC survey does not
constrain the $\Sigma$ parameter better than the Euclid survey; this
is due to the fact that in the early time parametrization, deviations
from $\Lambda$CDM are present also at high redshift, therefore we
do expect the information present at small redshift to be as relevant
as the one coming from higher $z$, where Euclid performs significantly
better than SKA1-SUR. The marginalized contours for the $\mu$-$\eta$ plane, 
comparing all these surveys, can be seen in the right panel of Fig.\ref{fig:combined_surveys}.

\subsection{Testing the effect of the Hu-Sawicki non-linear prescription on parameter estimation
\label{sub:Testing-the-effect-of-Zhao}}

In this section we show the effect of changing the parameters $\cnl$
and $s$ used in the HS prescription (specified in Eq.\ \ref{eq:PHSDefinition})
for the mildly non-linear matter power spectrum. As mentioned already
in Section \ref{sub:Prescription-HS}, previous works (see
\cite{zhao_modeling_2014,zhao_n-body_2011,koyama_non-linear_2009}),
have fitted the values of $\cnl$ and $s$ to match N-body simulations
in specific Modified Gravity models. In all these cases the HS parameters
$\cnl$ and $s$ have been found to be of order unity, with $\cnl$
ranging usually from $0.1$ to $3$ and $s$ from about $1/3$ to
a value of around $2$. In the absence of N-body simulations for our
models, we selected our benchmark HS parameters
to be $\cnl=1$ and $s=1$, as discussed in Section \ref{sub:Prescription-HS}, which we used
for all the analysis presented above. However, in order to test
the effect of a change of $c_{nl}$ and $s$ non-linear parameters on our estimation of errors
on the cosmological parameters, we perform our GC and WL forecasts on the
MG late-time model (Section \ref{sub:MG-DE}) also changing one at
a time the values of both HS parameters. We use the following values
for our test: $\cnl=\{0.1,0.5,\,1,\,3\}$ and $s=\{0,\,1/3,\,2/3,\,1\}$.

\begin{figure}[htbp]
\centering{}
\includegraphics[width=0.35\textwidth]{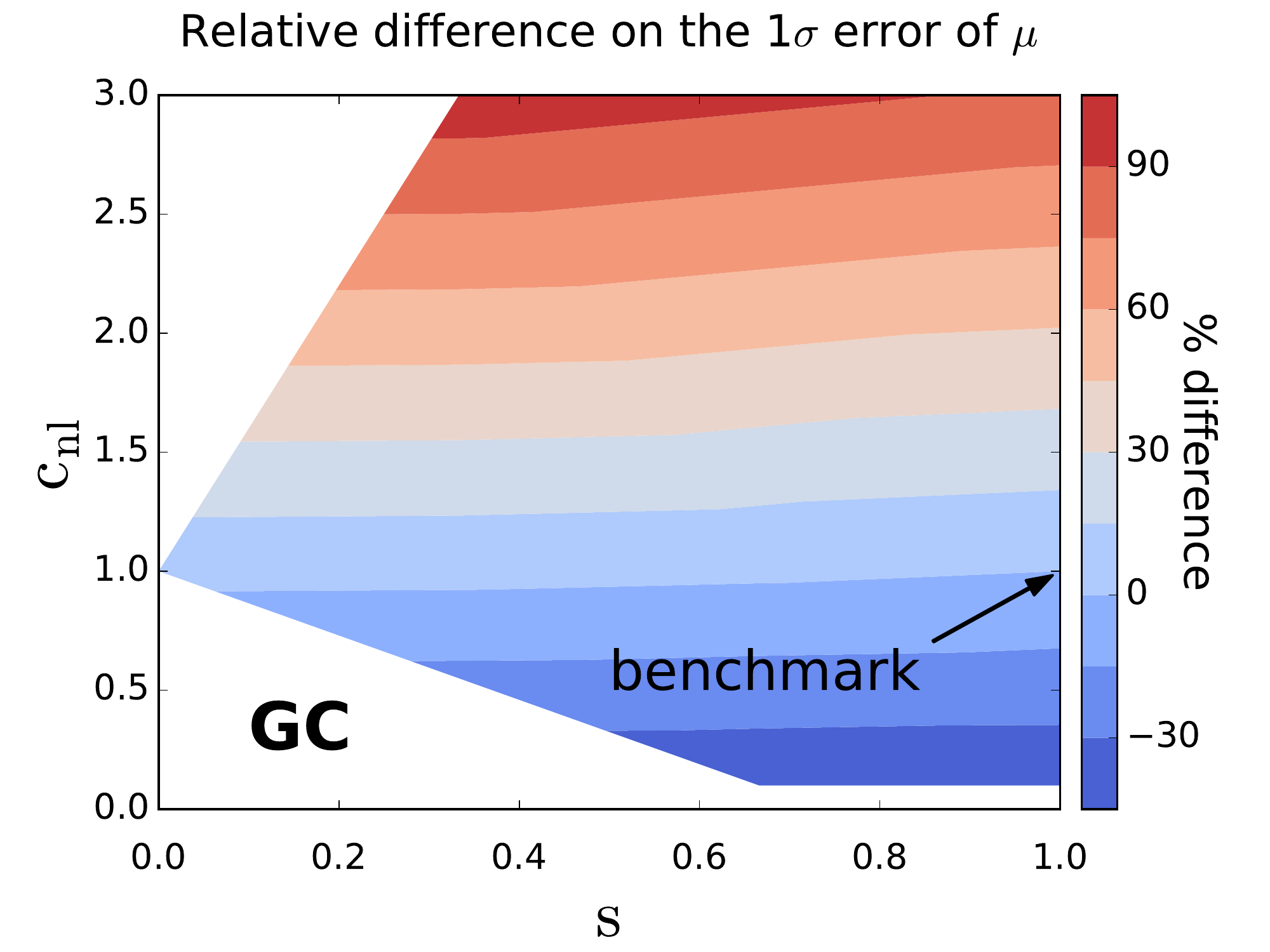}
\includegraphics[width=0.35\textwidth]{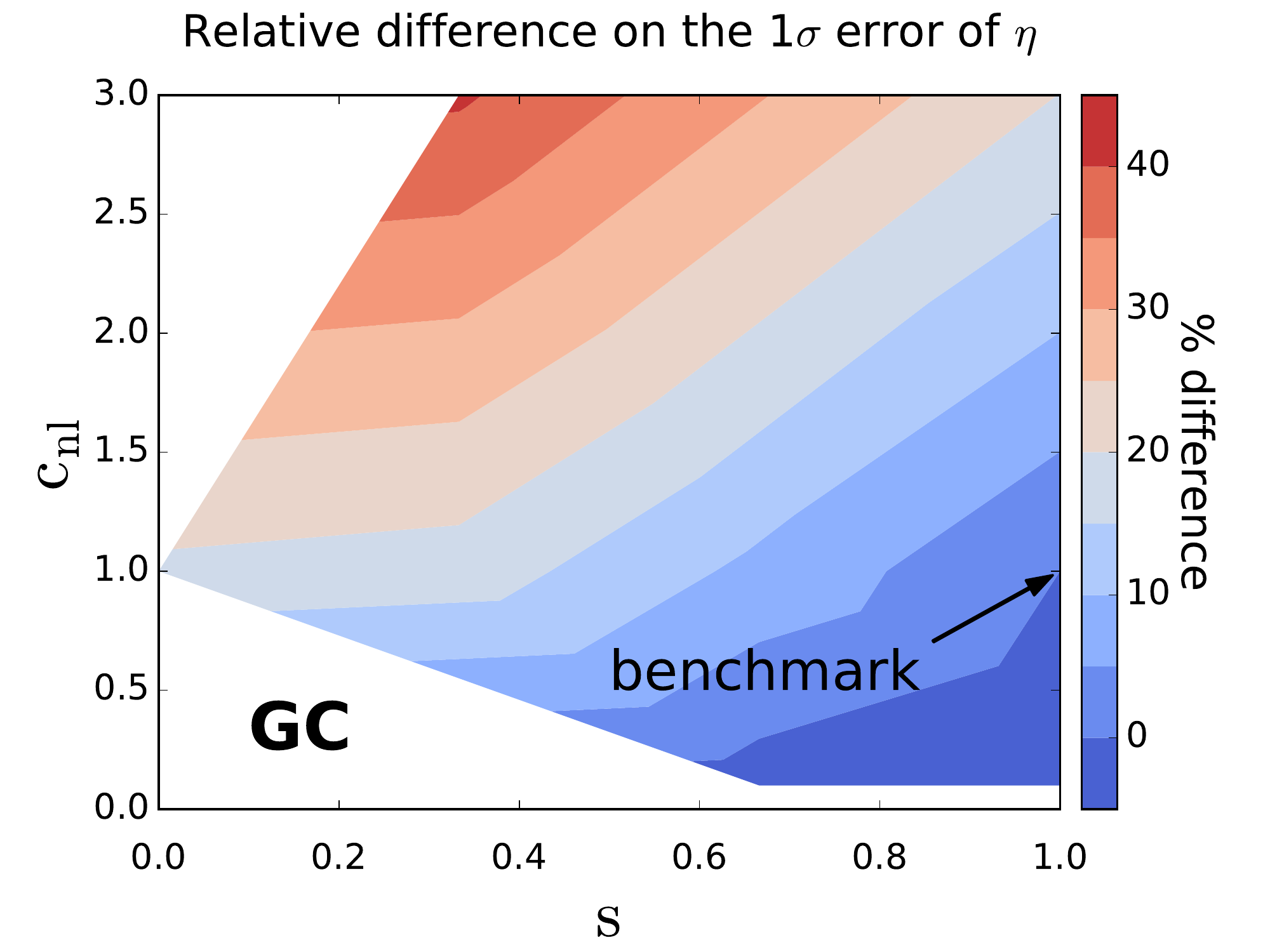}
\caption{\label{fig:Density-GC-HSpars}Effect
of the $\cnl$ and $s$ parameters
on the 1$\sigma$ marginalized error of the $\mu$ (left panel) and the
$\eta$ (right panel) parameters
in the MG late-time parametrization for a Euclid Galaxy Clustering
forecast with Redbook specifications. The colored contours show the percentage discrepancy when
departing from the benchmark case $\cnl=1$ and $s=1$ (marked with
a black arrow) to all other points in the $\cnl$-$s$
space. The red (blue) contours signal the regions of maximum positive (negative) discrepancy.
For example in the left panel, choosing $\cnl$ and $s$ in the
red region, will yield a 1$\sigma$ marginalized error on $\mu$
which is 90\% larger than in the benchmark case (see Table
\ref{tab:errors-Euclid-GC-WL-late_time}
for the benchmark forecast). For the standard $\lcdm$ cosmological
parameters (not shown here) the discrepancy is smaller than 4\% for all choices of $\cnl$ and $s$.
}
\end{figure}

\begin{figure}[htbp]
\centering{}
\includegraphics[width=0.35\textwidth]{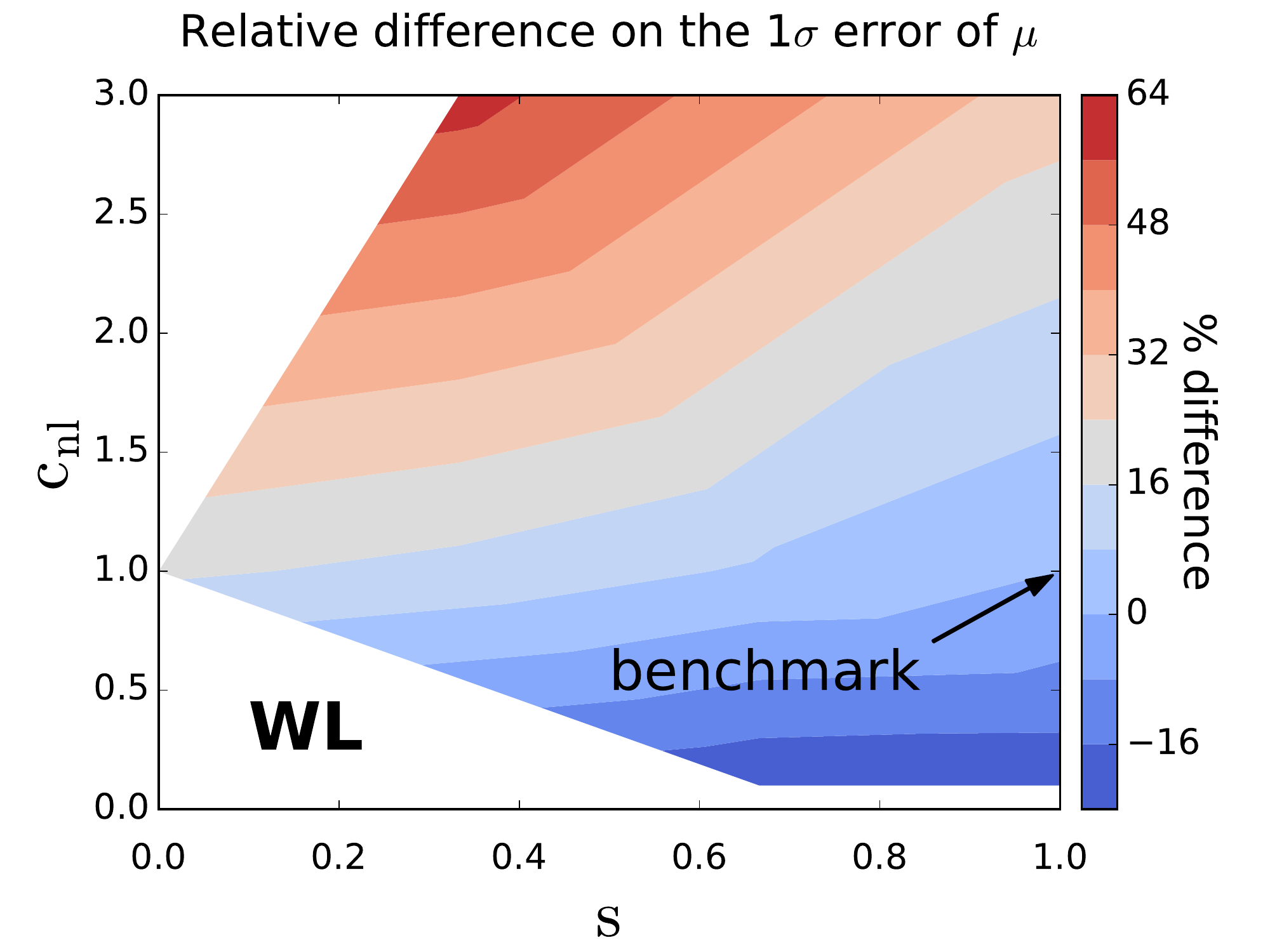}
\includegraphics[width=0.35\textwidth]{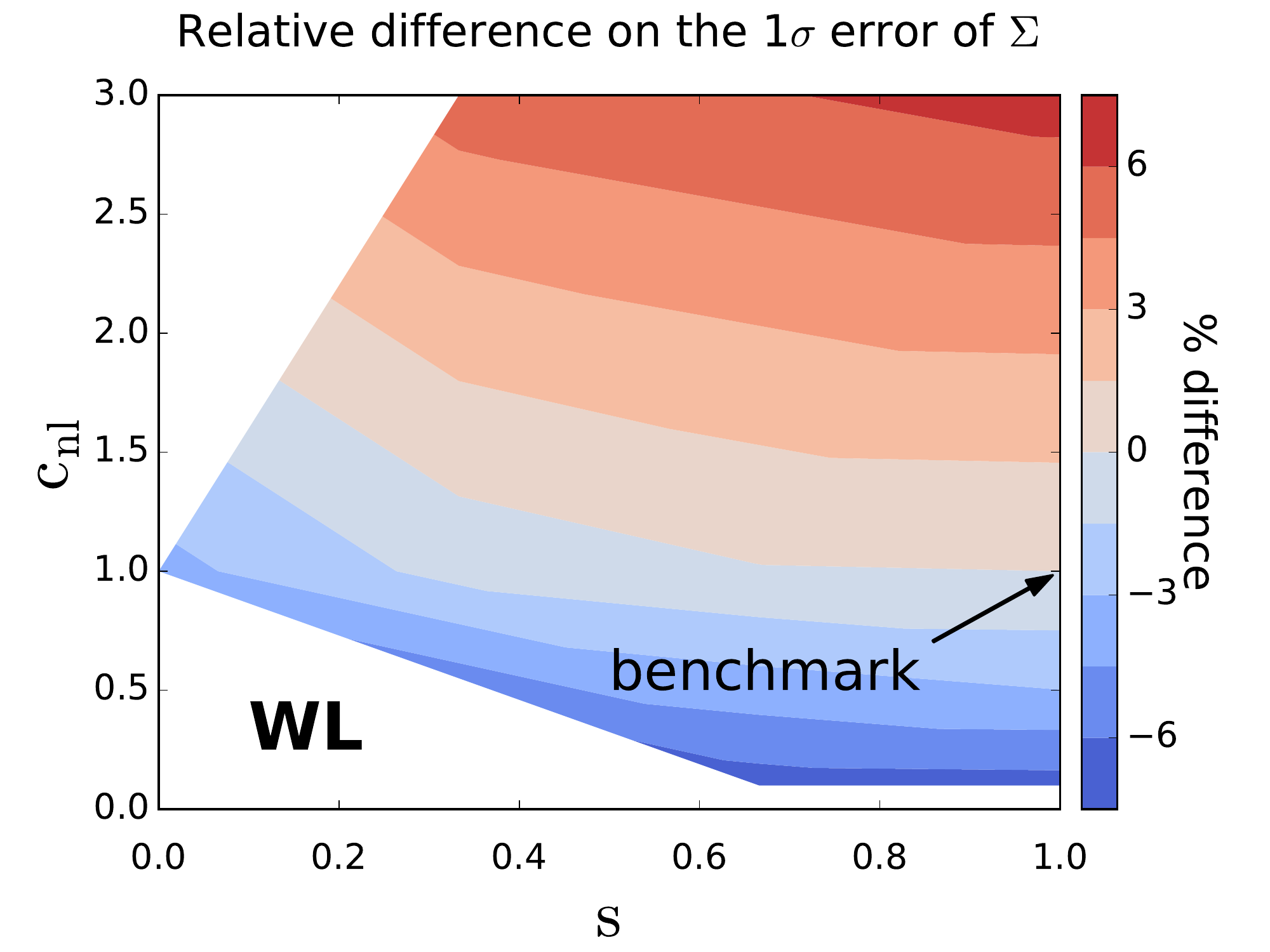}
\caption{\label{fig:Density-WL-HSpars}
Same as Fig.\ref{fig:Density-GC-HSpars} but for a Weak Lensing Euclid forecast using Redbook specifications.
In the left panel, choosing $\cnl$
and $s$ in the red region, will yield a 1$\sigma$ marginalized
error on $\mu$ which is 60\% larger than in the benchmark case
(see Table \ref{tab:errors-Euclid-GC-WL-late_time}
for the benchmark forecast). In the right panel we see that for the MG parameter $\Sigma$ the maximum positive and negative discrepancy
is only of about 6\%. The maximum and minimun discrepancy for the MG parameter $\eta$ is of -15\% and 50\%. This means that
the parameter $\Sigma$ (defined as the lensing Weyl potential, and therefore directly constrained by Weak Lensing) 
is much less sensitive to changes in the non-linear prescription parameters.}
\end{figure}

In Figure \ref{fig:Density-GC-HSpars} we show
the percentage discrepancy between the 1$\sigma$ marginalized error
obtained on parameters in the GC non-linear HS benchmark case (see Table
\ref{tab:errors-Euclid-GC-WL-late_time}
for the exact values) and the corresponding error obtained by performing
the same forecast with a different value of the $\cnl$-$s$ parameters.
In general terms we see that for the MG parameter $\mu$
(left panel of Figure \ref{fig:Density-GC-HSpars}) the relative difference
in the estimation of the 1$\sigma$ forecasted errors can lie between
90\% (at $\cnl=3$, $s=0.33$) and -30\% for $\cnl=0.1$ and $s=1.0$.
The behavior of the contour lines shows us that for a fixed value of $\cnl$, 
the forecasted error on the parameter $\mu$ remains unaffected.
For $\eta$ (right panel) the relative discrepancy lies between
40\% and -2\%. Here, to get the same 1$\sigma$ errors on $\eta$, one would have to vary 
both $\cnl$ and $s$.
We also tested the effect on the standard $\lcdm$ parameters and found it to be smaller than
4\% for all choices of $\cnl$ and $s$.
In the case of WL forecasts, we perform the same tests, which are shown in Figure \ref{fig:Density-WL-HSpars}.
We can observe that the relative discrepancies in the case of the
$\mu$ parameter lie between $\sim$60\% and $\sim$-15\%, while for $\Sigma$ the discrepancy is considerably smaller. 
The 1$\sigma$ error on $\Sigma$ varies only between $\pm 6$\%.
This is however a particular effect of choosing $\Sigma$, which is the true WL observable.
If we perform this test on $\eta$ using WL, we will find a stronger discrepancy,
that lies in between $\sim$50\% and $\sim$-15\%, similar to the one found for $\mu$.

We can also test the effect of adding these two HS parameters as extra
nuisance parameters to our model of the non-linear power spectrum;
therefore, by taking derivatives of the observed power spectrum with
respect to these parameters, we can forecast what would be the estimated
error on $\cnl$ and $s$. Then, marginalizing over $\cnl$ and
$s$ would yield more realistic constraints on our cosmological parameters,
and take into account our ignorance on the correct parameters
of the HS prescription. In Table
\ref{tab:errors-GC+WL-Marginalize-HS-Zhao-Euclid-DEpar-muetasigma}
we list the 1$\sigma$ marginalized constraints on $\cnl$ and
$s$ for our benchmark fiducial ( $\cnl=1$, $s=1$) and using the
standard fiducial for the cosmological parameters used previously
for the MG late time parametrization.

Taking these HS parameters into account in our Fisher forecast, will
automatically change the constraints on the other parameters. For our
method to be consistent, we would like this effect to remain as small
as possible. In Table
\ref{tab:errors-GC+WL-Marginalize-HS-Zhao-Euclid-DEpar-muetasigma},
we list the same constraints reported in Table
\ref{tab:errors-Euclid-GC-WL-late_time},
but obtained marginalizing over $\cnl$ and $s$ at the benchmark
fiducial $\cnl=1$, $s=1$. Comparing the two tables, we can
see that for GC the errors on the cosmological parameters remain quite
stable, except for
$\eta$ and $\Sigma$; this is understandable since those parameters
are not well constrained by GC alone. For WL, we see a difference
of 4 to 7 percent points in the errors on $h$, $n_{s}$ and $\Omega_{b}$,
while all other errors remain stable, with $\Sigma$ varying less
than 2 percent points.
Remarkably, the combined constraints from GC+WL are even less affected by the two nuisance parameters.
Comparing the Modified Gravity FoM between the two tables, shows the expected behavior, the MG FoM is reduced
when adding two extra parameters, but the change is very small, of just 0.3-0.5 nits.

\begin{table}[htbp]
\centering{}%
\begin{tabular}{|c|c|c|c|c|c||c|c|c||c|c|c|}
\hline 
\textbf{Euclid} (Redbook)  & $\Omega_{c}$  & $\Omega_{b}$  & $n_{s}$  &
$\ell\mathcal{A}_{s}$  & $h$  & $\mu$  & $\eta$  & $\Sigma$ \Tstrut &
$\cnl$  & $s$ & MG FoM \tabularnewline
\hline 
\Tstrut Fiducial  & {0.254}  & {0.048}  & {0.969}  & {3.060}  & {0.682}  & {1.042}  & {1.719}  & {1.416}  & {1}  & {1} & relative \tabularnewline
\hline
\hline
\Tstrut \textbf{GC(nl-HS)}  & 1.0\% & 2.8\% & 1.3\% & 1.1\% & 2.0\% & 1.7\% & 784\% & 480\% & 372\% & 236\% 
& 2.4 \tabularnewline 
\Tstrut \textbf{WL(nl-HS)}   & 6.5\% & 25\% & 8.3\% & 9.1\% & 19\% & 25\% & 46\% & 6.0\% & 1680\% & 899\% 
& 4.2 \tabularnewline 
\Tstrut \textbf{GC+WL(nl-HS)} & 1\% & 2.8\% & 1.2\% & 1\% & 1.9\% & 1.6\% & 2.6\% & 1.2\% & 333\% & 166\%  
& 8.5 \tabularnewline
\hline 
\end{tabular}\protect\caption{\label{tab:errors-GC+WL-Marginalize-HS-Zhao-Euclid-DEpar-muetasigma}
1-$\sigma$ fully marginalized errors on the cosmological parameters
and the two HS parameters $\cnl$ and $s$ for a Euclid Galaxy
Clustering forecast, a Weak Lensing forecast and the combination of
both in the late-time parameterization of Modified Gravity using non-linear
scales and the HS prescription. In contrast to Table
\ref{tab:errors-Euclid-GC-WL-late_time} (where $c_{nl}$ and $s$ had been fixed to the benchmark value)
here we include $\cnl$ and $s$ as free parameters
and marginalize over them. The MG FoM is computed relative to the same Euclid Redbook GC 
linear case,
used previously. The errors and the MG FoM show the expected behavior
of adding two nuisance parameters, and remain quite stable. All other naming conventions are the
same as for Table \ref{tab:errors-Euclid-GC-WL-late_time}. Remarkably, the combination of GC and WL is still able to constrain all Modified Gravity parameters at the level of 1-2 $\%$ after marginalizing over the non-linear parameters.}
\end{table}

\section{Conclusions}

We study in this paper the constraining power of upcoming large scale surveys on Modified Gravity theories, choosing a phenomenological approach that does not require to specify any particular model. To this purpose we consider the two functions $\mu$ and $\eta$ that encode general modifications to the Poisson equation and the anisotropic stress.
We study three different approaches to MG: redshift-binning, where we discretize the functions $\mu(z)$ and $\eta(z)$ in 5 redshift bins and 
let the values of $\mu$ and $\eta$ in each of the bins vary independently with respect to the others; an early-time parameterization, where $\mu$ and $\eta$ are allowed to vary at early times and their amplitude
can be different from unity today; a late-time parameterization where  $\mu$ and $\eta$ are linked to the energy density of dark energy and therefore they are very
close to unity in the past, but they can vary considerably at small redshifts. For convenience, we summarize all results in Table \ref{table_results}, with a direct link to the section and tables related to each scenario.

\begin{table}
\begin{tabular}{|l|l|l|} 
\hline 
\Tstrut \textbf{Model} & \textbf{Description} & \textbf{Results} \\ 
\hline 
\Tstrut Redshift Binned $\mu$ and $\eta$ & Section \ref{sub:param-z-bins-th}, Eqns.\ (\ref{eq:MGbin-mu-parametrization})-(\ref{eq:MGbin-muderiv-parametrization})  & Table \ref{tab:errors-all-MGBin3}; \,\,\,\,\,\,\,Figs.\ \ref{fig:GCcorr}-\ref{fig:GC+WL+Planck-bestconst-errspq} \\ 
\hline 
\Tstrut Late-time parameterization & Section \ref{sub:param-smooth-funct}, Eqns.\ (\ref{eq:DE-mu-parametrization})-(\ref{eq:DE-eta-parametrization}) & Tables \ref{tab:errors-Euclid-GC-WL-late_time},\ref{tab:errors-GC-SKAcompare-MG-DE-mu-eta-sigma}; Figs.\ \ref{fig:DE+Planck-ellipses-mu-sig-eta}, \ref{fig:BarPlot-DE-GC-1}, \ref{fig:combined_surveys} \\ 
\hline 
\Tstrut Early-time parameterization & Section \ref{sub:param-smooth-funct}, Eqns.\ (\ref{eq:TR-mu-parametrization})-(\ref{eq:TR-eta-parametrization}) & Tables \ref{tab:errors-Euclid-GC-WL-early_time},\ref{tab:errors-GC-SKAcompare-MG-TR-mu-eta-sigma-Zhao-1}; Figs.\ \ref{fig:T-related-ellipses-mu-omegac}, \ref{fig:BarPlot-MGTR-Surveys}, \ref{fig:combined_surveys} \\ 
\hline 
\Tstrut Effect of non-linear prescription & Section \ref{sub:Prescription-HS}, Eqns.\ (\ref{eq:PHSDefinition}),(\ref{eq:prescription_sigma_def}) & Table \ref{tab:errors-GC+WL-Marginalize-HS-Zhao-Euclid-DEpar-muetasigma}; \,\,\,\,\, Figs.\ \ref{fig:Density-GC-HSpars}, \ref{fig:Density-WL-HSpars} \\ 
\hline 
\end{tabular}
\caption{Summary of results for this work. For each studied model, we indicate where to find the explanations and where to find the main Fisher forecast results.} \label{table_results}
\end{table}

We use the predictions of linear perturbation theory to compute the linear power spectrum in Modified Gravity and then
use a prescription to add the non-linearities, by interpolating between Halofit non-linear corrections computed for the linear power spectrum for the MG model and for the corresponding GR model ($\eta=\mu=1$). We find that the non-linear power spectrum is sensitive to 
changes in $\mu$ and $\eta$; limiting the analysis to linear scales significantly reduces the constraining power on the anisotropic stress.
Using this prescription, we perform Fisher forecasts for Galaxy Clustering and Weak Lensing, taking into account linear and non-linear scales. 
We use the specifications for Euclid (using Redbook specifications), SKA1 \& SKA2 and DESI (only ELG). In addition to these surveys we also include
\planck\ priors obtained by performing an MCMC analysis with \planck\ data for the MG parametrizations considered here.

In the redshift-binned case, we find that in the linear case the $\mu_i$ and $\eta_i$ parameters are strongly correlated, while including the information coming from non linear 
scales reduces this correlation. We compute a figure of merit (FoM), given by the determinant of the $\mu$-$\eta$ part of the Fisher matrix, for the cases examined, finding that 
the combination of Galaxy Clustering and Weak Lensing is able to break the degeneracies among Modified Gravity parameters; as an example the error obtained with the non-linear prescription on $\mu$ ($\eta$) in the first redshift bin changes 
from $7\%$ ($20\%$) for Galaxy Clustering to $2.2\%$ ($3.6\%$) when this is combined with Weak Lensing, even if Weak Lensing alone is not very constraining for the same parameters. 
Overall, constraints are stronger at low redshifts, with the first two bins (0 < z < 1) being constrained at better than $5\%$ for both $\mu$ and $\eta$ if non-linearities are included (while the constraints are half as good, $<10\%$, if we only consider linear scales).

Given the significant correlation between the $\mu_i$ and $\eta_i$ parameters, we apply the ZCA decorrelation method, in order to find a set of uncorrelated 
variables, which gives us information on which redshift dependence of $\mu$ and $\eta$ will be best constrained by future surveys. If one combines GC+WL (Euclid Redbook)+{\it Planck}, the best constrained combinations of parameters (effectively $2\mu+\eta$ in the lowest redshift bin) will be measured with a precision of better than 1\%.
In the linear case, the errors on the decorrelated $q_i$ parameters are about 2 orders of magnitude smaller than for the primary parameters, 
while in the non-linear HS case, the improvement in the errors is of one order of magnitude. This also shows that applying a decorrelation procedure is worth even when non-linearities are considered.

In addition to binning Modified Gravity functions in redshift, we also forecast the constraining power of the same probes in the case where we assume a specific time evolution for the $\mu$ and $\eta$ functions. We choose two different and complementary time evolutions, used in \cite{planck_collaboration_planck_2016} and to which we refer as late-time and early-time evolution. We investigate also in this case the impact of the non-linear prescription interpolating between Halofit and the MG power spectrum. For these parameterizations we 
extract constraints on the present reconstructed value of $\mu$,  $\eta$ and $\Sigma$, where the latter is the parameter actually measured directly by Weak Lensing.  
In the late time parameterization, in the linear case, $\mu$ is mainly constrained by GC (although poorly, at the level of $17\%$) while WL constrains directly $\Sigma$, the modification of the lensing potential, at the level of $9\%$. Adding non-linear scales allows to significantly improve constraints down to less than $2\%$ for GC (on $\mu$) and to less than $5\%$ on $\Sigma$ for those two probes. Combining probes allows to reach $1-2\%$ on all Modified Gravity values of the $\mu$,  $\eta$ and $\Sigma$ functions at z = 0.

In the early-time parameterization, we find that including non-linearities allows to constrain also the $\eta$ and $\Sigma$ functions with GC alone at the level of 8\% and 5\%, respectively. This is related to the early time 
deviations from GR allowed by this parameterization, which are not present in the late time case: a variation in $\eta$ can
yield a variation of the amplitude of the power spectrum, which can then be measured in the mildly non-linear regime.
Overall, also in this case the combination of Weak Lensing and Galaxy Clustering leads to errors of the order of $1\%$ on the
values of these functions at present.

Finally, we test the impact on the forecasts given by uncertainties appearing in the non linear HS prescription, related in particular to the parameters $c_{nl}$ and $s$. We find that
the errors on the parameters $\mu$ and $\eta$ can vary by up to 
$90\%$ for the Galaxy Clustering case and up to $65\%$ for Weak Lensing when we change the fiducial values of the HS parameters in a region between 0 and 3 for $c_{nl}$
and 0 and 1 for $s$. The effect on $\Sigma$ is quite small, with a discrepancy of $\pm 6 \%$ compared to the benchmark case.
Interestingly, when we include these two parameters as extra nuisance parameters in our forecast formalism and marginalize over them, 
the effect is very small and the errors found previously remain stable both for GC, WL and their combination.

It is clear that limiting the analysis to linear scales discards important information encoded in structure formation. On the other hand, a realistic analysis of non-linear scales would have to include several further effects (baryonic effects, higher order RSD's, damping of BAO peaks, corrections
to peculiar velocity perturbations, higher order perturbation theory in Modified Gravity, just to name a few), which make our non-linear case an optimistic limit.
Therefore, the quantitive true constraints given by a survey like Euclid will probably lie in between these two limiting cases.

\section*{Acknowledgments}
MM, SC and VP thank the COST Action (CANTATA/CA15117), supported by COST (European Cooperation in Science and Technology). SC and VP acknowledge support from the Heidelberg Graduate School for Fundamental Physics (HGSFP) and from the SFB-Transregio TR33 "The Dark Universe".
MM is supported by the Foundation for Fundamental Research on Matter (FOM) and the Netherlands Organization for Scientific Research / Ministry of Science and Education (NWO/OCW).
MK acknowledges financial support by the Swiss National Science Foundation. MM also thanks Alessandra Silvestri for useful
discussion. 

\bibliographystyle{unsrtnat}
\bibliography{1-MG-CAMB-Cosmo}

\begin{thebibliography}{82}
\providecommand{\natexlab}[1]{#1}
\providecommand{\url}[1]{\texttt{#1}}
\expandafter\ifx\csname urlstyle\endcsname\relax
  \providecommand{\doi}[1]{doi: #1}\else
  \providecommand{\doi}{doi: \begingroup \urlstyle{rm}\Url}\fi

\bibitem[{Planck Collaboration} et~al.(2016){Planck Collaboration}, Ade, and
  {others}]{planck_collaboration_planck_2016}
{Planck Collaboration}, P.~A.~R. Ade, and {others}.
\newblock Planck 2015 results. {XIV}. {Dark} energy and modified gravity.
\newblock \emph{Astronomy and Astrophysics}, 594:\penalty0 A14, September 2016.
\newblock \doi{10.1051/0004-6361/201525814}.
\newblock URL \url{http://arxiv.org/abs/1502.01590}.

\bibitem[Amendola and {others}(2013)]{amendola_cosmology_2013}
Luca Amendola and {others}.
\newblock Cosmology and fundamental physics with the {Euclid} satellite.
\newblock \emph{Living Reviews in Relativity}, 16, 2013.
\newblock ISSN 1433-8351.
\newblock \doi{10.12942/lrr-2013-6}.
\newblock URL \url{http://arxiv.org/abs/1206.1225}.

\bibitem[Casas et~al.(2015)Casas, Amendola, Baldi, Pettorino, and
  Vollmer]{casas_fitting_2015}
Santiago Casas, Luca Amendola, Marco Baldi, Valeria Pettorino, and Adrian
  Vollmer.
\newblock Fitting and forecasting non-linear coupled dark energy.
\newblock \emph{arXiv:1508.07208 [astro-ph]}, August 2015.
\newblock URL \url{http://arxiv.org/abs/1508.07208}.

\bibitem[Bielefeld et~al.(2014)Bielefeld, Huterer, and
  Linder]{bielefeld_cosmological_2014}
Jannis Bielefeld, Dragan Huterer, and Eric~V. Linder.
\newblock Cosmological {Leverage} from the {Matter} {Power} {Spectrum} in the
  {Presence} of {Baryon} and {Nonlinear} {Effects}.
\newblock \emph{SciRate}, November 2014.
\newblock URL \url{https://scirate.com/arxiv/1411.3725}.

\bibitem[Kunz(2012)]{kunz_phenomenological_2012}
Martin Kunz.
\newblock The phenomenological approach to modeling the dark energy.
\newblock \emph{Comptes Rendus Physique}, 13\penalty0 (6-7):\penalty0 539--565,
  July 2012.
\newblock ISSN 16310705.
\newblock \doi{10.1016/j.crhy.2012.04.007}.
\newblock URL \url{http://arxiv.org/abs/1204.5482}.

\bibitem[Amendola et~al.(2013)Amendola, Kunz, Motta, Saltas, and
  Sawicki]{amendola_observables_2013}
Luca Amendola, Martin Kunz, Mariele Motta, Ippocratis~D. Saltas, and Ignacy
  Sawicki.
\newblock Observables and unobservables in dark energy cosmologies.
\newblock \emph{Physical Review D}, 87:\penalty0 023501, January 2013.
\newblock ISSN 0556-2821.
\newblock \doi{10.1103/PhysRevD.87.023501}.
\newblock URL \url{http://arxiv.org/abs/1210.0439}.

\bibitem[Alonso et~al.(2016)Alonso, Bellini, Ferreira, and
  Zumalacarregui]{Alonso2016}
David Alonso, Emilio Bellini, Pedro~G. Ferreira, and Miguel Zumalacarregui.
\newblock The {Observational} {Future} of {Cosmological} {Scalar}-{Tensor}
  {Theories}.
\newblock \emph{arXiv:1610.09290 [astro-ph, physics:gr-qc, physics:hep-th]},
  October 2016.
\newblock URL \url{http://arxiv.org/abs/1610.09290}.
\newblock arXiv: 1610.09290.

\bibitem[Hojjati et~al.(2012)Hojjati, Zhao, Pogosian, Silvestri, Crittenden,
  and Koyama]{hojjati_cosmological_2012}
Alireza Hojjati, Gong-Bo Zhao, Levon Pogosian, Alessandra Silvestri, Robert
  Crittenden, and Kazuya Koyama.
\newblock Cosmological tests of {General} {Relativity}: a principal component
  analysis.
\newblock \emph{Phys. Rev.}, D85:\penalty0 043508, 2012.
\newblock \doi{10.1103/PhysRevD.85.043508}.
\newblock URL \url{https://arxiv.org/abs/1111.3960}.

\bibitem[Baker and Bull(2015)]{baker_observational_2015}
Tessa Baker and Philip Bull.
\newblock Observational signatures of modified gravity on ultra-large scales.
\newblock \emph{The Astrophysical Journal}, 811\penalty0 (2):\penalty0 116,
  September 2015.
\newblock ISSN 1538-4357.
\newblock \doi{10.1088/0004-637X/811/2/116}.
\newblock URL \url{http://arxiv.org/abs/1506.00641}.

\bibitem[Bull(2015)]{bull_extending_2015}
Philip Bull.
\newblock Extending cosmological tests of {General} {Relativity} with the
  {Square} {Kilometre} {Array}.
\newblock \emph{arXiv:1509.07562 [astro-ph, physics:gr-qc]}, September 2015.
\newblock URL \url{http://arxiv.org/abs/1509.07562}.

\bibitem[Gleyzes et~al.(2016)Gleyzes, Langlois, Mancarella, and
  Vernizzi]{Gleyzes2016}
Jérôme Gleyzes, David Langlois, Michele Mancarella, and Filippo Vernizzi.
\newblock Effective {Theory} of {Dark} {Energy} at {Redshift} {Survey}
  {Scales}.
\newblock \emph{Journal of Cosmology and Astroparticle Physics}, 2016\penalty0
  (02):\penalty0 056--056, February 2016.
\newblock ISSN 1475-7516.
\newblock \doi{10.1088/1475-7516/2016/02/056}.
\newblock URL \url{http://arxiv.org/abs/1509.02191}.
\newblock arXiv: 1509.02191.

\bibitem[Hojjati et~al.(2011)Hojjati, Pogosian, and Zhao]{hojjati_testing_2011}
Alireza Hojjati, Levon Pogosian, and Gong-Bo Zhao.
\newblock Testing gravity with {CAMB} and {CosmoMC}.
\newblock \emph{JCAP}, 1108:\penalty0 005, 2011.
\newblock \doi{10.1088/1475-7516/2011/08/005}.

\bibitem[Smith et~al.(2003)Smith, Peacock, Jenkins, White, Frenk, Pearce,
  Thomas, Efstathiou, and Couchman]{smith_stable_2003}
R.~E. Smith, J.~A. Peacock, A.~Jenkins, S.~D.~M. White, C.~S. Frenk, F.~R.
  Pearce, P.~A. Thomas, G.~Efstathiou, and H.~M.~P. Couchman.
\newblock Stable clustering, the halo model and non-linear cosmological power
  spectra.
\newblock \emph{MNRAS}, 341\penalty0 (4):\penalty0 1311--1332, June 2003.
\newblock ISSN 0035-8711, 1365-2966.
\newblock \doi{10.1046/j.1365-8711.2003.06503.x}.
\newblock URL \url{http://mnras.oxfordjournals.org/content/341/4/1311}.

\bibitem[Takahashi et~al.(2012)Takahashi, Sato, Nishimichi, Taruya, and
  Oguri]{takahashi_revising_2012}
Ryuichi Takahashi, Masanori Sato, Takahiro Nishimichi, Atsushi Taruya, and
  Masamune Oguri.
\newblock Revising the {Halofit} {Model} for the {Nonlinear} {Matter} {Power}
  {Spectrum}.
\newblock \emph{arXiv:1208.2701}, August 2012.
\newblock URL \url{http://arxiv.org/abs/1208.2701}.

\bibitem[Hu and Sawicki(2007)]{hu_parameterized_2007}
Wayne Hu and Ignacy Sawicki.
\newblock A {Parameterized} {Post}-{Friedmann} {Framework} for {Modified}
  {Gravity}.
\newblock \emph{Phys.Rev.}, D76:\penalty0 104043, 2007.
\newblock \doi{10.1103/PhysRevD.76.104043}.

\bibitem[Saltas et~al.(2014)Saltas, Sawicki, Amendola, and
  Kunz]{saltas_anisotropic_2014}
Ippocratis~D. Saltas, Ignacy Sawicki, Luca Amendola, and Martin Kunz.
\newblock Anisotropic {Stress} as a {Signature} of {Nonstandard} {Propagation}
  of {Gravitational} {Waves}.
\newblock \emph{Phys. Rev. Lett.}, 113\penalty0 (19):\penalty0 191101, 2014.
\newblock \doi{10.1103/PhysRevLett.113.191101}.

\bibitem[Sawicki et~al.(2016)Sawicki, Saltas, Motta, Amendola, and
  Kunz]{sawicki_non-standard_2016}
Ignacy Sawicki, Ippocratis~D. Saltas, Mariele Motta, Luca Amendola, and Martin
  Kunz.
\newblock Non-standard gravitational waves imply gravitational slip: on the
  difficulty of partially hiding new gravitational degrees of freedom.
\newblock 2016.

\bibitem[{Pogosian} et~al.(2010){Pogosian}, {Silvestri}, {Koyama}, and
  {Zhao}]{pogosian_how_2010}
L.~{Pogosian}, A.~{Silvestri}, K.~{Koyama}, and G.-B. {Zhao}.
\newblock How to optimally parametrize deviations from general relativity in
  the evolution of cosmological perturbations.
\newblock \emph{Phys. Rev.}, D81\penalty0 (10):\penalty0 104023, May 2010.
\newblock \doi{10.1103/PhysRevD.81.104023}.

\bibitem[Gubitosi et~al.(2013)Gubitosi, Piazza, and Vernizzi]{Gubitosi2013}
Giulia Gubitosi, Federico Piazza, and Filippo Vernizzi.
\newblock The {Effective} {Field} {Theory} of {Dark} {Energy}.
\newblock \emph{Journal of Cosmology and Astroparticle Physics}, 2013\penalty0
  (02):\penalty0 032--032, February 2013.
\newblock ISSN 1475-7516.
\newblock \doi{10.1088/1475-7516/2013/02/032}.
\newblock URL \url{http://arxiv.org/abs/1210.0201}.
\newblock arXiv: 1210.0201.

\bibitem[Bellini et~al.(2016)Bellini, Cuesta, Jimenez, and Verde]{Bellini2016}
Emilio Bellini, Antonio~J. Cuesta, Raul Jimenez, and Licia Verde.
\newblock Constraints on deviations from \$\{{\textbackslash}{Lambda}\}\${CDM}
  within {Horndeski} gravity.
\newblock \emph{Journal of Cosmology and Astroparticle Physics}, 2016\penalty0
  (02):\penalty0 053--053, February 2016.
\newblock ISSN 1475-7516.
\newblock \doi{10.1088/1475-7516/2016/02/053}.
\newblock URL \url{http://arxiv.org/abs/1509.07816}.
\newblock arXiv: 1509.07816.

\bibitem[Bellini and Sawicki(2014)]{Bellini2014}
Emilio Bellini and Ignacy Sawicki.
\newblock Maximal freedom at minimum cost: linear large-scale structure in
  general modifications of gravity.
\newblock \emph{arXiv:1404.3713 [astro-ph, physics:gr-qc]}, April 2014.
\newblock URL \url{http://arxiv.org/abs/1404.3713}.
\newblock arXiv: 1404.3713.

\bibitem[Zhao et~al.(2009)Zhao, Pogosian, Silvestri, and
  Zylberberg]{zhao_searching_2009}
Gong-Bo Zhao, Levon Pogosian, Alessandra Silvestri, and Joel Zylberberg.
\newblock Searching for modified growth patterns with tomographic surveys.
\newblock \emph{Phys. Rev.}, D79:\penalty0 083513, 2009.
\newblock \doi{10.1103/PhysRevD.79.083513}.

\bibitem[Lewis et~al.(2000)Lewis, Challinor, and Lasenby]{lewis_efficient_2000}
Antony Lewis, Anthony Challinor, and Anthony Lasenby.
\newblock Efficient {Computation} of {CMB} anisotropies in closed {FRW} models.
\newblock \emph{Astrophys. J.}, 538:\penalty0 473--476, 2000.
\newblock \doi{10.1086/309179}.

\bibitem[Bernardeau et~al.(2001)Bernardeau, Colombi, Gaztanaga, and
  Scoccimarro]{bernardeau_large-scale_2001}
F.~Bernardeau, S.~Colombi, E.~Gaztanaga, and R.~Scoccimarro.
\newblock Large-{Scale} {Structure} of the {Universe} and {Cosmological}
  {Perturbation} {Theory}.
\newblock \emph{arXiv:astro-ph/0112551}, December 2001.
\newblock \doi{10.1016/S0370-1573(02)00135-7}.
\newblock URL \url{http://arxiv.org/abs/astro-ph/0112551}.

\bibitem[Bernardeau(2013)]{bernardeau_evolution_2013}
Francis Bernardeau.
\newblock The evolution of the large-scale structure of the universe: beyond
  the linear regime.
\newblock \emph{arXiv:1311.2724 [astro-ph]}, November 2013.
\newblock URL \url{http://arxiv.org/abs/1311.2724}.

\bibitem[Crocce and Scoccimarro(2006)]{crocce_renormalized_2006}
Martín Crocce and Román Scoccimarro.
\newblock Renormalized cosmological perturbation theory.
\newblock \emph{Physical Review D}, 73\penalty0 (6):\penalty0 063519, March
  2006.
\newblock \doi{10.1103/PhysRevD.73.063519}.
\newblock URL \url{http://link.aps.org/doi/10.1103/PhysRevD.73.063519}.

\bibitem[Blas et~al.(2016)Blas, Garny, Ivanov, and
  Sibiryakov]{blas_time-sliced_2016}
Diego Blas, Mathias Garny, Mikhail~M. Ivanov, and Sergey Sibiryakov.
\newblock Time-sliced perturbation theory {II}: baryon acoustic oscillations
  and infrared resummation.
\newblock \emph{Journal of Cosmology and Astroparticle Physics}, 2016\penalty0
  (07):\penalty0 028, 2016.
\newblock ISSN 1475-7516.
\newblock \doi{10.1088/1475-7516/2016/07/028}.
\newblock URL \url{http://stacks.iop.org/1475-7516/2016/i=07/a=028}.

\bibitem[Taruya and Hiramatsu(2008)]{taruya_closure_2008}
Atsushi Taruya and Takashi Hiramatsu.
\newblock A {Closure} {Theory} for {Non}-linear {Evolution} of {Cosmological}
  {Power} {Spectra}.
\newblock \emph{The Astrophysical Journal}, 674\penalty0 (2):\penalty0
  617--635, February 2008.
\newblock ISSN 0004-637X, 1538-4357.
\newblock \doi{10.1086/526515}.
\newblock URL \url{http://arxiv.org/abs/0708.1367}.

\bibitem[Pietroni(2008)]{pietroni_flowing_2008}
Massimo Pietroni.
\newblock Flowing with {Time}: a {New} {Approach} to {Nonlinear} {Cosmological}
  {Perturbations}.
\newblock \emph{JCAP}, 0810:\penalty0 036, 2008.
\newblock \doi{10.1088/1475-7516/2008/10/036}.

\bibitem[Anselmi and Pietroni(2012)]{anselmi_nonlinear_2012}
Stefano Anselmi and Massimo Pietroni.
\newblock Nonlinear power spectrum from resummed perturbation theory: a leap
  beyond the {BAO} scale.
\newblock \emph{Journal of Cosmology and Astroparticle Physics}, 2012\penalty0
  (12):\penalty0 013, 2012.
\newblock URL \url{http://iopscience.iop.org/1475-7516/2012/12/013}.

\bibitem[Carrasco et~al.(2012)Carrasco, Hertzberg, and
  Senatore]{carrasco_effective_2012}
John Joseph~M. Carrasco, Mark~P. Hertzberg, and Leonardo Senatore.
\newblock The effective field theory of cosmological large scale structures.
\newblock \emph{Journal of High Energy Physics}, 2012\penalty0 (9):\penalty0
  1--40, 2012.
\newblock URL \url{http://link.springer.com/article/10.1007/JHEP09(2012)082}.

\bibitem[Baumann et~al.(2012)Baumann, Nicolis, Senatore, and
  Zaldarriaga]{baumann_cosmological_2012}
Daniel Baumann, Alberto Nicolis, Leonardo Senatore, and Matias Zaldarriaga.
\newblock Cosmological non-linearities as an effective fluid.
\newblock \emph{Journal of Cosmology and Astroparticle Physics}, 2012\penalty0
  (07):\penalty0 051, 2012.
\newblock URL \url{http://iopscience.iop.org/1475-7516/2012/07/051}.

\bibitem[Pietroni et~al.(2011)Pietroni, Mangano, Saviano, and
  Viel]{pietroni_coarse-grained_2011}
Massimo Pietroni, Gianpiero Mangano, Ninetta Saviano, and Matteo Viel.
\newblock Coarse-{Grained} {Cosmological} {Perturbation} {Theory}.
\newblock \emph{arXiv:1108.5203 [astro-ph, physics:hep-th]}, August 2011.
\newblock URL \url{http://arxiv.org/abs/1108.5203}.

\bibitem[Manzotti et~al.(2014)Manzotti, Peloso, Pietroni, Viel, and
  Villaescusa-Navarro]{manzotti_coarse_2014}
Alessandro Manzotti, Marco Peloso, Massimo Pietroni, Matteo Viel, and Francisco
  Villaescusa-Navarro.
\newblock A coarse grained perturbation theory for the {Large} {Scale}
  {Structure}, with cosmology and time independence in the {UV}.
\newblock \emph{arXiv:1407.1342 [astro-ph, physics:gr-qc, physics:hep-th]},
  July 2014.
\newblock URL \url{http://arxiv.org/abs/1407.1342}.

\bibitem[Springel(2005)]{springel_cosmological_2005}
Volker Springel.
\newblock The cosmological simulation code {GADGET}-2.
\newblock \emph{Mon. Not. Roy. Astron. Soc.}, 364:\penalty0 1105--1134, 2005.

\bibitem[Fosalba et~al.(2013)Fosalba, Crocce, Gaztanaga, and
  Castander]{fosalba_mice_2013}
P.~Fosalba, M.~Crocce, E.~Gaztanaga, and F.~J. Castander.
\newblock The {MICE} {Grand} {Challenge} {Lightcone} {Simulation} {I}: {Dark}
  matter clustering.
\newblock \emph{arXiv preprint arXiv:1312.1707}, 2013.
\newblock URL \url{http://arxiv.org/abs/1312.1707}.

\bibitem[Lawrence et~al.(2010)Lawrence, Heitmann, White, Higdon, Wagner, Habib,
  and Williams]{lawrence_coyote_2010}
Earl Lawrence, Katrin Heitmann, Martin White, David Higdon, Christian Wagner,
  Salman Habib, and Brian Williams.
\newblock The coyote universe. {III}. simulation suite and precision emulator
  for the nonlinear matter power spectrum.
\newblock \emph{The Astrophysical Journal}, 713\penalty0 (2):\penalty0 1322,
  2010.
\newblock URL \url{http://iopscience.iop.org/0004-637X/713/2/1322}.

\bibitem[Heitmann et~al.(2014)Heitmann, Lawrence, Kwan, Habib, and
  Higdon]{heitmann_coyote_2014}
Katrin Heitmann, Earl Lawrence, Juliana Kwan, Salman Habib, and David Higdon.
\newblock The {Coyote} {Universe} {Extended}: {Precision} {Emulation} of the
  {Matter} {Power} {Spectrum}.
\newblock \emph{The Astrophysical Journal}, 780\penalty0 (1):\penalty0 111,
  2014.
\newblock URL \url{http://iopscience.iop.org/0004-637X/780/1/111}.

\bibitem[Heymans and {others}(2013)]{heymans_cfhtlens_2013}
Catherine Heymans and {others}.
\newblock {CFHTLenS} tomographic weak lensing cosmological parameter
  constraints: {Mitigating} the impact of intrinsic galaxy alignments.
\newblock \emph{Mon. Not. Roy. Astron. Soc.}, 432:\penalty0 2433, 2013.
\newblock \doi{10.1093/mnras/stt601}.

\bibitem[Kitching and {others}(2014)]{kitching_3d_2014}
T.~D. Kitching and {others}.
\newblock 3d {Cosmic} {Shear}: {Cosmology} from {CFHTLenS}.
\newblock \emph{Mon. Not. Roy. Astron. Soc.}, 442\penalty0 (2):\penalty0
  1326--1349, 2014.
\newblock \doi{10.1093/mnras/stu934}.

\bibitem[Zhao(2014)]{zhao_modeling_2014}
Gong-Bo Zhao.
\newblock Modeling the nonlinear clustering in modified gravity models {I}: {A}
  fitting formula for matter power spectrum of f({R}) gravity.
\newblock \emph{The Astrophysical Journal Supplement Series}, 211\penalty0
  (2):\penalty0 23, April 2014.
\newblock ISSN 0067-0049, 1538-4365.
\newblock \doi{10.1088/0067-0049/211/2/23}.
\newblock URL \url{http://arxiv.org/abs/1312.1291}.

\bibitem[Taruya et~al.(2014)Taruya, Nishimichi, Bernardeau, Hiramatsu, and
  Koyama]{taruya_regularized_2014}
Atsushi Taruya, Takahiro Nishimichi, Francis Bernardeau, Takashi Hiramatsu, and
  Kazuya Koyama.
\newblock Regularized cosmological power spectrum and correlation function in
  modified gravity models.
\newblock \emph{Physical Review D}, 90\penalty0 (12), December 2014.
\newblock ISSN 1550-7998, 1550-2368.
\newblock \doi{10.1103/PhysRevD.90.123515}.
\newblock URL \url{http://arxiv.org/abs/1408.4232}.

\bibitem[Saracco et~al.(2010)Saracco, Pietroni, Tetradis, Pettorino, and
  Robbers]{saracco_non-linear_2010}
F.~Saracco, M.~Pietroni, N.~Tetradis, V.~Pettorino, and G.~Robbers.
\newblock Non-linear {Matter} {Spectra} in {Coupled} {Quintessence}.
\newblock \emph{Physical Review D}, 82\penalty0 (2), July 2010.
\newblock ISSN 1550-7998, 1550-2368.
\newblock \doi{10.1103/PhysRevD.82.023528}.
\newblock URL \url{http://arxiv.org/abs/0911.5396}.

\bibitem[Vollmer et~al.(2014)Vollmer, Amendola, and
  Catena]{vollmer_efficient_2014}
Adrian Vollmer, Luca Amendola, and Riccardo Catena.
\newblock Efficient implementation of the {Time} {Renormalization} {Group}.
\newblock \emph{arXiv:1412.1650 [astro-ph, physics:physics]}, December 2014.
\newblock URL \url{http://arxiv.org/abs/1412.1650}.

\bibitem[Brouzakis et~al.(2011)Brouzakis, Pettorino, Tetradis, and
  Wetterich]{brouzakis_nonlinear_2011}
N.~Brouzakis, V.~Pettorino, N.~Tetradis, and C.~Wetterich.
\newblock Nonlinear matter spectra in growing neutrino quintessence.
\newblock \emph{JCAP}, 1103:\penalty0 049, 2011.
\newblock \doi{10.1088/1475-7516/2011/03/049}.

\bibitem[Bird et~al.(2011)Bird, Viel, and Haehnelt]{bird_massive_2011}
Simeon Bird, Matteo Viel, and Martin~G. Haehnelt.
\newblock Massive {Neutrinos} and the {Non}-linear {Matter} {Power} {Spectrum}.
\newblock \emph{arXiv:1109.4416}, September 2011.
\newblock \doi{10.1111/j.1365-2966.2011.20222.x}.
\newblock URL \url{http://arxiv.org/abs/1109.4416}.

\bibitem[Zhao et~al.(2011)Zhao, Li, and Koyama]{zhao_n-body_2011}
Gong-Bo Zhao, Baojiu Li, and Kazuya Koyama.
\newblock N-body {Simulations} for f({R}) {Gravity} using a {Self}-adaptive
  {Particle}-{Mesh} {Code}.
\newblock \emph{Physical Review D}, 83\penalty0 (4), February 2011.
\newblock ISSN 1550-7998, 1550-2368.
\newblock \doi{10.1103/PhysRevD.83.044007}.
\newblock URL \url{http://arxiv.org/abs/1011.1257}.

\bibitem[Koyama et~al.(2009)Koyama, Taruya, and
  Hiramatsu]{koyama_non-linear_2009}
Kazuya Koyama, Atsushi Taruya, and Takashi Hiramatsu.
\newblock Non-linear {Evolution} of {Matter} {Power} {Spectrum} in {Modified}
  {Theory} of {Gravity}.
\newblock \emph{Physical Review D}, 79\penalty0 (12), June 2009.
\newblock ISSN 1550-7998, 1550-2368.
\newblock \doi{10.1103/PhysRevD.79.123512}.
\newblock URL \url{http://arxiv.org/abs/0902.0618}.

\bibitem[Tegmark et~al.(1998)Tegmark, Hamilton, Strauss, Vogeley, and
  Szalay]{tegmark_measuring_1998}
Max Tegmark, Andrew Hamilton, Michael Strauss, Michael Vogeley, and Alexander
  Szalay.
\newblock Measuring the galaxy power spectrum with future redshift surveys.
\newblock \emph{The Astrophysical Journal}, 499\penalty0 (2):\penalty0
  555--576, June 1998.
\newblock ISSN 0004-637X, 1538-4357.
\newblock \doi{10.1086/305663}.
\newblock URL \url{http://arxiv.org/abs/astro-ph/9708020}.

\bibitem[Seo and Eisenstein(2007)]{seo_improved_2007}
Hee-Jong Seo and Daniel~J. Eisenstein.
\newblock Improved forecasts for the baryon acoustic oscillations and
  cosmological distance scale.
\newblock \emph{The Astrophysical Journal}, 665\penalty0 (1):\penalty0 14--24,
  August 2007.
\newblock ISSN 0004-637X, 1538-4357.
\newblock \doi{10.1086/519549}.
\newblock URL \url{http://arxiv.org/abs/astro-ph/0701079}.

\bibitem[Seo and Eisenstein(2005)]{seo_baryonic_2005}
Hee-Jong Seo and Daniel~J. Eisenstein.
\newblock Baryonic acoustic oscillations in simulated galaxy redshift surveys.
\newblock \emph{The Astrophysical Journal}, 633\penalty0 (2):\penalty0
  575--588, November 2005.
\newblock ISSN 0004-637X, 1538-4357.
\newblock \doi{10.1086/491599}.
\newblock URL \url{http://arxiv.org/abs/astro-ph/0507338}.

\bibitem[Mukherjee et~al.(2008)Mukherjee, Kunz, Parkinson, and
  Wang]{mukherjee_planck_2008}
Pia Mukherjee, Martin Kunz, David Parkinson, and Yun Wang.
\newblock Planck priors for dark energy surveys.
\newblock \emph{Physical Review D}, 78\penalty0 (8), October 2008.
\newblock ISSN 1550-7998, 1550-2368.
\newblock \doi{10.1103/PhysRevD.78.083529}.
\newblock URL \url{http://arxiv.org/abs/0803.1616}.

\bibitem[Laureijs and {others}(2011)]{laureijs_euclid_2011}
R.~Laureijs and {others}.
\newblock Euclid {Definition} {Study} {Report}.
\newblock \emph{arXiv:1110.3193 [astro-ph]}, October 2011.
\newblock URL \url{http://arxiv.org/abs/1110.3193}.

\bibitem[Yahya et~al.(2015)Yahya, Bull, Santos, Silva, Maartens, Okouma, and
  Bassett]{yahya_cosmological_2015}
S.~Yahya, P.~Bull, M.~G. Santos, M.~Silva, R.~Maartens, P.~Okouma, and
  B.~Bassett.
\newblock Cosmological performance of {SKA} {HI} galaxy surveys.
\newblock \emph{Monthly Notices of the Royal Astronomical Society},
  450\penalty0 (3):\penalty0 2251--2260, May 2015.
\newblock ISSN 0035-8711, 1365-2966.
\newblock \doi{10.1093/mnras/stv695}.
\newblock URL \url{http://arxiv.org/abs/1412.4700}.

\bibitem[Santos et~al.(2015)Santos, Alonso, Bull, Silva, and
  Yahya]{santos_hi_2015}
Mario~G. Santos, David Alonso, Philip Bull, Marta Silva, and Sahba Yahya.
\newblock {HI} galaxy simulations for the {SKA}: number counts and bias.
\newblock \emph{arXiv:1501.03990 [astro-ph]}, January 2015.
\newblock URL \url{http://arxiv.org/abs/1501.03990}.

\bibitem[Raccanelli et~al.(2015)Raccanelli, Bull, Camera, Bacon, Blake, Dore,
  Ferreira, Maartens, Santos, Viel, and Zhao]{raccanelli_measuring_2015}
Alvise Raccanelli, Philip Bull, Stefano Camera, David Bacon, Chris Blake,
  Olivier Dore, Pedro Ferreira, Roy Maartens, Mario Santos, Matteo Viel, and
  Gong-bo Zhao.
\newblock Measuring redshift-space distortions with future {SKA} surveys.
\newblock 2015.

\bibitem[Bull et~al.(2015)Bull, Camera, Raccanelli, Blake, Ferreira, Santos,
  and Schwarz]{bull_measuring_2015}
Philip Bull, Stefano Camera, Alvise Raccanelli, Chris Blake, Pedro~G. Ferreira,
  Mario~G. Santos, and Dominik~J. Schwarz.
\newblock Measuring baryon acoustic oscillations with future {SKA} surveys.
\newblock \emph{arXiv:1501.04088 [astro-ph]}, January 2015.
\newblock URL \url{http://arxiv.org/abs/1501.04088}.

\bibitem[Harrison et~al.(2016)Harrison, Camera, Zuntz, and
  Brown]{harrison_ska_2016}
Ian Harrison, Stefano Camera, Joe Zuntz, and Michael~L. Brown.
\newblock {SKA} {Weak} {Lensing} {I}: {Cosmological} {Forecasts} and the
  {Power} of {Radio}-{Optical} {Cross}-{Correlations}.
\newblock \emph{arXiv:1601.03947 [astro-ph]}, January 2016.
\newblock URL \url{http://arxiv.org/abs/1601.03947}.

\bibitem[{DESI Collaboration} et~al.(2016{\natexlab{a}}){DESI Collaboration},
  Aghamousa, and {others}]{desi_collaboration_desi_2016-1}
{DESI Collaboration}, Amir Aghamousa, and {others}.
\newblock The {DESI} {Experiment} {Part} {I}: {Science},{Targeting}, and
  {Survey} {Design}.
\newblock \emph{arXiv:1611.00036 [astro-ph]}, October 2016{\natexlab{a}}.
\newblock URL \url{http://arxiv.org/abs/1611.00036}.

\bibitem[{DESI Collaboration} et~al.(2016{\natexlab{b}}){DESI Collaboration},
  Aghamousa, and {others}]{desi_collaboration_desi_2016}
{DESI Collaboration}, Amir Aghamousa, and {others}.
\newblock The {DESI} {Experiment} {Part} {II}: {Instrument} {Design}.
\newblock \emph{arXiv:1611.00037 [astro-ph]}, October 2016{\natexlab{b}}.
\newblock URL \url{http://arxiv.org/abs/1611.00037}.

\bibitem[Levi et~al.(2013)Levi, Bebek, Beers, Blum, Cahn, Eisenstein, Flaugher,
  Honscheid, Kron, Lahav, McDonald, Roe, Schlegel, and
  collaboration]{levi_desi_2013}
Michael Levi, Chris Bebek, Timothy Beers, Robert Blum, Robert Cahn, Daniel
  Eisenstein, Brenna Flaugher, Klaus Honscheid, Richard Kron, Ofer Lahav,
  Patrick McDonald, Natalie Roe, David Schlegel, and representing the~DESI
  collaboration.
\newblock The {DESI} {Experiment}, a whitepaper for {Snowmass} 2013.
\newblock \emph{arXiv:1308.0847 [astro-ph]}, August 2013.
\newblock URL \url{http://arxiv.org/abs/1308.0847}.

\bibitem[de~la Torre and Guzzo(2012)]{de_la_torre_modelling_2012}
Sylvain de~la Torre and Luigi Guzzo.
\newblock Modelling non-linear redshift-space distortions in the galaxy
  clustering pattern: systematic errors on the growth rate parameter.
\newblock \emph{arXiv:1202.5559 [astro-ph]}, February 2012.
\newblock URL \url{http://arxiv.org/abs/1202.5559}.

\bibitem[Alcock and Paczynski(1979)]{alcock_evolution_1979}
C.~Alcock and B.~Paczynski.
\newblock An evolution free test for non-zero cosmological constant.
\newblock \emph{Nature}, 281:\penalty0 358, October 1979.
\newblock ISSN 0028-0836.
\newblock \doi{10.1038/281358a0}.
\newblock URL \url{http://adsabs.harvard.edu/abs/1979Natur.281..358A}.

\bibitem[Ballinger et~al.(1996)Ballinger, Peacock, and
  Heavens]{ballinger_measuring_1996}
W.~E. Ballinger, J.~A. Peacock, and A.~F. Heavens.
\newblock Measuring the cosmological constant with redshift surveys.
\newblock \emph{Monthly Notices of the Royal Astronomical Society},
  282:\penalty0 877, October 1996.
\newblock ISSN 0035-8711.
\newblock \doi{10.1093/mnras/282.3.877}.
\newblock URL \url{http://adsabs.harvard.edu/abs/1996MNRAS.282..877B}.

\bibitem[Feldman et~al.(1994)Feldman, Kaiser, and Peacock]{feldman_power_1994}
Hume~A. Feldman, Nick Kaiser, and John~A. Peacock.
\newblock Power {Spectrum} {Analysis} of {Three}-{Dimensional} {Redshift}
  {Surveys}.
\newblock \emph{The Astrophysical Journal}, 426:\penalty0 23, May 1994.
\newblock ISSN 0004-637X, 1538-4357.
\newblock \doi{10.1086/174036}.
\newblock URL \url{http://arxiv.org/abs/astro-ph/9304022}.

\bibitem[Amendola et~al.(2012)Amendola, Pettorino, Quercellini, and
  Vollmer]{amendola_testing_2012}
Luca Amendola, Valeria Pettorino, Claudia Quercellini, and Adrian Vollmer.
\newblock Testing coupled dark energy with next-generation large-scale
  observations.
\newblock \emph{Physical Review D}, 85\penalty0 (10), May 2012.
\newblock ISSN 1550-7998, 1550-2368.
\newblock \doi{10.1103/PhysRevD.85.103008}.
\newblock URL \url{http://link.aps.org/doi/10.1103/PhysRevD.85.103008}.

\bibitem[Bailoni et~al.(2016)Bailoni, Spurio~Mancini, and
  Amendola]{bailoni_improving_2016}
Alberto Bailoni, Alessio Spurio~Mancini, and Luca Amendola.
\newblock Improving {Fisher} matrix forecasts for galaxy surveys: window
  function, bin cross-correlation, and bin redshift uncertainty.
\newblock \emph{ArXiv e-prints}, 1608:\penalty0 arXiv:1608.00458, August 2016.
\newblock URL \url{http://adsabs.harvard.edu/abs/2016arXiv160800458B}.

\bibitem[Lewis and Bridle(2002)]{lewis_cosmological_2002}
Antony Lewis and Sarah Bridle.
\newblock Cosmological parameters from {CMB} and other data: {A} {Monte}
  {Carlo} approach.
\newblock \emph{Phys. Rev.}, D66:\penalty0 103511, 2002.
\newblock \doi{10.1103/PhysRevD.66.103511}.

\bibitem[Lewis(2013)]{lewis_efficient_2013}
Antony Lewis.
\newblock Efficient sampling of fast and slow cosmological parameters.
\newblock \emph{Phys. Rev.}, D87\penalty0 (10):\penalty0 103529, 2013.
\newblock \doi{10.1103/PhysRevD.87.103529}.

\bibitem[Hu and Sugiyama(1996)]{hu_small_1996}
Wayne Hu and Naoshi Sugiyama.
\newblock Small {Scale} {Cosmological} {Perturbations}: {An} {Analytic}
  {Approach}.
\newblock \emph{The Astrophysical Journal}, 471\penalty0 (2):\penalty0
  542--570, November 1996.
\newblock ISSN 0004-637X, 1538-4357.
\newblock \doi{10.1086/177989}.
\newblock URL \url{http://arxiv.org/abs/astro-ph/9510117}.

\bibitem[Simpson and Peacock(2010)]{Simpson2010}
Fergus Simpson and John~A. Peacock.
\newblock Difficulties {Distinguishing} {Dark} {Energy} from {Modified}
  {Gravity} via {Redshift} {Distortions}.
\newblock \emph{Physical Review D}, 81\penalty0 (4), February 2010.
\newblock ISSN 1550-7998, 1550-2368.
\newblock \doi{10.1103/PhysRevD.81.043512}.
\newblock URL \url{http://arxiv.org/abs/0910.3834}.
\newblock arXiv: 0910.3834.

\bibitem[Leonard et~al.(2015)Leonard, Baker, and Ferreira]{Leonard2015}
C.~Danielle Leonard, Tessa Baker, and Pedro~G. Ferreira.
\newblock Exploring degeneracies in modified gravity with weak lensing.
\newblock \emph{Physical Review D}, 91\penalty0 (8), April 2015.
\newblock ISSN 1550-7998, 1550-2368.
\newblock \doi{10.1103/PhysRevD.91.083504}.
\newblock URL \url{http://arxiv.org/abs/1501.03509}.
\newblock arXiv: 1501.03509.

\bibitem[Bell and Sejnowski(1997)]{Bell19973327}
Anthony~J. Bell and Terrence~J. Sejnowski.
\newblock The “independent components” of natural scenes are edge filters.
\newblock \emph{Vision Research}, 37\penalty0 (23):\penalty0 3327 -- 3338,
  1997.
\newblock ISSN 0042-6989.
\newblock \doi{http://dx.doi.org/10.1016/S0042-6989(97)00121-1}.
\newblock URL
  \url{http://www.sciencedirect.com/science/article/pii/S0042698997001211}.

\bibitem[Kessy et~al.(2015)Kessy, Lewin, and Strimmer]{kessy_optimal_2015}
Agnan Kessy, Alex Lewin, and Korbinian Strimmer.
\newblock Optimal whitening and decorrelation.
\newblock \emph{arXiv:1512.00809 [stat]}, December 2015.
\newblock URL \url{http://arxiv.org/abs/1512.00809}.

\bibitem[Lacasa and Rosenfeld(2016)]{Lacasa2016}
Fabien Lacasa and Rogerio Rosenfeld.
\newblock Combining cluster number counts and galaxy clustering.
\newblock \emph{JCAP}, 1608:\penalty0 005, March 2016.
\newblock \doi{10.1088/1475-7516/2016/08/005, 10.1088/1475-7516/2016/08/005}.
\newblock URL \url{https://arxiv.org/abs/1603.00918}.

\bibitem[Friedman(1987)]{Friedman-PCA}
Jerome~H. Friedman.
\newblock Exploratory projection pursuit.
\newblock \emph{Journal of the American Statistical Association}, 82\penalty0
  (397):\penalty0 249--266, 1987.
\newblock \doi{10.1080/01621459.1987.10478427}.
\newblock URL
  \url{http://amstat.tandfonline.com/doi/abs/10.1080/01621459.1987.10478427}.

\bibitem[Kullback and Leibler(1951)]{kullback_information_1951}
S.~Kullback and R.~A. Leibler.
\newblock On {Information} and {Sufficiency}.
\newblock \emph{The Annals of Mathematical Statistics}, 22\penalty0
  (1):\penalty0 79--86, March 1951.
\newblock ISSN 0003-4851, 2168-8990.
\newblock \doi{10.1214/aoms/1177729694}.
\newblock URL \url{http://projecteuclid.org/euclid.aoms/1177729694}.

\bibitem[Kunz et~al.(2006)Kunz, Aghanim, Cayon, Forni, Riazuelo, and
  Uzan]{kunz_constraining_2006}
M.~Kunz, N.~Aghanim, L.~Cayon, O.~Forni, A.~Riazuelo, and J.~P. Uzan.
\newblock Constraining topology in harmonic space.
\newblock \emph{Physical Review D}, 73\penalty0 (2), January 2006.
\newblock ISSN 1550-7998, 1550-2368.
\newblock \doi{10.1103/PhysRevD.73.023511}.
\newblock URL \url{http://arxiv.org/abs/astro-ph/0510164}.

\bibitem[Raveri et~al.(2016)Raveri, Martinelli, Zhao, and
  Wang]{raveri_cosmicfish_2016}
Marco Raveri, Matteo Martinelli, Gongbo Zhao, and Yuting Wang.
\newblock {CosmicFish} {Implementation} {Notes} {V}1.0.
\newblock \emph{arXiv:1606.06268 [astro-ph, physics:gr-qc]}, June 2016.
\newblock URL \url{http://arxiv.org/abs/1606.06268}.

\bibitem[Seehars et~al.(2014)Seehars, Amara, Refregier, Paranjape, and
  Akeret]{seehars_information_2014}
Sebastian Seehars, Adam Amara, Alexandre Refregier, Aseem Paranjape, and Joël
  Akeret.
\newblock Information {Gains} from {Cosmic} {Microwave} {Background}
  {Experiments}.
\newblock \emph{Physical Review D}, 90\penalty0 (2), July 2014.
\newblock ISSN 1550-7998, 1550-2368.
\newblock \doi{10.1103/PhysRevD.90.023533}.
\newblock URL \url{http://arxiv.org/abs/1402.3593}.

\bibitem[Verde et~al.(2013)Verde, Protopapas, and Jimenez]{verde_planck_2013}
Licia Verde, Pavlos Protopapas, and Raul Jimenez.
\newblock Planck and the local {Universe}: {Quantifying} the tension.
\newblock \emph{Physics of the Dark Universe}, 2\penalty0 (3):\penalty0
  166--175, September 2013.
\newblock ISSN 2212-6864.
\newblock \doi{10.1016/j.dark.2013.09.002}.
\newblock URL
  \url{http://www.sciencedirect.com/science/article/pii/S2212686413000319}.

\bibitem[Zhao et~al.(2017)Zhao, Raveri, Pogosian, Wang, Crittenden, Handley,
  Percival, Brinkmann, Chuang, Cuesta, Eisenstein, Kitaura, Koyama, L'Huillier,
  Nichol, Pieri, Rodriguez-Torres, Ross, Rossi, Sánchez, Shafieloo, Tinker,
  Tojeiro, Vazquez, and Zhang]{Zhao2017}
Gong-Bo Zhao, Marco Raveri, Levon Pogosian, Yuting Wang, Robert~G. Crittenden,
  Will~J. Handley, Will~J. Percival, Jonathan Brinkmann, Chia-Hsun Chuang,
  Antonio~J. Cuesta, Daniel~J. Eisenstein, Francisco-Shu Kitaura, Kazuya
  Koyama, Benjamin L'Huillier, Robert~C. Nichol, Matthew~M. Pieri, Sergio
  Rodriguez-Torres, Ashley~J. Ross, Graziano Rossi, Ariel~G. Sánchez, Arman
  Shafieloo, Jeremy~L. Tinker, Rita Tojeiro, Jose~A. Vazquez, and Hanyu Zhang.
\newblock The clustering of galaxies in the completed {SDSS}-{III} {Baryon}
  {Oscillation} {Spectroscopic} {Survey}: {Examining} the observational
  evidence for dynamical dark energy.
\newblock 2017.

\end{thebibliography}

\appendix

\section{Transformation of primary variables in Modified Gravity}\label{sec:appjac}

\subsection{Late time parameterization}
Our main observables are parameterized
in terms of the primary variables $E_{11}$ and $E_{22}$ from Equation
\ref{eq:TR-mu-parametrization}; we are however interested in forecasting
the constraints on the pair of secondary variables $\{\mu$, $\eta\}$
or on the pair $\{\mu$, $\Sigma$\}.

Therefore we need to transform the variables using a Jacobian
$J_{ij}=\frac{\partial\theta_{i}}{\partial\tilde{\theta}_{j}}$,
where $\theta_{i}$ is the set of primary variables and $\tilde{\theta}_{i}$
is the vector of secondary variables. 
Eqns. (\ref{eq:SigmaofMuEta},\ref{eq:DE-mu-parametrization},\ref{eq:DE-eta-parametrization})
allow us to express the $\tilde{\theta}_{i}$ as a function of the variables $\theta_{i}$ and 
to obtain the non-vanishing derivatives
of $\mu$ and $\eta$ w.r.t to all cosmological parameters:

\begin{alignat}{2}
\frac{\partial\mu}{\partial\Omega_{c}} & =-E_{11},\qquad &
\frac{\partial\eta}{\partial\Omega_{c}} & =-E_{22}\\
\frac{\partial\mu}{\partial\Omega_{b}} & =-E_{11},\qquad &
\frac{\partial\eta}{\partial\Omega_{b}} & =-E_{22}\\
\frac{\partial\mu}{\partial E_{11}} &
=1-\Omega_{b}-\Omega_{c}-\Omega_{\nu},\qquad & \frac{\partial\eta}{\partial
E_{22}} & =1-\Omega_{b}-\Omega_{c}-\Omega_{\nu} \,\,\, .
\end{alignat}
With these derivatives we can construct the inverse of the Jacobian
$J_{ij}^{-1}=\frac{\partial\tilde{\theta}_{j}}{\partial\theta_{i}}$.
The Fisher matrix in the secondary variables $\tilde{F}{}_{ij}$ is
then given by 
\begin{equation}
\tilde{F}=J^{T}F\,J
\end{equation}
For the parameter set containing $\tilde{\theta}_{i}=\{\mu,\Sigma\}$
we obtain the following non-vanishing derivatives

\begin{alignat}{2}
\frac{\partial\mu}{\partial\Omega_{c}} & =-E_{11},\qquad &
\frac{\partial\Sigma}{\partial\Omega_{c}} &
=-\frac{1}{2}E_{22}(1+E_{11}(1-\Omega_{b}-\Omega_{c}-\Omega_{\nu}))\\
 &  &  & -\frac{1}{2}E_{11}(2+E_{22}(1-\Omega_{b}-\Omega_{c}-\Omega_{\nu}))\\
\frac{\partial\mu}{\partial\Omega_{b}} & =-E_{11},\qquad &
\frac{\partial\Sigma}{\partial\Omega_{b}} &
=-\frac{1}{2}E_{22}(1+E_{11}(1-\Omega_{b}-\Omega_{c}-\Omega_{\nu}))\\
 &  &  & -\frac{1}{2}E_{11}(2+E_{22}(1-\Omega_{b}-\Omega_{c}-\Omega_{\nu}))\\
\frac{\partial\mu}{\partial E_{11}} &
=1-\Omega_{b}-\Omega_{c}-\Omega_{\nu},\qquad & \frac{\partial\Sigma}{\partial
E_{11}} &
=\frac{1}{2}(2+E_{22}(1-\Omega_{b}-\Omega_{c}-\Omega_{\nu}))(1-\Omega_{b}-\Omega_{c}-\Omega_{\nu})\\
 &  & \frac{\partial\Sigma}{\partial E_{22}} &
=\frac{1}{2}(1+E_{11}(1-\Omega_{b}-\Omega_{c}-\Omega_{\nu}))(1-\Omega_{b}-\Omega_{c}-\Omega_{\nu})
\end{alignat}

\subsection{Early time parameterization}

In the early time case, we have two parameters for each $\mu(a)$
and $\eta(a)$ function as in Eq.\ (\ref{eq:TR-mu-parametrization}).
If we are interested in the parameters $\mu$, $\eta$ and $\Sigma$
today ($a=1$), then the parameters $E_{12}$ and $E_{21}$ are not
important anymore and we can simply marginalize over them in our Fisher
matrix. Then the relation between $\mu$ and $\eta$ is very simple
$\{\mu,\eta\}=1+\{E_{11},\,E_{22}\}$. The corresponding Jacobian
is simply a $7\times7$ identity matrix and we can apply it to the
Fisher matrix after we marginalize over the two unimportant parameters.

For the transformation to the pair $\mu$-$\Sigma$ we use the definition
of $\Sigma$ of Eq.\ (\ref{eq:SigmaofMuEta}) and then find the derivatives
with respect to $E_{ij}$. We obtain

\begin{align}
\frac{\partial\Sigma}{\partial E_{11}} & =\frac{1}{2}(2+E_{22})\\
\frac{\partial\Sigma}{\partial E_{22}} & =\frac{1}{2}(2+E_{11})\quad,
\end{align}
while all other derivatives remain the same.

\section{Derivatives of the Power Spectrum with respect to $\mu$ and $\eta$}\label{sec:appder}

In this section we investigate the derivatives of the power spectrum with respect to the MG parameters. 
Using the Jacobians from the previous Appendix \ref{sec:appjac}, we can convert 
our fundamental derivatives $\partial P(k,z)/\partial E_{ij}$ to derivatives of $\partial P(k,z)/\partial (\mu, \eta)$ evaluated at $z=0$.

\begin{figure}[htbp]
\includegraphics[width=0.4\textwidth]{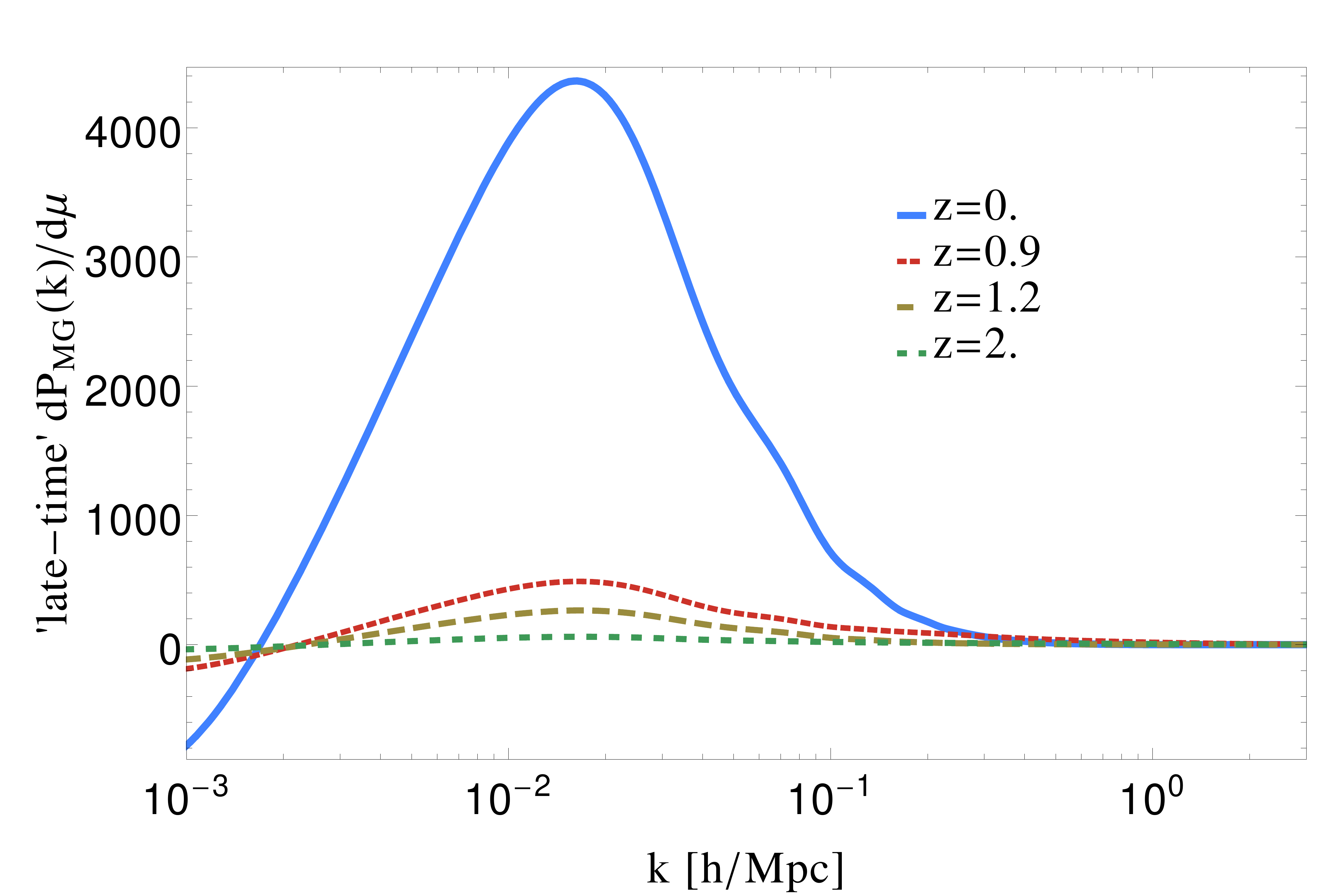}
\includegraphics[width=0.4\textwidth]{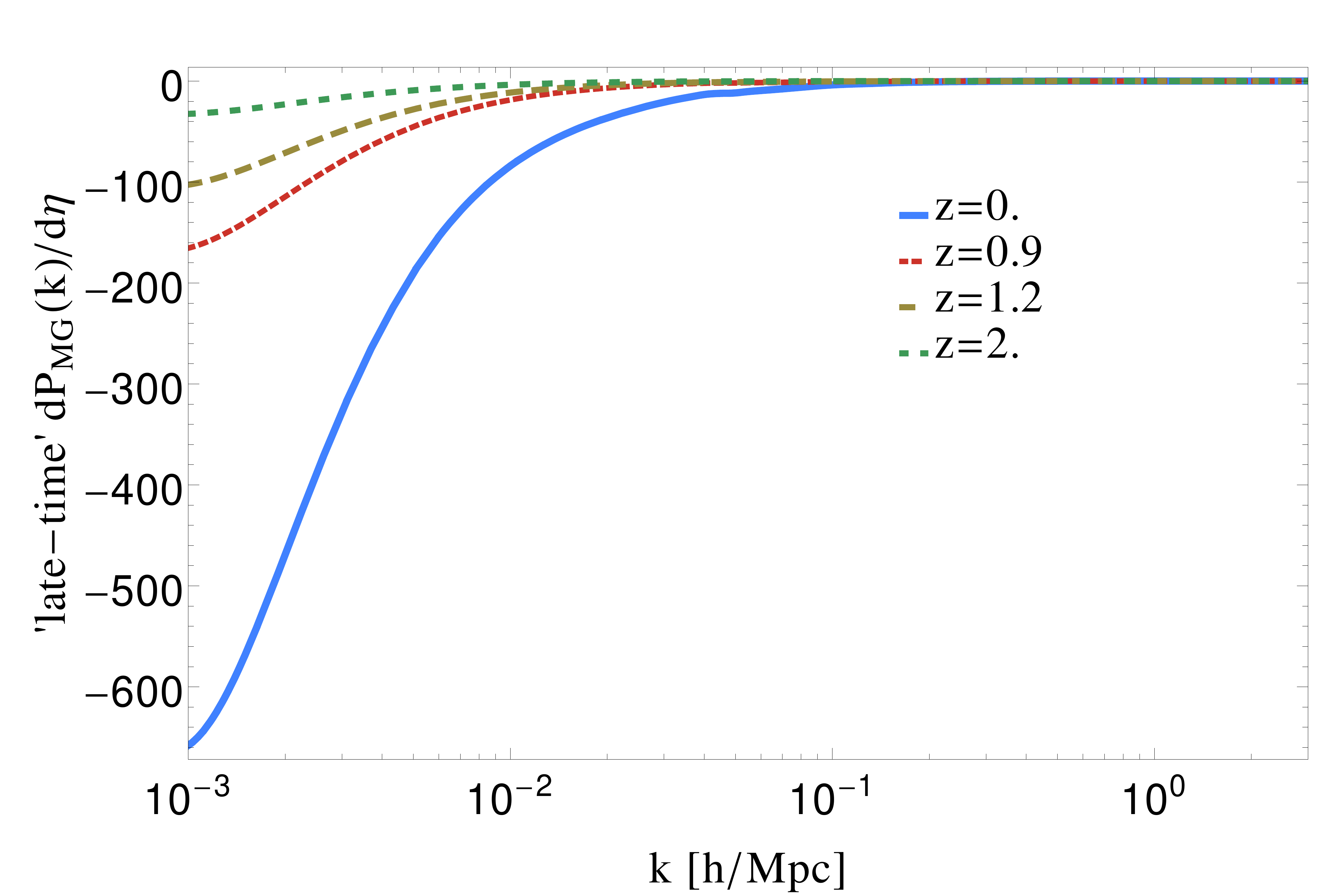}
\caption{ Derivatives of the matter power spectrum $P(k,z)$ w.r.t. the MG parameters $\mu$ and $\eta$ in the late-time parametrization.
}\label{fig:Pkderivs-latetime}
\end{figure}

\begin{figure}[htbp]
\includegraphics[width=0.4\textwidth]{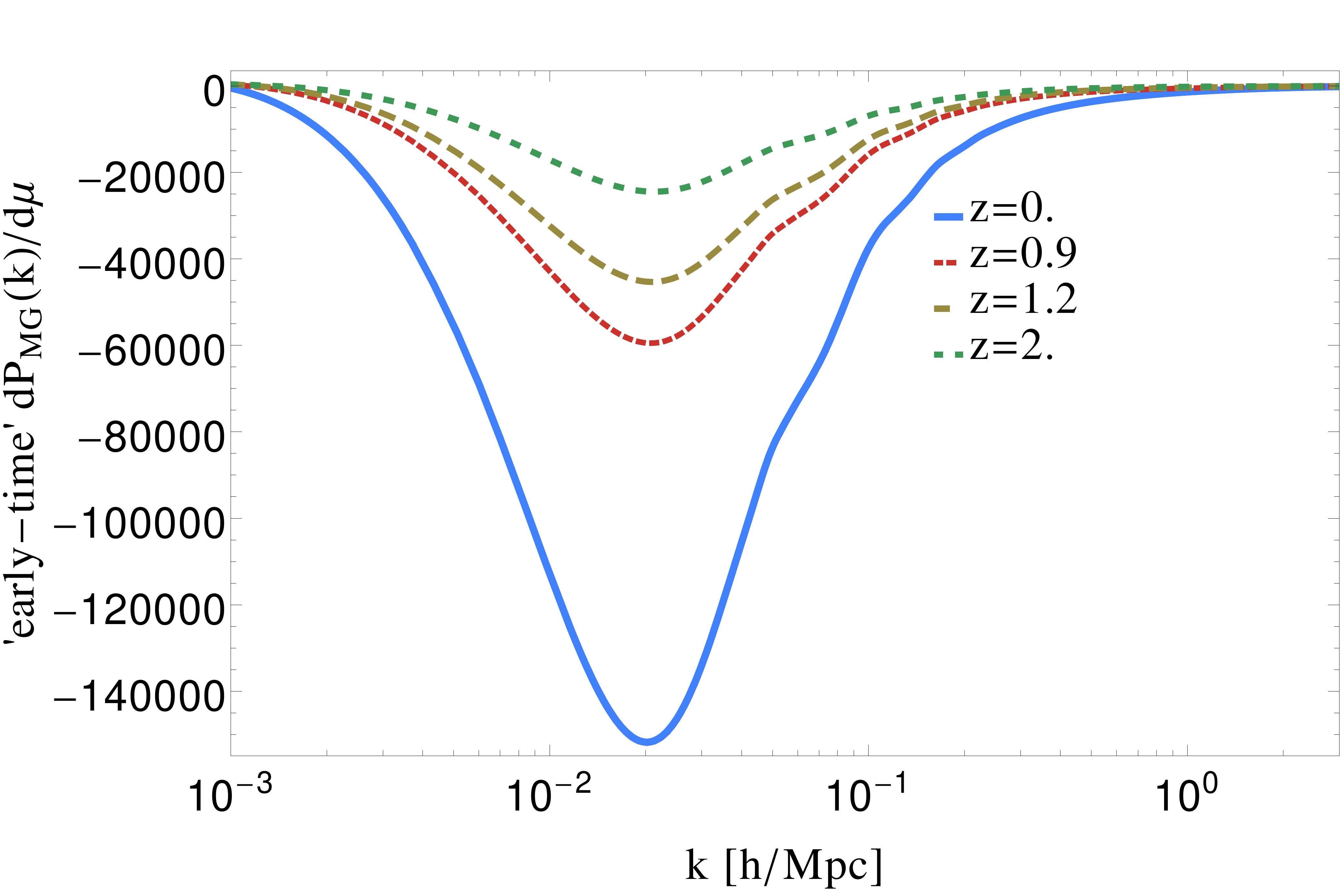}
\includegraphics[width=0.4\textwidth]{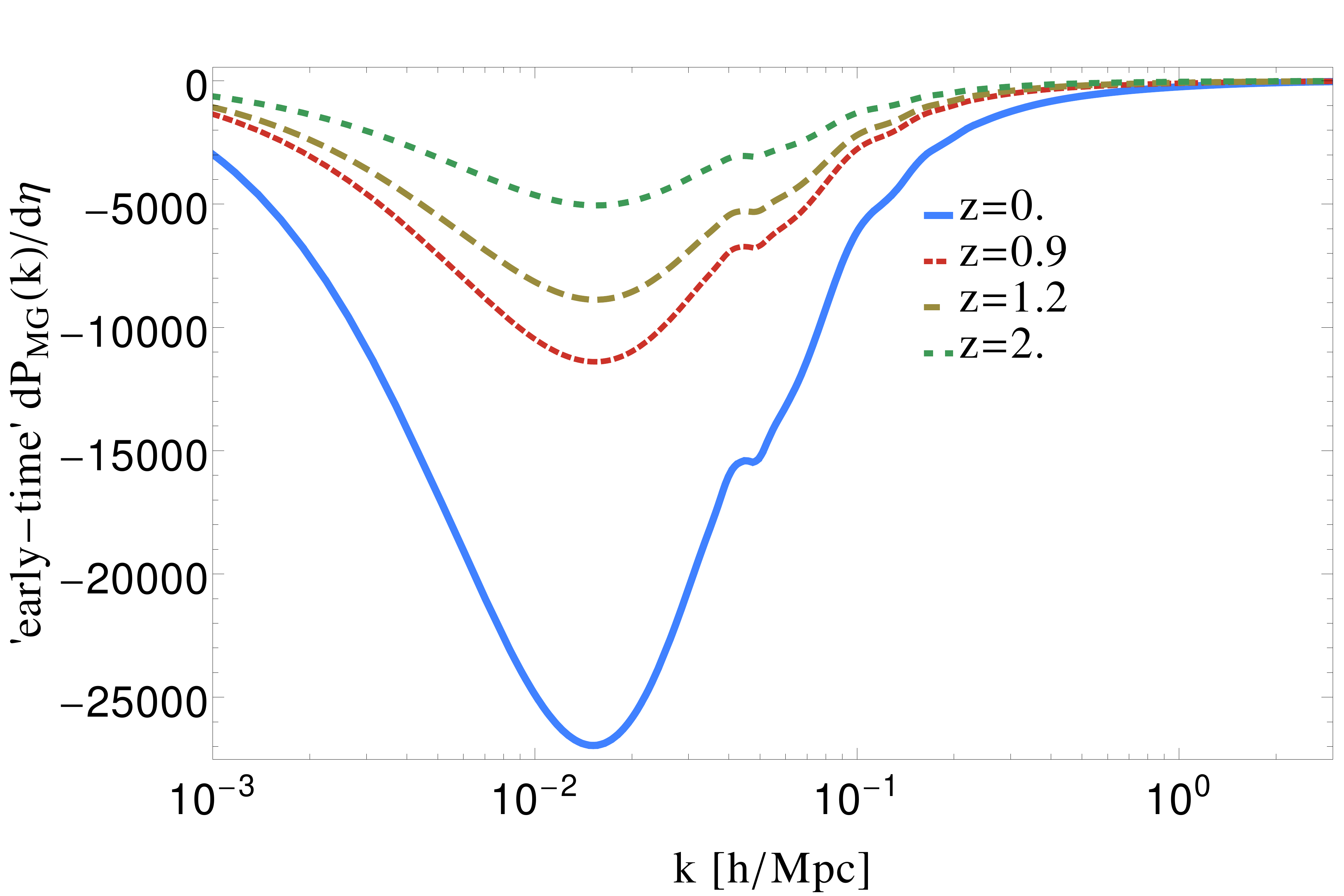}
\caption{ Derivatives of the matter power spectrum $P(k,z)$ w.r.t. the MG parameters $\mu$ and $\eta$ in the early-time parametrization.
}\label{fig:Pkderivs-latetime}
\end{figure}

We can see from the previous figures that in the early-time parametrization the derivative of the power spectrum
with respect to $\eta$ has a similar shape as the derivative with respect to $\mu$, making $\eta$ detectable by a Galaxy Clustering survey.
To explain this, we derive the equation governing the evolution of density fluctuations for a cold dark matter (CDM) species, based on the equations
implemented on the code MGCAMB presented in \cite{hojjati_testing_2011}, expressed here in the conformal Newtonian gauge.
In the following, a dot represents derivative with respect to conformal time $\tau$:
\begin{align}
\dot \delta &= - (1+w)(\theta - 3 \dot \Phi) - 3 \mathcal{H}(\frac{\delta P}{\delta \rho}-w)\delta\\
\dot \theta &= - \curH (1-3w)\theta - \frac{\dot w}{1+w} \theta + \frac{\delta P / \delta \rho}{1+w} k^2 \delta - k^2 \sigma + k^2 \Psi
\end{align}
For CDM we have $\sigma = w = c^2_s= \delta P / \delta \rho  = 0$,  then:
\begin{align}
\dot \delta &= -(\theta - 3 \dot \Phi) \label{eq:deltadot}\\
\dot \theta &= -\curH \theta + k^2 \Psi \label{eq:thetadot}
\end{align}
We have parameterized the solution to $\Psi$ as:
\begin{equation}\label{eq:Psipoisson}
k^2 \Psi = -4 \pi G a^2 \mu(\tau) \rho(\tau) \delta(\tau) \quad ,
\end{equation}
and since we are also requiring gravitational slip $\eta = \Phi / \Psi$, we then have the Poisson equation for $\Phi$:
\begin{equation}
k^2 \Phi = -4 \pi G a^2 \mu(\tau) \eta(\tau) \rho(\tau) \delta(\tau) \label{eq:Phipoisson}
\end{equation}

Taking the time derivative of (\ref{eq:deltadot}) and using (\ref{eq:thetadot}) and (\ref{eq:deltadot}) to 
replace $\dot \theta$ and $\theta$, and substituting $\Psi$ from (\ref{eq:Psipoisson}), we obtain:
\begin{align}
\ddot{\delta} + \curH \dot{\delta} &= 3\curH \dot{\Phi} + 3\ddot{\Phi} + 4 \pi G a^2 \mu \rho \delta
\label{eq-delta}
\end{align}

In general, derivatives of $\Phi$ appearing on the right hand side will depend on both $\mu$ and $\eta$. Their contribution is larger in the early-time parameterization with respect to the late-time one.

\section{Other Decorrelation Methods}\label{sec:appdec}

In section \ref{sub:Combined-GC-WL-Planck-Binned} we have worked with a special decorrelation method, ZCA (Zero-phase component analysis, first 
introduced by \cite{Bell19973327} in the context of image processing), which
allows us to find a new vector of decorrelated variables $q$ that is as similar as possible to the original
vector of variables $p$.
Other decorrelation methods do not share this property,
so in this section we want to illustrate their difference 
with respect to ZCA.
In the next subsections we show for the Principal Component Analysis and Cholesky decomposition methods, 
a subset of our previous results, namely the Galaxy Clustering non-linear case, 
using the HS prescription for a Euclid survey with Redbook specifications.

\subsubsection{Principal Component Analysis }

Principal Component Analysis (PCA) \cite{Friedman-PCA} is a well known method, which rotates
the vector of variables $p$ into a new basis, using the eigenmatrix of the 
covariance matrix $C$. At the same time, it is the method that maximizes the compression of all components
of $p$ into the components of $q$ using as measure the cross-covariance
between $q$ and $p$ (see \cite{kessy_optimal_2015} for more details and references).
This method is useful for dimensional reduction or data compression, 
since the information is stored in as few components as possible.
This is achieved by using the $W$ matrix

\begin{equation}
W=\Lambda^{-1/2}U^{T}
\end{equation}
where $\Lambda$ and $U$ represent the eigensystem of $C$ (defined in Eqn.\ \ref{eq:eigensystemofC}).

\begin{figure}[htbp]
\includegraphics[width=0.35\textwidth]{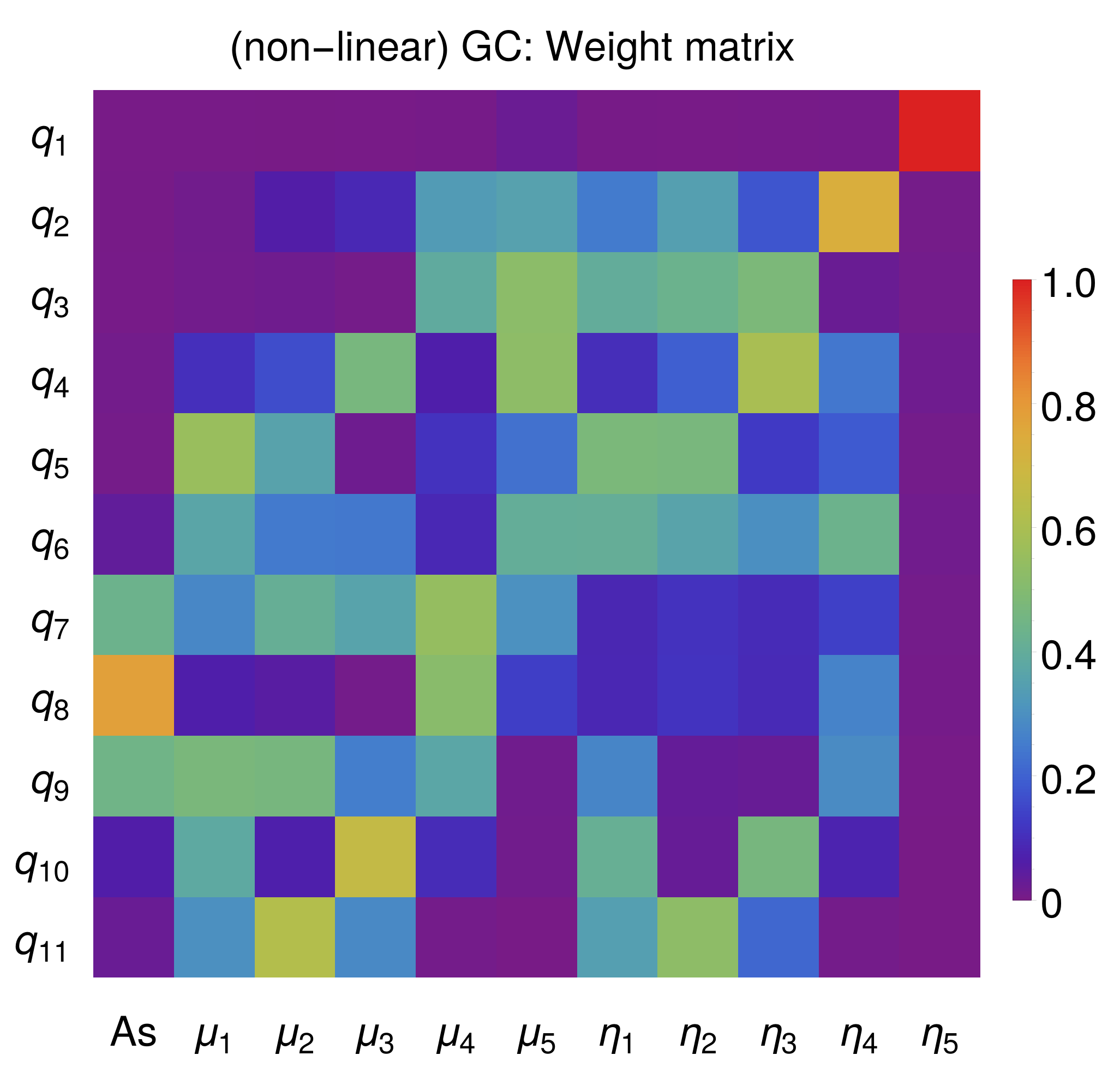}
\includegraphics[width=0.45\textwidth]{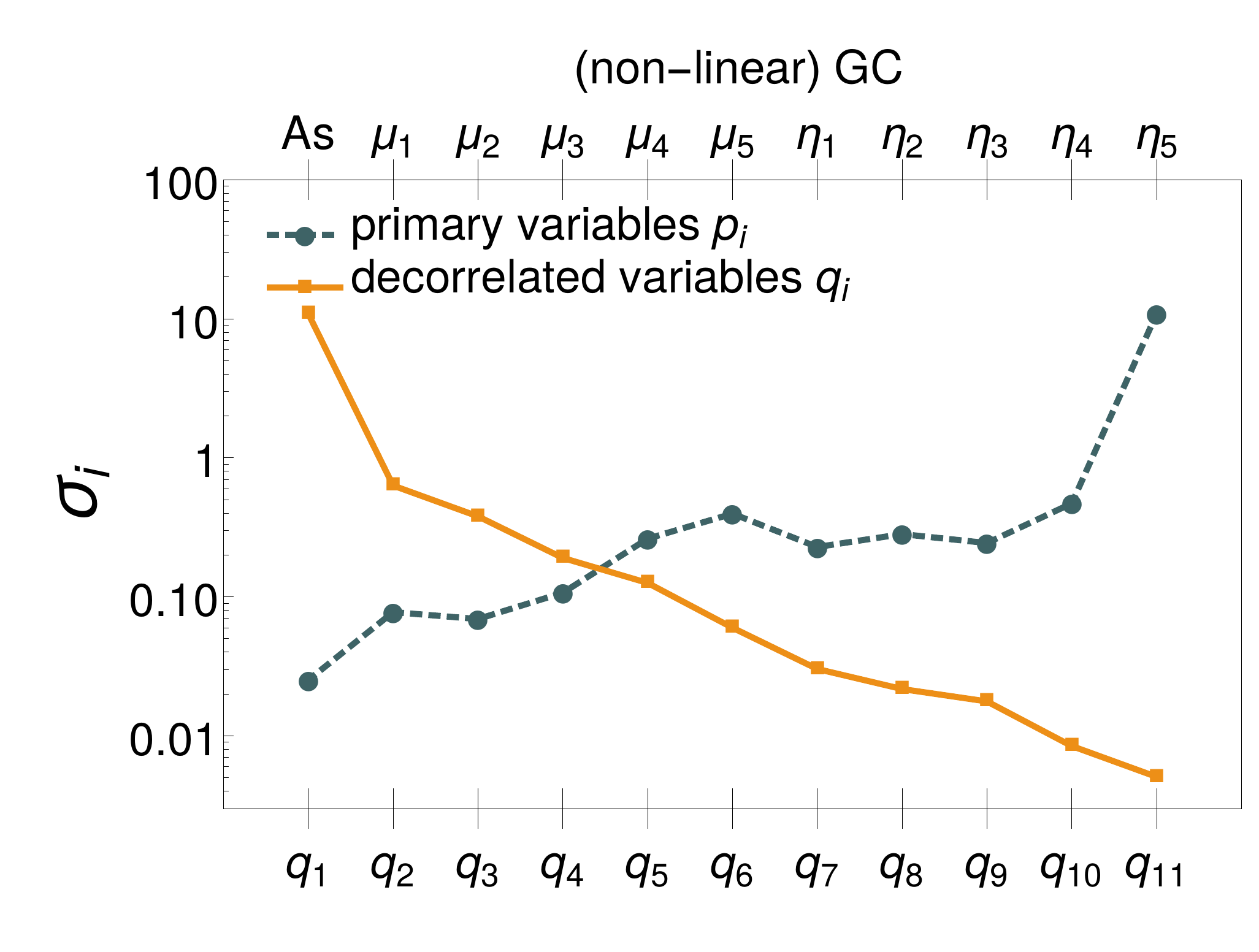}
\caption{\textbf{Left}: weight matrix $W$ for PCA. \textbf{Right}: 1$\sigma$ fully maximized errors on
the primary parameters $p$ (blue lines) and the errors on the uncorrelated
derived parameters $q$ (orange lines). Notice how all the important information is constrained in as few variables as possible,
namely the last elements of $q_i$.
}\label{fig:PCA-GCnlhs}
\end{figure}

Then, it follows that the transformed covariance matrix $\tilde{C}$ is whitened:
\begin{align}
\tilde{C} & =\Lambda^{-1/2}U^{T}C\Lambda^{-1/2}U^{T}\\
 & =\Lambda^{-1/2}U^{T}U\Lambda U^{T}\Lambda^{-1/2}U^{T}\\
 & =\mathbb{1}
\end{align}
As done previously, we renormalize $W$ in such a way that the sum of the square of the elements of each row sum up to
unity.
In Figure \ref{fig:PCA-GCnlhs} we show in the left panel the weight matrix $W$ and in the right panel the 1$\sigma$ errors for the original
and decorrelated variables. We see that the most constrained $q$ variables are the last components $q_8$-$q_{11}$, where most
of the information has been compressed into. These 4 variables are complicated linear combinations of $A_s$, $\mu_{1,2,3}$ 
and $\eta_{1,2,3}$. This is the same we found for ZCA (cf.\ Fig.\ \ref{fig:GCbestconst} and Eqns.\ (\ref{fig:GCbestconst})). However, the interpretation
in terms of the old variables $p$ is not so simple anymore. We can see from the weight matrix and the 1$\sigma$
errors, that the most unconstrained parameter is $q_1$, which is basically equivalent to $\eta_5$. This means that this parameter can be
eliminated from the analysis if one wants to do dimensional reduction.

\subsubsection{Cholesky decomposition}

\begin{figure}[htbp]
\includegraphics[width=0.35\textwidth]{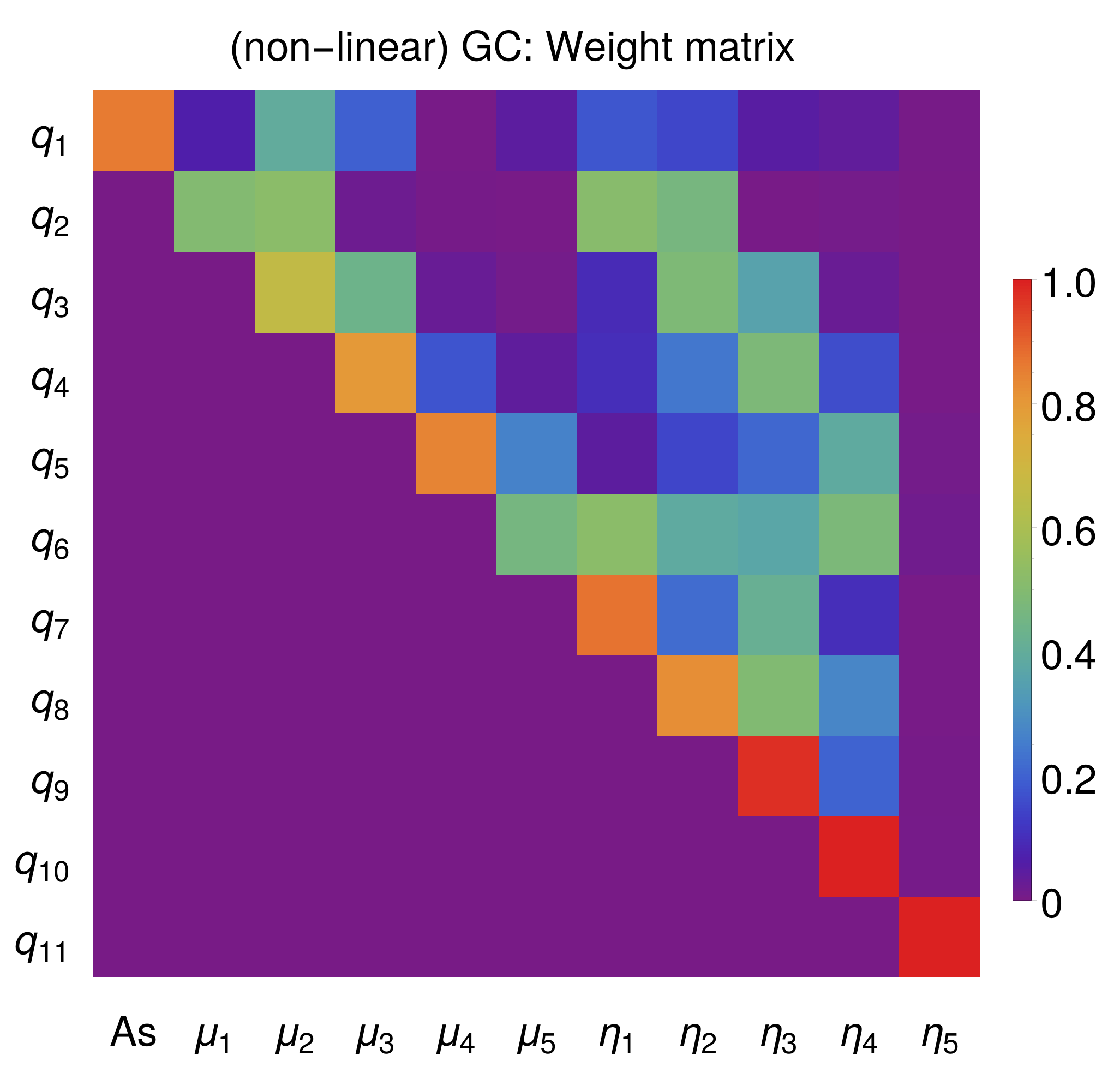}
\includegraphics[width=0.45\textwidth]{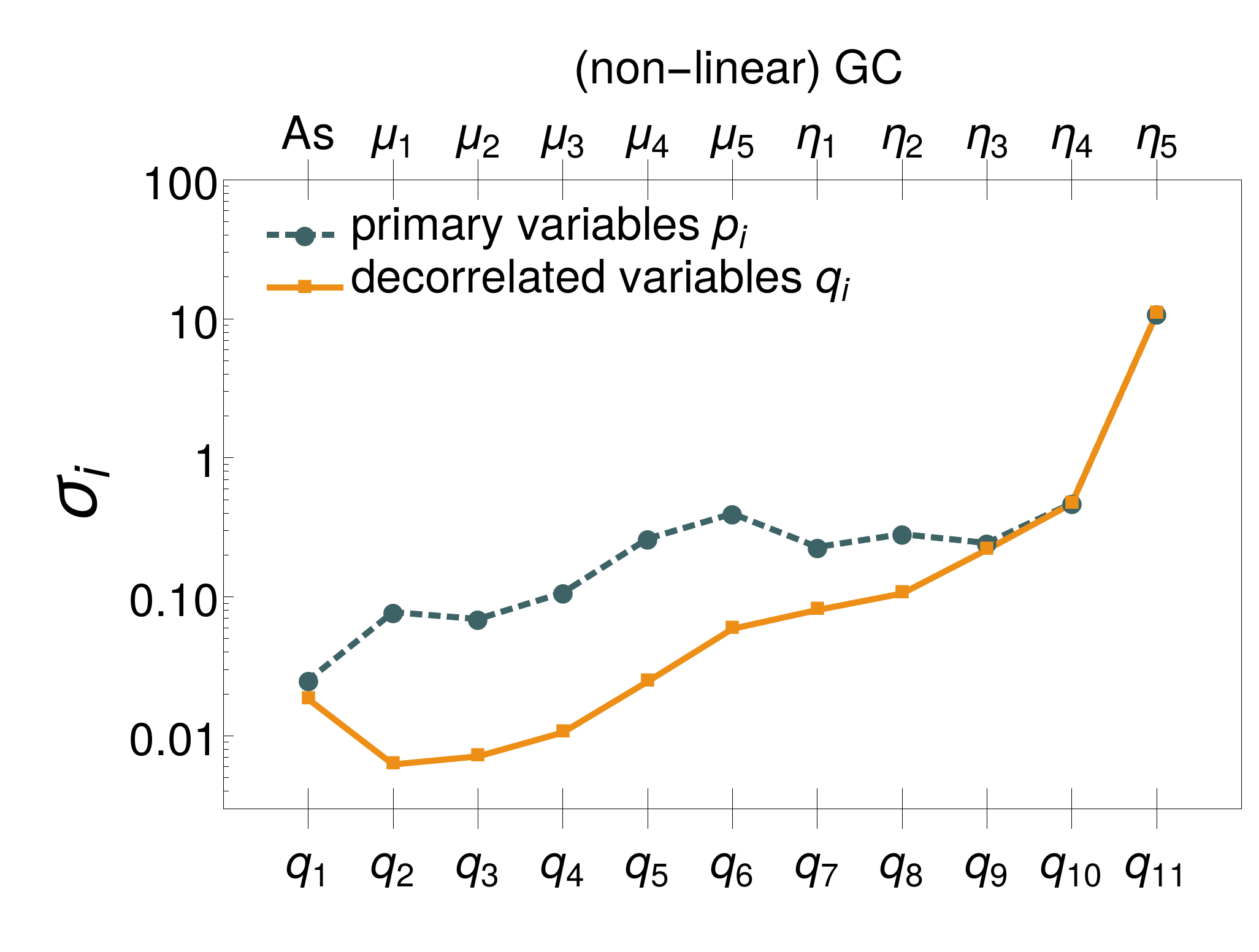}
\caption{\textbf{Left:} weight matrix $W$ for the Cholesky decorrelation. \textbf{Right:} 1$\sigma$ fully maximized errors on
the primary parameters $p$ (blue lines) and the errors on the uncorrelated
derived parameters $q$ (orange lines). Notice how in this case, because of the upper triangular construction, the new variables
$q_{9,10,11}$ are equivalent to $\eta_{3,4,5}$. The parameter $A_s$ is decorrelated in the non-linear HS case for GC, therefore
it is almost equivalent to $q_1$ as was also the case in ZCA (compare with Fig.\ \ref{fig:Wmat-ZCA-GC})
}\label{fig:Chol-GCnlhs}
\end{figure}

Cholesky decomposition of the Fisher matrix $F=LL^{T}$, allows us to
define a decorrelation method that compresses
all components of $p$ into an upper triangular matrix of components
of $q$. This is achieved by using the $W$ matrix:

\begin{equation}
W=L^{T}
\end{equation}
Then the covariance matrix will be whitened:
\begin{align}
\tilde{C} & =L^{T} (L L^{T})^{-1} L\\
 & =L^{T} (L^{T})^{-1} L^{-1} L\\
 & =\mathbb{1}
\end{align}
As done previously, we renormalize $W$ in such a way that the sum of the square of the elements of each row sum up to
unity.
In Cholesky decomposition, since we are constructing it via an upper triangular matrix (see left panel of Fig.\ \ref{fig:Chol-GCnlhs}), 
the new parameter $q_{11}$ will be identical 
to the parameter $\eta_5$, which is, as we have seen before (section \ref{subsub:ZCA-GC}), the less constrained parameter. This decorrelation method
is useful if one wants to have an ordering of the variables (see \cite{kessy_optimal_2015} and references therein).
On the other hand $q_1$ is almost identical to $A_s$, since as we have seen in section \ref{subsub:ZCA-GC},
it becomes decorrelated from the MG parameters in the non-linear case.

\section{Weight matrices coefficients}
\label{sec:Wmatrices}

Here we explicitly show the coefficient of the $W$ matrices illustrated in Figures
\ref{fig:Wmat-ZCA-GC}, \ref{fig:Wmat-ZCA-WL}
and \ref{fig:GC+WL+Planck-corr-Wmat}.
Tables \ref{tab:Wcoeff-lin-GC} and \ref{tab:Wcoeff-nlHS-GC} contain the coefficient matrices obtained for 
Galaxy Clustering in the linear and non-linear HS prescription cases respectively, while 
Tables \ref{tab:Wcoeff-lin-WL} and \ref{tab:Wcoeff-nlHS-WL} contain the coefficients obtained for WL. 
Finally, Table \ref{tab:Wcoeff-nlHS-WL+GC+Planck} reports the coefficients for the combination of GC, 
WL and \planck\ priors, in the non-linear HS case.

\begin{table}[H]
	\[
	q_i = 
	\footnotesize
	\left(
	\begin{array}{rrrrrrrrrrr}
	\color{red} 0.899 & 0.003 & -0.020 & 0.109 & \color{red} 0.315 & 0.250 & 0.097 & 0.021 & -0.083 & -0.022 & -0.007 \\
	0.001 & \color{red} 0.695 & \color{red} -0.304 & -0.089 & -0.084 & -0.026 & \color{red} 0.519 & \color{red} -0.362 & -0.082 & -0.033 & 0.001 \\
	-0.003 & -0.186 & \color{red} 0.745 & -0.211 & -0.090 & -0.027 & \color{red} -0.285 & \color{red} 0.491 & -0.181 & -0.036 & 0.001 \\
	0.022 & -0.065 & \color{red} -0.252 & \color{red} 0.737 & -0.174 & -0.019 & -0.075 & \color{red} -0.320 & \color{red} 0.486 & -0.116 & 0.001 \\
	0.148 & -0.142 & -0.250 & \color{red} -0.405 & \color{red} 0.666 & 0.070 & -0.094 & -0.102 & \color{red} -0.268 & \color{red} 0.439 & -0.002 \\
	\color{red} 0.654 & -0.246 & \color{red} -0.420 & -0.240 & \color{red} 0.388 & \color{red} 0.290 & 0.098 & 0.086 & -0.159 & 0.003 & -0.008 \\
	0.026 & \color{red} 0.494 & \color{red} -0.444 & -0.098 & -0.053 & 0.010 & \color{red} 0.621 & \color{red} -0.375 & -0.129 & -0.055 & -0.000 \\
	0.004 & \color{red} -0.265 & \color{red} 0.588 & \color{red} -0.320 & -0.044 & 0.007 & \color{red} -0.288 & \color{red} 0.578 & -0.248 & -0.045 & -0.000 \\
	-0.023 & -0.082 & \color{red} -0.295 & \color{red} 0.664 & -0.158 & -0.017 & -0.135 & \color{red} -0.338 & \color{red} 0.545 & -0.102 & 0.001 \\
	-0.015 & -0.082 & -0.147 & \color{red} -0.399 & \color{red} 0.651 & 0.001 & -0.146 & -0.155 & \color{red} -0.257 & \color{red} 0.525 & -0.000 \\
	\color{red} -0.519 & \color{red} 0.308 & \color{red} 0.510 & \color{red} 0.346 & \color{red} -0.348 & -0.239 & -0.061 & -0.070 & 0.181 & -0.004 & 0.193 \\
	\end{array}
	\right)
	\cdot 
    \footnotesize	
	\begin{pmatrix} \ln(10^{10}A_s) \\ \mu_{1}  \\ \mu_{2}   \\ \mu_{3}   \\ \mu_{4}   \\ \mu_{5}  
	\\ \eta_{1}   \\ \eta_{2}  \\ \eta_{3}   \\\eta_{4}  \\ \eta_{5}
	\end{pmatrix}
	\]
	\caption{\label{tab:Wcoeff-lin-GC}
		\normalfont Coefficients of the Matrix $W$ for the Euclid Redbook GC forecasted covariance matrix, using linear power spectra.
		The entries marked with red are those whose absolute values are larger than $0.25$.}
\end{table}
\begin{table}[H]
	\[
	q_i = 
	\footnotesize
	\left(
	\begin{array}{rrrrrrrrrrr}
	\color{red} 0.986 & -0.064 & -0.112 & -0.093 & -0.037 & -0.038 & -0.002 & 0.005 & -0.002 & 0.005 & 0.002 \\
	-0.021 & \color{red} 0.677 & \color{red} -0.349 & -0.091 & -0.042 & -0.008 & \color{red} 0.516 & \color{red} -0.367 & -0.086 & -0.021 & 0.001 \\
	-0.025 & -0.233 & \color{red} 0.731 & -0.194 & -0.051 & -0.011 & \color{red} -0.317 & \color{red} 0.492 & -0.166 & -0.021 & 0.001 \\
	-0.030 & -0.088 & \color{red} -0.281 & \color{red} 0.755 & -0.061 & 0.036 & -0.073 & \color{red} -0.331 & \color{red} 0.465 & -0.086 & -0.001 \\
	-0.032 & -0.108 & -0.197 & -0.161 & \color{red} 0.820 & 0.238 & -0.023 & -0.092 & -0.240 & \color{red} 0.357 & -0.007 \\
	-0.086 & -0.058 & -0.110 & \color{red} 0.252 & \color{red} 0.637 & \color{red} 0.659 & 0.224 & -0.001 & -0.143 & -0.046 & -0.018 \\
	-0.001 & \color{red} 0.492 & \color{red} -0.452 & -0.071 & -0.008 & 0.031 & \color{red} 0.617 & \color{red} -0.391 & -0.113 & -0.038 & -0.001 \\
	0.001 & \color{red} -0.296 & \color{red} 0.593 & \color{red} -0.275 & -0.029 & -0.000 & \color{red} -0.331 & \color{red} 0.575 & -0.209 & -0.031 & -0.000 \\
	-0.001 & -0.115 & \color{red} -0.329 & \color{red} 0.635 & -0.124 & -0.028 & -0.157 & \color{red} -0.344 & \color{red} 0.558 & -0.065 & 0.001 \\
	0.008 & -0.092 & -0.136 & \color{red} -0.390 & \color{red} 0.614 & -0.030 & -0.176 & -0.167 & -0.216 & \color{red} 0.581 & 0.001 \\
	0.140 & 0.126 & 0.180 & -0.194 & \color{red} -0.645 & \color{red} -0.608 & -0.218 & -0.029 & 0.137 & 0.049 & 0.199 \\
	\end{array}
	\right)
		\cdot 
	\footnotesize	
    \begin{pmatrix} \ln(10^{10}A_s) \\ \mu_{1}  \\ \mu_{2}   \\ \mu_{3}   \\ \mu_{4}   \\ \mu_{5}  
	\\ \eta_{1}   \\ \eta_{2}  \\ \eta_{3}   \\\eta_{4}  \\ \eta_{5}
	\end{pmatrix}
	\]
	\caption{\label{tab:Wcoeff-nlHS-GC}
		\normalfont Coefficients of the Matrix $W$ for the Euclid Redbook GC forecasted covariance matrix, using the non-linear HS prescription.
		The entries marked with red are those whose absolute values are larger than $0.25$.}
\end{table}
\begin{table}[H]
	\[
	q_i = 
	\footnotesize
	\left(
	\begin{array}{rrrrrrrrrrr}
	\color{red} 0.760 & -0.005 & \color{red} 0.475 & 0.131 & 0.005 & 0.034 & 0.247 & \color{red} 0.329 & 0.095 & 0.004 & -0.000 \\
	-0.000 & \color{red} 0.673 & \color{red} -0.592 & -0.055 & 0.013 & 0.008 & \color{red} 0.328 & \color{red} -0.290 & -0.034 & -0.001 & -0.000 \\
	0.033 & \color{red} -0.594 & \color{red} 0.674 & 0.013 & -0.003 & -0.000 & \color{red} -0.302 & \color{red} 0.318 & 0.010 & 0.000 & 0.000 \\
	0.075 & \color{red} -0.457 & 0.112 & \color{red} 0.754 & 0.002 & 0.004 & \color{red} -0.310 & 0.039 & \color{red} 0.326 & 0.001 & 0.000 \\
	0.022 & \color{red} 0.892 & -0.187 & 0.017 & 0.102 & 0.064 & \color{red} 0.326 & -0.208 & -0.064 & -0.000 & 0.000 \\
	\color{red} 0.251 & \color{red} 0.859 & -0.030 & 0.048 & 0.098 & 0.090 & \color{red} 0.414 & -0.078 & -0.029 & 0.004 & 0.009 \\
	0.034 & \color{red} 0.654 & \color{red} -0.600 & -0.074 & 0.009 & 0.008 & \color{red} 0.356 & \color{red} -0.280 & -0.037 & -0.000 & 0.000 \\
	0.047 & \color{red} -0.606 & \color{red} 0.662 & 0.010 & -0.006 & -0.002 & \color{red} -0.293 & \color{red} 0.324 & 0.013 & 0.001 & 0.001 \\
	0.108 & \color{red} -0.576 & 0.167 & \color{red} 0.655 & -0.015 & -0.005 & \color{red} -0.311 & 0.105 & \color{red} 0.304 & 0.003 & 0.002 \\
	\color{red} 0.256 & \color{red} -0.718 & \color{red} 0.380 & 0.136 & -0.003 & 0.032 & -0.010 & \color{red} 0.448 & 0.197 & 0.111 & 0.061 \\
	-0.029 & \color{red} -0.734 & \color{red} 0.253 & 0.068 & 0.000 & 0.166 & 0.084 & \color{red} 0.506 & 0.218 & 0.122 & 0.197 \\
	\end{array}
	\right)
	\cdot 
	\footnotesize
	\begin{pmatrix} \ln(10^{10}A_s) \\ \mu_{1}  \\ \mu_{2}   \\ \mu_{3}   \\ \mu_{4}   \\ \mu_{5}  
	\\ \eta_{1}   \\ \eta_{2}  \\ \eta_{3}   \\\eta_{4}  \\ \eta_{5}
	\end{pmatrix}
	\]
	\caption{\label{tab:Wcoeff-lin-WL}
		\normalfont Coefficients of the Matrix $W$ for the Euclid Redbook WL forecasted covariance matrix, using linear power spectra.
		The entries marked with red are those whose absolute values are larger than $0.25$.}
\end{table}
\begin{table}[H]
	\[
	q_i = \left(
	\footnotesize
	\begin{array}{rrrrrrrrrrr}
	\color{red} 0.926 & -0.004 & \color{red} -0.323 & -0.039 & -0.040 & -0.066 & -0.043 & -0.168 & -0.004 & -0.000 & 0.001 \\
	-0.000 & \color{red} 0.669 & \color{red} -0.595 & -0.051 & 0.012 & 0.013 & \color{red} 0.332 & \color{red} -0.291 & -0.031 & -0.001 & -0.000 \\
	-0.021 & \color{red} -0.552 & \color{red} 0.711 & 0.029 & 0.005 & 0.016 & \color{red} -0.270 & \color{red} 0.339 & 0.014 & 0.001 & 0.000 \\
	-0.025 & \color{red} -0.460 & \color{red} 0.288 & \color{red} 0.725 & 0.016 & 0.040 & -0.238 & 0.150 & \color{red} 0.312 & 0.002 & 0.000 \\
	-0.159 & \color{red} 0.657 & \color{red} 0.289 & 0.102 & \color{red} 0.476 & 0.156 & \color{red} 0.418 & 0.155 & -0.007 & 0.006 & 0.001 \\
	-0.168 & \color{red} 0.455 & \color{red} 0.622 & 0.158 & 0.100 & \color{red} 0.486 & 0.223 & 0.236 & -0.016 & -0.001 & -0.002 \\
	-0.006 & \color{red} 0.667 & \color{red} -0.585 & -0.053 & 0.015 & 0.012 & \color{red} 0.366 & \color{red} -0.274 & -0.026 & 0.000 & -0.000 \\
	-0.023 & \color{red} -0.560 & \color{red} 0.704 & 0.032 & 0.005 & 0.013 & \color{red} -0.263 & \color{red} 0.345 & 0.017 & 0.001 & 0.000 \\
	-0.005 & \color{red} -0.576 & \color{red} 0.277 & \color{red} 0.636 & -0.002 & -0.008 & -0.238 & 0.165 & \color{red} 0.319 & 0.004 & 0.001 \\
	-0.026 & \color{red} -0.506 & \color{red} 0.670 & 0.239 & 0.110 & -0.037 & 0.114 & \color{red} 0.359 & 0.209 & 0.169 & 0.096 \\
	0.180 & \color{red} -0.799 & 0.053 & 0.033 & 0.069 & -0.140 & -0.055 & 0.044 & 0.134 & \color{red} 0.271 & \color{red} 0.452 \\
	\end{array}
	\right)
	\cdot
	\footnotesize 
	\begin{pmatrix} \ln(10^{10}A_s) \\ \mu_{1}  \\ \mu_{2}   \\ \mu_{3}   \\ \mu_{4}   \\ \mu_{5}  
	\\ \eta_{1}   \\ \eta_{2}  \\ \eta_{3}   \\\eta_{4}  \\ \eta_{5}
	\end{pmatrix}
	\]
	\caption{\label{tab:Wcoeff-nlHS-WL}
		\normalfont Coefficients of the Matrix $W$ for the Euclid Redbook WL forecasted covariance matrix, using the non-linear HS prescription.
		The entries marked with red are those whose absolute values are larger than $0.25$.}
\end{table}
\begin{table}[H]
	\[
	q_i = \left(
	\footnotesize
\begin{array}{rrrrrrrrrrr}
\color{red} 0.932 & 0.033 & 0.149 & 0.220 & 0.202 & 0.138 & -0.005 & -0.026 & -0.001 & 0.010 & -0.003 \\
0.029 & \color{red} 0.839 & -0.214 & -0.085 & -0.035 & -0.021 & \color{red} 0.476 & -0.119 & -0.021 & 0.002 & -0.000 \\
0.138 & -0.222 & \color{red} 0.800 & -0.083 & 0.016 & 0.028 & \color{red} -0.256 & \color{red} 0.452 & -0.118 & -0.011 & -0.002 \\
\color{red} 0.284 & -0.123 & -0.115 & \color{red} 0.791 & 0.076 & 0.108 & -0.106 & \color{red} -0.286 & \color{red} 0.387 & -0.067 & -0.010 \\
\color{red} 0.506 & -0.097 & 0.043 & 0.148 & \color{red} 0.753 & 0.248 & -0.059 & -0.089 & -0.144 & 0.222 & -0.032 \\
\color{red} 0.545 & -0.092 & 0.120 & \color{red} 0.331 & \color{red} 0.390 & \color{red} 0.626 & -0.027 & -0.088 & -0.089 & -0.009 & 0.104 \\
-0.006 & \color{red} 0.621 & \color{red} -0.322 & -0.096 & -0.028 & -0.008 & \color{red} 0.656 & \color{red} -0.261 & -0.048 & -0.009 & -0.001 \\
-0.032 & -0.166 & \color{red} 0.605 & \color{red} -0.276 & -0.044 & -0.028 & \color{red} -0.277 & \color{red} 0.654 & -0.148 & -0.006 & 0.000 \\
-0.003 & -0.049 & \color{red} -0.270 & \color{red} 0.634 & -0.122 & -0.048 & -0.088 & \color{red} -0.252 & \color{red} 0.659 & -0.023 & 0.010 \\
0.067 & 0.015 & -0.082 & \color{red} -0.358 & \color{red} 0.614 & -0.017 & -0.055 & -0.034 & -0.074 & \color{red} 0.688 & 0.027 \\
-0.072 & -0.002 & -0.060 & -0.179 & \color{red} -0.286 & \color{red} 0.591 & -0.027 & 0.008 & 0.108 & 0.088 & \color{red} 0.713 \\
\end{array}
	\right)
	\cdot 
	\footnotesize
	\begin{pmatrix} \ln(10^{10}A_s) \\ \mu_{1}  \\ \mu_{2}   \\ \mu_{3}   \\ \mu_{4}   \\ \mu_{5}  
	\\ \eta_{1}   \\ \eta_{2}  \\ \eta_{3}   \\\eta_{4}  \\ \eta_{5}
	\end{pmatrix}
	\]
	\caption{\label{tab:Wcoeff-nlHS-WL+GC+Planck}
		\normalfont Coefficients of the Matrix $W$ for the Euclid Redbook GC+WL case using the non-linear HS prescription and adding a CMB \planck\ prior.
	     The entries marked with red are those whose absolute values are larger than $0.25$.}
\end{table}
  
\section{The Kullback-Leibler divergence \label{sec:KL}}

In Section \ref{sec:covcorr} we exploited the determinant of the covariance matrix $C$ as a Figure of Merit for our forecasts. Here we summarize a possible alternative, the Kullback-Leibler divergence \cite{kullback_information_1951}, also called relative entropy or information gain. It has been used in the field of cosmology for model selection, experiment design and forecasting, see among others \cite{kunz_constraining_2006, raveri_cosmicfish_2016, seehars_information_2014, verde_planck_2013, Zhao2017}.
The KL-divergence $\mathcal{D}(p_2||p_1)$ measures for a continuous, $d$-dimensional random variable $\theta$, the relative entropy between two probability density functions $p_1(\theta)$ and $p_2(\theta)$ and it is given by
\begin{equation}
\mathcal{D}(p_2||p_1) \equiv \int p_2(\theta)\ln\left(\frac{p_2(\theta)}{p_1(\theta)} \right) \mathrm{d}\theta .
\end{equation}
Although it is not symmetric in $p_1$ and $p_2$ it can be interpreted as a distance between the two distributions and measures the information gain since it is non-negative ($\mathcal{D}(p_2||p_1)\geq0$), non-degenerate ($\mathcal{D}(p_2||p_1)=0$ if and only if $p_1=p_2$) and it is invariant 
under re-parametrizations of the distributions as $p_1(\theta)\mathrm{d}\theta = p_1(\tilde{\theta})\mathrm{d}\tilde{\theta}$. In the form given here the information gain is measured in nits as in section \ref{sec:covcorr}, to convert nits to bits it is enough to divide the result by $\ln(2)$.

For the special case of $p_1(\theta)$ and $p_2(\theta)$ being multivariate Gaussian distributions, with the same mean values and covariance matrices $\mathcal{A}$ and $\mathcal{B}$ respectively, we obtain
\begin{equation}
\mathcal{D}(p_2||p_1) = -\frac{1}{2} \left[\ln\left(\frac{\det(\mathcal{A})}{\det(\mathcal{B})} \right) +\mathrm{Tr}\left[\mathbb{1}-\mathcal{B}^{-1}\mathcal{A}   \right] \right].
\end{equation}

\begin{figure}[h]
	\centering
	\includegraphics[width=0.45\linewidth]{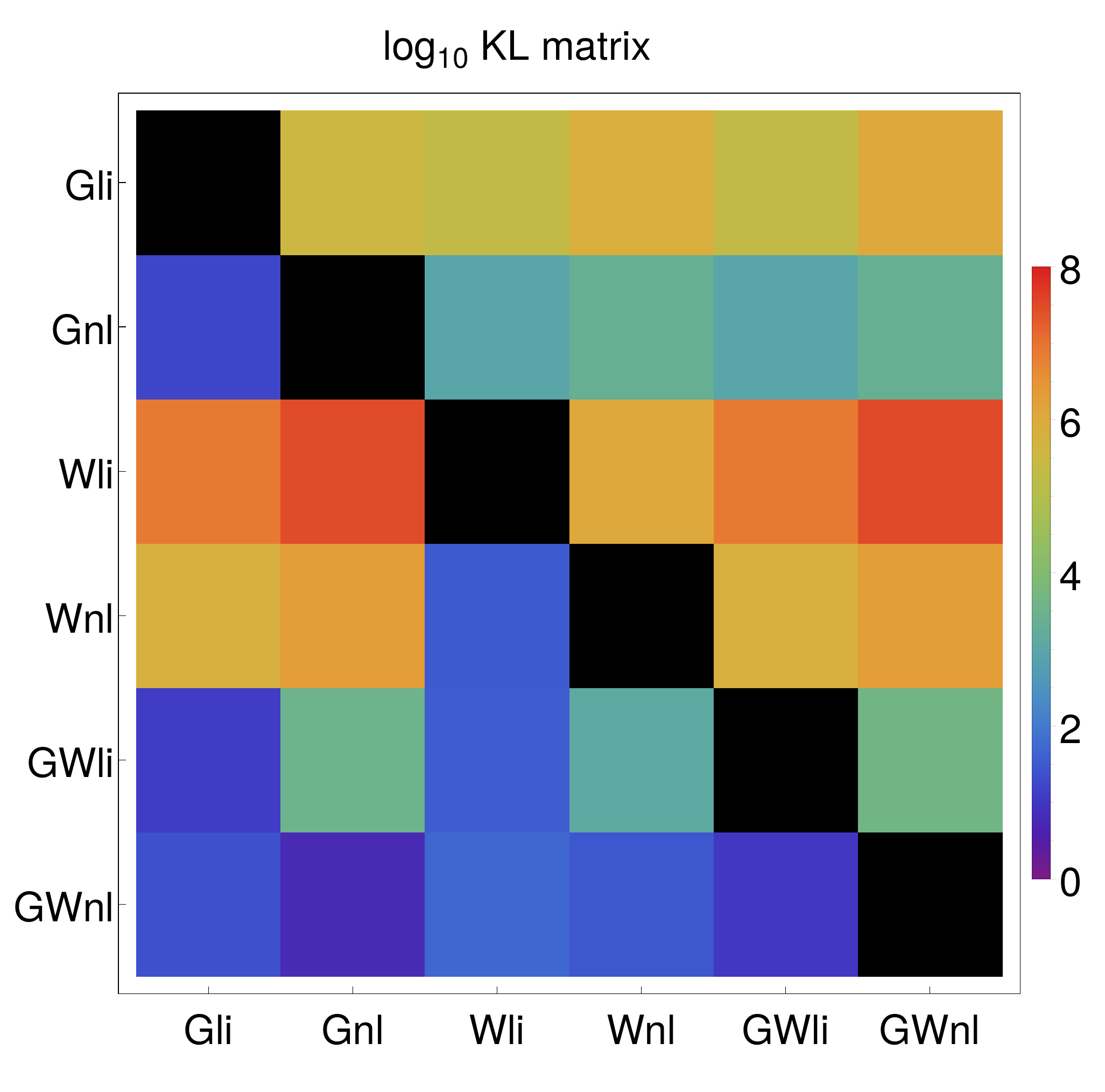}
	\includegraphics[width=0.45\linewidth]{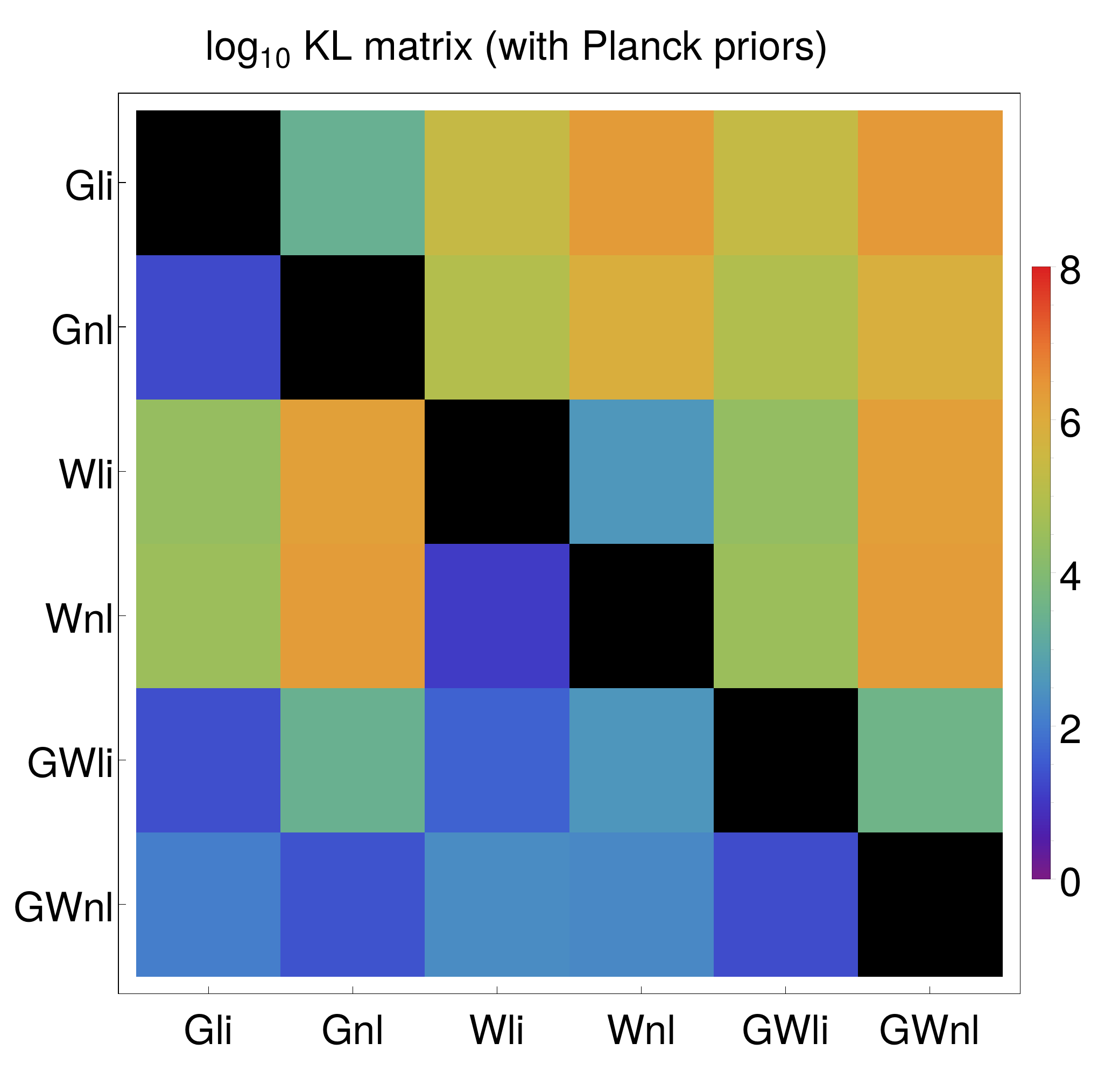}
	\caption[KL div mat]{Kullback-Leibler divergence matrices $\mathcal{K}_{ij}$. This matrix represents graphically
	the information gain between all possible observables in the redshift
binned parameterization of section \ref{sec:Results:-Redshift-Binned}. 
We have plotted here the logarithm of the KL-divergence matrix, for illustrative purposes. 
Therefore the diagonal is $-\infty$ and it is represented by a black color. 
\textbf{Left:} KL matrix without \planck\ priors. 
The maximum gain is about $10^7$ when going from WL(lin) to GC+WL(non-linear) 
and we can observe that GC+WL does not gain extra information when complemented with the other observables, which is expected.
\textbf{Right:} In this case we compare the observables, when a \planck\ prior is added beforehand. 
The overall information gain is now smaller, with a maximum of about $10^6$. 
The maximum gain comes when comparing WL (linear and non-linear) to GC and GC+WL (non-linear). 
    }
	\label{fig:kl-matrices}
\end{figure}

We then define a Kullback-Leibler matrix $\mathcal{K}$ composed of the KL-divergence measure among our observables, defined as:
\begin{equation}
\mathcal{K}_{ij}=\mathcal{D}(p_j||p_i).
\end{equation}
where $p_i$ and $p_j$ represent our observables: GC(lin), GC(nl-HS), WL(lin), WL(nl-HS), GC+WL(lin) and GC+WL(nl-HS) in the redshift
binned parameterization of section \ref{sec:Results:-Redshift-Binned}.
In this way we can quantitatively see how much information is gained when going from one probe like Galaxy Clustering 
to another probe like Weak Lensing. 
In Figure \ref{fig:kl-matrices} we plot in the left panel the KL matrix $\mathcal{K}_{ij}$ when no \planck\ priors are added and 
in the right panel the KL matrix when \planck\ priors are added to all probes. For visualization purposes, 
we plot the logarithm of the KL matrix: $\log_{10} \mathcal{K}_{ij}$. Blue (red) colours represent small (large) information gain, while
black represents no information gain at all, which is by construction the case on the diagonal, $\mathcal{K}_{ii}=0$. The rows of the matrix represent the reference observable of $p_1$ and the columns,
the new observable $p_2$.
The information gain when going from a linear WL observable to a non-linear GC observable or to a combined GC+WL (linear and non-linear) 
observable is considerably high at about $\approx 10^6$-$10^7$. 
However, when adding a \planck\ prior, this gets reduced by at least 2 orders of magnitude, 
since the prior is strong and then there is not as much new information gained in the new observables. 
On the opposite case, we can see that the row corresponding to GC+WL(nl-HS) is mostly blue, 
meaning that there is little gain of information when going to one of the other observables.

\end{document}